\documentclass[a4paper]{article}
\usepackage[USenglish]{babel}

\usepackage{fullpage}

\usepackage{amsmath} 
\usepackage{amssymb}
\usepackage{latexsym}
\usepackage{graphicx}
\usepackage{textcomp}
\usepackage{verbatim} 
\usepackage{xspace} 
\usepackage{array, booktabs, longtable, makecell} 
\usepackage{afterpage}
\usepackage{cancel}
\usepackage{tikz}
\usetikzlibrary{arrows,automata,calc,decorations.text,shapes,cd}
\usepackage{xxcolor}
\usepackage{multirow}
\usepackage{stmaryrd}
\usepackage{bbm} 
\usepackage{algorithm}
\usepackage{algpseudocode}
\usepackage{multicol}
\usepackage{hyperref}
\usepackage{amsthm}
\usepackage{xr-hyper}

\usepackage{ifoddpage}
\usepackage{marginnote}
\usepackage{skull}

\usepackage{csquotes}

\usepackage{subcaption}

\usepackage{booktabs}   
\usepackage{dashbox}
\usepackage{xargs}

\newcounter{WarnCounts}

\algblockdefx[Record]{Record}{EndRecord}%
[1]{\textbf{record} {#1}:}{\textbf{end record}}

\algblockdefx[Proc]{Proc}{EndProc}%
[2]{\textbf{proc} {#1}(#2)}{\textbf{end proc}}

\algblockdefx[Type]{Type}{EndType}%
[1]{\textbf{type} {#1}:}{\textbf{end type}}

\algblockdefx[Enum]{Enum}{EndEnum}%
[1]{\textbf{enum} {#1}:}{\textbf{end enum}}

\algblockdefx[EnumLine]{EnumLine}{EndEnumLine}%
[2]{\textbf{enum} {#1}: {#2}}{\textbf{end enum}}

\algblockdefx[ForEach]{ForEach}{EndForEach}%
[2]{\textbf{for each} {#1} \textbf{in} {#2}:}{\textbf{end for each}}

\algblockdefx{Match}{EndMatch}%
[1]{\textbf{match} {#1} \textbf{with}}{\textbf{end match}}

\algcblockdefx[Match]{Match}{MatchCase}{EndMatch}%
[1]{$\mid$ {#1} $\Rightarrow$}{\textbf{end match}}

\algblockdefx[Ghost]{Ghost}{EndGhost}[1]{}{}

\algrenewcommand\algorithmicprocedure{\textbf{proc}}
\newcommand{\returnCmd}{\textbf{return\ }}

\floatname{algorithm}{Listing}
\algtext*{EndRecord}
\algtext*{EndType}
\algtext*{EndEnum}
\algtext*{EndEnumLine}
\algtext*{EndForEach}
\algtext*{EndIf}
\algtext*{EndProcedure}
\algtext*{EndProc}

\theoremstyle{definition}
\newtheorem{defn}{Definition}[section]

\theoremstyle{plain}
\newtheorem{lem}{Lemma}[section]
\newtheorem{thm}{Theorem}[section]
 
\theoremstyle{remark}
\newtheorem*{claim}{Claim}
\newtheorem*{notation}{Notation}

\newenvironment{prf}{\begin{proof}}{\end{proof}}
\newenvironment{sketch}{\begin{proof}[Proof sketch]}{\end{proof}}

\NewDocumentEnvironment{subfigwrap}{mmO{\textwidth}}
   {\begin{subfigure}{#3}}
   {\caption{#1}
    \label{#2}
    \end{subfigure}
   }

\newcommand{\rdcssDesc}{\textsc{rdesc}} 
\newcommand{\mcasDesc}{\textsc{mdesc}} 

\newcommand{\rdcssAlg}{\textit{rdcss}}
\newcommand{\rdcssLoopAlg}{\textit{rdcss}'} 
\newcommand{\mcasAlg}{\textit{mcas}}
\newcommand{\mcasHelpAlg}{\textit{mcas}'} 
\newcommand{\completeAlg}{\textit{complete}} 
\newcommand{\rdcssReadAlg}{\textit{rread}}
\newcommand{\rdcssWriteAlg}{\textit{rwrite}}
\newcommand{\rdcssCasAlg}{\textit{rCAS}}
\newcommand{\rdcssReadCtlAlg}{\textit{rread-c}}
\newcommand{\rdcssWriteCtlAlg}{\textit{rwrite-c}}
\newcommand{\rdcssCasCtlAlg}{\textit{rCAS-c}}
\newcommand{\rdcssAllocAlg}{\textit{ralloc}}
\newcommand{\mcasReadAlg}{\textit{mread}}
\newcommand{\mcasWriteAlg}{\textit{mwrite}}

\newcommand{\mcasAllocAlg}{\textit{malloc}}
\newcommand{\writeAllDescsAlg}{\textit{writeall}} 

\newcommand{\readRepAlg}{!}
\newcommand{\writeRepAlg}{{:=}}
\newcommand{\allocRepAlg}{\textit{Alloc}}
\newcommand{\casRepAlg}{\textit{CAS}}
\newcommand{\isRdcssDescRepAlg}{\textit{is\_rdesc}}
\newcommand{\isMcasDescRepAlg}{\textit{is\_mdesc}}

\newcommand{\UNDECIDED}{\textsc{Undec}}
\newcommand{\SUCCEEDED}{\textsc{Succ}}
\newcommand{\FAILED}{\textsc{Fail}}
\newcommand{\CONTROL}{\textsc{cptr}}
\newcommand{\DATA}{\textsc{dptr}}
\newcommand{\rdcssNamespace}[1]{#1} 

\newcommand{\defini}{\eqdef}
\newcommand{\natorderSymbol}{<_{\mathbb{N}}}
\newcommand{\natorderEqSymbol}{\leq_{\mathbb{N}}}
\newcommand{\natorderSymbolRight}{>_{\mathbb{N}}}
\newcommand{\natorderEqSymbolRight}{\geq_{\mathbb{N}}}

\newcommand{\ValType}{\textsc{Val}}
\newcommand{\PtsType}{\textsc{pts}}
\newcommand{\BoolType}{\textsc{Bool}}

\newcommand{\projFirst}[1]{\pi_1({#1})}
\newcommand{\projSecond}[1]{\pi_2({#1})}
\newcommand{\ControlPtType}{\textsc{cptr}}
\newcommand{\ptKind}{\textsc{ptKind}}
\newcommand{\DataPtType}{\textsc{dptr}}
\newcommand{\mapEntry}[2]{{#1} \mapsto {#2}}
\newcommand{\mapExt}[2]{{#1}[{#2}]}
\newcommand{\mapExtThree}[3]{{#1}[{#2}]_{#3}}
\newcommand{\unitValue}{tt}

\newcommand{\absEvent}{\textsc{Ev}}
\newcommand{\repEvent}{\textsc{Rep}}

\newcommand{\ldot}{.\,}
\newcommand{\eqdef}{\mathrel{\:\widehat{=}\:}}
\newcommand{\inputValName}{\textsc{writer}}
\newcommand{\inputVal}[3]{\inputValName\ {#1}\ {#2}\ {#3}}
\newcommand{\ninputVal}[3]{\neg\inputVal {#1} {#2} {#3}}
\newcommand{\oportunisticPred}[2]{\mathsf{OpAcc}\ {#1}\ {#2}}
\newcommand{\refleTransCl}[1]{\mathrel{{#1}^{*}}}
\newcommand{\transCl}[1]{\mathrel{{#1}^{+}}}

\newcommand{\listvar}[1]{\overline{#1}}

\newcommand{\RDCSS}{\textsc{R}}
\newcommand{\MCAS}{\textsc{M}}
\newcommand{\MCASFam}{\genStructName{\MCAS}}
\newcommand{\RDCSSFam}{\genStructName{\RDCSS}}
\newcommand{\genStructName}[1]{V^{#1}}
\newcommand{\genSpanStructName}[1]{S^{#1}}
\newcommand{\genOportSpanStructName}[1]{O^{#1}}
\newcommand{\MCASSpanStruct}{\genSpanStructName{\MCAS}}

\newcommand{\terminatedEvent}{T}
\newcommand{\closedEventSymbol}{\overline{\terminatedEvent}}

\newcommand{\closedEvent}{\closedEventSymbol}
\newcommand{\postPredSymbol}{\mathcal{Q}}
\newcommand{\postPred}[2]{\postPredSymbol_{{#1},{#2}}}

\newcommand*{\sepEq}{%
\mathrel{\vcenter{\offinterlineskip
\hbox{\phantom{$\ltimes$}}\vskip-.35ex\hbox{$\ltimes$}\vskip-.35ex\hbox{$-$}}}}

\newcommand{\visObsSYMBOL}{\lessdot}
\newcommand{\visSepSYMBOL}{\ltimes}

\newcommand{\visSepEqSYMBOL}{\sepEq}

\newcommand{\visObsSymbol}[1]{\mathrel{\visObsSYMBOL}_{#1}}
\newcommand{\visSepSymbol}[1]{\mathrel{\visSepSYMBOL}_{#1}}

\newcommand{\visSepEqSymbol}[1]{\mathrel{\visSepEqSYMBOL}_{#1}}
\newcommand{\visObsIndxSymbol}[2]{\mathrel{\visObsSYMBOL}_{#1}^{#2}}
\newcommand{\visSepIndxSymbol}[2]{\mathrel{\visSepSYMBOL}_{#1}^{#2}}

\newcommand{\visObs}[3]{{#2} \visObsSymbol {#1} {#3}}
\newcommand{\visSep}[3]{{#2} \visSepSymbol {#1} {#3}}

\newcommand{\visSepEq}[3]{{#2} \visSepEqSymbol {#1} {#3}}
\newcommand{\visObsIndx}[4]{{#3} \visObsIndxSymbol {#1} {#2} {#4}}
\newcommand{\visSepIndx}[4]{{#3} \visSepIndxSymbol {#1} {#2} {#4}}

\newcommand{\genVisSymbol}{\prec}
\newcommand{\genVis}[2]{{#1} \genVisSymbol {#2}}
\newcommand{\genVisTransSymbol}{\genVisSymbol^{+}}
\newcommand{\genVisTrans}[2]{{#1} \genVisTransSymbol {#2}}

\newcommand{\triagVisSymbol}{\vartriangleleft}
\newcommand{\triagVis}[2]{{#1} \triagVisSymbol {#2}}

\newcommand{\triagVisEqSymbol}{\trianglelefteq}

\newcommand{\writesAbsSymbol}{\mathcal{W}}
\newcommand{\writesAbs}[1]{\writesAbsSymbol_{#1}}

\newcommand{\allocsAbsSymbol}{\mathcal{A}}
\newcommand{\allocsAbs}[1]{\allocsAbsSymbol_{#1}}

\newcommand{\precedesAbsSymbol}{\sqsubset}
\newcommand{\precedesAbs}[2]{{#1} \precedesAbsSymbol {#2}}
\newcommand{\precedesAbsEqSymbol}{\sqsubseteq}
\newcommand{\precedesAbsEq}[2]{{#1} \precedesAbsEqSymbol {#2}}

\newcommand{\nprecedesAbsEqSymbol}{\not\sqsubseteq}
\newcommand{\nprecedesAbsEq}[2]{{#1} \nprecedesAbsEqSymbol {#2}}

\newcommand{\spansSymbol}{\mathcal{S}}
\newcommand{\spans}[1]{\spansSymbol_{#1}}
 
\newcommand{\runFuncSymbol}{\left\llbracket{\cdot}\right\rrbracket}

\newcommand{\runFunc}[1]{\left\llbracket{#1}\right\rrbracket}
\newcommand{\outputRunFunc}[1]{o({#1})}

\newcommand{\writesSpansSymbol}{\mathcal{S}^W}
\newcommand{\writesSpans}[1]{\writesSpansSymbol_{#1}}
\newcommand{\descriptorSpansSymbol}{\mathcal{S}^D}
\newcommand{\descriptorSpans}[1]{\descriptorSpansSymbol_{#1}}
\newcommand{\opportunisticSpansSymbol}{\mathcal{S}^O}
\newcommand{\opportunisticSpans}[1]{\opportunisticSpansSymbol_{#1}}

\newcommand{\hspans}[2]{\spans{#1}(#2)}
\newcommand{\dspans}[2]{\descriptorSpans{#1}(#2)}
\newcommand{\ospans}[2]{\opportunisticSpans{#1}(#2)}

\newcommand{\allocsSpansSymbol}{\mathcal{S}^A}
\newcommand{\allocsSpans}[1]{\allocsSpansSymbol_{#1}}

\newcommand{\precedesSpansSymbol}{\sqsubset^S}
\newcommand{\precedesSpans}[2]{{#1} \precedesSpansSymbol {#2}}
\newcommand{\precedesSpansEqSymbol}{\sqsubseteq^S}
\newcommand{\precedesSpansEq}[2]{{#1} \precedesSpansEqSymbol {#2}}

\newcommand{\LastBlockRel}[2]{{\lastRep {#1}} \linRepsSymbol {\lastRep {#2}}}

\newcommand{\repsCompl}{[\linRepsEqSymbol]}

\newcommand{\linRepsSymbol}{<^R}
\newcommand{\linReps}[2]{{#1} \linRepsSymbol {#2}}
\newcommand{\linRepsEqSymbol}{\leq^R}
\newcommand{\linRepsEq}[2]{{#1} \linRepsEqSymbol {#2}}

\newcommand{\generatorEvent}[1]{\left\llbracket{#1}\right\rrbracket}

\newcommand{\outputPropName}{\textit{out}}
\newcommand{\outputProp}[1]{{#1}.\outputPropName}
\newcommand{\STimePropName}{\textit{start}}
\newcommand{\STimeProp}[1]{{#1}.\STimePropName}
\newcommand{\ETimePropName}{\textit{end}}
\newcommand{\ETimeProp}[1]{{#1}.\ETimePropName}
\newcommand{\inPropName}{\textit{in}}
\newcommand{\inProp}[1]{{#1}.\inPropName}

\newcommand{\linePropName}{\textit{line}}
\newcommand{\lineProp}[1]{{#1}.\linePropName}

\newcommand{\firstRepFuncSymbol}{f}
\newcommand{\lastRepFuncSymbol}{l}
\newcommand{\firstRep}[1]{\firstRepFuncSymbol({#1})}
\newcommand{\lastRep}[1]{\lastRepFuncSymbol({#1})}

\newcommand{\pointOne}[1]{{#1}.pt_1}
\newcommand{\pointTwo}[1]{{#1}.pt_2}
\newcommand{\expOne}[1]{{#1}.exp_1}
\newcommand{\expTwo}[1]{{#1}.exp_2}
\newcommand{\newTwo}[1]{{#1}.new_2}
\newcommand{\pointGen}[1]{{#1}.pt}
\newcommand{\expGen}[1]{{#1}.exp}
\newcommand{\newGen}[1]{{#1}.new}
\newcommand{\statusProp}[1]{{#1}.\textit{status}}
\newcommand{\entriesProp}[1]{{#1}.\textit{entries}}

\newcommand{\updateEntry}[2]{{#1}_{#2}}
\newcommand{\pointGenEntry}[1]{\updateEntry{pt}{#1}}
\newcommand{\expGenEntry}[1]{\updateEntry{exp}{#1}}
\newcommand{\newGenEntry}[1]{\updateEntry{new}{#1}}

\newcommand{\pointerIndx}[1]{{#1}}
\newcommand{\descIndx}{\mathsf{D}}
\newcommand{\opporIndx}{\mathsf{O}}

\newcounter{axiom}[figure]
\newcommand{\axiomHLabel}[1]{\refstepcounter{axiom}\label{#1}($C_{\theaxiom}$)}
\newcommand{\axiomCLabel}[1]{\refstepcounter{axiom}\label{#1}($C_{\theaxiom}$)}
\newcommand{\axiomHRef}[1]{$C_{\ref{#1}}$}
\newcommand{\axiomDLabel}[1]{\refstepcounter{axiom}\label{#1}($S_{\theaxiom}$)}
\newcommand{\axiomDRef}[1]{$S_{\ref{#1}}$}
\newcommand{\axiomOLabel}[1]{\refstepcounter{axiom}\label{#1}($O_{\theaxiom}$)}
\newcommand{\axiomORef}[1]{$O_{\ref{#1}}$}

\usepackage{authblk}  

\title{Declarative Linearizability Proofs for Descriptor-Based Concurrent Helping Algorithms}

\author[1,2]{Jes\'{u}s Dom\'{i}nguez}
\author[1]{Aleksandar Nanevski}
\affil[1]{IMDEA Software Institute, Spain}
\affil[2]{Universidad Polit\'{e}cnica de Madrid, Spain}
\date{}

\newcommand{\citet}[1]{\cite{#1}}

\begin{document}

\maketitle

\begin{abstract}
Linearizability is a standard correctness criterion for concurrent
algorithms, typically proved by establishing the algorithms'
linearization points. However, relying on linearization points leads
to proofs that are implementation-dependent, and thus hinder
abstraction and reuse.
In this paper we show that one can develop more declarative proofs by
foregoing linearization points and instead relying on a technique of
axiomatization of visibility relations. While visibility relations
have been considered before, ours is the first study where the
challenge is to formalize the helping nature of the algorithms.
In particular, we show that by axiomatizing the properties of 
separation between events that contain bunches of help requests, we
can extract what is common for high-level understanding of several
descriptor-based helping algorithms of Harris et al. (RDCSS, MCAS, and
optimizations), and produce novel proofs of their linearizability that
share significant components.

\end{abstract}


\section{Introduction}
 
\emph{Helping} is a design principle for concurrent algorithms in
which a process carries out work for other processes.
Helping is useful for achieving lock-free and wait-free
implementations, because whenever a process $p$ becomes stuck, other
processes can help $p$ make progress, or $p$ itself can help the
process that is in its way, so that $p$ can
continue~\cite{HelpCensor,AttiyaCastaneda,Aksenov}.
Helping algorithms are often equipped with \emph{descriptors}, which
are structures that a process uses to signal that it requests help
with its task. Whenever a process $p$ finds a descriptor in some
pointer $x$, it means that some other process $q$ has requested help
with computing the value of $x$. Process $p$ then extracts $q$'s task
information from the descriptor and $p$ \emph{fully} helps until $q$'s
task terminates; only then, $p$ attempts its own task.

In this paper, we focus on proving correct two descriptor-based helping algorithms:
RDCSS and MCAS~\cite{Harris}. 
\emph{Restricted Double-Compare Single Swap (RDCSS)} is
a generalization of the compare-and-swap operation CAS. Whereas
$\casRepAlg\,(pt_2,exp_2, new_2)$ updates $pt_2$ with $new_2$ if the
old value of $pt_2$ is $exp_2$, RDCSS adds another pointer $pt_1$ and
value $exp_1$ into the decision.  More precisely, RDCSS receives as
input a descriptor that is a record with five pieces of data: two
pointers $pt_1$, $pt_2$ and three values $exp_1$, $exp_2$, $new_2$.
To the invoking client, RDCSS gives the impression that it
\emph{atomically} carries out the following update: if $pt_1$ has
expected value $exp_1$ and $pt_2$ has expected value $exp_2$, then
$pt_2$ is updated to $new_2$.
\emph{Multiple Compare-And-Swap (MCAS)} also generalizes
CAS, but it updates an arbitrary number of pointers at once.
More precisely, MCAS receives a list of update entries, each being a
record with three pieces of data: a pointer $\textit{pt}$ and two
values $\textit{exp}$, $\textit{new}$. To a client, MCAS gives the
impression that it atomically carries out the following conditional
multiple update: if for every update entry $i$, the pointer
$\updateEntry{\textit{pt}}{i}$ has expected value
$\updateEntry{\textit{exp}}{i}$, then for every update entry $j$, the
pointer $\updateEntry{\textit{pt}}{j}$ is updated to the new value
$\updateEntry{\textit{new}}{j}$.

These algorithms have recently become somewhat of a verification
benchmark~\cite{vafeiadis,IrisFuture,LiangFeng} due to their tricky
\emph{linearizability}\footnote{%
  A structure is \emph{linearizable}~\cite{herlihy:90} if in every
  concurrent execution history of the structure's exportable methods,
  the method invocations can be ordered linearly just by permuting
  overlapping invocations, so that the obtained history is
  \emph{sequentially sound}; that is, executing the methods
  sequentially in the linear order produces the same outputs that the
  methods had in the concurrent history. In other words, every
  concurrent history is equivalent to a sequential one where methods
  execute without interference, i.e., \emph{atomically}. The linear
  order must include all terminated methods, but may also include
  selected non-terminated ones, if their partial execution influenced
  (i.e., was \emph{visible} to)
  others.\label{footnote::linearizability}} proofs.
The existing linearizability proofs of RDCSS and MCAS all employ the
\emph{linearization point} (LP) approach. Given a method (henceforth,
\emph{event}), its LP is the moment in the event's duration at which the
event's effect can be considered to have occurred \emph{abstractly},
in the sense that the linearization order of the events is determined
by the real-time order of the chosen LPs.  The LPs are always
described \emph{operationally}, by indicating a line in the code
together with a run-time condition under which the line applies (thus,
the code line chosen as the LP may vary). The linearizability proof
then shows that the effect of the invocation abstractly occurs at the
declared line.
The operational nature of the LP description leads to proofs that are
implementation-dependent. Concretely, RDCSS and MCAS
are operationally tricky because, in addition to helping, they exhibit
so called ``future-dependent LPs''. These are LPs whose position
depends on a run-time condition in the future, potentially even after
the considered event has terminated.

In this paper, we develop a novel approach to proving linearizability
of RDCSS and MCAS (and optimizations) that elides LPs altogether. Our
approach is \emph{declarative} in nature, rather than operational. In
particular, we show that RDCSS and MCAS can be explained in an
implementation-independent manner that \emph{brings to light the key
  high-level abstract design principles that they both share}. This in
turn leads to abstract proofs of linearizability that reuse
significant portions of the formal development.

More specifically, our first contribution is proposing that the kind
of helping employed by RDCSS and MCAS induces two natural notions that
can be used to capture the essence of both algorithms.
The first notion is that of \emph{separation} between events. Since an
event that finds a descriptor carries out its task only after
completing the task of the event that requested help, there is a
\emph{gap} in real time between tasks carried out by the two events.
This behavior also induces the effect that help requests on the same
pointer are \emph{disjoint} in time, since a process must help first
(hence, cannot make a help request) if there is a descriptor present.
In addition, RDCSS and MCAS collect their help requests into
``bunches'' so that the bunched help requests of one event can easily
be separated from the bunched help requests of another event.
The second notion is that of \emph{observation}. It arises because
RDCSS and MCAS modules contain procedures that modify the state, and
procedures that read the state. Thus, a reader event $A$ may observe a
value written by writer event $B$, leading to a dependence of $A$'s
result on $B$.  
In a sequentially sound reordering of events, no event sequenced
between $B$ and $A$ may overwrite $B$'s write.

Our second contribution is \emph{axiomatizing} separation and
observation,
which formally exposes the common semantic structure behind RDCSS and
MCAS and makes it possible to decompose their proofs into independent
components. As common in axiomatic systems, a proof of a property
developed out of axioms applies without change to any algorithm that
satisfies the axioms.
We thus proceed to prove linearizability of RDCSS and
MCAS out of the axioms about separation and observation alone. The axioms are much
easier to establish and understand than attempting the linearizability
proof for either of the algorithms from scratch and in its
totality. 
Certain aspects of the abstract linearizability proof (e.g., existence
of a linear order) are common to MCAS and RDCSS, while other aspects
(e.g., sequential soundness) are unique to each data structure, but
are still abstract in the sense that the proof can be reused by other
implementations. 

To state our axioms, we encode the notions of separation and
observation as binary relations between events. Each relation
describes a different notion of dependence between events. The
axioms then state properties or behaviors expected to hold whenever
these order dependences are present. A prior event on which a latter
event may depend, is said to be ``visible'' to the latter event,
lending the name of \emph{visibility relations}~\cite{Viotti} to our
notions of separation and observation.
Visibility relations are clearly relevant to linearizability, where a
notion of one event influencing others is crucial, as evident from our
description of linearizability in
Footnote~\ref{footnote::linearizability}.

\begin{figure}[t]
\includegraphics[scale=0.7]{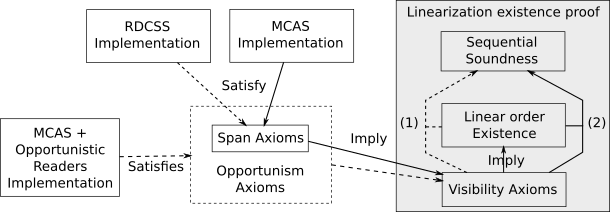}
\centering
\caption{Steps in the linearizability proof for the RDCSS and MCAS
  implementations.  (1) and (2) represent distinct proofs of
  sequential soundness for RDCSS and MCAS, respectively, which
  share the construction of the linearization order. Dashed  lines indicate the material 
  in the appendices.
}
\label{fig::proof-diagram}
\end{figure}

Axiomatizing the visibility between events via binary relations was
first proposed as an alternative to the LP approach by
Henzinger et al.~\citet{henzinger:concur13}, who applied visibility to axiomatize
concurrent queues, and termed ``aspects'' the various properties that
implied the axiomatization.
Subsequently, axioms for event visibility were developed for
the concurrent timestamped stack~\cite{Dodds}, for sophisticated
concurrent snapshot algorithms~\cite{Joakim}, and to formalize
relaxations of linearizability and 
of memory consistency~\cite{EmmiEnea,RaadAzalea}. 
Our paper, however, is the first to apply visibility relations to a study
of helping algorithms, which poses very different challenges. In
particular, whereas axioms for concurrent queues, stacks, and
snapshots mostly capture how the contents of the structure evolve
extensionally (e.g., queues are FIFO, stacks are LIFO, etc.), with
RDCSS and MCAS the challenge is in describing intensional aspects of
event interaction, such as temporal bunching of help requests and the
resulting separation of events.
In summary, our contributions are:
\begin{itemize}
\item We propose separation and observation as foundational notions to
  formalize the properties of RDCSS and MCAS algorithms in a
  declarative style.
\item We provide the first axiomatization of these two algorithms (or
  of any helping algorithm) in the style of visibility relations
  (Section~\ref{subsect::overview::separable-before-relations}). 
  The axiomatization captures the
  common semantic structure underpinning the two algorithms, and
  substantiates that visibility relations are a powerful method for
  verification of concurrent algorithms. We show that separation and
  observation axioms (henceforth, visibility axioms) imply
  linearizability (Section~\ref{sect::axiomatization}).
%
\item To 
  relate separation, observation and descriptors, we formulate another
  set of axioms, called span axioms
  (Section~\ref{sect::key-concepts-spans}).
  These mathematically describe, in implementation-independent way,
  how descriptors in RDCSS and MCAS collect the help requests into
  disjoint bunches. We prove that the span axioms imply the visibility
  axioms (Section~\ref{sect::key-concepts-spans}). Thus, linearizability is
  further reduced to establishing the span axioms, allowing for even
  more proof reuse.
\item We show that RDCSS and MCAS satisfy the span axioms, concluding
  that both algorithms are linearizable
  (Section~\ref{sect::span-structure-mcas} for MCAS,
  Appendix~\ref{appendix::sub::sect::impl::RDCSS} for 
  RDCSS).
\item We also show that separation and observation generalize beyond
 helping. In particular, we consider
 (Section~\ref{subsubsect::helping-vs-oppor}) an optimization of the
 read procedure in MCAS, introduced by Harris et al.~\cite{Harris}, which we call
 ``opportunistic reading''. Opportunistic readers do not help on a
 descriptor; they merely determine their return value from the
 descriptor, but improve their own efficiency by leaving the actual
 helping to others (proofs in Appendix~\ref{sect::appendix::opportunistic-impl-proof}).
\end{itemize}
Figure \ref{fig::proof-diagram} summarizes the structure of our
linearizability proof and of our paper.  Solid boxes and lines
indicate what we discuss in the main body of the paper; these include
common considerations, and detailed treatment of the proof of
MCAS. Dashed boxes and lines indicate the material about RDCSS, MCAS
with opportunistic readers, and the generalization of the span axioms
that deal with opportunism, found in the
appendices.

\section{Algorithm Description}
\label{sect::overview}
 
\subsection{RDCSS}\label{subsect::overview::rdcss-algorithms}

\begin{figure}[t]
\begin{multicols*}{2}

\begin{algorithmic}[1]
\Record{$\rdcssDesc$}
\State $pt_1$ : $\ControlPtType$
\State $pt_2$ : $\DataPtType$
\State $exp_1$, $exp_2$, $new_2$ : $\ValType$
\EndRecord
%
\State
%
\Proc{$\rdcssAlg$\,}{$desc: \rdcssDesc$}
\State \label{alg-alloc-desc-RDCSS} $d \gets \allocRepAlg\,(\textit{desc})$ as $\DataPtType$
\State \returnCmd $\rdcssLoopAlg\,(d, \textit{desc})$
\EndProc
%
\State 
%
\Proc{$\rdcssLoopAlg$\,}{$d: \DataPtType$, $\textit{desc} : \rdcssDesc$}
\State \label{alg-CAS-RDCSS} $\textit{old} \gets \casRepAlg\,(\pointTwo{desc}, \expTwo{desc}, d)$
\If{$\isRdcssDescRepAlg\,(\textit{old})$} \label{alg-is-desc-RDCSS}
  \State \label{alg-help-complete-invoke-RDCSS} $\completeAlg\,(\textit{old})$
  \State \label{alg-help-try-again-rdcssLoop-RDCSS} \returnCmd $\rdcssLoopAlg\,(d, \textit{desc})$
\Else
  \If {$\textit{old} = \textit{desc}.\textit{exp}_2$} \label{alg-is-cas-successful-RDCSS}
  	\State \label{alg-self-complete-invoke-RDCSS} $\completeAlg\,(d)$
  \EndIf
  \State \label{alg-return-old-value-RDCSS} \returnCmd{$\textit{old}$}
\EndIf
\State
\EndProc
%
%
\Proc{$\completeAlg$\,}{$d: \DataPtType$}
\State \label{alg-desc-read-RDCSS} $\textit{desc} \gets \,\readRepAlg d$
\State \label{alg-pt1-read-RDCSS} $x \gets \,\readRepAlg \pointOne{\textit{desc}}$
\If {$x = \expOne{\textit{desc}}$}
  \State \label{alg-pt2-write-success-RDCSS} $\casRepAlg\,(\pointTwo{\textit{desc}}, d, \newTwo{\textit{desc}})$
\Else
  \State \label{alg-pt2-write-fail-RDCSS} $\casRepAlg\,(\pointTwo{\textit{desc}}, d, \expTwo{\textit{desc}})$
\EndIf
\EndProc
%
\columnbreak
%
\Proc{$\rdcssReadAlg$\,}{$\textit{pt} : \DataPtType$}
\State \label{alg-access-Read-RDCSS} $\textit{old} \gets \,\readRepAlg \textit{pt}$
\If {$\isRdcssDescRepAlg\,(\textit{old})$} \label{alg-is-desc-Read-RDCSS}
  \State \label{alg-help-complete-invoke-Read-RDCSS} $\completeAlg\,(\textit{old})$
  \State \label{alg-help-try-again-read-RDCSS} \returnCmd $\rdcssReadAlg\,(\textit{pt})$
\Else
  \State \label{alg-return-old-value-Read-RDCSS} \returnCmd{$\textit{old}$}
\EndIf
\EndProc
%
\State 
%
\Proc{$\rdcssCasAlg$\,}{$\textit{pt} : \DataPtType$, $exp$, $new: \ValType$}
\State \label{alg-access-CAS-RDCSS} $\textit{old} \gets \casRepAlg\,(\textit{pt}, \textit{exp}, \textit{new})$
\If {$\isRdcssDescRepAlg\,(\textit{old})$} \label{alg-is-desc-CAS-RDCSS}
  \State \label{alg-help-complete-invoke-CAS-RDCSS} $\completeAlg\,(\textit{old})$
  \State \label{alg-help-try-again-cas-RDCSS} \returnCmd $\rdcssCasAlg\,(\textit{pt}, \textit{exp}, \textit{new})$
\Else
  \State \label{alg-return-old-value-CAS-RDCSS} \returnCmd{$\textit{old}$}
\EndIf
\EndProc
%
\State 
%
\Proc{$\rdcssWriteAlg$\,}{$\textit{pt} : \DataPtType$, $v: \ValType$}
\State \label{alg-access-Write-RDCSS} $\textit{old} \gets \,\readRepAlg \textit{pt}$
\If {$\isRdcssDescRepAlg\,(\textit{old})$} \label{alg-is-desc-Write-RDCSS}
  \State \label{alg-help-complete-invoke-Write-RDCSS} $\completeAlg\,(\textit{old})$
  \State \label{alg-help-try-again-write-RDCSS} $\rdcssWriteAlg\,(\textit{pt}, v)$
\Else 
  \State \label{alg-attempt-write-Write-RDCSS} $x \gets \casRepAlg\,(\textit{pt}, \textit{old}, v)$
  \If {$x \neq \textit{old}$}
    \State $\rdcssWriteAlg\,(\textit{pt}, v)$
  \EndIf 
\EndIf
\State
\EndProc
\end{algorithmic}
%
%
\end{multicols*}
\caption{C-like pseudo code of the RDCSS implementation (essential fragment). $\readRepAlg$,
  $\allocRepAlg$ and $\casRepAlg$ are the system calls for memory
  dereference, allocation, and compare-and-swap.
}
\label{alg-RDCSS}
\end{figure}

Figure~\ref{alg-RDCSS} shows the 
implementation of the RDCSS data structure ($\rdcssAlg$ and associated
methods, and the descriptor type $\rdcssDesc$). $\ValType$ denotes the
set of all possible input values; it excludes descriptors and
descriptor-storing pointers.
Every pointer exposed to the clients is classified as a control
($\ControlPtType$) or data ($\DataPtType$) pointer. Data pointers may
store $\ValType$ values, descriptors, and pointers, while control
pointers only store $\ValType$ values. Along with the $\rdcssAlg$
method, the implementation exports data pointer methods
$\rdcssReadAlg$, $\rdcssWriteAlg$, and $\rdcssCasAlg$. The latter
replace the system calls for pointer dereference, update, and
$\textit{CAS}$, which they adapt to the presence of control and data
pointers.  Not shown, but also exported, are the methods for
dereference, update and $\textit{CAS}$ over control pointers,
implemented simply as respective system calls. Methods $\rdcssLoopAlg$
and $\completeAlg$ are internal, and not exported.
The implementation further assumes: (1) an exported method
$\rdcssAllocAlg\,(v,t)$ for allocating a pointer of type
$t \in \{ \CONTROL, \DATA \}$ with initial value $v$, which replaces
the system operation $\allocRepAlg$, and (2) an internal Boolean
predicate $\isRdcssDescRepAlg\,(p)$ that returns true iff pointer $p$
stores an RDCSS descriptor. We show their implementation in
Appendix~\ref{appendix::sub::sect::impl::RDCSS}, but elide here as it
is not essential for understanding.

We next give a high-level description of $\rdcssAlg$.  Any thread $T$
invoking $\rdcssAlg$ allocates a fresh address $d$ for the input
descriptor $\textit{desc}$ (line~\ref{alg-alloc-desc-RDCSS}); $d$ will
serve as a unique identifier for the $\rdcssAlg$ invocation. Next, $T$
calls the recursive procedure $\rdcssLoopAlg$. This first CASs on
$\pointTwo{\textit{desc}}$ to read the $\textit{old}$ value and write
$d$ if the old value is the expected $\expTwo{\textit{desc}}$
(line~\ref{alg-CAS-RDCSS}). By this write, $T$ essentially requests
help with $d$, and signals to other threads that:
\begin{itemize}
\item Descriptor $\textit{desc}$ (which is stored in $d$) is currently
  active in $\pointTwo{\textit{desc}}$.
\item $T$ is invoking $\rdcssAlg$ in which pointer
  $\pointOne{\textit{desc}}$ still needs to be read.
\item While $desc$ remains active in $\pointTwo{\textit{desc}}$, value
  $\expTwo{\textit{desc}}$ is stored in $\pointTwo{\textit{desc}}$
  \emph{indirectly}, as it can be reached by following the descriptor.
\end{itemize}
After the CAS in line~\ref{alg-CAS-RDCSS}, if $\textit{old}$ is a
descriptor, then $T$ first \emph{helps} by invoking
$\completeAlg{(\textit{old})}$
(line~\ref{alg-help-complete-invoke-RDCSS}), and then $T$ recurses to
reattempt its task. If $\textit{old}$ is a value not matching
$\expTwo{\textit{desc}}$, then no modification to
$\pointTwo{\textit{desc}}$ is performed and $T$ 
returns $\textit{old}$. If $\textit{old}$ is the expected value (i.e.,
the CAS succeeded), then $T$ invokes $\completeAlg{(d)}$ to help itself
(line~\ref{alg-self-complete-invoke-RDCSS}).

We describe $\completeAlg$ from the point of view of another thread
$T'$ that reads $\pointTwo{\textit{desc}}$ and finds $d$ (lines
\ref{alg-is-desc-RDCSS}, \ref{alg-is-desc-Read-RDCSS},
\ref{alg-is-desc-CAS-RDCSS}, \ref{alg-is-desc-Write-RDCSS}). Before
doing anything else, $T'$ helps on $d$ by invoking $\completeAlg$ as
follows (lines \ref{alg-help-complete-invoke-RDCSS},
\ref{alg-help-complete-invoke-Read-RDCSS},
\ref{alg-help-complete-invoke-CAS-RDCSS},
\ref{alg-help-complete-invoke-Write-RDCSS}). $T'$ first reads pointer
$\pointOne{\textit{desc}}$ (line~\ref{alg-pt1-read-RDCSS}).  If it
finds the expected value $\expOne{\textit{desc}}$, and since pointer
$\pointTwo{\textit{desc}}$ indirectly stores the expected value, then
$\rdcssAlg$ can succeed; thus, $T'$ attempts to replace the descriptor
in $\pointTwo{\textit{desc}}$ with the new value
$\newTwo{\textit{desc}}$ (line~\ref{alg-pt2-write-success-RDCSS}). The
replacement in line~\ref{alg-pt2-write-success-RDCSS} is performed by
$\casRepAlg$, which may fail if some other thread managed to help $T$
before $T'$. Either way, $T$ has been helped after the call to
$\completeAlg$.
%
%
Otherwise, if $T'$ does not find the expected value
$\expOne{\textit{desc}}$, it attempts to undo $T$'s modification to
$\pointTwo{\textit{desc}}$ (line~\ref{alg-pt2-write-fail-RDCSS}) by
replacing $d$ in $\pointTwo{\textit{desc}}$ back to
$\expTwo{\textit{desc}}$, via $\casRepAlg$. This undoing may fail if
some other thread managed to help $T$ before $T'$.  Either way, $T$
has again been helped after the call to $\completeAlg$.

Procedures $\rdcssReadAlg$, $\rdcssCasAlg$ and $\rdcssWriteAlg$ follow
the same helping strategy, i.e., whenever they find a descriptor in
their input pointer, they invoke $\completeAlg$ to help the pending
$\rdcssAlg$ before recursing to reattempt their own task.

\subsection{MCAS}\label{subsect::overview::mcas-algorithms}

\begin{figure}[t]
\begin{multicols}{2}

\begin{algorithmic}[1]
\Record{$\textsc{update\_entry}$} 
	\State $\textit{pt}$ : $\DataPtType$
	\State $\textit{exp}$, $\textit{new}$ : $\ValType$
\EndRecord
\EnumLine{$\textsc{status}$}{$\UNDECIDED$, $\SUCCEEDED$, $\FAILED$}
\EndEnumLine
\Record{$\mcasDesc$}
\State $\textit{status}$ : $\ControlPtType$ $\textsc{status}$
\State $\textit{entries}$ : $\textsc{list}\,\textsc{update\_entry}$
\EndRecord
%
\State
%
\Proc{$\mcasHelpAlg$\,}{$d : \DataPtType$}
\State \label{read-desc-MCAS} $\textit{desc} \gets \rdcssNamespace{\rdcssReadAlg\,(d)}$
\State \label{read-phase1-status-MCAS} $\textit{phase}_1 \gets \,\rdcssNamespace{\readRepAlg \statusProp{\textit{desc}}}$
\If{$\textit{phase}_1 = \UNDECIDED$} \label{is-phase1-still-undecided}
	\State \label{write-all-descs-MCAS} $s \gets \writeAllDescsAlg\,(d,\textit{desc})$
	\State \label{resolve-status-MCAS} $\rdcssNamespace{\casRepAlg\,(\statusProp{\textit{desc}}, \UNDECIDED, s)}$
\EndIf
\State \label{read-phase2-status-MCAS} $\textit{phase}_2 \gets \,\rdcssNamespace{\readRepAlg \statusProp{\textit{desc}}}$
\State $b \gets (\textit{phase}_2 = \SUCCEEDED)$
\ForEach{$e$}{$\entriesProp {\textit{desc}}$} \label{remove-all-descs-loop-MCAS}
	\State \label{remove-all-descs-MCAS} $\rdcssNamespace{\rdcssCasAlg\,(\pointGen e, d, b\ ?\ \newGen e : \expGen e)}$
\EndForEach 
\State \returnCmd{$b$}
\EndProc
\State
\Proc{$\mcasAlg$\,}{$\listvar u: \textsc{list}\,\textsc{update\_entry}$}
\State \label{alloc-status-MCAS} $s \gets \rdcssNamespace{\rdcssAllocAlg\,(\UNDECIDED, \CONTROL)}$
\State \label{create-mcas-desc-MCAS} $\textit{desc} \gets \mcasDesc\ (s, \listvar u)$
\State \label{alloc-desc-MCAS} $d \gets \rdcssNamespace{\rdcssAllocAlg\,(\textit{desc}, \DATA)}$
\State \returnCmd $\mcasHelpAlg\,(d)$
\EndProc
\State 
\Proc{$\writeAllDescsAlg$\,}{$d : \DataPtType, \textit{desc}: \mcasDesc$}
\ForEach{$e$}{$\entriesProp {\textit{desc}}$}
	\State \label{create-rdcss-desc-MCAS} $\textit{rd} \gets \rdcssNamespace{\rdcssDesc}\ ( \statusProp {\textit{desc}},$ 
	\State $\phantom{\textit{rd} \gets \quad}\pointGen e, \UNDECIDED, \expGen e, d )$
	\State \label{invoke-rdcss-in-MCAS} $\textit{old} \gets \rdcssNamespace{\rdcssAlg\,(\textit{rd})}$
	\If{$\isMcasDescRepAlg\,(\textit{old})$} \label{alg-is-desc-MCAS-MCAS}
	  \If{$\textit{old} \neq d$} \label{is-my-desc-MCAS}
	     \State \label{alg-help-complete-invoke-MCAS-MCAS} $\mcasHelpAlg\,(\textit{old})$
	     \State \label{alg-help-restart-MCAS} \returnCmd{$\writeAllDescsAlg\,(d, \textit{desc})$}
	  \EndIf
	\ElsIf{$\textit{old} \neq \expGen e$} \label{rdcss-failed-MCAS}
	  \State \returnCmd{$\FAILED$}
    \EndIf
\EndForEach
\State \returnCmd{$\SUCCEEDED$}
\EndProc
\end{algorithmic}
\end{multicols}
\caption{C-like pseudo code of the MCAS implementation (essential fragment). The command \returnCmd finishes the procedure execution where it is used.}
\label{alg-MCAS}
\end{figure}

Figure~\ref{alg-MCAS} shows the implementation of the MCAS data structure
($\mcasAlg$ and associated methods, and the descriptor type
$\mcasDesc$) atop RDCSS.
Procedures $\mcasHelpAlg$ and $\writeAllDescsAlg$ are internal. The
figure elides the exported procedures $\mcasReadAlg$, $\mcasWriteAlg$,
and $\mcasAllocAlg$, which can be found in 
Appendix~\ref{appendix::sub::sect::impl::MCAS}.
Again, $\ValType$ is the set of input values that excludes descriptors
and descriptor-storing pointers.
We use the notation $\pointGenEntry i$, $\expGenEntry i$, and 
$\newGenEntry i$ to refer to the components of an $\textsc{update\_entry}$
$i$.

High-level description of $\mcasAlg$ is as follows.  Thread $T$
invoking $\mcasAlg$ first creates an MCAS descriptor $\textit{desc}$
containing the non-empty list of update entries and a status pointer
(line~\ref{create-mcas-desc-MCAS}).  The status pointer starts in an
undecided state ($\UNDECIDED$) and evolves into either a success
($\SUCCEEDED$) or failed ($\FAILED$) state.  $\SUCCEEDED$ indicates
that the descriptor was written into all input pointers, while
$\FAILED$ indicates the failure of at least one such write. Next, $T$
allocates a fresh address $d$ for $\textit{desc}$
(line~\ref{alloc-desc-MCAS}), which serves as the unique identifier
for the $\mcasAlg$ invocation.  Eventually, through a call to
$\mcasHelpAlg$, $T$ invokes $\writeAllDescsAlg$
(line~\ref{write-all-descs-MCAS}) to attempt storing $d$ into each
input pointer, via $\rdcssAlg$.  If any individual $\rdcssAlg$ of
$\writeAllDescsAlg$ fails, then $T$ attempts to mark the status
pointer of $desc$ as failed (line~\ref{resolve-status-MCAS}). In
general, $T$ writing $d$ into pointer $\pointGenEntry i$ of entry $i$
signals that writing $d$ into previous entries has succeeded and:
\begin{itemize}
\item Descriptor $\textit{desc}$ (stored in $d$) is currently active
  in $\pointGenEntry j$ for every $j \leq i$.
\item $T$ has an ongoing $\mcasAlg$ in which pointers
  $\pointGenEntry j$ for $j > i$ still need to be updated with $d$.
\item Pointers $\pointGenEntry j$ for $j \leq i$ indirectly have
  their expected values.
\end{itemize}
If another thread $T'$ attempts to access $\pointGenEntry i$ and finds
$d$, it will \emph{help} $T$ complete $\mcasAlg$ before doing anything
else, by invoking $\mcasHelpAlg$ as follows.  $T'$ first attempts to
write the descriptor into all remaining pointers
(line~\ref{write-all-descs-MCAS}).  If $T'$ succeeds, then all
pointers had the expected values. The variable $s$ in line 
\ref{write-all-descs-MCAS} is set to
$\SUCCEEDED$, and $T'$ attempts changing the status field of $desc$
accordingly (line \ref{resolve-status-MCAS}) and replacing all the
descriptors with the new values (line~\ref{remove-all-descs-MCAS}). These
attempts utilize $\textit{CAS}$ which may fail if some other thread
already helped $T$ to either succeed or fail in its $\mcasAlg$.
Alternatively, if $T'$ fails to write $d$ into some pointer, it means
that not all pointers had the expected values. The variable $s$ in
line \ref{write-all-descs-MCAS} is bound to $\FAILED$, $T'$ attempts to change the status field
of $desc$ accordingly (line~\ref{resolve-status-MCAS}), and to undo
$T$'s writing of $d$
(line~\ref{remove-all-descs-MCAS}). As
before, the $\textit{CAS}$'s in these attempts may fail if some other
thread already helped $T$ to either succeed or fail in its $\mcasAlg$.

Notice that $\mcasHelpAlg$ and $\writeAllDescsAlg$ are mutually
recursive. This is necessary, because while a thread $T$ is writing
the descriptor into all the input pointers using $\writeAllDescsAlg$,
$T$ may encounter other descriptors that force it to help by invoking
$\mcasHelpAlg$ (lines
\ref{alg-is-desc-MCAS-MCAS}-\ref{alg-help-complete-invoke-MCAS-MCAS}),
after which $T$ must reattempt writing the descriptor again
(line~\ref{alg-help-restart-MCAS}).

\subsubsection{Helping reading vs. opportunistic reading}
\label{subsubsect::helping-vs-oppor}

\begin{figure}[t]
\begin{subfigwrap}{Helping reading}{fig::sub::standard-mcas-read-impl}[0.5\textwidth]
\begin{algorithmic}[1]
\Proc{$\mcasReadAlg$\,}{$pt : \DataPtType$}
\State \label{alg-access-Read-MCAS} $old \gets \rdcssReadAlg(pt)$
\If {$\isMcasDescRepAlg(old)$} \label{alg-is-desc-Read-MCAS}
  \State \label{alg-help-complete-invoke-Read-MCAS} $\mcasHelpAlg(old)$
  \State \label{alg-restart-Read-MCAS} \returnCmd{$\mcasReadAlg(pt)$}
\Else 
  \State \label{alg-return-old-value-Read-MCAS} \returnCmd{$old$}
\EndIf
\EndProc
\end{algorithmic}
\end{subfigwrap}
\begin{subfigwrap}{Opportunistic reading}{fig::sub::optimal-mcas-read-impl}[0.5\textwidth]
\begin{algorithmic}[1]
\Proc{$\mcasReadAlg$\,}{$pt : \DataPtType$}
\State \label{alg-access-Read-Oport-MCAS} $old \gets \rdcssReadAlg(pt)$
\If {$\isMcasDescRepAlg(old)$} \label{alg-is-desc-Read-Oport-MCAS}
  \State \label{alg-Read-Desc-Oport-MCAS} $d \gets \rdcssReadAlg(old)$
  \State \label{alg-Read-status-Oport-MCAS} $s \gets \readRepAlg(\statusProp{d})$
  \State \label{alg-Read-Entry-Oport-MCAS} $e \gets$ entry for $pt$ in $\entriesProp{d}$
  \If {$s = \SUCCEEDED$}
     \State \label{alg-return-new-Oport-MCAS} \returnCmd{$\newGen e$}
  \Else
     \State \label{alg-return-exp-Oport-MCAS} \returnCmd{$\expGen e$}
  \EndIf
\Else 
  \State \label{alg-return-old-value-Read-Oport-MCAS} \returnCmd{$old$}
\EndIf
\EndProc
\end{algorithmic}
\end{subfigwrap}
\caption{Different implementations for $\mcasReadAlg$.}
\label{alg-MCAS-Read}
\end{figure}

Figure \ref{alg-MCAS-Read} shows two possible implementations of the
$\mcasReadAlg$ procedure, realizing two different reading strategies:
\emph{helping reading} (Figure
\ref{fig::sub::standard-mcas-read-impl}), and its optimization
\emph{opportunistic reading} (Figure
\ref{fig::sub::optimal-mcas-read-impl}). Both have been introduced
by Harris et al.~\citet{Harris}; the second one informally.
In the rest of the paper, we assume helping reading, delegating
opportunistic reading to
Appendix~\ref{sect::appendix::opportunistic-impl-proof}.
Nevertheless, we show both implementations here, to emphasize that our
approach applies to both implementations.

The helping reader works as follows. After reading the input pointer
(line~\ref{alg-access-Read-MCAS}), if an MCAS descriptor is found, it
helps the ongoing $\mcasAlg$ by invoking $\mcasHelpAlg$ (line
\ref{alg-help-complete-invoke-Read-MCAS}).  Then, it starts again
(line~\ref{alg-restart-Read-MCAS}).
In contrast, an opportunistic reader optimizes by using the following
strategy. If an MCAS descriptor is found (line
\ref{alg-is-desc-Read-Oport-MCAS}), instead of helping, $\mcasReadAlg$
reads the descriptor's status (line \ref{alg-Read-status-Oport-MCAS}).
If the status is $\SUCCEEDED$, the descriptor was successfully written
into all input pointers in the ongoing $\mcasAlg$ (including pointer $pt$,
i.e. the input pointer to $\mcasReadAlg$).  In this case,
$\mcasReadAlg$ returns the new value for $pt$ (as found in the
descriptor) because this is the value $\mcasAlg$ will write eventually
anyway.  If the status is $\FAILED$, $\mcasReadAlg$ returns the
expected value for $pt$ because $\mcasAlg$ will leave $pt$ unmodified, and
the presence of the descriptor means that $pt$ indirectly has the
expected value.  If the status is $\UNDECIDED$, $\mcasReadAlg$ also
returns the expected value for $pt$ because $\mcasReadAlg$ can pretend
that the read to $pt$ occurred before any modification to $pt$, since
$\mcasAlg$ is still attempting to write the descriptor into all the input
pointers.  We say that the optimized implementation is an
\emph{opportunistic} reader because instead of helping, it takes
advantage of what helpers have registered in the descriptor.
Opportunistic reading is an optimization over helping reading because
it avoids the need to wait for a value to return, as such a wait
increases the response time in the common case when the number of
$\mcasReadAlg$'s is greater than the number of $\mcasAlg$
invocations.

\section{Introducing and Axiomatizing Visibility}

\begin{figure}[t]
  \centering 
  \begin{subfigwrap}{State-based sequential specification.
  $\mapExtThree{H}{\mapEntry {\pointGenEntry i} {\newGenEntry
      i}}{i\in\listvar{u}}$ is the heap (i.e., memory) obtained when
  pointers $\pointGenEntry i$ ($i \in \listvar u$) in the heap $H$ are
  mutated into $\newGenEntry i$, while leaving the rest of $H$
  unchanged.}{subfig:atomic-spec-state}
\centering
\begin{tabular}{ll}
$(A_1)$ Successful $\mcasAlg$ & \\
\quad $H \xrightarrow{\mcasAlg(\listvar{u})\ \langle true \rangle} \mapExtThree{H}{\mapEntry {\pointGenEntry i} {\newGenEntry i}}{i\in\listvar{u}}$ &
if $\forall i \in \listvar{u}.\ H(\pointGenEntry i) = \expGenEntry i$\\
$(A_2)$ Failing $\mcasAlg$ & \\
\quad $H \xrightarrow{\mcasAlg(\listvar{u})\ \langle false \rangle} H$ &
if $\exists i \in \listvar{u}.\ H(\pointGenEntry i) \neq \expGenEntry i$ \\
\end{tabular}
\end{subfigwrap}

\begin{subfigwrap}{History-based sequential specification. Relation $\visObsSymbol {\pointerIndx p} : \absEvent \times \absEvent$ is
abstract.}{subfig:atomic-spec-history}
\centering
\begin{tabular}{c}
\begin{tabular}{l}
$(B_1)$ No in-between \\
\quad $(\visObs {\pointerIndx p} w r \wedge w' \in \writesAbs{\pointerIndx p}) \implies 
  (\precedesAbsEq {w'} w \vee \precedesAbsEq r {w'})$ \\
$(B_2)$ Observed events are writes \\
\quad $\visObs {\pointerIndx p} w {\_} \implies w \in \writesAbs{\pointerIndx p}$ \\
$(B_3)$ Dependences occur in the past \\
\quad $\visObs {\pointerIndx p} w r \implies \precedesAbs w r$\\
$(B_{4.1})$ Successful $\mcasAlg$ \\
\quad $r = \mcasAlg(\listvar{u})\left<true\right> \implies 
\forall i \in \listvar{u}.\ \exists w.\ \visObs {\pointerIndx {\pointGenEntry i}} w r \wedge
\inputVal w {\pointGenEntry i} {\expGenEntry i}$ \\
$(B_{4.2})$ Failing $\mcasAlg$ \\
\quad $r = \mcasAlg(\listvar{u})\left<false\right> \implies 
\exists i \in \listvar{u}.\ \exists w.\ \exists v'.\ \visObs {\pointerIndx {\pointGenEntry i}} w r \wedge {}
\inputVal w {\pointGenEntry i} {v'} \wedge v' \neq \expGenEntry i$\\
\end{tabular}
\\
\begin{tabular}{c}
\ \\
$\begin{aligned}
{\inputVal x {p} {v}} \defini & \ x = \mcasAlg([\ldots,\{ \pointGenEntry{} : p,\ \expGenEntry{} : \_,\ \newGenEntry{} : v \},\ldots]) \left< \_ \right> \vee {}\\
                            & \ x = \mcasWriteAlg(p,v)\left< \_ \right> \vee x = \mcasAllocAlg(v)\left< p \right>\\
\end{aligned}$
\\
\ \\
$\begin{aligned}
x \in \writesAbs{p} \Longleftrightarrow & \ x = \mcasAlg([\ldots,\{ \pointGenEntry{} : p,\ \ldots\},\ldots]) \left< true \right> \vee
x = \mcasWriteAlg(p,\_)\left< \_ \right> \vee x = \mcasAllocAlg(\_)\left< p \right>\\
\end{aligned}$
\end{tabular}
\end{tabular}
\end{subfigwrap}
\caption{State-based and history-based sequential specifications for $\mcasAlg$.}
\label{fig::atomic-specs-mcas}
\end{figure}

\subsection{Sequential History Specifications and Observation Relations}
\label{subsect::overview::observation-relations-and-seq-specs}

Following Henzinger et al.~\citet{henzinger:concur13}, we start the development of
visibility relations by introducing \emph{history-based
  specifications} for our data structures.  
The idea is that a history-based specification in terms of visibility
relations will be more concrete---and thus more direct to prove---than
merely asserting linearizability of the involved structures.
At the same time, the specification will still be sufficiently
abstract and high-level that: (1) it is largely \emph{shared} by RDCSS
and MCAS, thus providing a unifying explanation of both (and of the
opportunistic reading optimization), and (2) it generically implies
linearizability. We next illustrate the intuition behind history-based
specifications, focusing on the $\mcasAlg$ procedure.
 
History-based specifications describe relationships between the data
structure's procedures in an execution history. They are significantly
different (and more involved) from the perhaps more customary
state-based specifications that describe the actions of a procedure in
terms of input and output state. However, history-based specifications
scale better to the concurrent setting, which is why concurrent
consistency criteria such as linearizability are invariably defined in
terms of execution histories (e.g., see Footnote~\ref{footnote::linearizability} of Introduction).

In this section we focus on \emph{sequential} histories in order to
introduce the idea of \emph{observation relation} in a simple way,
before generalizing to concurrent histories in
Section~\ref{subsect::overview::separable-before-relations}.
A sequential history is a sequence of the form
$[proc(in_1) \langle out_1 \rangle,\ \ldots,\ proc(in_n) \langle out_n
\rangle]$, where $proc(in_i) \langle out_i \rangle$ means that
$proc(in_i)$ executed \emph{atomically} and produced output
$out_i$. We term \emph{event} each element in a sequential history $h$,
and $\absEvent$ denotes the set of all events in $h$.

Figure~\ref{fig::atomic-specs-mcas} illustrates the distinction
between sequential state-based and history-based specifications for
$\mcasAlg$.  The state-based specification in
Figure~\ref{subfig:atomic-spec-state} says in axiom $A_1$ that a
successful $\mcasAlg$ occurs when all input pointers contain the
expected values; the pointers are then mutated to their new
values. Axiom $A_2$ says that a failing $\mcasAlg$ occurs when some
input pointer does not have the expected value, leaving the heap
unchanged.


Figure~\ref{subfig:atomic-spec-history} shows the history sequential
specification for $\mcasAlg$.  The specification utilizes the
\emph{observation relation} $\visObsSymbol {\pointerIndx p}$ to
capture a read-write causal dependence between events. In particular,
$\visObs {\pointerIndx p} w r$ means that ``event $r$ reads a value
that event $w$ wrote into pointer $p$''. 
Under this interpretation,
axioms $B_1,...,B_{4.2}$ state the following 
expected properties.\footnote{Our paper will make heavy use of several different
  relations. To help the reader keep track of them, we shall denote
  the relations by symbols that graphically associate to the
  relation's meaning. For example, we use $\visObsSYMBOL$ for the
  observation relation, because the symbol graphically resembles an
  eye.}

Axiom $B_1$ (No in-between) says that if $r$ reads from $w$ in $p$,
then no other successful $p$-write can occur between $w$ and $r$
(otherwise, such a write would overwrite $w$).  Relation
$\precedesAbsSymbol$ is the \emph{returns-before} relation (with
$\precedesAbsEqSymbol$ its reflexive closure), where
$\precedesAbs x y$ means that $x$ terminated before $y$ started. Note that $\precedesAbsSymbol$ is a total order on events, as
in a sequential setting the executions of different events cannot
overlap. Set $\writesAbs p$ collects the successful $p$-writes, e.g.,
writes of the form $\mcasWriteAlg(p,\_)$, $\mcasAllocAlg$ events 
returning $p$, and $\mcasAlg(\listvar u)$ events returning $true$ and having $p$ in
$\listvar u$.

Axiom $B_2$ (Observed events are writes) says that any observed event
at $p$ must be a successful $p$-write, i.e., if $r$ reads from $w$ in
$p$, then $w$ must have actually written into $p$.  Axiom $B_3$
(Dependences occur in the past) says that if a read depends on a
write, then the write executes before the read.

Axioms $B_{4.1}$ (Successful $\mcasAlg$) and $B_{4.2}$ (Failing
$\mcasAlg$) essentially encode the state-based sequential
specification for $\mcasAlg$.  For example, axiom $B_{4.1}$ directly
says that a successful $\mcasAlg(\listvar{u})\left<true\right>$ event
$r$ observes---for each of its input entries $\listvar u$---a successful write
event ($\mcasAlg$, $\mcasWriteAlg$ or $\mcasAllocAlg$, as per axiom $B_2$) that
wrote the expected value into the
appropriate pointer. Because the write events are
\emph{observed} by $r$, axiom $B_3$ ensures that they execute before $r$, while 
axiom $B_1$ guarantees the none of them is
overwritten before $r$ executes. Thus, diagrammatically, 
for each entry $i \in \listvar u$, the execution
looks as follows, where the write into pointer $\pointGenEntry {i}$
persists until the heap $H_n$ at which $r$
executes.

\begin{center}
\begin{tikzcd}[ampersand replacement=\&, column sep=small, row sep=small]
\ldots 
\arrow[r]\&
H_k
\arrow[r, "{w_{i}}"{name=Wi,yshift=2pt}] \& 
H_{k+1} 
\arrow[r] \&
\ldots 
\arrow[r] \&
H_n
\arrow[r, "{\mcasAlg(\listvar{u})\left< true \right>}"{name=MOne}] \&[30pt] 
\ldots
\\
{} \& {} \& {} \arrow[rr, mapsfrom, maps to, "\pointGenEntry{i} \mapsto \expGenEntry{i} \text{ persists}"] \& {} \& {} \\
\arrow[from=MOne, to=Wi, bend right, dashed, no head, "{\visObsSymbol{\pointerIndx {\pointGenEntry {i}}}}"{below}]
\end{tikzcd}
\end{center}\vspace{-5mm}

\begin{figure}[t]
\begin{subfigwrap}{Concurrent specification. Relations $\visObsSymbol {\pointerIndx p}, \visSepSymbol {\pointerIndx p} : \absEvent \times \absEvent$ are
existentially quantified.}{subfig:concurrent-spec-history}
\centering
\begin{tabular}{l}
\axiomCLabel{vis-ax::cc-no-in-between} No in-between \\
\quad $(\visObs {\pointerIndx p} w r \wedge w' \in \writesAbs{\pointerIndx p}) \implies 
  (\visSepEq {\pointerIndx p} {w'} w \vee \visSepEq {\pointerIndx p} r {w'})$ \\
\axiomCLabel{vis-ax::cc-observed-are-writes} Observed events are writes \\
\quad $\visObs {\pointerIndx p} w {\_} \implies w \in \writesAbs{\pointerIndx p}$ \\
\end{tabular}
\begin{tabular}{l}
\axiomCLabel{vis-ax::cc-no-future-dependence} No future dependences \\
\quad $x \transCl{\genVisSymbol} y \implies \nprecedesAbsEq y x$\\
\axiomCLabel{vis-ax::cc-return-completion} Return value completion \\
\quad $\exists v.\ \postPred x v \wedge (x \in \terminatedEvent \implies v = \outputProp x)$ \\
\end{tabular} 
\end{subfigwrap}

\begin{subfigwrap}{Defined notions.}{subfig:defined-notions-concurrent-spec-history}
\centering
\begin{tabular}{c}
\begin{tabular}{l}
General visibility relation \\
\quad ${\genVisSymbol} \defini \bigcup_p (\visObsSymbol {\pointerIndx p} \cup \visSepSymbol {\pointerIndx p})$\\
Returns-before relation\\
\quad $\precedesAbs e {e'} \defini \ETimeProp e \natorderSymbol \STimeProp {e'}$\\
\end{tabular}
\begin{tabular}{l}
Set of terminated events\\
\quad $\terminatedEvent \defini \{ e \mid \ETimeProp e \neq \bot \}$\\
Closure of terminated events\\
\quad $\closedEvent \defini \{ e \mid \exists t \in \terminatedEvent.\ e \refleTransCl{\genVisSymbol} t \} $\\
\end{tabular}
\\
\begin{tabular}{c}
  \\
$\postPred {\mcasAlg(\listvar{u})} v \defini v \in \BoolType \wedge
\begin{cases}
\forall i \in \listvar{u}.\ \exists w.\ \visObs {\pointerIndx {\pointGenEntry i}} w {\mcasAlg(\listvar{u})} \wedge
\inputVal w {\pointGenEntry i} {\expGenEntry i} &
\text{if $v = true$} \\
\exists i \in \listvar{u}.\ \exists w.\ \exists v'.\ \visObs {\pointerIndx {\pointGenEntry i}} w {\mcasAlg(\listvar{u})} \wedge {} & \text{if $v = false$} \\
\hspace{30pt} \inputVal w {\pointGenEntry i} {v'} \wedge v' \neq \expGenEntry i & \\
\end{cases}$
\\
\ \\
$\begin{aligned}
\mcasAlg(\listvar{u}) \in \writesAbs{p} \Longleftrightarrow
(\exists j \in \listvar{u}.\ p = \pointGenEntry j) \wedge 
\forall i \in \listvar{u}.\ \exists w.\ \visObs {\pointerIndx {\pointGenEntry i}} w {\mcasAlg(\listvar{u})} \wedge
\inputVal w {\pointGenEntry i} {\expGenEntry i}
\end{aligned}$
\end{tabular}
\end{tabular}
\end{subfigwrap}
\caption{Concurrent history-based specification (fragment) for $\mcasAlg$. Variables $w$, $w'$, $r$, $x$, $y$ range over $\closedEvent$.
Variables $e$, $e'$ range over $\absEvent$.
Full definitions of $\postPredSymbol$, $\writesAbs p$, and $\inputValName$ for the MCAS module are found in Appendix~\ref{sect::appendix::lin::MCAS}.
Two more axioms involving allocs are elided (also elided from Figure~\ref{subfig:atomic-spec-history}). The full list of axioms is shown in 
Figure~\ref{fig:visibility-relation-axioms} of Appendix~\ref{sect::appendix::lin::full-appendix}. }
\label{fig::concurrent-spec-mcas}
\end{figure}

\subsection{Concurrent Specifications and Separable-Before Relations}
\label{subsect::overview::separable-before-relations}

Concurrent execution histories do not satisfy the sequential axioms in
Figure~\ref{subfig:atomic-spec-history} for two main reasons.  First,
concurrent events can \emph{overlap} in real time. As a consequence,
the axioms $B_1$ (No in-between) and $B_3$ (Dependencies occur in the
past) are too restrictive, as they force events to be disjoint, due to
the use of the returns-before relation $\precedesAbsSymbol$.
Second, events can no longer be treated as atomic; thus event's start
and end times (if the event is terminated) must be taken into
account. As a consequence, axioms $B_{4.1}$ and $B_{4.2}$ and set
$\writesAbs p$ must be modified to account for the output of an
unfinished event not being available yet.  We continue using
$\absEvent$ for the set of events in the concurrent history. We denote
by $\STimeProp {e}$, $\ETimeProp e$, the start and end time of event
$e$, respectively. We use the standard order relation on natural
numbers $\natorderSymbol$ to compare start and end times.

Figure~\ref{fig::concurrent-spec-mcas} shows the concurrent
specification that addresses the above issues. Importantly, in
addition to the observation relation, the specification utilizes the
\emph{separable-before relation} $\visSepSymbol {\pointerIndx p}$ to
capture an ordering dependence between events. In particular,
$\visSep {\pointerIndx p} x y$ means that ``event $x$ is separable
before $y$ because of some logical gap when using shared pointer
$p$''.
We now explain how the concurrent specification of
Figure~\ref{fig::concurrent-spec-mcas} is obtained from the sequential
one in Figure~\ref{subfig:atomic-spec-history}.

Axiom \axiomHRef{vis-ax::cc-no-in-between} is obtained from $B_1$ by
replacing $\precedesAbsSymbol$ with $\visSepSymbol p$.  The intuition
is that we want to relax the real-time strong separation imposed by
$\precedesAbsSymbol$ into a more permissive separation
$\visSepSymbol p$. The latter captures that a gap can be viewed as
existing between possibly overlapping events, forcing one event to be
ordered before another.\footnote{The symbol $\visSepSYMBOL$ twists
  $\precedesAbsSymbol$, suggesting that the separable-before relation
  $\visSepSYMBOL$ relaxes (i.e., is a twist on) returns-before
  relation $\precedesAbsSymbol$.} The
definition of $\visSepSymbol p$ is different for different
implementations (thus, the gap is described by logically different
properties in different implementations), but they all satisfy axiom
\axiomHRef{vis-ax::cc-no-in-between}. We shall see in
Section~\ref{subsubsect::visibility-mcas} the concrete definition of
$\visSepSymbol p$ (i.e., the description of the logical gap) for the
$\mcasAlg$ implementation of Figure~\ref{alg-MCAS}.

Axiom \axiomHRef{vis-ax::cc-observed-are-writes} is unchanged compared to $B_2$.

Axiom \axiomHRef{vis-ax::cc-no-future-dependence} is obtained from $B_3$ as follows. In the sequential
specification, $\visObsSymbol p$ was the only relation encoding
dependences between events, but now we have two relations encoding
dependences, $\visObsSymbol p$ and $\visSepSymbol p$. The relation
${\genVisSymbol} \defini {\cup_p {(\visObsSymbol p \cup \visSepSymbol
    p)}}$ thus collects all the dependences, and we call
$\genVisSymbol$ the \emph{general} visibility relation.\footnote{The
  symbol $\genVisSymbol$ reminds us of an eye without an iris,
  suggesting that the general visibility relation is ``blinder'' than
  $\visObsSYMBOL$, since $\genVisSymbol$ also contains
  $\visSepSYMBOL$.} We can consider modifying
Axiom $B_3$ into $\genVis x y \implies \precedesAbs x y$ to say that
any dependence $x$ of $y$ must terminate before $y$ starts.
However, such a modification of $B_3$ is too stringent, as it
does not allow $x$ to overlap with $y$. Instead, we relax the
conclusion to say that an event cannot depend on itself or events from
the future, i.e., $\genVis x y \implies \nprecedesAbsEq y x$. Finally,
we get axiom \axiomHRef{vis-ax::cc-no-future-dependence} by replacing $\genVisSymbol$ with its transitive
closure $\transCl{\genVisSymbol}$ to account for \emph{indirect}
dependences of $y$; e.g., in $x_1 \genVisSymbol x_2 \genVisSymbol y$,
event $x_1$ is an indirect dependence of $y$.
Hence, \axiomHRef{vis-ax::cc-no-future-dependence} reads ``any direct
or indirect dependence does not execute in the future, and events do
not depend on themselves''.

To understand Axiom \axiomHRef{vis-ax::cc-return-completion}, we need to consider the set $T$ of all
\emph{terminated} events and its closure under the general visibility
relation
$\closedEvent \defini \{ e \in \absEvent \mid \exists t \in
\terminatedEvent.\ e \refleTransCl{\genVisSymbol} t \}$. As usual,
$\refleTransCl{\genVisSymbol}$ is the reflexive-transitive closure of
$\genVisSymbol$.
The idea here is that $\closedEvent$ contains all the events that have
contributed to producing values in the execution history, either by
being terminated (and thus directly returning a value), or by being an
event on which some terminated event depends. The dependence may be
indirect, and the depended event may be unterminated. For example, a
terminated $\mcasAlg$ $r$ may observe an unterminated $\mcasAlg$ $w$,
i.e., $\visObs p w r$ as follows: $w$ mutates $p$ but is preempted
just before termination, thus allowing $r$ to read. At any rate,
$\closedEvent$ contains all the events required to ``explain'' the
values obtained in the concurrent execution, and thus establish that
the linearized execution is \emph{sequentially sound}, as per
Footnote~\ref{footnote::linearizability}.  In our example with $r$ and
$w$, we must consider $w$ if we want to explain the return value of
$r$, which is why $w$ is included in $\closedEvent$.

Axiom \axiomHRef{vis-ax::cc-return-completion} first coalesces axioms $B_{4.1}$ and $B_{4.2}$ into the
postcondition predicate $\postPredSymbol$, as shown in
Figure~\ref{subfig:defined-notions-concurrent-spec-history}.  Then
\axiomHRef{vis-ax::cc-return-completion} requires that every event $x \in \closedEvent$ satisfies
$\postPredSymbol$. In case $x$ is a terminated $\mcasAlg$ event, it is
easy to see that this requirement, along with the requirement that $v$
equals the output of $x$, directly corresponds to the sequential
axioms $B_{4.1}$ and $B_{4.2}$. In case $x$ is an unterminated
$\mcasAlg$ event, the axiom posits that some value $v$ can be found to
make $x$ appear as if it has terminated in a manner coherent with the
other events. The latter ``completion'' of $x$ with $v$ is a common
pattern that we inherit from the work on linearizability.

We further emphasize that in the concurrent setting we also need to
redefine the set $\writesAbs p$ of successful writes into pointer $p$.
It does not suffice to consider an $\mcasAlg$ as a successful write if
it returns $true$; we need a criterion when an unfinished $\mcasAlg$
is a successful write as well. Thus, we define in
Figure~\ref{subfig:defined-notions-concurrent-spec-history} that an
$\mcasAlg(\listvar u)$ is a successful $p$-write if $p$ is a pointer
in $\listvar u$, and $\mcasAlg(\listvar u)$ observes events writing
the expected values for each entry in $\listvar u$.

\begin{figure}[t]
\centering
\begin{tabular}{c}
$\postPred {\rdcssAlg(d)} v \defini 
\begin{aligned}[t]
\exists w_2\ldot
  \visObs {\pointerIndx {\pointTwo d}} {w_2} {\rdcssAlg(d)} \wedge 
   \inputVal {w_2} {\pointTwo d} v \wedge {} \\
   (v = \expTwo d \implies \exists w_1\ldot \visObs {\pointerIndx {\pointOne d}} {w_1} {\rdcssAlg(d)})
\end{aligned}$
\\
\ \\
$\rdcssAlg(d) \in \writesAbs{p} \Longleftrightarrow
\begin{aligned}[t]
p = \pointTwo d \wedge \exists w_1, w_2.\ \visObs
{\pointerIndx {\pointOne d}} {w_1} {\rdcssAlg(d)} \wedge 
\visObs {\pointerIndx {\pointTwo d}} {w_2} {\rdcssAlg(d)} \wedge {} \\
\inputVal {w_1} {\pointOne d} {\expOne d} \wedge 
\inputVal {w_2} {\pointTwo d} {\expTwo d}
\end{aligned}$
\end{tabular}
\caption{Together with axioms in
  Figure~\ref{subfig:concurrent-spec-history}, this is a (fragment of) concurrent specification for $\rdcssAlg$.
Full definitions of $\postPredSymbol$, $\writesAbs p$, and $\inputValName$ for the RDCSS module are in Appendix~\ref{sect::appendix::lin::RDCSS}.}
\label{fig::concurrent-spec-rdcss}
\end{figure}

Section~\ref{sect::axiomatization} will show that the axioms in
Figure~\ref{subfig:concurrent-spec-history}, henceforth called
\emph{visibility axioms}, suffice to prove the existence of a
sequentially sound total order. Thus, proving linearizability for an
MCAS implementation reduces to finding $\visObsSymbol p$
and $\visSepSymbol p$ that satisfy the visibility axioms.

One can repeat the steps for $\rdcssAlg$, and obtain again the axioms in
Figure~\ref{subfig:concurrent-spec-history}, but with $\writesAbs p$ and $\postPredSymbol$ defined
as in Figure~\ref{fig::concurrent-spec-rdcss}.
The postcondition for $\rdcssAlg(d)$ states that $\rdcssAlg(d)$ must observe 
an event that wrote $\rdcssAlg(d)$'s output $v$ into $\pointTwo d$,
and in case $v$ is the expected value, $\rdcssAlg(d)$ must have read
pointer $\pointOne d$ also. An $\rdcssAlg(d)$ successfully writes
if it observes two events that wrote the expected values into
pointers $\pointOne d$ and $\pointTwo d$.

\begin{figure}[t]
\includegraphics[scale=0.5]{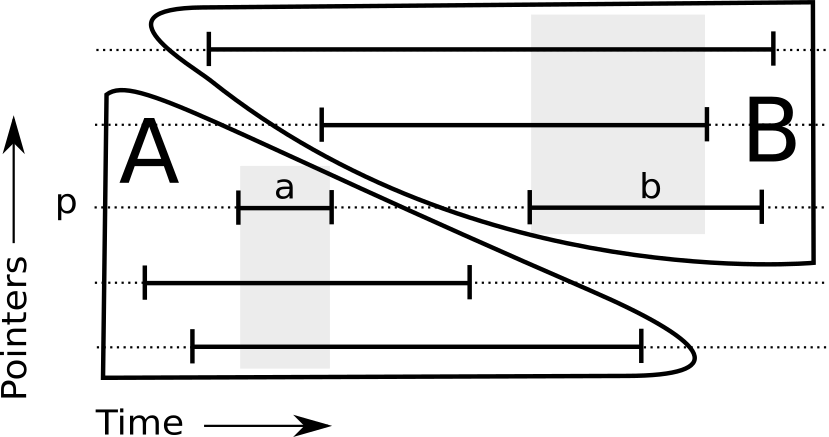}
\centering
\caption{An execution of $\mcasAlg$ events $A$ and $B$.  Each event
  encircles the spans belonging to it (i.e., helping it).  The
  intersection of spans for each event are shown as gray
  columns. Spans $a$ and $b$, both accessing pointer $p$, are named
  for later reference.}
\label{fig::spans-example}
\end{figure}

\subsection{Visibility Relations for MCAS}
\label{subsubsect::visibility-mcas}

\subsubsection{Definitions of Visibility Relations}
\label{subsubsect::visibility-definitions-mcas}

We now proceed to explain how relations $\visObsSymbol p$ and 
$\visSepSymbol p$ are defined for the implementation of $\mcasAlg$ in Figure 
\ref{alg-MCAS}. 
The key idea is to analyze the behavior of the descriptors during the
execution.  Figure~\ref{fig::spans-example} shows the execution of two
overlapping invocations of $\mcasAlg$, which we refer to as events $A$
and $B$, respectively.\footnote{We focus on the successful
  $\mcasAlg$ case. Section~\ref{sub::sect::lifespan-axiomatization} will
  explain the failing $\mcasAlg$ case.} During the execution of
$\mcasAlg$, threads write the descriptor into a pointer
(line~\ref{invoke-rdcss-in-MCAS} in Figure~\ref{alg-MCAS}), and then
remove it (line~\ref{remove-all-descs-MCAS}). We term \emph{span} the
time interval between writing the descriptor and removing it. We also
say that the removed descriptor is
\emph{resolved}. Figure~\ref{fig::spans-example} depicts spans as
bounded line segments. 

There are two key invariants of $\mcasAlg$ that give rise to 
its visibility relations.
The first key
invariant is that all the spans of an event \emph{must overlap} (we
call this the \emph{bunching invariant}). Concretely for $\mcasAlg$
events, any descriptor write must occur before any descriptor
removal. This is so because of the following two properties related to
how control flows through $\mcasAlg$. First, threads can reach
line~\ref{remove-all-descs-MCAS} (descriptor removal) only if some
thread changed the descriptor status at
line~\ref{resolve-status-MCAS}.\footnote{Recall that all descriptors
  are $\UNDECIDED$ initially, and can evolve to $\SUCCEEDED$ or
  $\FAILED$ only.}  This holds because
line~\ref{remove-all-descs-MCAS} (descriptor removal) is in the
$\mathit{false}$ branch of the conditional at
line~\ref{is-phase1-still-undecided} that checks if the descriptor
status is $\UNDECIDED$. Therefore, if no thread has changed the
descriptor status at line~\ref{resolve-status-MCAS}, threads starting
the $\mcasHelpAlg$ procedure will keep entering the $\mathsf{true}$
branch at line~\ref{is-phase1-still-undecided}.  The second property
is that any descriptor write at line~\ref{invoke-rdcss-in-MCAS} must
occur before some thread changes the descriptor status at
line~\ref{resolve-status-MCAS}.  This holds because the $\rdcssAlg$ at
line~\ref{invoke-rdcss-in-MCAS} (descriptor write attempt) fails if
the descriptor status is no longer $\UNDECIDED$.  These two properties
imply that any descriptor write must occur before the status change,
and any descriptor removal must occur after the status change, i.e.,
any span starts before any span finishes in the event; hence, span
intersection is non-empty. Figure~\ref{fig::spans-example} depicts the
spans' intersection for each event as gray columns. Since the presence
of a descriptor signals that a pointer indirectly has the expected
value, inside the gray column all the event's pointers simultaneously
have the expected values. Therefore, the event can be considered as
abstractly occurring (i.e., having its LP) anywhere inside the gray
column.

The second key invariant is that spans writing into \emph{the same
  pointer} (e.g., spans $a$ and $b$ in
Figure~\ref{fig::spans-example}) \emph{must be disjoint} (we call this
the \emph{disjointness invariant}).
This is so because $\mcasAlg$ uses helping; thus an event writes into
a pointer only if there is no descriptor currently present. More
specifically, the $\rdcssAlg$ at line \ref{invoke-rdcss-in-MCAS} in
Figure~\ref{alg-MCAS} writes the descriptor only if the expected value
is present directly (i.e., not via a descriptor). Since spans start 
by writing a descriptor, no span can start while there is a descriptor present, or
equivalently, when another span is already active in the pointer. The span
can start only after the currently present descriptor is removed.

These two invariants imply, as also apparent from
Figure~\ref{fig::spans-example}, that gray columns of
different events $A$ and $B$ accessing a common pointer must be
disjoint. Thus, we can \emph{separate} event $A$ before event $B$ in
time, because $A$'s gray column
executes before $B$'s. In the figure, we can describe the separation
(i.e., the gap between the gray columns) by saying that
there is a span in $A$ (namely $a$) and a span in $B$ (namely $b$),
both accessing a common pointer, and $a$ completes before $b$ starts.
Spans $a$ and $b$, being disjoint, induce a gap in the
gray columns. 

We define the separable-before relation $\visSepSymbol {\pointerIndx p}$ so that 
it directly formalizes the above description of the separation between $A$ and $B$,
\begin{align}
\label{eq::overview::helpers-separability}
\visSep{\pointerIndx p} A B \defini \exists x \in \hspans p A, y \in \hspans p B.\ \precedesSpans x y 
\end{align}
In the definition, $x \in \hspans p A$ means that $x$ is a span
accessing pointer $p$ in event $A$ (similarly for
$y \in \hspans p B$), and $\precedesSpans x y$ means that span $x$
terminates before $y$ starts.
Relation $\precedesSpansSymbol$ is a partial order on spans that
totally orders spans accessing the same pointer due to span
disjointness. $\precedesSpansEqSymbol$ denotes its reflexive closure.

We now focus on the observation relation
$\visObsSymbol {\pointerIndx p}$.  We previously informally explained
that $\visObs {\pointerIndx p} A B$ intuitively captures that event
$B$ reads a value written by the $p$-write $A$, with no other
intervening writes between.
To illustrate how this intuition can be expressed using spans, consider the
span $b$ in Figure~\ref{fig::spans-example}.
When this span starts, it reads the input pointer (this is so because the $\rdcssAlg$ at
line \ref{invoke-rdcss-in-MCAS} in Figure \ref{alg-MCAS} obtains the
pointer's value, writes the descriptor and returns the read
value). This obtained value must have been written by the most recent
span that successfully wrote into the pointer, which, in
Figure~\ref{fig::spans-example} is span $a$.

To formally capture the described situation, we say that event $A$ is
\emph{observed} by $B$ at pointer $p$ if there are spans $x$ and $y$ in
$A$ and $B$, respectively, such that $x$ is the most recent span that terminated before
$y$ started, and that successfully wrote into $p$,
\begin{align}
\label{eq::overview::helpers-observation}
\visObs {\pointerIndx p} A B \defini \exists x \in 
\hspans p A, y \in \hspans p B.\  x = \max_{\precedesSpansEqSymbol} \{ z \in \writesSpans p \mid \precedesSpans z y \}
\end{align}
In the definition, the maximum is taken under
$\precedesSpansEqSymbol$, and set $\writesSpans p$ collects the
spans that \emph{successfully} write into $p$. We differentiate
successful writes because some spans do not carry out changes. For
example, line \ref{remove-all-descs-MCAS} in Figure
\ref{alg-MCAS} produces a span that writes the new value only if the
descriptor's status was set to $\SUCCEEDED$, but produces a span that
undoes changes (as if the descriptor was not written) if the status
was set to $\FAILED$.  
Notice that it makes sense 
to take the maximum under $\precedesSpansEqSymbol$
because spans accessing the same pointer are disjoint, hence,
linearly ordered under $\precedesSpansEqSymbol$.

\subsubsection{Spans and Visibility Axioms}
\label{subsubsect::spans-and-visibility-axioms}

Section~\ref{sub::sect::lifespan-axiomatization} will substantiate more how
definitions~\eqref{eq::overview::helpers-separability}
and~\eqref{eq::overview::helpers-observation} satisfy the visibility
axioms of Figure~\ref{subfig:concurrent-spec-history}.  Here, we
briefly illustrate how the definitions satisfy the no in-between axiom
\axiomHRef{vis-ax::cc-no-in-between} using Figure~\ref{fig::spans-example}.  In
Figure~\ref{fig::spans-example}, assertion
$\visObs {\pointerIndx p} A B$ means that span $b$ reads a value
written by span $a$ in pointer $p$.  Since $a$ is the most recent
$p$-writing span before $b$ in real-time, and due to span
disjointness, any other $p$-writing span $c$ must execute either
completely before $a$ or completely after $b$. Therefore, if $c$
belongs to some $p$-write $C$, there will be a gap between the gray
columns of either $C$ and $A$ or $B$ and $C$, depending on whether $c$
executes before $a$ or after $b$.

\subsubsection{Alternative Span Definitions}
\label{subsubsect::span-alternative-definitions-mcas}

As we just saw, the definitions
\eqref{eq::overview::helpers-separability} and
\eqref{eq::overview::helpers-observation} satisfy axiom \axiomHRef{vis-ax::cc-no-in-between}. 
Notice, however, that the argument relied solely on the span disjointness
invariant, and not on a specific definition of spans.  This is actually
an overarching property that we shall utilize when proving each of the
visibility axioms from Figure~\ref{subfig:concurrent-spec-history}.
Neither proof will depend on the particular definition of spans, but
only on high-level abstract span invariants such as bunching,
disjointness and a small number of additional structural ones
that we shall introduce in
Section~\ref{sub::sect::lifespan-axiomatization}.

This abstraction affords some freedom to deviate from the operational
definition of spans as the time intervals between writing the descriptor
and resolving it,
so long as the high-level span invariants remain satisfied by the
programs.

To illustrate, bunching and disjointness remain satisfied if the
successful CAS at line~\ref{resolve-status-MCAS} is chosen as the
ending point of a span, instead of line~\ref{remove-all-descs-MCAS}
that we previously considered. Recall that while explaining the bunching
invariant in Section~\ref{subsubsect::visibility-definitions-mcas},
we stated that any descriptor write must occur before the status change
at line~\ref{resolve-status-MCAS}. Therefore, the successful CAS at line~\ref{resolve-status-MCAS} is perfectly 
fine to end spans, because any span must start before it, i.e., bunching 
holds because any span starts before the CAS, which is the ending point for all the spans. 
Disjointness still holds because of the same reason as in 
Section~\ref{subsubsect::visibility-definitions-mcas}: no descriptor can be 
written if there is currently a descriptor present. Therefore, if we end the span
prematurely at line~\ref{resolve-status-MCAS}, still no span can start until the descriptor
is removed at line~\ref{remove-all-descs-MCAS} much later.

Additionally, single reads that return a $\ValType$ value can be treated as 
spans that write a descriptor and instantaneously resolve it (i.e. a ``collapsed span'').
For example, the read at line~\ref{alg-access-Read-MCAS} of Figure~\ref{fig::sub::standard-mcas-read-impl}.
It is essential that we treat as spans only those reads that return $\ValType$ values
to ensure the span disjointness invariant: if we treat reads that
return descriptors as spans, then the read will occur while a descriptor is present,
i.e., inside another span, violating disjointness.

\subsubsection{Visibility Relations for RDCSS and Opportunistic Readers}

Perhaps surprisingly, definitions
\eqref{eq::overview::helpers-separability} and
\eqref{eq::overview::helpers-observation} apply without change not
only to $\mcasAlg$, but to all events using helping in RDCSS and MCAS
implementations such as $\mcasAlg$, $\mcasWriteAlg$, $\mcasReadAlg$,
$\rdcssAlg$, $\rdcssReadAlg$, etc.

For example, $\rdcssAlg$ events can be seen as a special case of
Figure \ref{fig::spans-example}, in which each $\rdcssAlg$ event
executes at most two spans. The first span of $\rdcssAlg$ is generated
by writing the descriptor at line \ref{alg-CAS-RDCSS} in Figure
\ref{alg-RDCSS}, and resolved at either line
\ref{alg-pt2-write-success-RDCSS} or line
\ref{alg-pt2-write-fail-RDCSS}.  The second span consists of a single
read of the control pointer at line \ref{alg-pt1-read-RDCSS} (thus, it
is a ``collapsed'' span). Trivially, $\rdcssAlg$ satisfies the bunching
invariant because the thread that resolves the descriptor at either
line \ref{alg-pt2-write-success-RDCSS} or line
\ref{alg-pt2-write-fail-RDCSS}, must have also previously read the
control pointer at line \ref{alg-pt1-read-RDCSS} while the descriptor
was present. In other words, the collapsed span overlaps the first
span.  Also, $\rdcssAlg$ satisfies span disjointness: an $\rdcssAlg$
invocation cannot write a new descriptor into a pointer if there is
currently another descriptor present.

On the other hand, opportunistic reading requires definitions
different from~\eqref{eq::overview::helpers-separability}
and~\eqref{eq::overview::helpers-observation} (see
Definition~\ref{defn::appendix::valid::visibility-relations-opor} in
Appendix~\ref{sect::appendix::opportunistic-impl-proof}).
Nevertheless, the concepts of separable-before and observation relations 
satisfying the axioms in 
Figure~\ref{subfig:concurrent-spec-history}, are general
high-level characteristics of all these algorithms.

\subsection{Comparing Visibility Relations and Linearization Points}
  
The standard way of proving a concurrent algorithm linearizable is to
exhibit the algorithm's LPs.
For example, the LPs for $\rdcssAlg$ from Figure~\ref{alg-RDCSS} have
been described operationally in previous
work~\cite{Harris,vafeiadis,IrisFuture}: one LP occurs at
line~\ref{alg-CAS-RDCSS} whenever the $\casRepAlg$ in
line~\ref{alg-CAS-RDCSS} fails and the value returned by the
$\casRepAlg$ is not a descriptor; another LP occurs at 
line~\ref{alg-pt1-read-RDCSS}, as long as the thread that executes this
line can successfully execute one of the $\casRepAlg$s at lines
\ref{alg-pt2-write-success-RDCSS} and \ref{alg-pt2-write-fail-RDCSS},
\emph{later on}.
While a proof that this is a correct description of the LPs of
$\rdcssAlg$ exists (e.g., in Jung et. al.~\cite{IrisFuture}), the LP assignment
alone does not really provide much intuition as to the underlying
principles behind $\rdcssAlg$, or how these principles are shared with
$\mcasAlg$.

For comparison, we can recover the LP approach by defining the
visibility relations in terms of LPs. For example, defining the
separable-before relation as,
\begin{align}
\label{eq::overview::separable-relation-with-LPs}
\visSep {\pointerIndx p} x y \defini LP(x) \precedesAbsSymbol LP(y) 
\end{align}
where $LP(x)$ and $LP(y)$ are the linearization points of $x$ and $y$,
respectively. However,
\eqref{eq::overview::separable-relation-with-LPs} leads to a less
abstract and less modular proof than ours.
Regarding abstraction,
\eqref{eq::overview::separable-relation-with-LPs} does not bring out
that span bunching and disjointness are the two high-level invariants
that $\rdcssAlg$ and $\mcasAlg$ share, and that suffice to describe
separation between events for both algorithms.
Regarding modularity,
\eqref{eq::overview::separable-relation-with-LPs} defines a total
order on events \emph{right away}. But notice that the specification
in Figure~\ref{subfig:concurrent-spec-history} does not require the
relation $\visSepSymbol {\pointerIndx p}$ to be total, not even over
events sharing pointer $p$; it \emph{only} requires to somehow
separate the events that share $p$. In particular,
\eqref{eq::overview::helpers-separability} is not a total order over
events sharing pointer $p$. Instead, the visibility approach divides the
burden of proving linearizability into two stages that are easier to
conquer than directly proving linearizability. In the first stage, we
define the visibility relations and prove the visibility axioms. In the
second stage, the (total) linearization order is constructed out of
the visibility relations. Importantly, the second stage is generic and
independent of the first, as it can be developed out of the visibility
axioms alone, without having the definitions of the visibility
relations in hand.

Where the LP approach very concretely and operationally locates the LP
of each event, the visibility approach is more concerned with
capturing the semantic structure of the algorithms, such as, in the
example of RDCSS and MCAS, recognizing that each event is
characterized by the disjoint bunches of its spans.
In fact, this semantic structure
appears to capture how Harris et al.~\citet{Harris} themselves thought about these
algorithms, as they reveal in this quote from their Section
4.2:
\begin{displayquote}
  Each invocation of RDCSS can be considered separately. This
  surprising observation follows by examining the memory
  accesses. [...]  Different RDCSS operations acting on the same
  location are thereby serialized by the order of their active
  periods, so we can consider them individually.
\end{displayquote}
Clearly, ``active periods'' corresponds to spans, and ``serialization
by the order of active periods'' corresponds to the separable-before
relation.  On the other hand, Harris et al.~\citet{Harris} do not pursue
axiomatization, which is why no concept analogous to the observation
relation appears for them. Be it as it may, the quote already
contrasts with the (arguably, not very intuitive) description of the
LPs of $\rdcssAlg$ we gave at the beginning of this section.
That visibility relations can formalize this (shared) intuition behind
algorithms such as RDCSS and MCAS, is the point of our paper.

\section{Proving Linearizability out of Visibility Axioms} 
\label{sect::axiomatization}

Having provided a high-level overview, we now proceed with the
technical development of the proofs of linearizability that are
\emph{abstract}, i.e., they proceed from the visibility axioms, and
apply to arbitrary implementations of RDCSS and MCAS (the gray box in
Figure~\ref{fig::proof-diagram}). In this section we exhibit the
abstract proof for MCAS. The one for RDCSS is in
Appendix~\ref{sect::appendix::lin::RDCSS}.  Later on, we shall
instantiate the abstract proofs to conclude that the \emph{concrete}
implementations are linearizable. In particular, the helping MCAS
implementation is shown linearizable in
Section~\ref{sub::sect::lifespan-axiomatization}, the RDCSS
implementation in Appendix~\ref{appendix::sub::sect::impl::RDCSS}, and
the opportunistic MCAS implementation in
Appendix~\ref{sect::appendix::opportunistic-impl-proof}.

\begin{defn}[Linearizability] 
\label{sect::defn-linearizability}
We say that a data structure's implementation $D$ is
\emph{linearizable} if for any set of events $\absEvent$
generated from an arbitrary execution history in $D$, there is a
binary relation
${\genVisSymbol} \subseteq {\absEvent \times \absEvent}$
(\emph{general visibility relation}) and a linear order
${\leq} \subseteq {\closedEvent \times
  \closedEvent}$ (\emph{linearization})\footnote{$\closedEvent$ is the closure of the terminated events $\terminatedEvent$ under $\genVisSymbol$, as in Figure~\ref{subfig:defined-notions-concurrent-spec-history}.} s.t.,
\begin{itemize}
\item $\leq$ respects the real-time ordering of events, 
i.e., $\precedesAbsSymbol$ restricted to $\closedEvent$ is contained in $\leq$.
\item $\leq$ respects the ordering constraints in $\genVisSymbol$,
i.e., $\genVisSymbol$ restricted to $\closedEvent$ is 
contained in $\leq$.
\item $\leq$ is \emph{sequentially sound}, i.e., when events are executed in
  the order determined by $\leq$, the execution output $o_i$ of the
  $i$-th event $x_i$ matches $\outputProp {x_i}$ if
  $x_i \in \terminatedEvent$ or it is consistent to define
  $\outputProp {x_i}$ as $o_i$ if $x_i \notin \terminatedEvent$.  The
  operational semantics of executing events in the order of $\leq$ is
  data-structure specific.
\end{itemize}
\end{defn}

Our definition of linearizability differs slightly from the usual
formulation~\cite{herlihy:90} in that it emphasizes working with
visibility relations, as follows. When judging a history $h$
linearizable, the standard definition starts by requiring an
\emph{arbitrary} completion $\hat{h}$, which closes up some
non-terminated events in $h$. For us, the completion is not arbitrary,
but is generated by the specific visibility relation that we work
with, as captured by the set $\closedEvent$. It is thus easy to see
that our definition implies the standard one. Our definition is
justified, because in practice of proving linearizability, the choice
of the completion is always guided by an implicit notion of dependence
between events; we merely make this notion explicit by means of
$\genVisSymbol$.

\subsection{Constructing the linearization order}
\label{subsec::linearizability-from-axioms}

Given arbitrary $\visObsSymbol {\pointerIndx p}$, $\visSepSymbol{\pointerIndx p}$, we denote by 
$\MCASFam(\visObsSymbol {\pointerIndx p}, \visSepSymbol{\pointerIndx p})$ the specification for MCAS 
in Figure~\ref{fig::concurrent-spec-mcas}.
We first give an overview of the structure of the proof. 

The proof is divided into two steps.
The first step constructs a linear order $\leq$ containing both
$\genVisSymbol$ and $\precedesAbsSymbol$ (the first two requirements in Definition~\ref{sect::defn-linearizability}) and
satisfying:
\begin{align}
\label{eq::sequential-recency}
\visObs {\pointerIndx p} x y \implies \neg \exists z \in \writesAbs {p} \cap \closedEvent.\ x < z < y
\end{align}
i.e., there are no $p$-writes in between a $p$-observation in the
linear order.
The second step consists in proving that $\leq$ is sequentially sound
(the third requirement in Definition~\ref{sect::defn-linearizability})
by using~\eqref{eq::sequential-recency}, which ensures that
the order of any mutation performed by $x$ remains unchanged in $\leq$
up to the point where $y$ reads. We assume throughout that
the visibility axioms of $\MCASFam(\visObsSymbol {\pointerIndx p}, \visSepSymbol{\pointerIndx p})$ 
hold.  

The first step of the proof is independent of the particular definitions for the postcondition predicate 
$\postPredSymbol$ and the successful writes set $\writesAbs{p}$ in $\MCASFam(\visObsSymbol {\pointerIndx p}, \visSepSymbol{\pointerIndx p})$.
Hence, to generalize the first step, we introduce the following definition. Given $\visObsSymbol {p},
\visSepSymbol{p}$, we say that a \emph{specification}, denoted $\genStructName{}(\visObsSymbol{\pointerIndx p},
\visSepSymbol{\pointerIndx p})$, is a pair $(\postPredSymbol,\writesAbs{p})$.\footnote{This definition is extended in Definition~\ref{defn-visibility-specification} of Appendix~\ref{sect::appendix::lin::generic}
	to include allocation events. We elide such events here, as they do not modify the intuition behind the presentation.}
A \emph{valid} specification $\genStructName{}(\visObsSymbol{\pointerIndx p},
\visSepSymbol{\pointerIndx p})$ satisfies the axioms in Figure~\ref{subfig:concurrent-spec-history}. 

We now focus on the first step, i.e., showing the existence of the
linear order, for a valid specification. 
We sketch the construction here, with the full proof in
Appendix~\ref{sect::appendix::lin::generic}.
\begin{lem}
\label{lma::existence-linear-order}
Given relations $\visObsSymbol{\pointerIndx p}$,
$\visSepSymbol{\pointerIndx p}$ and \emph{valid}
specification
$\genStructName{}(\visObsSymbol{\pointerIndx p},
\visSepSymbol{\pointerIndx p})$, there is a linear order
$\leq$ over $\closedEvent$, such that
${({\genVisSymbol} \cup {\precedesAbsSymbol})} \subseteq {\leq}$ and property
\eqref{eq::sequential-recency} holds.
\end{lem}

\begin{sketch}
Define relation ${\triagVisSymbol} \defini {(\precedesAbsSymbol \cup
\genVisSymbol)^{+}}$, i.e., the transitive closure of the union of the general visibility
and returns-before relations. Let $\triagVisEqSymbol$ be its reflexive closure.
We show that $\triagVisEqSymbol$ is a partial order and can be extended into a linear order.

First, axiom \axiomHRef{vis-ax::cc-no-future-dependence} ensures that
$\triagVisEqSymbol$ is a partial order over $\closedEvent$ (Lemma
\ref{lem::appendix::lin::triag-poset} in Appendix
\ref{sect::appendix::lin::generic}).  Intuitively,
\axiomHRef{vis-ax::cc-no-future-dependence} states that events do not look
for dependencies into their real-time future, so that cycles will not
be inserted when $\genVisSymbol$ is combined with
$\precedesAbsSymbol$, ensuring irreflexivity of $\triagVisSymbol$.

Second, given an \emph{arbitrary} partial order extension $\leq$ of
$\triagVisEqSymbol$ (over domain $\closedEvent$), axiom
\axiomHRef{vis-ax::cc-no-in-between} ensures property
\eqref{eq::sequential-recency}.  Indeed, if there is such write $z$, then
axiom \axiomHRef{vis-ax::cc-no-in-between} says that $z$ is separable
before $x$ or after $y$. But $\leq$ contains $\triagVisEqSymbol$
(hence the separable-before relation as well), which means that $z$
must also occur in $\leq$ before $x$ or after $y$ (Contradiction).

Third, since any partial order can be extended to a linear
order,\footnote{See topological sorting
  algorithms~\cite{topoSorting}.} choose some linear order $\leq$
extending $\triagVisEqSymbol$. Notice that $\leq$ will satisfy
\eqref{eq::sequential-recency} because it is a partial order extension
of $\triagVisEqSymbol$.  Also, $\leq$ contains 
${\genVisSymbol}$ and ${\precedesAbsSymbol}$, since $\leq$ contains 
$\triagVisEqSymbol$ by construction.
\end{sketch}

Now that we have a linear order $\leq$ that respects both
$\genVisSymbol$ and $\precedesAbsSymbol$, 
for the second step in the linearizability proof
we show that $\leq$ is sequentially sound.

We need to show that executing the events in the order of $\leq$
produces outputs that match the outputs registered in each event.  To
define the notion of execution of events in MCAS, we introduce a
state-based sequential operational semantics.  States for MCAS are
memory heaps. Let us denote by
$H \xrightarrow{proc(in)\ \langle out \rangle} H'$ the statement that
procedure $proc$ with input $in$ executes atomically on heap $H$,
produces output $out$ and modifies the heap into $H'$. We can then
define the effect of the $\mcasAlg$ procedure on a heap by the
following operational semantics (we will focus only on $\mcasAlg$, for
the rest of procedures in MCAS, see
Definition~\ref{defn::appendix::lin::mcas-legality} in
Appendix~\ref{sect::appendix::lin::MCAS}),

\begin{itemize}
\item $H \xrightarrow{\mcasAlg(\listvar u)\ 
\langle false \rangle} H$, if for some $j \in \listvar u$,
$H(\pointGenEntry j) \neq \expGenEntry j$.
\item $H \xrightarrow{\mcasAlg(\listvar u)\ \langle true \rangle} 
\mapExtThree{H}{\mapEntry
{\pointGenEntry j} {\newGenEntry j}}{j \in \listvar u}$, if for every $j \in \listvar u$, $H(\pointGenEntry j) = \expGenEntry j$.
\end{itemize}

The first case corresponds to a failing execution in which some input
pointer does not have the expected value, while the second case
corresponds to a successful execution in which all input pointers have
their expected values and the pointers mutate into their new
values.

Executing $\leq$ then translates to proving the existence of a path in
the operational semantics such that the $i$-th event $x_i$ in $\leq$
matches the $i$-th procedure in the path, and if
$x_i \in \terminatedEvent$, then $\outputProp {x_i}$ matches the $i$-th
output in the path. Notice that the existence of the path implicitly
assigns an output to those events in
$\closedEvent \setminus \terminatedEvent$.
 
The hypothesis of validity of the specification $\MCASFam$
for MCAS, will ensure the existence of such a path, as the following
lemma shows (the full proof is in 
%
Appendix~\ref{sect::appendix::lin::MCAS}).

\begin{lem}
  \label{lma::linear-order-sequential-soundness}
  Given relations $\visObsSymbol{\pointerIndx p}$,
  $\visSepSymbol{\pointerIndx p}$ such that
  $\MCASFam(\visObsSymbol{\pointerIndx p},
  \visSepSymbol{\pointerIndx p})$ is valid, the linear
  order $\leq$ of Lemma~\ref{lma::existence-linear-order} is
  sequentially sound.
\end{lem}

\begin{sketch}
  We prove by induction on $n$ the following claim, so that sequential
  soundness follows by applying the claim with
  $n = \lvert \closedEvent \rvert$.
\begin{claim}
  For any
  $1 \natorderEqSymbol n \natorderEqSymbol \lvert \closedEvent
  \rvert$, there is a path of length $n$ starting from the empty heap
  that matches $\leq$ from the $1$-th event to the $n$-th event.
\end{claim}
We focus on the inductive case, as the base case $n=1$ must have an
alloc as first event in $\leq$ 
(see Lemma \ref{lem::appendix::lin::mcas-legality-lemma} in 
Appendix \ref{sect::appendix::lin::MCAS}
for details).  The inductive hypothesis provides a path
of length $n$ that matches $\leq$ up to the $n$-th event,
\begin{align}
\label{eqn::lin-proof-path-example}
H_0 = \emptyset \xrightarrow{proc_1(in_1)\ \left< out_1 \right>} H_1 \xrightarrow{proc_2(in_2)\ \left< out_2 \right>} H_2 \ldots 
\xrightarrow{proc_n(in_n)\ \left< out_n \right>} H_n
\end{align}

We need to show that we can extend the path with a $(n+1)-st$ step
that matches the $(n+1)-st$ event in $\leq$. We case analyze the
$(n+1)-st$ event.  Let us focus on the $\mcasAlg(\listvar u)$ case (see Lemma
\ref{lem::appendix::lin::mcas-legality-lemma} in 
Appendix \ref{sect::appendix::lin::MCAS} for the other cases).

From Axiom \axiomHRef{vis-ax::cc-return-completion} and the definition of 
$\MCASFam(\visObsSymbol{\pointerIndx p},
  \visSepSymbol{\pointerIndx p})$, we have that event
$\mcasAlg(\listvar u)$ must satisfy for some
$v \in \BoolType$,
\begin{align*}
\begin{cases}
\forall i \in \listvar u.\ \exists z.\ \visObs {\pointerIndx {\pointGenEntry i}} z {\mcasAlg(\listvar u)} \wedge \inputVal z {\pointGenEntry i} {\expGenEntry i} & \text{ if }v=true\\
\exists i \in \listvar u,v'\in\ValType.\ \exists z.\ \visObs {\pointerIndx {\pointGenEntry i}} z {\mcasAlg(\listvar u)} \wedge \inputVal z {\pointGenEntry i} {v'} \wedge v' \neq {\expGenEntry i} & \text{ if }v=false
\end{cases}
\end{align*}
Focusing on the $v = true$ case, event $\mcasAlg(\listvar u)$ must be observing
at every input pointer $\pointGenEntry i$ an event $z$ that wrote the
expected value $\expGenEntry i$ into $\pointGenEntry i$. Notice that
$z$ must be a write thanks to Axiom
\axiomHRef{vis-ax::cc-observed-are-writes}.  Since path
\eqref{eqn::lin-proof-path-example} matches $\leq$ and $z$ is an event
occurring before $\mcasAlg(\listvar u)$ (i.e.,
$\visObs {\pointerIndx {\pointGenEntry i}} z {\mcasAlg(\listvar u)}$
implies $z < \mcasAlg(\listvar u)$), $z$ must be one of the $proc_k$'s in the
path, which means that $H_k(\pointGenEntry i) = \expGenEntry i$.  In
addition, property \eqref{eq::sequential-recency} ensures that there
is no other write to pointer $\pointGenEntry i$ between $z$ and
$\mcasAlg(\listvar u)$, which means that there is no $\pointGenEntry i$-write
in path \eqref{eqn::lin-proof-path-example} from $proc_k$ up to
$proc_n$, having as consequence that
$H_n(\pointGenEntry i) = \expGenEntry i$ (see Lemma
\ref{lem::appendix::lin::mcas-writers-change-heap} in Appendix
\ref{sect::appendix::lin::MCAS}).  Since this happens with every input
pointer, heap $H_n$ contains the expected values in the input
pointers, so that we can augment path
\eqref{eqn::lin-proof-path-example} with the step
$H_n \xrightarrow{\mcasAlg(\listvar u)\ \langle true \rangle}
\mapExtThree{H_n}{\mapEntry {\pointGenEntry i} {\newGenEntry i}}{i \in
  \listvar u}$.  In case $\mcasAlg(\listvar u) \in \terminatedEvent$, Axiom
\axiomHRef{vis-ax::cc-return-completion} also ensures that
$true = v = \outputProp {\mcasAlg(\listvar u)}$, so that the new step matches
$\leq$.  The case $v = false$ is similar.
\end{sketch}

Together, Lemmas~\ref{lma::existence-linear-order}
and~\ref{lma::linear-order-sequential-soundness} imply the following
theorem.

\begin{thm}
\label{thm::mcas-linearizability-from-vis-structure}
Given an implementation of MCAS, suppose that for any set of
events $\absEvent$ generated from an arbitrary execution history in
the implementation, there are relations
$\visObsSymbol{\pointerIndx p}$,
$\visSepSymbol{\pointerIndx p}$ such that
$\MCASFam(\visObsSymbol {\pointerIndx p},
{\visSepSymbol{\pointerIndx p}})$ is valid. Then, the
implementation is linearizable.
\end{thm}

\begin{prf}
  Let $\absEvent$ be a set of events generated from an arbitrary
  execution history in the implementation. 
  From the hypothesis, $\MCASFam(\visObsSymbol {\pointerIndx p},
  {\visSepSymbol{\pointerIndx p}})$ is valid 
  for some $\visObsSymbol {\pointerIndx p}$ and
  $\visSepSymbol{\pointerIndx p}$.
  Let $\leq$ be the linear order of Lemma
  \ref{lma::existence-linear-order}.  Then, we take $\genVisSymbol$ and $\leq$ to be the
  relations required by Definition~\ref{sect::defn-linearizability}.  
  By construction, $\leq$ respects both $\genVisSymbol$ and
  $\precedesAbsSymbol$. Also,
  $\leq$ is sequentially sound by Lemma
  \ref{lma::linear-order-sequential-soundness}.
\end{prf}

\section{Axiomatizing Spans and Proving Visibility}
\label{sub::sect::lifespan-axiomatization}

This section fills the remaining gap in our proofs of linearizability
and concludes the technical development. Again, we illustrate with the
helping MCAS implementation, relegating RDCSS and opportunistic MCAS
to Appendices~\ref{appendix::sub::sect::impl::RDCSS}
and~\ref{sect::appendix::opportunistic-impl-proof}, respectively.

To prove that the implementation in Figure~\ref{alg-MCAS} is
linearizable,
Theorem~\ref{thm::mcas-linearizability-from-vis-structure} requires
exhibiting two relations $\visObsSymbol{\pointerIndx p}$,
$\visSepSymbol{\pointerIndx p}$ that make the MCAS specification
$\genStructName{\MCAS}(\visObsSymbol{\pointerIndx p},
\visSepSymbol{\pointerIndx p})$ valid.
We will show that the separable-before
\eqref{eq::overview::helpers-separability} and observation
\eqref{eq::overview::helpers-observation} relations from Section
\ref{subsubsect::visibility-mcas} suffice. In this section,
$\visSepSymbol{\pointerIndx p}$ and $\visObsSymbol{\pointerIndx p}$
refer to \eqref{eq::overview::helpers-separability} and
\eqref{eq::overview::helpers-observation}, respectively.
As apparent, $\visSepSymbol{\pointerIndx p}$ and
$\visObsSymbol{\pointerIndx p}$ are defined in terms of spans. While
Section~\ref{subsubsect::visibility-mcas} describes spans informally,
here we make the definitions formal and abstract, via a notion of
\emph{span structure}. 

We shall postulate a number of axioms that a span structure must
satisfy.  Our proof obligations for validity of
$\genStructName{\MCAS}(\visObsSymbol{\pointerIndx p},
\visSepSymbol{\pointerIndx p})$ then split into two parts. First, we
show that if the span structure we define for MCAS satisfies the
axioms, then
$\genStructName{\MCAS}(\visObsSymbol{\pointerIndx p},
\visSepSymbol{\pointerIndx p})$ is valid, and thus, by
Theorem~\ref{thm::mcas-linearizability-from-vis-structure}, MCAS is
linearizable. Second, we show that the span structure for MCAS
satisfies the axioms. We remark that the first part of the proof is
actually more general, as it also applies to RDCSS; indeed, it only
uses the span axioms and definitions
\eqref{eq::overview::helpers-separability} and
\eqref{eq::overview::helpers-observation}, which are shared with RDCSS
(see Appendix~\ref{appendix::sect::impl::proof-of-validity}).
 
\subsection{Intuition} 
\label{subsect::motivation-spans}

\begin{figure}[t]
	\begin{subfigwrap}{Successful execution.}
		{subfig:successful-execution}[0.47\textwidth]
		\centering
		\includegraphics[scale=0.6]{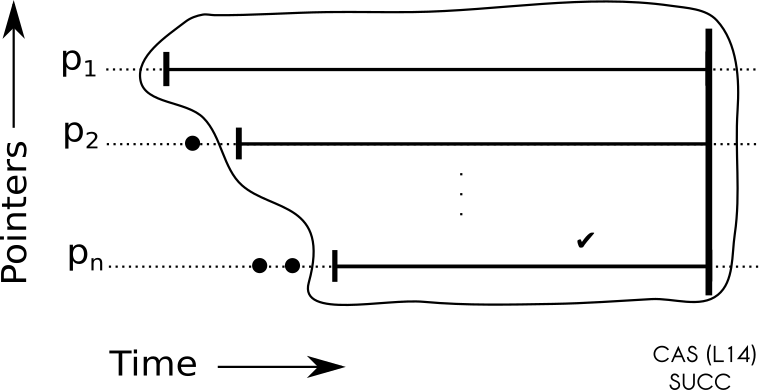}
	\end{subfigwrap}\qquad
	\begin{subfigwrap}{Failing execution.}
		{subfig:failing-execution}[0.47\textwidth]
		\centering
		\includegraphics[scale=0.6]{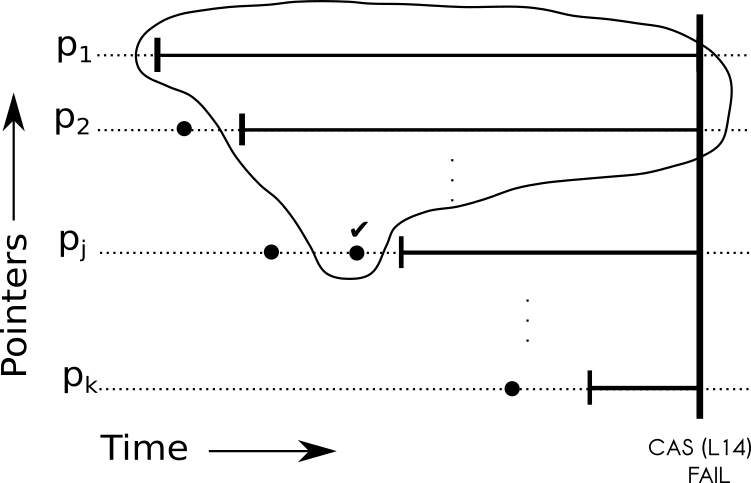}
	\end{subfigwrap}
	\caption{Spans of $\mcasAlg(\listvar u)$. Full spans represent
          successful help attempts, and are shown as line
          segments. Collapsed spans represent failed help attempts,
          and are represented as dots. Encircled spans are the ones
          chosen to formally denote the execution of the
          $\mcasAlg(\listvar u)$ event. All spans write into the same
          descriptor. All full spans share the same ending rep
          event---the successful CAS at line~\ref{resolve-status-MCAS}
          that changes the descriptor status---represented as a
          vertical line on the right. The span with $\checkmark$
          indicates that the thread that started the span,
          successfully executed the CAS at
          line~\ref{resolve-status-MCAS}.}
	\label{fig::executions-of-mcas}
\end{figure}

As mentioned in
Section~\ref{subsubsect::span-alternative-definitions-mcas}, we can
use many different definitions of spans, as long as they satisfy the
span invariants (i.e., span axioms, to be presented in
Section~\ref{sect::key-concepts-spans}). In particular, we shall use
two different kinds of spans: full and collapsed spans, illustrated in
Figure~\ref{fig::executions-of-mcas} as line segments and dots,
respectively.
A \emph{full span} starts when the successful $\rdcssAlg$ at
line~\ref{invoke-rdcss-in-MCAS} in Figure~\ref{alg-MCAS} writes the
descriptor, and finishes when the descriptor status is updated at
line~\ref{resolve-status-MCAS}, as explained in
Section~\ref{subsubsect::span-alternative-definitions-mcas}. A
\emph{collapsed span} is any failing $\rdcssAlg$ at
line~\ref{invoke-rdcss-in-MCAS} that returns the non-expected $\ValType$ value; thus,
a collapsed span corresponds to a failed help attempt.
Figure~\ref{fig::executions-of-mcas} shows how full and collapsed
spans interact in two kinds of executions of $\mcasAlg(\listvar u)$.

{\textbf {Successful execution}} of $\mcasAlg(\listvar u)$
(Figure~\ref{subfig:successful-execution}) occurs when a descriptor
$d$ is created at line~\ref{alloc-desc-MCAS} containing the entries in
$\listvar u$, each full span writes the descriptor $d$ into each entry
pointer of $\listvar u$, and the thread that executed the last full
span sets the descriptor status to $\SUCCEEDED$ (indicated in the
figure with a $\checkmark$).
	
While the actual successful execution of $\mcasAlg(\listvar u)$ may
witness both kinds of spans (as evident in
Figure~\ref{subfig:successful-execution}), it is only the full spans
writing $d$ that explain the successful outcome.  Indeed, a collapsed
span may arise if the considered pointer did not contain the expected
value when help was attempted. However, such a failed help attempt
must be followed by a full span at the same pointer, or else the
execution of $\mcasAlg(\listvar u)$ would not be successful. Such a
full span overrides any prior collapsed span at the same pointer, and,
along with other full spans, ultimately explains the overall outcome
of the $\mcasAlg(\listvar u)$.

This is why we use the set of full spans as the semantic abstraction
(i.e., \emph{denotation}) of the successful execution of
$\mcasAlg(\listvar u)$, and represent this in
Figure~\ref{subfig:successful-execution} by encircling the full spans
only.
It is this denotation that our span axioms apply
to. For example, we introduced the bunching invariant in
Section~\ref{subsubsect::visibility-definitions-mcas} to hold
of a number of spans if they share a common subinterval. As evident
from Figure~\ref{subfig:successful-execution}, bunching holds of the
encircled spans. However, it would fail if, for example, the collapsed
span (dot) on $p_2$ were to be encircled as well, since this dot does
not overlap with the segment on $p_n$.

{\textbf {Failing execution}} of $\mcasAlg(\listvar u)$
(Figure~\ref{subfig:failing-execution}) occurs when a descriptor $d$
is created at line~\ref{alloc-desc-MCAS} containing the entries
$\listvar u$, and:
\begin{itemize}
\item there is an entry $j \in \listvar u$ such that some
collapsed span failed to write descriptor $d$ into the $j$-th pointer,
\item there are full spans that write descriptor $d$ into each pointer of
entries $i < j$, and 
\item the thread that executed the collapsed span for
the $j$-th entry managed to set the descriptor status to $\FAILED$
using the CAS in line~\ref{resolve-status-MCAS} (the collapsed span
for the $j$-th entry is indicated in the figure with a $\checkmark$).
\end{itemize}
As encircled in Figure~\ref{subfig:failing-execution}, the full spans
writing $d$ for entries $i < j$, along with the $\checkmark$-ed
collapsed span, explain the failed outcome of $\mcasAlg(\listvar u)$,
and thus we use this set of spans as the denotation for the failed
$\mcasAlg(\listvar u)$ event.

The collapsed spans for entries $i < j$, the full spans following the
$\checkmark$-ed collapsed span, and the full and collapsed spans for
entries $i > j$, do not influence the $\FAILED$ outcome of
$\mcasAlg(\listvar u)$, because the $\checkmark$-ed collapsed span is
the one responsible for setting the status to $\FAILED$. In other
words, the thread executing the $\checkmark$-ed span ultimately
confirms the failure by setting the descriptor status to $\FAILED$ at
the CAS.

As in the successful case, we apply our span axioms to
the denotation of a failing $\mcasAlg(\listvar u)$. For example,
while bunching holds for the encircled spans in
Figure~\ref{subfig:failing-execution}, it would fail if we
additionally encircled the full span on $p_j$.

\subsection{Axiomatizing Spans}
\label{sect::key-concepts-spans}

We now formalize spans via the notion of span
structure, so that invariants (now called span axioms), like bunching
and disjointness, are stated independently of the span definitions
(e.g., as in
Section~\ref{subsubsect::spans-and-visibility-axioms}, where the
argument for \axiomHRef{vis-ax::cc-no-in-between} depended solely on
the disjointness invariant, and not on the span definition). We first
need a helper notion of rep events.

\subparagraph*{Rep events.}
\emph{Rep events} are generated by the invocation of a code line
inside a procedure.  For example, invoking line
\ref{resolve-status-MCAS} in Figure \ref{alg-MCAS} produces a rep
event. 
We do not assume that rep events are
atomic, because events in $\absEvent$ may become rep events from the point
of view of another implementation; for example, the exportable procedures in
the RDCSS implementation are used as primitive procedures in MCAS.
If $x$ is a rep event, $\STimeProp x$, $\ETimeProp x$, and $\outputProp x$
denotes its start time, ending time, and output, respectively. 
Since rep events are not atomic, we assume that we are given a
linearization $\linRepsEqSymbol$ on rep events, and use $\repsCompl$
to denote the domain of this relation. The set of all rep events in an
execution history is denoted as $\repEvent$.  The distinction between
events ($\absEvent$) and rep events is standard in the theory of
linearizability~\cite{herlihy:90}.

\subparagraph*{Span Structure.} 
A \emph{span structure}
$(\spans p, \writesSpans p,
\runFuncSymbol)$ consists
of:
\begin{itemize}
\item For every pointer $p$, a set $\spans p \subseteq \repEvent \times \repEvent$, called the \emph{spans} accessing $p$.
\item For every pointer $p$, a set $\writesSpans p \subseteq \spans p$, called the \emph{successful write spans} into $p$.
\item A function $\runFuncSymbol: \absEvent \rightarrow (\mathcal{P}(\bigcup_p \spans p) \times (\ValType \cup \{ \bot \}))$,
called \emph{event denotation}, written $\runFunc{x}$ for event $x$. Here, $\mathcal{P}(\bigcup_p \spans p)$ denotes the power set of $\bigcup_p \spans p$.
\end{itemize}

Each span in $\spans p$ is a pair of rep events, encoding the end
points of the span. We denote as $\firstRepFuncSymbol$ (``first'')
and $\lastRepFuncSymbol$ (``last'') the standard pair projection
functions. A collapsed span has coinciding end-points $(a,a)$, and is
abbreviated as $(a)$.

Set $\writesSpans p$ collects spans that mutate state, for example,
those full spans in Figure~\ref{subfig:successful-execution} that set
their descriptor status to $\SUCCEEDED$.

The denotation $\runFunc{x}$ for event $x \in \absEvent$
returns a pair, whose first component is the set of spans belonging to 
$x$, and whose second component is the output associated
with the denotation ($\bot$ if no output is defined).  Intuitively,
the denotation models the closed path encircling the spans in
Figure~\ref{fig::executions-of-mcas}, together with its
associated output.

We define $\hspans{}{x} \defini \projFirst {\runFunc{x}}$ to
be the set of spans in the denotation of $x$;
$\hspans{p}{x} \defini \hspans{}{x} \cap \spans p$ to be the set of
$p$-accessing spans in the denotation of $x$.\footnote{This is the
  notation we already used in Section
  \ref{subsubsect::visibility-mcas} and in definitions
  \eqref{eq::overview::helpers-separability} and
  \eqref{eq::overview::helpers-observation}.}  Also, we define
$\outputRunFunc x \defini \projSecond {\runFunc{x}}$
to be the denotation output of event $x$.\footnote{It is essential to not
  confuse $\outputRunFunc x$ with the event's output
  $\outputProp x$. An event $x$ may have $\outputProp x$ undefined
  because $x$ has not terminated, but still have
  $\outputRunFunc x$ defined because $x$ executed the
  corresponding spans and is ready to terminate. For terminated
  events, the two values coincide.}

\subparagraph*{Span returns-before relation.}
Relation
$\precedesSpans b c \defini \linReps {\lastRep b} {\firstRep c}$
states that span $b$ terminates before $c$ starts in the linearization
of rep events.  $\precedesSpans b c$ can be interpreted as ``$b$
resolves its descriptor before $c$ writes its own''.

\subparagraph*{Implementation by a span structure.}

Given arbitrary relations $\visObsSymbol {\pointerIndx p}$,
$\visSepSymbol{\pointerIndx p}$, we say that
$\genStructName{}(\visObsSymbol {\pointerIndx p},
\visSepSymbol{\pointerIndx p})$ (which is a pair
$(\writesAbs p, \postPredSymbol)$) is \emph{implemented by span
  structure} $(\spans p, \writesSpans p, \runFuncSymbol)$ if all the
span axioms in Figure~\ref{fig::span-descriptor-lifespans-axioms} are
satisfied.  We also say that
$\genStructName{}(\visObsSymbol {\pointerIndx p},
\visSepSymbol{\pointerIndx p})$ is \emph{implemented by spans} if
there exists some span structure that implements it.
The idea is that the span axioms imply the validity of
$\genStructName{\MCAS}(\visObsSymbol{\pointerIndx p},
\visSepSymbol{\pointerIndx p})$, as per the following theorem, which
thus, together with
Theorem~\ref{thm::mcas-linearizability-from-vis-structure},
reduces linearizability to proving the span axioms.

\begin{thm}
\label{thm::span-axioms-imply-visibility}
If
$\genStructName{\MCAS}(\visObsSymbol {\pointerIndx p},
\visSepSymbol{\pointerIndx p})$ is implemented by spans, then
$\genStructName{\MCAS}(\visObsSymbol {\pointerIndx p},
\visSepSymbol{\pointerIndx p})$ is valid.\footnote{This is
  theorem~\ref{thm::appendix::valid::span-axioms-imply-visibility}
  stated in a more general form in
  Appendix~\ref{appendix::sect::impl::proof-of-validity}, here
  instantiated with $\mu = \MCAS$.}
\end{thm}

\begin{figure}[t]
\begin{subfigwrap}{Defined notions for span axioms.}{fig::sub::span-defined-spans}
\centering
\begin{tabular}{l}
Span returns-before relation \\
\quad $\precedesSpans b c \defini \linReps {\lastRep b} {\firstRep c}$ \\
\end{tabular}
\begin{tabular}{l}
Set of spans in denotation \\
\quad $\hspans{}{x} \defini \projFirst {\runFunc{x}}$ \\
\end{tabular}
\begin{tabular}{l}
Denotation output \\
\quad $\outputRunFunc x \defini \projSecond {\runFunc{x}}$ \\
\end{tabular}
\end{subfigwrap}

\begin{subfigwrap}{Key Axioms}{fig::sub::span-spans-key-axioms}  
\centering
 \begin{tabular}{l}
 \begin{tabular}{ll}
    \axiomDLabel{span::descriptors-do-not-interfere} Disjointness &     \axiomDLabel{span::all-descriptors-are-written-before-any-resolution} Bunching \\
    \qquad $b, c \in \spans p \implies ({\precedesSpans b c} \vee {\precedesSpans c b} \vee {b = c})$ & 
    \qquad $b, c \in \hspans{}{x} \implies \linRepsEq {\firstRep b} {\lastRep c}$ \\
 \end{tabular}\\[1em]
 \begin{tabular}{l}
   \axiomDLabel{span::finished-operations-have-a-run} Adequacy \\
   \qquad $x \in \terminatedEvent \implies \outputRunFunc x = \outputProp x \wedge \hspans{}{x} \neq \emptyset$ \\
 \end{tabular}
 \end{tabular}
\end{subfigwrap}

\begin{subfigwrap}{Structural Axioms}{fig::sub::span-spans-structural-axioms} 
\centering 
 \begin{tabular}{c l}  
  \axiomDLabel{span::descriptor-write-precedes-resolution} & $\linRepsEq {\firstRep b} {\lastRep b}$ \\
        
  \axiomDLabel{span::runs-are-injective} & $\hspans{}{x} \cap \hspans{}{y} \neq \emptyset \implies x = y$ \\
  
  \axiomDLabel{span::writer-blocks-belong-to-runs} & $b \in \writesSpans p \implies \exists x.\ b \in \hspans{}{x}$ \\
  
  \axiomDLabel{span::postcondition-predicate-holds} & $\hspans{}{x} \neq \emptyset \implies 
  \outputRunFunc x \neq \bot \wedge \postPred x {\outputRunFunc x}$ \\
\end{tabular}
\begin{tabular}{c l}
  \axiomDLabel{span::writers-have-writer-blocks} & $\hspans{}{x} \neq \emptyset \implies (x \in \writesAbs p \Leftrightarrow \hspans{}{x} \cap \writesSpans p \neq \emptyset)$ \\
  
    \axiomDLabel{span::blocks-contained-in-abstract-time-interval} & (i) $b \in \hspans{}{x} \implies \STimeProp x \natorderEqSymbol \STimeProp {\firstRep b}$ \\
    
                                                                     & (ii) $(b \in \hspans{}{x} \wedge x \in \terminatedEvent) \implies {}$ \\ 
                                                                     & \qquad $\exists i \in \repsCompl.\ \ETimeProp i \natorderEqSymbol \ETimeProp x \wedge \linRepsEq {\lastRep b} i$ \\
\end{tabular}
\end{subfigwrap}
  \caption{Span axioms for span structure $(\spans p, \writesSpans p, \runFuncSymbol)$ implementing $\genStructName{}(\visObsSymbol {\pointerIndx p}, \visSepSymbol{\pointerIndx p}) = (\writesAbs p, \postPredSymbol)$. Variables $x$, $y$ range over $\absEvent$. Variables $b$, $c$ over $\bigcup_p \spans p$. Variable $p$ over $\PtsType$. The list omits three axioms involving allocs, which are shown in Figure~\ref{fig:descriptor-lifespans-axioms} of Appendix~\ref{appendix::sect::impl::proof-of-validity}.}
  \label{fig::span-descriptor-lifespans-axioms}
\end{figure}

We now explain the span axioms. The first two axioms formalize the
disjointness and bunching invariants from
Section~\ref{subsubsect::visibility-mcas}. The remaining axioms relate
the code to the span structure and formalize book-keeping invariants
required for the proofs.

\subparagraph*{Key Axioms.}
\emph{Disjointness Axiom 
	\axiomDRef{span::descriptors-do-not-interfere}} states that
spans accessing the same pointer $p$ do not overlap, i.e., 
any descriptor in $p$ must be resolved before another one is written
in $p$.
\emph{Bunching Axiom \axiomDRef{span::all-descriptors-are-written-before-any-resolution}} 
states that any two spans executed by the same event must
overlap. In other words, no descriptor can be resolved unless all
descriptors pertinent to the event have been written first (i.e.,
$\firstRep b \linRepsEqSymbol \lastRep c$, pairwise, for all spans
$b$, $c$ in the event's denotation).
\emph{Adequacy Axiom \axiomDRef{span::finished-operations-have-a-run}}
states that for terminated events, the denotation output coincides with
the event's output, and the event contains at least one span.  This
axiom is proved by showing that executing the actual code produces the
denotation. For example, proving adequacy for an $\mcasAlg$ event amounts 
to showing that the code does produce the spans encircled in Figures 
\ref{subfig:successful-execution} and \ref{subfig:failing-execution},
as shown in Theorem~\ref{thm::span-axioms-satisfied-general} of
Section~\ref{sect::span-structure-mcas}.
  
\subparagraph*{Structural Axioms.}
\emph{Axiom
  \axiomDRef{span::descriptor-write-precedes-resolution}}
justifies calling functions $\firstRepFuncSymbol$,
$\lastRepFuncSymbol$ the first and last rep events in a span.  The
axiom states that writing a descriptor precedes its resolution.
\emph{Axiom \axiomDRef{span::runs-are-injective}} 
states that denotations of different events do not share spans.
\emph{Axiom \axiomDRef{span::writer-blocks-belong-to-runs}} states
that successful writer spans must belong to the denotation of some event
(i.e. mutations do not occur ``out of thin air'').  Although it is
possible to strengthen the axiom to \emph{all} spans, this
is not required for correctness since non-writer spans can 
spuriously occur in the execution history as they do not have an effect 
that could be discovered by other spans.
\emph{Axiom \axiomDRef{span::postcondition-predicate-holds}} considers
an event $x$, which has completed its execution, though it need not
have terminated yet. We identify the completion of the execution with
the set of spans in $x$'s denotation being nonempty
(i.e.~$\hspans{}{x} \neq \emptyset$). The axiom then says that $x$
produced a denotation output and $x$'s postcondition predicate must hold at
$x$'s denotation output, as if $x$ has terminated.
\emph{Axiom \axiomDRef{span::writers-have-writer-blocks}} states that
successful $p$-write events must contain a successful $p$-writer span in their denotation.
\emph{Axiom
  \axiomDRef{span::blocks-contained-in-abstract-time-interval}}
describes the relation between an event's duration and the spans in
its denotation.  The first part of Axiom
\axiomDRef{span::blocks-contained-in-abstract-time-interval} says that
spans in a denotation start (in real time) after the event starts.  The
second part of Axiom
\axiomDRef{span::blocks-contained-in-abstract-time-interval} says that
spans in a denotation finish (according to the rep linearization) before
some rep event $i$ that witnesses the event's end in real-time. For
the second part, we cannot require the stronger property
$\ETimeProp {\lastRep b} \natorderEqSymbol \ETimeProp x$ because spans
may be ended by threads different from the one invoking $x$, which
means that $\lastRep b$ could still be executing by the time $x$
finishes.\footnote{Recall that rep events are not required to be atomic.}

\subsection{Proving the Span Axioms for MCAS}
\label{sect::span-structure-mcas}

By Theorem~\ref{thm::span-axioms-imply-visibility}, the problem of
proving linearizability of MCAS reduces to defining a span
structure $\MCASSpanStruct$ that implements the MCAS specification
$\genStructName{\MCAS}(\visObsSymbol {\pointerIndx p},
\visSepSymbol{\pointerIndx p})$.
Here we define $\MCASSpanStruct$ by using the code for $\mcasAlg$ in
Figure~\ref{alg-MCAS}. Appendix~\ref{appendix::sub::sect::impl::MCAS}
gives the full definition of $\MCASSpanStruct$ using the full code in
Figure~\ref{appendix::alg-MCAS}, which contains procedures elided from
Figure~\ref{alg-MCAS}.  Then, we show that
$\genStructName{\MCAS}(\visObsSymbol {\pointerIndx p},
\visSepSymbol{\pointerIndx p})$ is implemented by span structure
$\MCASSpanStruct$, i.e, that $\mcasAlg$ satisfies the span axioms.

Intuitively, structure $\MCASSpanStruct$ encodes the ideas presented in 
Section~\ref{subsect::motivation-spans}. 
Set $\spans p$ contains all full and collapsed spans 
accessing pointer $p$, defined as follows,

\begin{itemize}
	\item A \emph{full} span at pointer $p$ is a tuple $(a,b)$ of rep events
	$a$, $b$, where $a$ is the successful $\rdcssAlg$ at
	line~\ref{invoke-rdcss-in-MCAS} writing some descriptor $d$ in $p$,
	and $b$ is the successful CAS at line~\ref{resolve-status-MCAS}
	updating $d$'s status to either $\SUCCEEDED$ or $\FAILED$.
	
	\item A \emph{collapsed} span at pointer $p$ is $(c)$, where rep event $c$
	is an $\rdcssAlg$ at line~\ref{invoke-rdcss-in-MCAS} that fails to
	write a descriptor into $p$ (i.e., $c$ returns a $\ValType$ value
	different from the expected value).
\end{itemize}

Set $\writesSpans p$ contains all spans $(a,b) \in \spans p$ where 
$b$ updates the status to
$\SUCCEEDED$. The denotation $\runFunc x$ for $x = \mcasAlg(\listvar u)$ is defined
by cases, encoding the two kinds of 
executions discussed in Section~\ref{subsect::motivation-spans},	
	
\begin{itemize}
	\item (Successful execution) 
	$
	\runFunc x \defini (\{ (a_i,b) \in \spans{\pointGenEntry i} \mid i \in \listvar u \}, \textit{true})
	$, 
	where each $a_i$ wrote in $\pointGenEntry i$ the same descriptor $d$ (created at line~\ref{alloc-desc-MCAS}) containing the entries $\listvar u$, 
	and $b$ is the unique CAS at line~\ref{resolve-status-MCAS} that successfully updated $d$'s status to $\SUCCEEDED$.
	
	\item (Failing execution)
	$
	\runFunc x \defini (\{ (a_i,b) \in \spans{\pointGenEntry i} \mid i < j \} \cup \{ (c) \}, \textit{false})
	$, 
	where each $a_i$ wrote in $\pointGenEntry i$ the same descriptor $d$ (created at line~\ref{alloc-desc-MCAS}) containing the entries $\listvar u$;
	$b$ is the unique CAS at line~\ref{resolve-status-MCAS} that successfully updated $d$'s status to $\FAILED$; $b$ and $c$
	are executed by some thread $T$ such that 
	$c$ is the failed $\rdcssAlg$ at line~\ref{invoke-rdcss-in-MCAS} that attempted to write $d$ and is the most recent such $\rdcssAlg$ preceding $b$ in the control
	flow of $T$; and $j$ is the entry in which $c$ failed to write $d$ (i.e., $c$ failed to write in pointer $\pointGenEntry j$).
		
	\item $\runFunc x \defini (\emptyset, \bot)$, if the history does not satisfy the previous two cases.
\end{itemize}

In both non-empty cases of the definition of $\runFunc x$, full spans
are uniquely determined, because the descriptor $d$ is unique to the
$\mcasAlg$ invocation. Moreover, once $b$ changes $d$'s status, $d$ cannot
be written nor its status be changed, since both the $\rdcssAlg$ and CAS at
lines~\ref{invoke-rdcss-in-MCAS}, \ref{resolve-status-MCAS} expect the $\UNDECIDED$
status. Hence, for each $i$, $\spans{\pointGenEntry i}$ contains at
most one full span writing $d$.  Additionally, in the failing
execution, $(c)$ is uniquely determined because it is executed by the same
thread $T$ that executed $b$. More specifically, when $T$ executes $c$ at line~\ref{invoke-rdcss-in-MCAS} 
(descriptor write attempt), immediately after $c$ fails, $T$ exits the $\writeAllDescsAlg$ procedure at 
line~\ref{write-all-descs-MCAS} and executes $b$ (descriptor status
change), meaning that $c$ is the instruction immediately preceding $b$ in the control
flow of $T$.

The empty case of the definition of $\runFunc x$ is when the denotation
is undefined. It is the role of the adequacy
axiom \axiomDRef{span::finished-operations-have-a-run} to ensure that
this case does not arise for terminated events, and therefore that the denotation
correctly captures the spans that are executed by the code of a terminated
$\mcasAlg$.

With this definition for $\MCASSpanStruct$, we have the following
theorem, which together with theorems
\ref{thm::mcas-linearizability-from-vis-structure} and
\ref{thm::span-axioms-imply-visibility} implies that $\mcasAlg$ is
linearizable.

\begin{thm}
\label{thm::span-axioms-satisfied-general}
$\genStructName{\MCAS}(\visObsSymbol {\pointerIndx p}, \visSepSymbol{\pointerIndx p})$ 
is implemented by
span structure $\genSpanStructName{\MCAS}$.
\end{thm}

\begin{sketch}
  We consider only the case of the $\mcasAlg$ procedure in
  Figure~\ref{alg-MCAS}, focusing on the proof of the adequacy axiom,
  in the case when $\mcasAlg$ returns $\mathit{true}$. The proof of the $\mathit{false}$ case and the other span axioms 
  are found in their respective lemmas in
  Appendix~\ref{appendix::sub::sect::impl::MCAS}, which uses the full
  code in Figure~\ref{appendix::alg-MCAS}.	
	
Since $\mcasAlg(\listvar u)$ returned $true$, line~\ref{read-phase2-status-MCAS} in Figure~\ref{alg-MCAS} must have read the status pointer of descriptor $d$ 
to be $\SUCCEEDED$ (where $d$ was created at line~\ref{alloc-desc-MCAS}).
This implies that some thread $T'$ (which could be $T$ as well)
must have set $d$'s status pointer to $\SUCCEEDED$ at line~\ref{resolve-status-MCAS}, and $T'$ must
have finished the execution of $\writeAllDescsAlg$ with a $\SUCCEEDED$ at line~\ref{write-all-descs-MCAS}.
Since $\writeAllDescsAlg$ is a recursive method which only recurses when it needs to try again, and we know that $T'$
finished executing $\writeAllDescsAlg$, $T'$ must have reached
an invocation of $\writeAllDescsAlg$ that did not recurse, i.e., an
invocation of $\writeAllDescsAlg$ where 
lines \ref{alg-help-complete-invoke-MCAS-MCAS}-\ref{alg-help-restart-MCAS} were not executed. So, we
can assume we are on such execution.

Since $\writeAllDescsAlg$ returned $\SUCCEEDED$,
thread $T'$ must have tried all the entries inside the loop. 
For each entry $i \in \listvar{u}$, $T'$ must have tried to write $d$ using the 
$\rdcssAlg$ at line~\ref{invoke-rdcss-in-MCAS}. 
The $\rdcssAlg$ must have returned either $d$ or the expected
value $\expGenEntry{i}$ (it cannot return a descriptor different
from $d$, because lines \ref{alg-help-complete-invoke-MCAS-MCAS}-\ref{alg-help-restart-MCAS} would be executed). 
If the returned
value was the expected value $\expGenEntry{i}$, then the $\rdcssAlg$ succeeded (since we
already know that $d$'s status was set to $\SUCCEEDED$ by $T'$ later, and $d$'s
status is $\UNDECIDED$ previous to the change). If the $\rdcssAlg$ at line~\ref{invoke-rdcss-in-MCAS}
returned $d$, then some other thread (or $T'$ in a previous recursive
$\writeAllDescsAlg$ invocation) already wrote the descriptor. In other words,
in both cases, the descriptor was written. Since $d$ was
written for each entry and its status pointer was set to $\SUCCEEDED$
later, this corresponds to the successful case in 
$\runFunc {\mcasAlg(\listvar u)}$, which has as output $true$ and the denotation is not empty.
\end{sketch}

\section{Related and Future Work}
\label{sect::related-work}

\subparagraph*{Proofs of RDCSS and MCAS} A number of papers have
considered linearizability of MCAS and RDCSS using LPs.  For MCAS, the
PhD thesis of Vafeiadis~\citet{vafeiadis} and the original paper of
Harris et al.~\citet{Harris} are the main sources for the LP proof.
For RDCSS, Vafeiadis' PhD thesis proposed the use of \emph{prophecy
  variables}~\cite{lamport} to model that the LPs of RDCSS depend on
future events. The argument was later mechanized in Coq by Liang and
Feng~\citet{LiangFeng} using \emph{speculations} (related to prophecy
variables, but less general), and eventually using prophecy variables
in Iris~\cite{IrisFuture}.
Both speculations and prophecies lead to operational proofs, as they
essentially codify the LPs of the methods. Our proofs, being
declarative, do not rely on speculations or prophecies, but use spans
and visibility as the common abstractions that unify RDCSS and
MCAS. That said, we have not mechanized our proofs yet.

\subparagraph*{Visibility relations in other contexts} Our
approach uses visibility relations to model the ordering dependencies
between events, by axiomatizing the notions of observation and
separation.  A general survey of the use of visibility relations in
concurrency and distributed systems is given by Viotti and
Vukoli\'c~\citet{Viotti}.  Visibility relations and declarative proofs
have also been utilized to specify consistency criteria weaker than
linearizability (Emmi and Enea~\cite{EmmiEnea}), to introduce a
specification framework for weak memory models (Raad et
al.~\cite{RaadAzalea}), and to specify the RC11 memory model (Lahav et
al.~\cite{LahavRC11}).  In particular, for the RC11
model~\cite{LahavRC11}, our observation relation is reminiscent of
the reads-from relation, and there may be a potential connection
between our separable-before and the (extended) coherence order in
RC11. However, the extended coherence order in RC11 is defined in
terms of reads-from, while in our approach, observation and
separable-before are independent abstract relations linked by the
visibility axioms.

In contrast to the above papers that focus on the semantics of
consistency criteria, our use of visibility relations focuses on
verifying specific algorithms and data structures, and is thus closer
to the following work where visibility relations are applied to
concurrent queues (Henzinger et
al.~\cite{henzinger:concur13,aspects}), concurrent stacks (Dodds et
al.~\cite{Dodds}), and memory snapshot algorithms~(\"Ohman and
Nanevski~\cite{Joakim}).  We differ from these in the addressed
structures, RDCSS and MCAS, which, unlike the related work, further
utilize helping.

\subparagraph*{Alternative RDCSS and MCAS implementations} The
implementations of RDCSS and MCAS that we used here were originally
developed by Harris et al.~\citet{Harris}. However, there are
alternative implementations as well. For example, Arbel-Raviv and
Brown~\citet{Trevor} present a variant of RDCSS where pointers to
descriptors are reused instead of being created at each RDCSS
invocation. Guerraoui et al.~\citet{Guerraoui} present an MCAS
implementation that does not use RDCSS as a subroutine, and that uses
a minimal amount of CAS calls. Guerraoui et al.~also survey a number
of other MCAS implementations. Feldman et
al.~\citet{waitFreeFeldman} present a wait-free implementation of MCAS
(the implementation we used here is only lock-free).

In the future, we will consider how our axioms apply to these
alternative implementations.
For example, our span axioms are agnostic in that they impose no
restriction on how many times a pointer to a descriptor can be used
across spans. Thus, we expect that our axiomatization will be able to
handle implementations with descriptor reuse, such as those of
Arbel-Raviv and Brown~\citet{Trevor}. Similarly, we expect that we
could handle MCAS of Guerraoui et al.~\citet{Guerraoui}, as it too
should satisfy the span axioms. Indeed, this implementation follows the
general outline of MCAS by writing the descriptor into all input
pointers before any descriptor resolution is attempted.

\subparagraph*{Observation and separation as a general methodology}
The pattern suggested by Sections~\ref{subsect::overview::observation-relations-and-seq-specs} and 
\ref{subsect::overview::separable-before-relations}, where 
one starts with a history-based sequential specification and transforms
it into a concurrent specification by replacing the returns-before relation
$\precedesAbsSymbol$ with a separable-before
relation $\visSepSymbol {\pointerIndx p}$ defined in terms of gaps 
between events,
points towards a general methodology for obtaining visibility
relations for a given data structure.
We have also attempted the pattern on queues, and found that it does
derive a variant of the queue axioms of Henzinger et
al.~\citet{henzinger:concur13,aspects} and applies to the queue of
Herlihy and Wing~\citet{herlihy:90}. In the future, we plan to study
if this pattern applies to other concurrent data structures (e.g.,
memory snapshots, trees, lists, sets, etc.) and if the induced notions
of observation and separation are usable in more general settings
(e.g., to address other flavors of
helping~\cite{Shavit}).

\section{Conclusions}
\label{sect::future-work}

In this paper, we show that axiomatization of visibility relations are
a powerful alternative to linearization points, when it comes to
verifying linearizability of a number of helping algorithms with
common semantic structure. In particular, we applied the technique to
the descriptor-based algorithms RDCSS and MCAS, identifying and
axiomatizing the relations of separation and observation between
events. We achieved further abstraction by axiomatizing the notion of
spans to model how help requests bunch together in these algorithms,
and to allow expressing the separation and disjointness of the
bunches.

\bibliography{bibmacros,references}

\appendix

\newpage

\section{Proof of Linearizability for RDCSS and MCAS}
\label{sect::appendix::lin::full-appendix}

\subsection{Common definitions and results}
\label{sect::appendix::lin::generic}

This section has results that apply to
both the RDCSS and MCAS data structures.
In particular, the section
focuses on the first step of the linearization proof
(see gray box in Figure \ref{fig::proof-diagram}), which
is proving the existence of the linear order.
Sections \ref{sect::appendix::lin::RDCSS}
and \ref{sect::appendix::lin::MCAS}
then prove sequential soundness of the linear order 
for RDCSS and MCAS, respectively.

\begin{defn}[Specification]
\label{defn-visibility-specification}
Given two pointer-indexed relations
${\visObsSymbol {\pointerIndx p}} \subseteq \absEvent
\times \absEvent$ (\emph{observation relation}) and
${\visSepSymbol {\pointerIndx p}} \subseteq \absEvent
\times \absEvent$ (\emph{separable-before relation}), a
\emph{specification}, denoted
$\genStructName{}(\visObsSymbol {\pointerIndx p},
\visSepSymbol {\pointerIndx p})$, is a triple
$(\writesAbs p, \allocsAbs p, \postPredSymbol)$ consisting
of:
\begin{itemize}
\item A pointer-indexed set ${\writesAbs p} \subseteq \absEvent$ (called \emph{successful write events}).
\item A pointer-indexed set ${\allocsAbs p} \subseteq \absEvent$ (called \emph{alloc events}).
\item A predicate ${\postPredSymbol} : \absEvent \times \ValType \rightarrow \BoolType$ (called \emph{postcondition predicate}). 
  We denote by
  $\postPred x v$ the application of $\postPredSymbol$ to
  $x\in\absEvent$ and $v\in\ValType$.  
\end{itemize}
Given relations $\visObsSymbol {\pointerIndx p}$, $\visSepSymbol {\pointerIndx p}$, we say that $\genStructName{}(\visObsSymbol {\pointerIndx p}, \visSepSymbol {\pointerIndx p}) = 
(\writesAbs p, \allocsAbs p, \postPredSymbol)$ is \emph{valid} if all
the \emph{visibility axioms} from Figure~\ref{fig:helpingstruct} are satisfied.
\end{defn}

\begin{figure}[t]
\centering
\begin{subfigwrap}{Defined notions for visibility axioms.}{fig:defined-notions-axioms}
\centering
\begin{tabular}[t]{l}
General visibility relation \\
\quad ${\genVisSymbol} \defini \bigcup_p (\visObsSymbol {\pointerIndx p} \cup \visSepSymbol {\pointerIndx p})$\\
Returns-before relation\\
\quad $\precedesAbs e {e'} \defini \ETimeProp e \natorderSymbol \STimeProp {e'}$\\
\end{tabular}
\qquad\qquad
\begin{tabular}[t]{l}
Set of terminated events\\
\quad $\terminatedEvent \defini \{ e \mid \ETimeProp e \neq \bot \}$\\
Closure of terminated events\\
\quad $\closedEvent \defini \{ e \mid \exists t \in \terminatedEvent.\ e \refleTransCl{\genVisSymbol} t \} $\\
\end{tabular}
\end{subfigwrap}

\begin{subfigwrap}{Visibility axioms for the validity of $V(\visObsSymbol {\pointerIndx p}, {\visSepSymbol{\pointerIndx p}}) = (\writesAbs p, \allocsAbs p, \postPredSymbol)$.}{fig:helpingstruct}
\centering
\begin{tabular}[t]{ll}
\axiomHLabel{help-focused::non-helpers} No in-between \\
\quad $(\visObs {\pointerIndx p} w r \wedge w' \in \writesAbs p) \implies 
       (\visSepEq {\pointerIndx p} {w'} w \vee \visSepEq {\pointerIndx p} r {w'})$ \\
\axiomHLabel{help-focused::helped-are-writers} Observed events are writes \\ 
\quad $\visObs {\pointerIndx p} w {-} \implies w \in \writesAbs p$\\
\axiomHLabel{help-focused::vis-acyclic} No future dependences  \\
\quad $\genVisTrans x y \implies \nprecedesAbsEq y x$\\
\end{tabular}
\quad
\begin{tabular}[t]{ll}
\axiomHLabel{help-focused::fin-predicate} Return value completion \\
\quad $\exists v.\ \postPred x v \wedge (x \in \terminatedEvent \implies v = \outputProp x)$ \\
\axiomHLabel{help-focused::allocs} Allocation uniqueness \\
\quad $w,w' \in \allocsAbs p \implies w = w'$\\
\axiomHLabel{help-focused::all-writers-are-willing-helpers} Written pointers are allocated \\
\quad $w \in \writesAbs p \implies \exists w' \in \allocsAbs p.\ \visSepEq {\pointerIndx p} {w'} w$ \\
\end{tabular}
\end{subfigwrap}
\caption{Visibility Axioms. Variables $w$, $w'$, $r$, $x$, $y$ range over $\closedEvent$.
	Variables $e$, $e'$ range over $\absEvent$. Variables $p$ and $v$ over $\PtsType$ and $\ValType$, respectively.}
  \label{fig:visibility-relation-axioms}
\end{figure}

Section~\ref{subsect::overview::separable-before-relations} already explained axioms \axiomHRef{help-focused::non-helpers} through \axiomHRef{help-focused::fin-predicate}. 
Here, we quickly mention the rest. Axiom \axiomHRef{help-focused::allocs} states that pointers are 
created by at most one alloc. Axiom \axiomHRef{help-focused::all-writers-are-willing-helpers}
states that for any successful $p$-write, there must
exist a $p$-alloc separable before the write. Intuitively, this just
expresses that input pointers must have been created by previous
allocs.

\begin{defn}[Section Hypotheses]
\label{defn::appendix::lin::hypotheses-for-existence-of-linear-order}
To shorten the statements of the lemmas and theorems, we will assume the following 
hypotheses along this entire section,
\begin{itemize}
\item $\absEvent$ is an arbitrary set of events.
\item $\visObsSymbol{\pointerIndx p}$ and $\visSepSymbol{\pointerIndx p}$
are two arbitrary pointer-indexed binary relations over $\absEvent$.
\item $\genStructName{}(\visObsSymbol{\pointerIndx p}, \visSepSymbol{\pointerIndx p})$
is an arbitrary valid specification.
\end{itemize}
\end{defn}

\begin{lem}
\label{lem::appendix::lin::vis-acyclic}
Suppose hypotheses \ref{defn::appendix::lin::hypotheses-for-existence-of-linear-order}.
Relation $\genVisSymbol$ is acyclic on $\closedEvent$.
\end{lem}

\begin{prf}
Suppose there is a cycle for some $x \in \closedEvent$.
Hence, $\genVisTrans x x$ holds. But, by 
Axiom \axiomHRef{help-focused::vis-acyclic}, 
$\nprecedesAbsEq x x$ must hold, which means 
$x \neq x$ (Contradiction).
\end{prf}

\begin{lem}
\label{lem::appendix::lin::precedes-poset}
Suppose hypotheses \ref{defn::appendix::lin::hypotheses-for-existence-of-linear-order}. 
Relation $\precedesAbsEqSymbol$ is a partial order on $\absEvent$. In addition, $\precedesAbsSymbol$ satisfies the interval 
order property,
\[
\forall w,x,y,z \in \absEvent.\ 
(\precedesAbs {w} {x} \wedge \precedesAbs {y} {z})
\rightarrow
(\precedesAbs {w} {z} \vee \precedesAbs {y} {x})
\]
\end{lem}

\begin{prf}
To show that $\precedesAbsEqSymbol$ is a partial order, 
it is enough to show that $\precedesAbsSymbol$
is irreflexive and transitive, because it is 
a standard result that the reflexive closure of
an irreflexive and transitive relation is a partial order.

\begin{itemize}
\item Irreflexivity. If $\precedesAbs x x$, then we would
have $\ETimeProp x \natorderSymbol \STimeProp x 
\natorderEqSymbol \ETimeProp x$, which is a contradiction.

\item Transitivity. If $\precedesAbs {x} {y}$ and
$\precedesAbs {y} {z}$, then 
$\ETimeProp x \natorderSymbol \STimeProp y 
\natorderEqSymbol \ETimeProp y 
\natorderSymbol \STimeProp z$, which means
$\ETimeProp x \natorderSymbol \STimeProp z$.
Hence, $\precedesAbs {x} {z}$.
\end{itemize}

We now show that $\precedesAbsSymbol$ satisfies the
interval order property.
 
Let $\precedesAbs {w} {x}$ and $\precedesAbs {y} {z}$.
Hence, $\ETimeProp w \natorderSymbol \STimeProp x$ 
and $\ETimeProp y \natorderSymbol \STimeProp z$.

Either $\ETimeProp w \natorderSymbol \STimeProp z$ or
$\STimeProp z \natorderEqSymbol \ETimeProp w$. The first
case leads to $\precedesAbs {w} {z}$.

For the second case, by using the hypotheses, we have
$\ETimeProp y \natorderSymbol \STimeProp z
\natorderEqSymbol \ETimeProp w 
\natorderSymbol \STimeProp x$.
Therefore, $\precedesAbs {y} {x}$.
\end{prf}

\begin{lem}
\label{lem::appendix::lin::no-cycles}
Suppose hypotheses \ref{defn::appendix::lin::hypotheses-for-existence-of-linear-order}.
Let $n \natorderSymbolRight 0$. If 
$a_0 R_1 a_1 R_2 \ldots R_{n-1} a_{n-1} R_n a_{n}$
is a sequence of $n$ steps where each $R_i$ is either 
$\genVisSymbol$ or $\precedesAbsSymbol$, and each $a_i \in \closedEvent$, 
then $\nprecedesAbsEq {a_n} {a_0}$.
\end{lem}

\begin{prf}
By strong induction on $n$.
\begin{itemize}
\item Case $n = 1$. So, we have $a_0 R_1 a_1$.

If $R_1 = {\genVisSymbol}$, then from
$\genVis {a_0} {a_1}$ and Axiom 
\axiomHRef{help-focused::vis-acyclic},
we get $\nprecedesAbsEq {a_1} {a_0}$.

If $R_1 = {\precedesAbsSymbol}$, then from
$\precedesAbs {a_0} {a_1}$, we cannot have
$\precedesAbsEq {a_1} {a_0}$, because we would get
$a_0 \precedesAbsSymbol a_1 \precedesAbsEqSymbol
a_0$ (Contradiction by Lemma \ref{lem::appendix::lin::precedes-poset}).

\item Inductive case. Let 
$a_0 R_1 a_1 R_2 \ldots R_n a_n R_{n+1} a_{n+1}$ be a 
sequence of $n+1$ steps.

If all $R_i$ are $\genVisSymbol$, then 
$\nprecedesAbsEq {a_{n+1}} {a_0}$ follows by Axiom
\axiomHRef{help-focused::vis-acyclic}.

Therefore, we can assume that for some 
$1 \natorderEqSymbol i \natorderEqSymbol n+1$,
we have $R_i = {\precedesAbsSymbol}$, i.e., 

\begin{align}
\label{eqn::appendix::lin::triag-irreflexive-1}
\overbrace{a_0 R_1 \ldots R_{i-1} a_{i-1}}^{i-1 \text{ steps}} 
\precedesAbsSymbol 
\overbrace{a_{i} R_{i+1} \ldots R_{n+1} a_{n+1}}^{n-i+1 \text{ steps}}
\end{align}

Suppose for a contradiction that 
$\precedesAbsEq {a_{n+1}} {a_0}$.

If $a_{n+1} = a_0$. Then,
\[
a_{i} R_{i+1} \ldots R_n a_n R_{n+1} a_0 R_1 \ldots R_{i-1} a_{i-1}
\]
is a sequence of $(n-i+1) + (i-1) = n$ steps. Therefore,
by the inductive hypothesis, 
$\nprecedesAbsEq {a_{i-1}} {a_{i}}$, which contradicts
\eqref{eqn::appendix::lin::triag-irreflexive-1}.

It remains to check the case 
$\precedesAbs {a_{n+1}} {a_0}$.

\begin{itemize}
\item Subcase $i = 1$. Hence, 
${a_{n+1}} \precedesAbsSymbol {a_0} 
\precedesAbsSymbol a_1$.

But ${a_{1}} R_2 \ldots R_{n+1} {a_{n+1}}$ is a sequence
of $n$ steps. Therefore, by 
the inductive hypothesis, 
$\nprecedesAbsEq {a_{n+1}} {a_{1}}$ (Contradiction).

\item Subcase $i = n+1$. Hence, 
${a_{n}} \precedesAbsSymbol {a_{n+1}} 
\precedesAbsSymbol a_0$.

But ${a_{0}} R_1 \ldots R_n {a_{n}}$ is a sequence
of $n$ steps. Therefore, by 
the inductive hypothesis, 
$\nprecedesAbsEq {a_{n}} {a_{0}}$ (Contradiction).

\item Subcase $1 \natorderSymbol i \natorderSymbol n+1$.

Since $\precedesAbs {a_{i-1}} {a_{i}}$ and 
$\precedesAbs {a_{n+1}} {a_0}$, by Lemma 
\ref{lem::appendix::lin::precedes-poset}, 
either $\precedesAbs {a_{i-1}} {a_0}$ or
$\precedesAbs {a_{n+1}} {a_{i}}$.

But $a_0 R_1 \ldots R_{i-1} a_{i-1}$ is a sequence with 
$0 \natorderSymbol i-1 \natorderSymbol n$
steps and 
$a_{i} R_{i+1} \ldots R_{n+1} a_{n+1}$ 
is a sequence with 
$0 \natorderSymbol n-i+1 \natorderSymbol n$ steps. 
Therefore, by the inductive hypothesis,
$\nprecedesAbsEq {a_{i-1}} {a_0}$ and
$\nprecedesAbsEq {a_{n+1}} {a_{i}}$ 
(Contradiction).
\end{itemize}
\end{itemize}
\end{prf}

\begin{defn}
Define the \emph{happens-before} relation ${\triagVisSymbol} \defini {(\genVisSymbol \cup \precedesAbsSymbol)^{+}}$, i.e.,
the transitive closure of the union of the general visibility and returns-before relations.
\end{defn}

\begin{lem}
\label{lem::appendix::lin::triag-irreflexive}
Suppose hypotheses \ref{defn::appendix::lin::hypotheses-for-existence-of-linear-order}.
The relation $\triagVisSymbol$ is irreflexive on
$\closedEvent$.
\end{lem}

\begin{prf}
If $\triagVis x x$, then for some 
$n \natorderSymbolRight 0$, there is a sequence
$a_0 R_1 \ldots R_n a_{n}$ of $n$ steps, where 
each $R_i$ is either 
$\genVisSymbol$ or $\precedesAbsSymbol$, each $a_i \in \closedEvent$ and
$a_0 = x$, and $a_n = x$.

But by Lemma 
\ref{lem::appendix::lin::no-cycles},
we would have $\nprecedesAbsEq x x$ 
which is a contradiction.
\end{prf}

\begin{lem}
\label{lem::appendix::lin::triag-poset}
Suppose hypotheses \ref{defn::appendix::lin::hypotheses-for-existence-of-linear-order}.
The relation $\triagVisEqSymbol$ is a partial order on $\closedEvent$.
\end{lem}

\begin{prf}
It is a standard result that the reflexive closure of
an irreflexive and transitive relation is a partial order.
Hence, $\triagVisEqSymbol$ is a partial order on
$\closedEvent$ by Lemma 
\ref{lem::appendix::lin::triag-irreflexive} and
the fact that $\triagVisSymbol$ is transitive by definition.
\end{prf}

\begin{lem}
\label{eq::appendix::lin::committed-downward-closed}
Suppose hypotheses \ref{defn::appendix::lin::hypotheses-for-existence-of-linear-order}.
Set $\closedEvent$ is $\genVisSymbol$-downward closed. In other words,
if $x \in \closedEvent$ and $\genVis y x$, then $y \in \closedEvent$.
\end{lem}

\begin{prf}
Since $x \in \closedEvent$, there is $z \in \terminatedEvent$ such that 
$x \refleTransCl{\genVisSymbol} z$. So $y \genVisSymbol x \refleTransCl{\genVisSymbol} z$
which means $y \in \closedEvent$.
\end{prf}

\begin{lem}
\label{lem::appendix::lin::recency-holds-in-any-extension}
Suppose hypotheses \ref{defn::appendix::lin::hypotheses-for-existence-of-linear-order}.
Given a partial order $\leq$ over $\closedEvent$ 
such that ${\triagVisEqSymbol} \subseteq {\leq}$, the following holds,
\begin{align}
\label{eq::appendix::lin::sequential-recency}
\forall x,y \in \closedEvent.\ \visObs {\pointerIndx p} x y \implies \neg \exists z \in \writesAbs {p} \cap \closedEvent.\ x < z < y
\end{align}
\end{lem}

\begin{prf}
Let $\visObs {\pointerIndx p} x y$. Suppose for a contradiction that $x < z < y$
for some $z \in \writesAbs p \cap \closedEvent$.

By Axiom \axiomHRef{help-focused::non-helpers},
either $\visSep {\pointerIndx p} y z$ or 
$\visSep {\pointerIndx p} z x$ or $y = z$ or $x = z$.

\begin{itemize}
\item Case $\visSep {\pointerIndx p} y z$.
Since ${\triagVisEqSymbol} \subseteq {\leq}$, we have $y < z$. 
But by hypothesis $z < y$ (Contradiction).

\item Case $\visSep {\pointerIndx p} z x$.
Since ${\triagVisEqSymbol} \subseteq {\leq}$, we have
$z < x$. But by hypothesis $x < z$ (Contradiction).

\item Case $y = z$. This contradicts hypothesis $z < y$.

\item Case $x = z$. This contradicts hypothesis $x < z$.
\end{itemize}
\end{prf}

\begin{lem}
\label{lem::appendix::lin::existence-basic}
Suppose hypotheses \ref{defn::appendix::lin::hypotheses-for-existence-of-linear-order}.
There is a linear order $\leq$ over $\closedEvent$ containing 
$\genVisSymbol$ and $\precedesAbsSymbol$, such that property \eqref{eq::appendix::lin::sequential-recency}
holds.
\end{lem}

\begin{prf}
Any finite partial order is contained in a linear order~\cite{topoSorting}.
Pick some linear order $\leq$ containing 
the partial order $\triagVisEqSymbol$ over $\closedEvent$.

By Lemma \ref{lem::appendix::lin::recency-holds-in-any-extension},
property \eqref{eq::appendix::lin::sequential-recency} holds.
\end{prf}

\subsection{Linearizability Proof for the RDCSS Data Structure}
\label{sect::appendix::lin::RDCSS}

This section focuses on proving, for the RDCSS data structure, sequential soundness 
of the linear order built in Lemma \ref{lem::appendix::lin::existence-basic}.
This section makes use of Definition \ref{defn-visibility-specification}.

\begin{defn}[Exportable Procedures in the RDCSS Data Structure]
The following are the exportable procedures in the RDCSS Data Structure,
\begin{center}
\begin{tabular}{ll}
$\rdcssAlg(d: \rdcssDesc)$ & (RDCSS for descriptor $d$) \\
$\rdcssAllocAlg(v: \ValType,\ k:\ptKind)$ & (Allocation of pointer kind $k \in \{ \CONTROL, \DATA \}$) \\
$\rdcssCasAlg(p: \DataPtType,\ e: \ValType,\ n: \ValType)$ & (CAS for data pointers) \\
$\rdcssWriteAlg(p: \DataPtType,\ v: \ValType)$ & (Write for data pointers) \\
$\rdcssReadAlg(p: \DataPtType)$ & (Read for data pointers) \\
$\rdcssCasCtlAlg(p: \ControlPtType,\ e: \ValType,\ n: \ValType)$ & (CAS for control pointers) \\
$\rdcssWriteCtlAlg(p: \ControlPtType,\ v: \ValType)$ & (Write for control pointers) \\
$\rdcssReadCtlAlg(p: \ControlPtType)$ & (Read for control pointers) \\
\end{tabular}
\end{center}
\end{defn}

For each $x \in \absEvent$ and pointer $p$, we define 
the \emph{writer predicate} $\inputVal x p v$.

\begin{defn}[Writer Predicate for the RDCSS Data Structure]
\label{defn::appendix::lin::rdcss-input-map}
The \emph{writer predicate} $\inputValName \subseteq \absEvent \times \PtsType \times \ValType$, 
denoted $\inputVal x p v$, is defined by cases,
\begin{align*}
\inputVal x p v \defini 
& (\exists d.\ x = \rdcssAlg(d) \wedge p = \pointTwo d \wedge v = \newTwo d) \vee {} \\
& (x = \rdcssAllocAlg(v, \_) \wedge \ETimeProp x \neq \bot \wedge p = \outputProp x) \vee{} \\
& x = \rdcssCasAlg(p,\_,v) \vee x = \rdcssWriteAlg(p, v) \vee {}\\
& x = \rdcssCasCtlAlg(p,\_,v) \vee x = \rdcssWriteCtlAlg(p, v) \\
\end{align*}
Predicate $\inputVal x p v$ denotes the value $v$
in $\inProp x$ that $x$ will attempt to write into pointer $p$.
\end{defn}

Notice that for the case of allocs, $\inputValName$
is defined only after the alloc has finished, since
$\inputValName$ depends on the pointer generated
by the alloc. We only need to consider terminated allocs
because any procedure $x$ requires in its input the pointers produced by allocs
at the moment $x$ is invoked.

We now define the specification for the RDCSS data structure.

\begin{defn}[Specification for the RDCSS Data Structure]
\label{defn::appendix::lin::rdcss-vis-spec}
Given relations $\visObsSymbol {\pointerIndx p}$, $\visSepSymbol {\pointerIndx p}$,
we define the specification for the RDCSS data structure, denoted 
$\RDCSSFam(\visObsSymbol {\pointerIndx p}, \visSepSymbol {\pointerIndx p})$, as follows,
\begin{align*}
x \in \allocsAbs p & \defini x = \rdcssAllocAlg(\_,\_) \wedge
\ETimeProp x \neq \bot \wedge p = \outputProp x \\
x \in \writesAbs p & \defini 
\begin{cases}
p = \pointTwo d \wedge \exists z_1, z_2.\ \visObs {\pointerIndx {\pointOne d}} {z_1} x \wedge \visObs {\pointerIndx {\pointTwo d}} {z_2} x \wedge {} & \text{if } x = \rdcssAlg(d) \\
\quad \inputVal {z_1} {\pointOne d} {\expOne d} \wedge \inputVal {z_2} {\pointTwo d} {\expTwo d} & \\
\ETimeProp x \neq \bot \wedge p = \outputProp x & \text{if } x = \rdcssAllocAlg(\_, \_) \\
p = q \wedge \exists z.\ \visObs {\pointerIndx q} z x \wedge \inputVal z q e & \text{if } x = \rdcssCasAlg(q,e,\_) \\
p = q & \text{if } x = \rdcssWriteAlg(q, \_) \\
p = q \wedge \exists z.\ \visObs {\pointerIndx q} z x \wedge \inputVal z q e & \text{if } x = \rdcssCasCtlAlg(q,e,\_) \\
p = q & \text{if } x = \rdcssWriteCtlAlg(q, \_) \\
\end{cases} \\
\postPred x v & \defini
\begin{cases}
\exists z_2.\ \visObs {\pointerIndx {\pointTwo d}} {z_2} x \wedge \inputVal {z_2} {\pointTwo d} v \wedge {} & \text{if } x = \rdcssAlg(d) \\
\quad (v = \expTwo d \implies \exists z_1.\ \visObs {\pointerIndx {\pointOne d}} {z_1} x) & \\
\ETimeProp x \neq \bot \wedge v = \outputProp x & \text{if } x = \rdcssAllocAlg(\_, \_) \\
\exists z.\ \visObs {\pointerIndx q} z x \wedge \inputVal z q v & \text{if } x = \rdcssCasAlg(q,\_,\_) \\
v = \unitValue & \text{if } x = \rdcssWriteAlg(\_, \_) \\
\exists z.\ \visObs {\pointerIndx q} z x \wedge \inputVal z q v & \text{if } x = \rdcssReadAlg(q) \\
\exists z.\ \visObs {\pointerIndx q} z x \wedge \inputVal z q v & \text{if } x = \rdcssCasCtlAlg(q,\_,\_) \\
v = \unitValue & \text{if } x = \rdcssWriteCtlAlg(\_, \_) \\
\exists z.\ \visObs {\pointerIndx q} z x \wedge \inputVal z q v & \text{if } x = \rdcssReadCtlAlg(q) \\
\end{cases}
\end{align*}
\end{defn}

We have the following immediate lemma,

\begin{lem}
\label{lem::appendic::lin::rdcss-writers-have-input-maps}
If $x \in \writesAbs p$, then there is 
$v \in \ValType$ such that $\inputVal x p v$.
\end{lem}

\begin{prf}
By cases on $x$.
\begin{itemize}
\item Case $x = \rdcssAlg(d)$. We have $p = \pointTwo d$ and
$\inputVal x p {\newTwo d}$ follows by definition of $\inputValName$.
\item Case $x = \rdcssAllocAlg(v,\_)$. We have 
$p = \outputProp x$ and
$\inputVal x p v$ follows by definition of $\inputValName$.
\item Case $x = \rdcssCasAlg(q,\_,n)$. We have 
$p = q$ and
$\inputVal x p n$ follows by definition of $\inputValName$.
\item Case $x = \rdcssWriteAlg(q,v)$. We have 
$p = q$ and
$\inputVal x p v$ follows by definition of $\inputValName$.
\item Cases when $x$ is one of $\rdcssCasCtlAlg(q,\_,n)$ or $\rdcssWriteCtlAlg(q,v)$
are similar.
\end{itemize}
\end{prf}

Towards defining the notion of sequential soundness for a linear order over 
events in RDCSS, we need to describe what it means for events to execute.
For that matter, we define the following state-based operational semantics,
so that intuitively, events execute by building 
a path in the operational semantics.

\begin{defn}[Operational semantics for events in the RDCSS Data Structure]
\label{defn::appendix::lin::rdcss-legality}
Let $Op^R$ denote the operational semantics generated by 
the following base steps, where states are heaps and labels
are of the form ``$proc(in)\ \langle out \rangle$''
where $proc$ is the procedure name, $in$ the procedure's input, and $out$ the procedure's output.
\begin{itemize}
\item $H \xrightarrow{\rdcssAlg(d)\ \langle v \rangle} H$, 
if $H(\pointTwo d) = v$ and $v \neq \expTwo d$.
\item $H \xrightarrow{\rdcssAlg(d)\ \langle \expTwo d \rangle} H$, 
if $H(\pointTwo d) = \expTwo d$ and $H(\pointOne d) \neq \expOne d$.
\item $H \xrightarrow{\rdcssAlg(d)\ \langle {\expTwo d} \rangle} \mapExt H {\mapEntry {\pointTwo d} {\newTwo d}}$, 
if $H(\pointTwo d) = \expTwo d$ and $H(\pointOne d) = \expOne d$.
\item $H \xrightarrow{\rdcssAllocAlg(v,\_)\ \langle p \rangle} \mapExt H {\mapEntry p v}$, 
if $p \notin dom(H)$.
\item $H \xrightarrow{\rdcssCasAlg(p,e,\_)\ \langle v \rangle} H$, 
if $H(p) = v$ and $v \neq e$.
\item $H \xrightarrow{\rdcssCasAlg(p,e,n)\ \langle e \rangle} \mapExt H {\mapEntry p n}$,
if $H(p) = e$.
\item $H \xrightarrow{\rdcssWriteAlg(p,v)\ \langle \unitValue \rangle} \mapExt H {\mapEntry p v}$,
if $p \in dom(H)$.
\item $H \xrightarrow{\rdcssReadAlg(p)\ \langle v \rangle} H$, if 
$H(p) = v$.
\item $H \xrightarrow{\rdcssCasCtlAlg(p,e,\_)\ \langle v \rangle} H$, 
if $H(p) = v$ and $v \neq e$.
\item $H \xrightarrow{\rdcssCasCtlAlg(p,e,n)\ \langle e \rangle} \mapExt H {\mapEntry p n}$,
if $H(p) = e$.
\item $H \xrightarrow{\rdcssWriteCtlAlg(p,v)\ \langle \unitValue \rangle} \mapExt H {\mapEntry p v}$,
if $p \in dom(H)$.
\item $H \xrightarrow{\rdcssReadCtlAlg(p)\ \langle v \rangle} H$, if 
$H(p) = v$.
\end{itemize}
\end{defn}

We now define the notion of sequential soundness of a linear order $\leq$.

\begin{defn}[Sequential soundness of a linear order]
Given a linear order $\leq$ over a set of events, and a path $P$ in the operational semantics $Op^R$, 
we say that $P$ \emph{matches} $\leq$ if for every step $i$ in $P$, 
\begin{itemize}
\item The $i$-th procedure's name in $P$ equals the procedure's name of the $i$-th event in $\leq$.
\item The $i$-th procedure's input in $P$ equals the procedure's input of the $i$-th event in $\leq$.
\item If the $i$-th event in $\leq$ is terminated, then the 
$i$-th procedure's output in $P$ equals the $i$-th event output in $\leq$.
\end{itemize}
We say that $\leq$ is \emph{sequentially sound} if there is a matching path in $Op^R$, starting from the empty heap and
having as many steps as the number of events in the domain of $\leq$.
\end{defn}

\begin{notation}
If path $P$ matches $\leq$, and $H \xrightarrow{proc(in)\ \langle out \rangle} H'$ is a step in $P$,
we denote by $\generatorEvent {proc}$ the corresponding event in $\leq$. 
\end{notation}

To prove sequential soundness, we require two lemmas.
The first lemma, called writes
lemma, expresses that successful writes modify the heap 
and as long as there are no successful
writes afterwards, such changes remain intact.
The second lemma, called allocs lemma, states that allocs create
pointers in the heap.

\begin{defn}[Section Hypotheses]
\label{defn::appendix::lin::hypotheses-for-sequential-soundness-rdcss}
To shorten the statements of lemmas from \ref{lem::appendix::lin::rdcss-writers-change-heap} 
to \ref{lem::appendix::lin::linear-order-rdcss-sequential-soundness}, the
following hypotheses will apply to those lemmas,
\begin{itemize}
\item $\absEvent$ is an arbitrary set of events.
\item $\visObsSymbol{\pointerIndx p}$ and $\visSepSymbol{\pointerIndx p}$
are two arbitrary pointer-indexed binary relations over $\absEvent$.
\item $\genStructName{\RDCSS}(\visObsSymbol{\pointerIndx p}, \visSepSymbol{\pointerIndx p})$
is valid, where $\genStructName{\RDCSS}$ is the specification for RDCSS defined in \ref{defn::appendix::lin::rdcss-vis-spec}.
\end{itemize}
\end{defn}

We can now prove the writes lemma,

\begin{lem}[Writes lemma]
\label{lem::appendix::lin::rdcss-writers-change-heap}
Suppose hypotheses \ref{defn::appendix::lin::hypotheses-for-sequential-soundness-rdcss}.
Let $\leq$ be the linear order of Lemma \ref{lem::appendix::lin::existence-basic}.
Suppose $\mathcal P$ is a path of length $1 \natorderEqSymbol n \natorderEqSymbol \vert \closedEvent \vert$ in $Op^R$ that matches
$\leq$,
\[
H_0 = \emptyset \xrightarrow{proc_1(in_1)\ \langle out_1 \rangle} H_1 \xrightarrow{proc_2(in_2)\ \langle out_2 \rangle} 
\ldots \xrightarrow{proc_n(in_n)\ \langle out_n \rangle} H_n 
\]

Let $1 \natorderEqSymbol i \natorderEqSymbol n$.
If $\generatorEvent {proc_i} \in \writesAbs p$
and $\inputVal {\generatorEvent {proc_i}} p v$ 
and for all $j \natorderSymbolRight i$, 
$\generatorEvent {proc_j} \notin \writesAbs p$, 
then $H_n(p) = v$.
\end{lem}

\begin{prf}
By induction on $n$.

\begin{itemize}
\item Case $n = 1$. 

Suppose $\generatorEvent {proc_1} \in \writesAbs p$ and 
$\inputVal {\generatorEvent {proc_1}} p v$. 

All non-alloc steps at $proc_1$ require that some 
pointer exists at $H_0 = \emptyset$, which is impossible.
Therefore,
the only applicable step is when $proc_1$ is an alloc.  
Hence, $proc_1(in_1) = \rdcssAllocAlg(v',\_)$
for some $v'$.

But,
$p = \outputProp {\generatorEvent {proc_1}} = out_1$, since
$\generatorEvent {proc_1} \in \writesAbs p$, 
$proc_1$ matches $\generatorEvent {proc_1}$,
and $\generatorEvent {proc_1} \in \terminatedEvent$ by definition of
$\writesAbs p$. 
Also, $v' = v$ by definition of $\inputValName$.

Therefore, $H_{1}(p) = v$,
since the alloc step creates $out_1 = p$ with initial value $v' = v$.

\item Inductive case. Let $n \natorderEqSymbolRight 1$ and 
suppose the statement holds for any path of length $n$.
Let $\mathcal P$ be a matching path of length $n+1$,
\[
H_0 = \emptyset \xrightarrow{proc_1(in_1)\ \langle out_1 \rangle} 
\ldots \xrightarrow{proc_n(in_n)\ \langle out_n \rangle} H_n \xrightarrow{proc_{n+1}(in_{n+1})\ \langle out_{n+1} \rangle} H_{n+1}
\]

Let $1 \natorderEqSymbol i \natorderEqSymbol n+1$
and $\generatorEvent {proc_i} \in \writesAbs p$
and $\inputVal {\generatorEvent{proc_i}} p v$ 
and for all $j \natorderSymbolRight i$, 
$\generatorEvent{proc_j} \notin \writesAbs p$. 
We need to show that $H_{n+1}(p) = v$.

\begin{itemize}
\item Case $i \natorderSymbol n + 1$, hence $i \natorderEqSymbol n$.
Since $\left[ proc_1 , \ldots , proc_n \right]$ is a subpath of length $n$ 
that matches $\leq$, the inductive hypothesis concludes
$H_n(p) = v$. 

By hypothesis,
$\generatorEvent{proc_{n+1}} \notin \writesAbs p$ must hold.
We do a case analysis on $proc_{n+1}$.

\begin{itemize}
\item Case $proc_{n+1}(in_{n+1}) = \rdcssAlg(d)$. 
Since $\generatorEvent{proc_{n+1}} \notin \writesAbs p$, we have either 
$p \neq \pointTwo d$ or,
\begin{align}
\label{eqn::appendix::lin::rdcss-non-writer}
\begin{split}
\forall z_1, z_2.\ (\visObs {\pointerIndx 
{\pointOne d}} {z_1} {\generatorEvent{proc_{n+1}}} \wedge \visObs 
{\pointerIndx {\pointTwo d}} {z_2} {\generatorEvent{proc_{n+1}}}) 
\implies {} \\
(\ninputVal {z_1} {\pointOne d} {\expOne d} \vee 
\ninputVal {z_2} {\pointTwo d} {\expTwo d})
\end{split}
\end{align} 

The case $p \neq \pointTwo d$ implies $H_{n+1}(p) = v$
since any of the RDCSS steps do not change 
pointers different from $\pointTwo d$. We now focus on the case 
\eqref{eqn::appendix::lin::rdcss-non-writer}.

By Axiom 
\axiomHRef{help-focused::fin-predicate}, we have for some $v'$,
\begin{align}
\label{eqn::appendix::lin::rdcss-post-1}
\begin{split}
\exists z_2.\ \visObs {\pointerIndx {\pointTwo d}} 
{z_2} {\generatorEvent {proc_{n+1}}} \wedge \inputVal {z_2} {\pointTwo d} {v'} \wedge {} \\
(v' = \expTwo d \implies \exists z_1.\ \visObs 
{\pointerIndx {\pointOne d}} {z_1} {\generatorEvent{proc_{n+1}}})
\end{split}
\end{align} 

\begin{itemize}
\item Case $v' = \expTwo d$. Hence, by 
\eqref{eqn::appendix::lin::rdcss-post-1} and Axiom \axiomHRef{help-focused::helped-are-writers},
$z_1 \in \writesAbs {\pointOne d}$ holds. So, by Lemma
\ref{lem::appendic::lin::rdcss-writers-have-input-maps} with
$z_1$, and
\eqref{eqn::appendix::lin::rdcss-non-writer} with $z_1$ and
$z_2$ we have
$\inputVal {z_1} {\pointOne d} {v''}$ for some
$v'' \neq \expOne d$ and 
$\inputVal {z_2} {\pointTwo d} {\expTwo d}$.

We know that $\left[ proc_1 , \ldots , proc_{n} \right]$ is a subpath 
of length $n$ that matches $\leq$, and $z_1$, $z_2$ must
appear before $\generatorEvent{proc_{n+1}}$ in $\leq$,
since $\leq$ contains $\genVisSymbol$.
So, there are $proc_l$ and $proc_m$ such that $l,m \natorderEqSymbol n$ and 
$\generatorEvent{proc_{l}} = z_1$ and
$\generatorEvent{proc_{m}} = z_2$. 

By Axiom 
\axiomHRef{help-focused::helped-are-writers} and
property \eqref{eq::appendix::lin::sequential-recency},
we also have the following facts:
$z_1 \in \writesAbs {\pointOne d}$,
$z_2 \in \writesAbs {\pointTwo d}$,
and for all $l \natorderSymbol j \natorderEqSymbol n$,
$\generatorEvent {proc_{j}} \notin \writesAbs {\pointOne d}$, 
and for all $m \natorderSymbol j \natorderEqSymbol n$, 
$\generatorEvent {proc_{j}} \notin \writesAbs {\pointTwo d}$.

Therefore, by the inductive hypothesis applied twice on the subpath $\left[ proc_1 , \ldots , proc_{n} \right]$
for $l$ and $m$, we must have
$H_n(\pointOne d) = v'' \neq \expOne d$ and $H_n(\pointTwo d) = \expTwo d$.

Hence, $proc_{n+1}$ must be a failing $\rdcssAlg$ step in the operational semantics.
This means that if either $p = \pointTwo d$
or not, we will have $H_{n+1}(p) = v$,
since $H_n(p) = v$ and the $n+1$ step does not modify pointers.

\item Case $v' \neq \expTwo d$. From \eqref{eqn::appendix::lin::rdcss-post-1}, $\visObs {\pointerIndx {\pointTwo d}} 
{z_2} {\generatorEvent {proc_{n+1}}}$ and $\inputVal {z_2} {\pointTwo d} {v'}$ hold.

By a similar argument as the previous case, $z_2$ must be one of the 
procedures in the subpath $\left[ proc_1 , \ldots , proc_{n} \right]$.
Therefore, by Axiom 
\axiomHRef{help-focused::helped-are-writers}, property \eqref{eq::appendix::lin::sequential-recency},
and the inductive hypothesis, $H_n(\pointTwo d) = v' \neq \expTwo d$ must hold.

Hence, $proc_{n+1}$ must be a failing $\rdcssAlg$ step in the operational semantics.
This means that if either $p = \pointTwo d$
or not, we will have $H_{n+1}(p) = v$,
since $H_n(p) = v$ and the $n+1$ step does not modify pointers.
\end{itemize}

\item Case $proc_{n+1}(in_{n+1}) = \rdcssAllocAlg(v,\_)$. 
Since $\generatorEvent{proc_{n+1}} \notin \writesAbs p$,
we have,
\[
\outputProp {\generatorEvent{proc_{n+1}}} \neq \bot \implies 
p \neq \outputProp{\generatorEvent{proc_{n+1}}}
\]

But by Axiom \axiomHRef{help-focused::fin-predicate}, 
$\outputProp {\generatorEvent{proc_{n+1}}} \neq \bot$.
Hence, $p \neq \outputProp{\generatorEvent{proc_{n+1}}}$.

But since $proc_{n+1}$ matches $\leq$ and 
$\generatorEvent{proc_{n+1}}$ has terminated,
we also have $\outputProp{\generatorEvent{proc_{n+1}}} = out_{n+1}$.
Therefore, $p \neq out_{n+1}$.

Since the alloc step only creates $out_{n+1}$, pointer $p$ remains
unmodified, hence $H_{n+1}(p) = v$.

\item Case $proc_{n+1}(in_{n+1}) = \rdcssCasAlg(r,e,m)$. 
Since $\generatorEvent{proc_{n+1}} \notin \writesAbs p$, we have either 
$p \neq r$ or,
\begin{align}
\label{eqn::appendix::lin::rdcss-cas-non-writer}
\begin{split}
\forall z.\ \visObs {\pointerIndx r} z {\generatorEvent{proc_{n+1}}} 
\implies \ninputVal z r e
\end{split}
\end{align} 

The case $p \neq r$ implies $H_{n+1}(p) = v$
since the $\rdcssCasAlg$ steps do not change 
pointers different from $r$. We now focus on the case 
\eqref{eqn::appendix::lin::rdcss-cas-non-writer}.
 
By Axiom \axiomHRef{help-focused::fin-predicate}, we have for some $v'$,
\begin{align}
\label{eqn::appendix::lin::rdcss-cas-post-1}
\exists z.\ \visObs {\pointerIndx r} 
z {\generatorEvent{proc_{n+1}}} \wedge \inputVal z r {v'}
\end{align} 

Hence, by 
\eqref{eqn::appendix::lin::rdcss-cas-post-1} and
\eqref{eqn::appendix::lin::rdcss-cas-non-writer} with $z$, 
we have $v' \neq e$.

By a similar argument as the $\rdcssAlg$ case, $z$ must be one of the 
procedures in the subpath $\left[ proc_1 , \ldots , proc_{n} \right]$.
Therefore, by Axiom 
\axiomHRef{help-focused::helped-are-writers}, property \eqref{eq::appendix::lin::sequential-recency},
and the inductive hypothesis, $H_n(r) = v' \neq e$ must hold.

Hence, $proc_{n+1}$ must be a failing $\rdcssCasAlg$ step in the operational semantics.
This means that if either $p = r$
or not, we will have $H_{n+1}(p) = v$,
since $H_n(p) = v$ and the $n+1$ step does not modify pointers.

\item Case $proc_{n+1}(in_{n+1}) = \rdcssWriteAlg(r,m)$. 
By definition, we have $\generatorEvent{proc_{n+1}} \in 
\writesAbs r$.
Therefore, $p \neq r$ (since $\generatorEvent{proc_{n+1}} \notin 
\writesAbs p$ by hypothesis), which means $H_{n+1}(p) = v$
since the $\rdcssWriteAlg$ step does not change 
pointers different from $r$.

\item Case $proc_{n+1}(in_{n+1}) = \rdcssReadAlg(r)$. 
We have $H_{n+1}(p) = v$ since the $\rdcssReadAlg$ step does not change pointers.

\item Cases for $\rdcssCasCtlAlg$, $\rdcssWriteCtlAlg$, and $\rdcssReadCtlAlg$
are identical to the cases for $\rdcssCasAlg$, $\rdcssWriteAlg$, and $\rdcssReadAlg$,
respectively.
\end{itemize}

\item Case $i = n+1$. So, we have $\generatorEvent{proc_{n+1}} \in 
\writesAbs p$ and
$\inputVal {\generatorEvent{proc_{n+1}}} p v$. 
We do a case analysis on
$proc_{n+1}$.

\begin{itemize}
\item Case $proc_{n+1}(in_{n+1}) = \rdcssAlg(d)$. 

Hence, $p = \pointTwo d$ (by definition of $\writesAbs p$),
$v = \newTwo d$ (by definition of $\inputValName$), and,
\begin{align*}
\begin{split}
\exists z_1, z_2.\ \visObs {\pointerIndx 
{\pointOne d}} {z_1} {\generatorEvent{proc_{n+1}}} \wedge 
\visObs {\pointerIndx {\pointTwo d}} {z_2} 
{\generatorEvent{proc_{n+1}}} \wedge {} \\
\inputVal {z_1} {\pointOne d} {\expOne d} \wedge 
\inputVal {z_2} {\pointTwo d} {\expTwo d}
\end{split}
\end{align*}

By a similar argument as the $\rdcssAlg$ case when $i \natorderSymbol n + 1$, $z_1$ and $z_2$ must be one of the 
procedures in the subpath $\left[ proc_1 , \ldots , proc_{n} \right]$.
Therefore, by Axiom 
\axiomHRef{help-focused::helped-are-writers}, property \eqref{eq::appendix::lin::sequential-recency},
and the inductive hypothesis, $H_n(\pointOne d) = \expOne d$ and $H_n(\pointTwo d) = \expTwo d$ must hold.

Therefore, the $proc_{n+1}$ step
must be the successful $\rdcssAlg$ step, which means 
$H_{n+1} = \mapExt {H_n} {\mapEntry {\pointTwo d} {\newTwo d}}$.
Hence $H_{n+1}(p) = H_{n+1}(\pointTwo d) = \newTwo d = v$.

\item Case $proc_{n+1}(in_{n+1}) = \rdcssAllocAlg(v',\_)$. 
Hence, $p = \outputProp {\generatorEvent{proc_{n+1}}}$
and $\ETimeProp{\generatorEvent{proc_{n+1}}} \neq \bot$ 
(by definition of $\writesAbs p$). 
Hence, $v = v'$ (by definition of $\inputValName$).

Since $\generatorEvent{proc_{n+1}}$ has terminated
and $proc_{n+1}$ matches $\leq$, $\outputProp {\generatorEvent{proc_{n+1}}} = out_{n+1}$.
Hence, $out_{n+1} = p$.

But the alloc step implies $H_{n+1} = \mapExt {H_n} {\mapEntry {out_{n+1}} {v'}}$,
which means $H_{n+1}(p) = H_{n+1}(out_{n+1}) = v' = v$. 

\item Case $proc_{n+1}(in_{n+1}) = \rdcssCasAlg(q,e,m)$. 

By definitions of $\writesAbs p$ and $\inputValName$, $p = q$, $v = m$, and,
\begin{align*}
\exists z.\ \visObs {\pointerIndx {q}} z 
{\generatorEvent{proc_{n+1}}} 
\wedge \inputVal z q e
\end{align*}

So, $z$ must be one of the procedures in the subpath $\left[ proc_1 , \ldots , proc_{n} \right]$.
Therefore, by Axiom 
\axiomHRef{help-focused::helped-are-writers}, property \eqref{eq::appendix::lin::sequential-recency},
and the inductive hypothesis, $H_n(q) = e$ must hold.

Therefore, the $proc_{n+1}$ step
must be the successful $\rdcssCasAlg$ step, which means 
$H_{n+1} = \mapExt {H_n} {\mapEntry {q} {e}}$.
Hence $H_{n+1}(p) = H_{n+1}(q) = e = v$.

\item Case $proc_{n+1}(in_{n+1}) = \rdcssWriteAlg(q,m)$. 
By definitions of $\writesAbs p$ and $\inputValName$, $p = q$ and $v = m$.

The only applicable transition for the $p_{n+1}$ step
is the $\rdcssWriteAlg$ step, which means 
$H_{n+1} = \mapExt {H_n} {\mapEntry {q} {m}}$.
Hence $H_{n+1}(p) = H_{n+1}(q) = m = v$.

\item Case $proc_{n+1}(in_{n+1}) = \rdcssReadAlg(q)$. 
This case is impossible, since reads are not elements of 
$\writesAbs p$.

\item Cases for $\rdcssCasCtlAlg$, $\rdcssWriteCtlAlg$, and $\rdcssReadCtlAlg$
are identical to the cases for $\rdcssCasAlg$, $\rdcssWriteAlg$, and $\rdcssReadAlg$,
respectively.

\end{itemize}
\end{itemize}
\end{itemize}
\end{prf}

\begin{lem}[Allocs lemma]
\label{lem::appendix::lin::rdcss-pointers-are-allocated}
Suppose hypotheses \ref{defn::appendix::lin::hypotheses-for-sequential-soundness-rdcss}.
Let $\leq$ be the linear order of Lemma \ref{lem::appendix::lin::existence-basic}.
Suppose $\mathcal P$ is a path of length $1 \natorderEqSymbol n \natorderEqSymbol \vert \closedEvent \vert$ in $Op^R$ that matches
$\leq$,
\[
H_0 = \emptyset \xrightarrow{proc_1(in_1)\ \langle out_1 \rangle} H_1 \xrightarrow{proc_2(in_2)\ \langle out_2 \rangle} 
\ldots \xrightarrow{proc_n(in_n)\ \langle out_n \rangle} H_n 
\]

Then, $p \in dom(H_n)$ if and only if 
there is $i \natorderEqSymbol n$, 
such that $\generatorEvent{proc_i} \in \allocsAbs p$.
\end{lem}

\begin{prf}
By induction on $n$.
\begin{itemize}
\item Case $n = 1$. 

\begin{itemize}
\item $\Rightarrow$. Let $p \in dom(H_1)$. 
Procedure $proc_1$ must be an alloc, 
since all other steps preserve the 
heap domain (i.e. it would be the case that 
$p \in dom(H_0) = \emptyset$ for non-alloc steps, which
is impossible).
Hence, $proc_1(in_1) = \rdcssAllocAlg(v,\_)$ for some $v$,
and $out_1 = p$, since $p$ is in the only pointer in the domain of $H_1$.

By Axiom \axiomHRef{help-focused::fin-predicate}, 
$\ETimeProp{\generatorEvent{proc_1}} \neq \bot$,
and since the path matches $\leq$, $p = out_1 = \outputProp{\generatorEvent{proc_1}}$.
Therefore, $\generatorEvent{proc_1} \in \allocsAbs p$ by definition of 
$\allocsAbs p$.

\item $\Leftarrow$. Suppose 
$\generatorEvent{proc_1} \in \allocsAbs p$. Therefore,
$p = \outputProp {\generatorEvent{proc_1}}$,
$\ETimeProp {\generatorEvent{proc_1}} \neq \bot$,
and $proc_1 = \rdcssAllocAlg$. 

Since the path matches $\leq$, $out_1 = \outputProp {\generatorEvent{proc_1}} = p$.
Therefore, the step creates pointer $p$ and $p \in dom(H_1)$.  
\end{itemize}

\item Inductive case. Let $n \natorderEqSymbolRight 1$ and 
suppose the statement holds for any path of length $n$.
Let $\mathcal P$ be a matching path of length $n+1$,
\[
H_0 = \emptyset \xrightarrow{proc_1(in_1)\ \langle out_1 \rangle} 
\ldots \xrightarrow{proc_n(in_n)\ \langle out_n \rangle} H_n \xrightarrow{proc_{n+1}(in_{n+1})\ \langle out_{n+1} \rangle} H_{n+1}
\]

\begin{itemize}

\item $\Rightarrow$. Let $p \in dom(H_{n+1})$. We consider the following cases,

\begin{itemize}
\item Case $p \in dom(H_n)$. By the inductive
hypothesis on the subpath $[proc_1,\ldots,proc_n]$, 
there is $proc_i$ such that $i \natorderEqSymbol n$ 
(hence, $i \natorderEqSymbol n+1$) such that 
$\generatorEvent{proc_i} \in \allocsAbs p$.

\item Case $p \notin dom(H_n)$. Procedure
$proc_{n+1}$ must be an alloc, since the rest
of the steps in the operational semantics preserve the heap domain,
and we know that $p \in dom(H_{n+1})$.
Hence, $proc_{n+1}(in_{n+1}) = \rdcssAllocAlg(v,\_)$ for some $v$,
and $out_{n+1} = p$, since an alloc step adds a single pointer to a heap and 
$p$ is a new pointer in the domain of $H_{n+1}$.

By Axiom \axiomHRef{help-focused::fin-predicate}, 
$\ETimeProp{\generatorEvent{proc_{n+1}}} \neq \bot$,
and since the path matches $\leq$, we have $p = out_{n+1} = \outputProp{\generatorEvent{proc_{n+1}}}$.
Therefore, $\generatorEvent{proc_{n+1}} \in \allocsAbs p$ by definition of 
$\allocsAbs p$.
\end{itemize}

\item $\Leftarrow$. Suppose there is $i \natorderEqSymbol n+1$ such 
that $\generatorEvent{proc_i} \in \allocsAbs p$. 

For the case $i \natorderSymbol n+1$ (hence, $i \natorderEqSymbol n$), the inductive 
hypothesis on the subpath $[proc_1,\ldots,proc_n]$ concludes $p \in dom(H_n)$. But the fact that 
steps in the operational semantics do not remove pointers implies that
$p \in dom(H_{n+1})$.

We now check the case $i = n + 1$. 
Since $\generatorEvent{proc_{n+1}} \in \allocsAbs p$,
we have $p = \outputProp {\generatorEvent{proc_{n+1}}}$,
$\ETimeProp {\generatorEvent{proc_{n+1}}} \neq \bot$,
and $proc_{n+1} = \rdcssAllocAlg$. 

Since the path matches $\leq$, we have $out_{n+1} = \outputProp {\generatorEvent{proc_{n+1}}} = p$.
Therefore, the step creates pointer $p$ and $p \in dom(H_{n+1})$. 
\end{itemize}
\end{itemize}
\end{prf}

With the writes and alloc lemmas, we can now prove the existence of a path.

\begin{lem}[Path Existence]
\label{lem::appendix::lin::rdcss-legality-lemma}
Suppose hypotheses \ref{defn::appendix::lin::hypotheses-for-sequential-soundness-rdcss}.
Let $\leq$ be the linear order of Lemma \ref{lem::appendix::lin::existence-basic}.
For any $1 \natorderEqSymbol n \natorderEqSymbol \lvert \closedEvent \rvert$, 
there is a path in $Op^R$ of length $n$ that matches $\leq$ and starts from the empty heap.
\end{lem}

\begin{prf}
By induction on $n$.

\begin{itemize}
\item Case $n = 1$. Denote by $x_1$ the first event in $\leq$. 

If $x_1$ is any of 
$\rdcssAlg$, $\rdcssCasAlg$, $\rdcssReadAlg$, $\rdcssCasCtlAlg$, and
$\rdcssReadCtlAlg$, Axiom \axiomHRef{help-focused::fin-predicate}
and the definition of the postcondition predicate force these events
to observe another event in at least one pointer. For example,
$\rdcssAlg(d)$ observes a $z_2$ in pointer $\pointTwo d$, and 
$\rdcssCasAlg(q,e,n)$ observes a $z$ in pointer $q$.
These observed events (like $z_2$ and $z$) 
must be in $\closedEvent$ because $\closedEvent$ is $\genVisSymbol$-downward closed
(Lemma \ref{eq::appendix::lin::committed-downward-closed}).
Therefore, there must exist an event occurring \emph{before}
$x_1$ in $\leq$, which is impossible.

Hence, $x_1$ must be either a $\rdcssWriteAlg$, $\rdcssWriteCtlAlg$, or
$\rdcssAllocAlg$. However, if $x_1$ is either  
$\rdcssWriteAlg$ or $\rdcssWriteCtlAlg$, Axiom \axiomHRef{help-focused::all-writers-are-willing-helpers}
forces the existence of an alloc $z$ such that $\visSep {\pointerIndx p} z {x_1}$ for some
$p$ (equality cannot hold because $x_1$ is not an alloc).
This $z$ must be in $\closedEvent$ because $\closedEvent$ is $\genVisSymbol$-downward closed
(Lemma \ref{eq::appendix::lin::committed-downward-closed}).
In other words, $z$ must occur \emph{before} $x_1$ in $\leq$, which
is impossible. 

Hence, $x_1$ must be an alloc of the form $\rdcssAllocAlg(v,\_)$
for some $v$. 

Now, Axiom \axiomHRef{help-focused::fin-predicate}
concludes $\ETimeProp{x_1} \neq \bot$. Define $q \defini \outputProp{x_1}$,
and $H_1 \defini \mapExt {\emptyset} {\mapEntry q v}$.
Then, $\emptyset \xrightarrow{\rdcssAllocAlg(v,\_)\ \left\langle q \right\rangle} H_1$ is a path
of length $1$ starting from the empty heap that matches $\leq$.

\item Inductive case. Let $n \natorderEqSymbolRight 1$. Suppose 
$n + 1 \natorderEqSymbol \vert \closedEvent \vert$. 
We need to show that there is a matching path of length $n+1$.
Since $1 \natorderEqSymbol n \natorderEqSymbol \vert \closedEvent \vert$, 
the inductive hypothesis implies that there is a matching path of length $n$,
\begin{align}
\label{eqn::appendix::lin::rdcss-legality-path-1}
H_0 = \emptyset \xrightarrow{proc_1(in_1)\ \langle out_1 \rangle} H_1 
\xrightarrow{proc_2(in_2)\ \langle out_2 \rangle} 
\ldots \xrightarrow{proc_n(in_n)\ \langle out_n \rangle} H_n
\end{align}

We need to show that we can extend this path with a matching $n+1$ step 
for the $n+1$ event in $\leq$. Denote the $n+1$ event in $\leq$ as $x_{n+1}$.

We do a case analysis on $x_{n+1}$,
\begin{itemize}
\item Case $x_{n+1} = \rdcssAlg(d)$. 

By Axiom 
\axiomHRef{help-focused::fin-predicate}, we have for some $v$,
\begin{align}
\label{eqn::appendix::lin::rdcss-post}
\begin{split}
\exists z_2.\ \visObs {\pointerIndx {\pointTwo d}} 
{z_2} {x_{n+1}} \wedge 
\inputVal {z_2} {\pointTwo d} v \wedge {} \\
(v = \expTwo d \implies \exists z_1.\ \visObs 
{\pointerIndx {\pointOne d}} {z_1} 
{x_{n+1}})
\end{split}
\end{align} 

Since path \eqref{eqn::appendix::lin::rdcss-legality-path-1} 
matches $\leq$, \eqref{eqn::appendix::lin::rdcss-post} implies 
that event $z_2$ must be one of the procedures in the path,
since it occurs before $x_{n+1}$ in $\leq$,
i.e. $z_2 = \generatorEvent{proc_i}$ for some $i$.

Notice that by Axiom 
\axiomHRef{help-focused::helped-are-writers},
$\generatorEvent{proc_i} \in \writesAbs {\pointTwo d}$.
Also, for every $j \natorderSymbolRight i$,  
$\generatorEvent{proc_j} \notin \writesAbs {\pointTwo d}$ 
by property \eqref{eq::appendix::lin::sequential-recency}, since 
$\visObs {\pointerIndx {\pointTwo d}} {z_2} 
{x_{n+1}}$. 
Therefore, by Lemma 
\ref{lem::appendix::lin::rdcss-writers-change-heap} 
applied on path 
\eqref{eqn::appendix::lin::rdcss-legality-path-1}, we have
$H_n(\pointTwo d) = v$.

We now case analyze $v$. 
\begin{itemize} 
\item Case $v = \expTwo d$. By 
\eqref{eqn::appendix::lin::rdcss-post}, 
$\visObs {\pointerIndx {\pointOne d}} {z_1} 
{x_{n+1}}$.
Hence, by Axiom 
\axiomHRef{help-focused::helped-are-writers}, 
$z_1 \in \writesAbs {\pointOne d}$.
By Lemma \ref{lem::appendic::lin::rdcss-writers-have-input-maps}, 
$\inputVal {z_1} {\pointOne d} {v'}$ for some $v'$.

By a similar argument as for 
$\visObs {\pointerIndx {\pointTwo d}} {z_2} 
{x_{n+1}}$ above, we will have
$H_n(\pointOne d) = v'$. 

We now case analyze $v'$.
\begin{itemize}
\item Case $v' = \expOne d$. Define $H_{n+1} \defini 
\mapExt {H_n} {\mapEntry {\pointTwo d} {\newTwo d}}$.
Hence, $H_n \xrightarrow{\rdcssAlg(d)\ \left\langle v \right\rangle} H_{n+1}$ is a valid step
consisting on the successful $\rdcssAlg$ step, since $v = \expTwo d$ and $v' = \expOne d$.

In case $x_{n+1} \in \terminatedEvent$, Axiom \axiomHRef{help-focused::fin-predicate} 
also concludes $v = \outputProp{x_{n+1}}$. Hence, the step
matches $\leq$.

\item Case $v' \neq \expOne d$. Define $H_{n+1} \defini H_n$.
Hence, $H_n \xrightarrow{\rdcssAlg(d)\ \left\langle v \right\rangle} H_{n+1}$ is a valid step
consisting on a failing $\rdcssAlg$ step, since $v = \expTwo d$ but $v' \neq \expOne d$.

In case $x_{n+1} \in \terminatedEvent$, Axiom \axiomHRef{help-focused::fin-predicate} 
also concludes $v = \outputProp{x_{n+1}}$. Hence, the step
matches $\leq$.
\end{itemize}

\item Case $v \neq \expTwo d$. Define $H_{n+1} \defini H_n$.
Hence, $H_n \xrightarrow{\rdcssAlg(d)\ \left\langle v \right\rangle} H_{n+1}$ is a valid step
consisting on a failing $\rdcssAlg$ step, since $v \neq \expTwo d$.

In case $x_{n+1} \in \terminatedEvent$, Axiom \axiomHRef{help-focused::fin-predicate} 
also concludes $v = \outputProp{x_{n+1}}$. Hence, the step
matches $\leq$.
\end{itemize}

\item Case $x_{n+1} = \rdcssAllocAlg(v,\_)$.

By Axiom \axiomHRef{help-focused::fin-predicate}, $\ETimeProp{x_{n+1}} \neq \bot$.
Define $q \defini \outputProp{x_{n+1}}$.

We claim $q \notin dom(H_n)$. For suppose 
$q \in dom(H_n)$. Then, by Lemma 
\ref{lem::appendix::lin::rdcss-pointers-are-allocated} applied
on path \eqref{eqn::appendix::lin::rdcss-legality-path-1}, there is
$proc_i$ such that $i \natorderEqSymbol n$ and 
$\generatorEvent{proc_i} \in \allocsAbs q$, which means
$\generatorEvent{proc_i} < x_{n+1}$ since
the path matches $\leq$.
But by Axiom 
\axiomHRef{help-focused::allocs}, 
$\generatorEvent{proc_i} = x_{n+1}$ (Contradiction).

Define $H_{n+1} \defini 
\mapExt {H_n} {\mapEntry q v}$.
Hence, $H_n \xrightarrow{\rdcssAllocAlg(v,\_)\ \left\langle q \right\rangle} H_{n+1}$ is a valid 
alloc step that matches $\leq$.

\item Case $x_{n+1} = \rdcssCasAlg(q,e,m)$. 

By Axiom 
\axiomHRef{help-focused::fin-predicate}, we have for some $v$,
\begin{align*}
\exists z.\ \visObs {\pointerIndx q} 
z {x_{n+1}} \wedge \inputVal z q v 
\end{align*} 

By a similar argument as in the $\rdcssAlg$ case, 
$H_n(q) = v$. 

We now case analyze $v$. 
\begin{itemize} 
\item Case $v = e$. Define $H_{n+1} \defini 
\mapExt {H_n} {\mapEntry q m}$.
Hence, $H_n \xrightarrow{\rdcssCasAlg(q,e,m)\ \left\langle v \right\rangle} H_{n+1}$ 
is a valid step corresponding to the successful $\rdcssCasAlg$ step.

In case $x_{n+1} \in \terminatedEvent$, Axiom \axiomHRef{help-focused::fin-predicate} 
also concludes $v = \outputProp{x_{n+1}}$. Hence, the step
matches $\leq$.

\item Case $v \neq e$. Define $H_{n+1} \defini H_n$.
Hence, $H_n \xrightarrow{\rdcssCasAlg(q,e,m)\ \left\langle v \right\rangle} H_{n+1}$ is a valid step 
corresponding to the failing $\rdcssCasAlg$ step.

In case $x_{n+1} \in \terminatedEvent$, Axiom \axiomHRef{help-focused::fin-predicate} 
also concludes $v = \outputProp{x_{n+1}}$. Hence, the step
matches $\leq$.
\end{itemize}

\item Case $x_{n+1} = \rdcssWriteAlg(q,m)$. 

By definition, we have $x_{n+1} \in 
\writesAbs q$. 
Hence, by Axiom 
\axiomHRef{help-focused::all-writers-are-willing-helpers} 
there is $z \in \allocsAbs q$ such that 
$\visSepEq {\pointerIndx q} z {x_{n+1}}$.
Case $z = x_{n+1}$ cannot hold, since $x_{n+1} \notin \allocsAbs q$
by definition of $\allocsAbs q$. Hence,
$\visSep {\pointerIndx q} z {x_{n+1}}$ must be true.
In addition, $z$ must be in $\closedEvent$ because $\closedEvent$ is $\genVisSymbol$-downward closed
(Lemma \ref{eq::appendix::lin::committed-downward-closed}).

Since the path matches $\leq$, $z$ must be one of the procedures
in the path, i.e., $\generatorEvent{proc_i} = z$ for some 
$i \natorderEqSymbol n$.
By Lemma \ref{lem::appendix::lin::rdcss-pointers-are-allocated}
applied on path
\eqref{eqn::appendix::lin::rdcss-legality-path-1},
$q \in dom(H_n)$ holds.

Define $H_{n+1} \defini \mapExt {H_n} {\mapEntry q m}$.
Hence, $H_n \xrightarrow{\rdcssWriteAlg(q,m)\ \left\langle tt \right\rangle} H_{n+1}$ is a valid step 
corresponding to the $\rdcssWriteAlg$ step.

In case $x_{n+1} \in \terminatedEvent$, Axiom \axiomHRef{help-focused::fin-predicate} 
also concludes $\outputProp{x_{n+1}} = tt$. Hence, the step
matches $\leq$.

\item Case $x_{n+1} = \rdcssReadAlg(q)$. 

By Axiom 
\axiomHRef{help-focused::fin-predicate}, we have for some $v$,
\begin{align*}
\exists z.\ \visObs {\pointerIndx q} 
z {x_{n+1}} \wedge \inputVal z q v 
\end{align*} 

By a similar argument as in the previous cases,
$H_n(q) = v$. 

Define $H_{n+1} \defini H_n$.
Hence, $H_n \xrightarrow{\rdcssReadAlg(q)\ \left\langle v \right\rangle} H_{n+1}$ is a valid step 
corresponding to the $\rdcssReadAlg$ step.

In case $x_{n+1} \in \terminatedEvent$, Axiom \axiomHRef{help-focused::fin-predicate} 
also concludes $v = \outputProp{x_{n+1}}$. Hence, the step
matches $\leq$.

\item Cases for $\rdcssCasCtlAlg$, $\rdcssWriteCtlAlg$, and $\rdcssReadCtlAlg$
are identical to the cases for $\rdcssCasAlg$, $\rdcssWriteAlg$, and $\rdcssReadAlg$,
respectively.
\end{itemize}
\end{itemize}
\end{prf}

\begin{lem}[Sequential Soundness]
\label{lem::appendix::lin::linear-order-rdcss-sequential-soundness}
Suppose hypotheses \ref{defn::appendix::lin::hypotheses-for-sequential-soundness-rdcss}.
The linear order $\leq$ of Lemma \ref{lem::appendix::lin::existence-basic} is sequentially sound.
\end{lem}

\begin{prf}
If $\closedEvent = \emptyset$, then the empty path matches $\leq$. 
If $\closedEvent \neq \emptyset$, then Lemma \ref{lem::appendix::lin::rdcss-legality-lemma}
applied with $n = \vert \closedEvent \vert$ ensures the existence of a matching path
for $\leq$.
\end{prf}

\begin{thm}
\label{thm::appendix::lin::rdcss-linearizability-from-vis-structure}
Given an implementation of the RDCSS, suppose that for any set of
abstract events $\absEvent$ generated from an arbitrary execution
history in the implementation, there are relations
$\visObsSymbol{\pointerIndx p}$,
$\visSepSymbol{\pointerIndx p}$ such that
$\RDCSSFam(\visObsSymbol {\pointerIndx p},
{\visSepSymbol{\pointerIndx p}})$ is valid. Then, the
implementation is linearizable.
\end{thm}

\begin{prf}
  Let $\absEvent$ be a set of events generated from an arbitrary
  execution history in the implementation. 
  From the hypothesis, $\RDCSSFam(\visObsSymbol {\pointerIndx p},
  {\visSepSymbol{\pointerIndx p}})$ is valid 
  for some $\visObsSymbol {\pointerIndx p}$ and
  $\visSepSymbol{\pointerIndx p}$.
  Let $\leq$ be the linear order of Lemma
  \ref{lem::appendix::lin::existence-basic}.  Then, we take the general
  visibility relation $\genVisSymbol$ and $\leq$ to be the
  relations required by the definition of linearizability.  
  By construction, $\leq$ respects both $\genVisSymbol$ and
  $\precedesAbsSymbol$. Also,
  $\leq$ is sequentially sound by Lemma
  \ref{lem::appendix::lin::linear-order-rdcss-sequential-soundness}.
\end{prf}

\subsection{Linearizability Proof for the MCAS Data Structure}
\label{sect::appendix::lin::MCAS}

This section focuses on proving, for the MCAS data structure, sequential soundness 
of the linear order built in Lemma \ref{lem::appendix::lin::existence-basic}.
This section makes use of Definition \ref{defn-visibility-specification}.

\begin{defn}[Exportable Procedures in the MCAS Data Structure]
The following are the exportable procedures in the MCAS Data Structure,
\begin{center}
\begin{tabular}{ll}
$\mcasAlg(\listvar u: \textsc{list}\,\textsc{update\_entry})$ & (MCAS for list of entries $\listvar u \neq \emptyset$) \\
$\mcasAllocAlg(v: \ValType)$ & (Allocation of pointer) \\
$\mcasWriteAlg(p: \DataPtType,\ v: \ValType)$ & (Write procedure) \\
$\mcasReadAlg(p: \DataPtType)$ & (Read procedure) \\
\end{tabular}
\end{center}
\end{defn}

As we did for the RDCSS data structure, we define the writer predicate,

\begin{defn}[Writer Predicate for the MCAS Data Structure]
\label{defn::appendix::lin::mcas-input-map}
The \emph{writer predicate} $\inputValName \subseteq \absEvent \times \PtsType \times \ValType$, 
denoted $\inputVal x p v$, is defined by cases,
\begin{align*}
\inputVal x p v \defini  
& (\exists \listvar u,i \in \listvar u.\ x = \mcasAlg(\listvar u) \wedge v = \newGenEntry i \wedge p = \pointGenEntry i) \vee {} \\
& (x = \mcasAllocAlg(v) \wedge \ETimeProp x \neq \bot \wedge p = \outputProp x) \vee {} \\
& x = \mcasWriteAlg(p, v)\\
\end{align*}
Predicate $\inputVal x p v$ returns the value $v$
in $\inProp x$ that $x$ will attempt to write into pointer $p$.
\end{defn}

We now define the specification for the MCAS data structure.

\begin{defn}[Specification for the MCAS Data Structure]
\label{defn::appendix::lin::mcas-vis-spec}
Given relations $\visObsSymbol {\pointerIndx p}$, $\visSepSymbol {\pointerIndx p}$,
we define the specification for the MCAS data structure, denoted 
$\MCASFam(\visObsSymbol {\pointerIndx p}, \visSepSymbol {\pointerIndx p})$, as follows,
\begin{align*}
x \in \allocsAbs p & \defini x = \mcasAllocAlg(\_) \wedge \ETimeProp x \neq \bot \wedge p = \outputProp x \\
x \in \writesAbs p & \defini 
\begin{cases}
(\exists j \in \listvar u.\ p = \pointGenEntry j) \wedge (\forall i \in \listvar u.\ \exists z.\ \visObs {\pointerIndx {\pointGenEntry i}} z x \wedge 
\inputVal z {\pointGenEntry i} {\expGenEntry i}) & \text{if } x = \mcasAlg(\listvar u) \\
\ETimeProp x \neq \bot \wedge p = \outputProp x & \text{if } x = \mcasAllocAlg(\_) \\
p = q & \text{if } x = \mcasWriteAlg(q, \_) \\
\end{cases} \\
\postPred x v & \defini
\begin{cases}
v \in \BoolType \wedge {} & \text{if } x = \mcasAlg(\listvar u) \\
\quad \begin{cases}
\forall i \in \listvar u.\ \exists z.\ \visObs {\pointerIndx {\pointGenEntry i}} z x \wedge 
\inputVal z {\pointGenEntry i} {\expGenEntry i} & \text{if } v = true \\
\begin{aligned}
\exists i \in \listvar u,v' \in \ValType.\ \exists z.\ \visObs {\pointerIndx {\pointGenEntry i}} z x \wedge {}\\
\inputVal z {\pointGenEntry i} {v'} \wedge v' \neq {\expGenEntry i}
\end{aligned}
 & \text{if } v = false \\
\end{cases} &  \\
\ETimeProp x \neq \bot \wedge v = \outputProp x & \text{if } x = \mcasAllocAlg(\_) \\
v = \unitValue & \text{if } x = \mcasWriteAlg(\_, \_) \\
\exists z.\ \visObs {\pointerIndx q} z x \wedge \inputVal z q v & \text{if } x = \mcasReadAlg(q) \\
\end{cases}
\end{align*}
\end{defn}

We have the following immediate lemma,

\begin{lem}
\label{lem::appendic::lin::mcas-writers-have-input-maps}
If $x \in \writesAbs p$, then there is 
$v \in \ValType$ such that $\inputVal x p v$.
\end{lem}

\begin{prf}
By cases on $x$.
\begin{itemize}
\item Case $x = \mcasAlg(\listvar u)$. We have $p = \pointGenEntry j$ for some 
$j \in \listvar u$. And $\inputVal x p {\newGenEntry j}$ follows by definition of $\inputValName$.
\item Cases when $x$ is one of $\mcasAllocAlg(v)$ or $\mcasWriteAlg(q,v)$ are as in the proof of Lemma
\ref{lem::appendic::lin::rdcss-writers-have-input-maps}.
\end{itemize}
\end{prf}

Towards defining the notion of sequential soundness for a linear order over 
events in MCAS, we need to describe what it means for events to execute.
For that matter, we define the following state-based operational semantics,
so that intuitively, events execute by building 
a path in the operational semantics.

\begin{defn}[Operational semantics for events in the MCAS Data Structure]
\label{defn::appendix::lin::mcas-legality}
Let $Op^M$ denote the operational semantics generated by 
the following base steps, where states are heaps and labels
are of the form ``$proc(in)\ \langle out \rangle$''
where $proc$ is the procedure name, $in$ the procedure's input, and $out$ the procedure's output.
\begin{itemize}
\item $H \xrightarrow{\mcasAlg(\listvar u)\ \langle false \rangle} H$, 
if for some $j \in \listvar u$, $H(\pointGenEntry j) \neq \expGenEntry j$.
\item $H \xrightarrow{\mcasAlg(\listvar u)\ \langle true \rangle} \mapExtThree{H}{\mapEntry {\pointGenEntry j} {\newGenEntry j}}{j \in \listvar u}$, 
if for every $j \in \listvar u$, $H(\pointGenEntry j) = \expGenEntry j$.
\item $H \xrightarrow{\mcasAllocAlg(v)\ \langle p \rangle} \mapExt H {\mapEntry p v}$, 
if $p \notin dom(H)$.
\item $H \xrightarrow{\mcasWriteAlg(p,v)\ \langle \unitValue \rangle} \mapExt H {\mapEntry p v}$,
if $p \in dom(H)$.
\item $H \xrightarrow{\mcasReadAlg(p)\ \langle v \rangle} H$, 
if $H(p) = v$.
\end{itemize}
\end{defn}

We now define the notion of sequential soundness of a linear order $\leq$.

\begin{defn}[Sequential soundness of a linear order]
Given a linear order $\leq$ over a set of events, and a path $P$ in the operational semantics $Op^M$, 
we say that $P$ \emph{matches} $\leq$ if for every step $i$ in $P$, 
\begin{itemize}
\item The $i$-th procedure's name in $P$ equals the procedure's name of the $i$-th event in $\leq$.
\item The $i$-th procedure's input in $P$ equals the procedure's input of the $i$-th event in $\leq$.
\item If the $i$-th event in $\leq$ is terminated, then the 
$i$-th procedure's output in $P$ equals the $i$-th event output in $\leq$.
\end{itemize}
We say that $\leq$ is \emph{sequentially sound} if there is a matching path in $Op^M$, starting from the empty heap and
having as many steps as the number of events in the domain of $\leq$.
\end{defn}

\begin{notation}
If path $P$ matches $\leq$, and $H \xrightarrow{proc(in)\ \langle out \rangle} H'$ is a step in $P$,
we denote by $\generatorEvent {proc}$ the corresponding event in $\leq$. 
\end{notation}

To prove sequential soundness, we require two lemmas.
The first lemma, called writes
lemma, expresses that successful writes modify the heap 
and as long as there are no successful
writes afterwards, such changes remain intact.
The second lemma, called allocs lemma, states that allocs create
pointers in the heap.

\begin{defn}[Section Hypotheses]
\label{defn::appendix::lin::hypotheses-for-sequential-soundness-mcas}
To shorten the statements of lemmas from \ref{lem::appendix::lin::mcas-writers-change-heap} 
to \ref{lem::appendix::lin::linear-order-mcas-sequential-soundness}, the
following hypotheses will apply to those lemmas,
\begin{itemize}
\item $\absEvent$ is an arbitrary set of events.
\item $\visObsSymbol{\pointerIndx p}$ and $\visSepSymbol{\pointerIndx p}$
are two arbitrary pointer-indexed binary relations over $\absEvent$.
\item $\genStructName{\MCAS}(\visObsSymbol{\pointerIndx p}, \visSepSymbol{\pointerIndx p})$
is valid, where $\genStructName{\MCAS}$ is the specification for MCAS defined in \ref{defn::appendix::lin::mcas-vis-spec}.
\end{itemize}
\end{defn}

We can now prove the writes lemma,

\begin{lem}[Writes lemma]
\label{lem::appendix::lin::mcas-writers-change-heap}
Suppose hypotheses \ref{defn::appendix::lin::hypotheses-for-sequential-soundness-mcas}.
Let $\leq$ be the linear order of Lemma \ref{lem::appendix::lin::existence-basic}.
Suppose $\mathcal P$ is a path of length $1 \natorderEqSymbol n \natorderEqSymbol \vert \closedEvent \vert$ in $Op^M$ that matches
$\leq$,
\[
H_0 = \emptyset \xrightarrow{proc_1(in_1)\ \langle out_1 \rangle} H_1 \xrightarrow{proc_2(in_2)\ \langle out_2 \rangle} 
\ldots \xrightarrow{proc_n(in_n)\ \langle out_n \rangle} H_n 
\]

Let $1 \natorderEqSymbol i \natorderEqSymbol n$.
If $\generatorEvent {proc_i} \in \writesAbs p$
and $\inputVal {\generatorEvent {proc_i}} p v$ 
and for all $j \natorderSymbolRight i$, 
$\generatorEvent {proc_j} \notin \writesAbs p$, 
then $H_n(p) = v$.
\end{lem}

\begin{prf}
By induction on $n$.

\begin{itemize}
\item Case $n = 1$. 

Suppose $\generatorEvent {proc_1} \in \writesAbs p$ and 
$\inputVal {\generatorEvent {proc_1}} p v$. 

All non-alloc steps at $proc_1$ in the operational semantics 
(including successful $\mcasAlg(\listvar u)$ since $\listvar u$ is not empty) 
require that some 
pointer exists at $H_0 = \emptyset$, which is impossible.
Therefore,
the only applicable case is when $proc_1$ is an alloc.  
Hence, $proc_1(in_1) = \rdcssAllocAlg(v',\_)$
for some $v'$.

But $p = \outputProp {\generatorEvent {proc_1}} = out_1$, since
$\generatorEvent {proc_1} \in \writesAbs p$, 
$proc_1$ matches $\generatorEvent {proc_1}$,
and $\generatorEvent {proc_1} \in \terminatedEvent$ by definition of
$\writesAbs p$. 
Also, $v' = v$ by definition of $\inputValName$.

Therefore, $H_{1}(p) = v$,
since the alloc step creates $out_1 = p$ with initial value $v' = v$.

\item Inductive case. Let $n \natorderEqSymbolRight 1$ and 
suppose the statement holds for any path of length $n$.
Let $\mathcal P$ be a matching path of length $n+1$,
\[
H_0 = \emptyset \xrightarrow{proc_1(in_1)\ \langle out_1 \rangle} 
\ldots \xrightarrow{proc_n(in_n)\ \langle out_n \rangle} H_n \xrightarrow{proc_{n+1}(in_{n+1})\ \langle out_{n+1} \rangle} H_{n+1}
\]

Let $1 \natorderEqSymbol i \natorderEqSymbol n+1$
and $\generatorEvent {proc_i} \in \writesAbs p$
and $\inputVal {\generatorEvent{proc_i}} p v$ 
and for all $j \natorderSymbolRight i$, 
$\generatorEvent{proc_j} \notin \writesAbs p$. 
We need to show that $H_{n+1}(p) = v$.

\begin{itemize}
\item Case $i \natorderSymbol n + 1$, hence $i \natorderEqSymbol n$.
Since $\left[ proc_1 , \ldots , proc_n \right]$ is a subpath of length $n$ 
that matches $\leq$, the inductive hypothesis concludes
$H_n(p) = v$. 

By hypothesis,
$\generatorEvent{proc_{n+1}} \notin \writesAbs p$ must hold.
We do a case analysis on $proc_{n+1}$.

\begin{itemize}
\item Case $proc_{n+1}(in_{n+1}) = \mcasAlg(\listvar u)$. 
Since $\generatorEvent{proc_{n+1}} \notin \writesAbs p$, we have either 
$\forall k \in \listvar u.\ p \neq \pointGenEntry k$ or, 
for some $m \in \listvar u$,
\begin{align}
\label{eqn::appendix::lin::mcas-non-writer}
\forall z.\ \visObs {\pointerIndx {\pointGenEntry m}} {z} {\generatorEvent{proc_{n+1}}} \implies  
\ninputVal z {\pointGenEntry m} {\expGenEntry m}
\end{align} 

The case $\forall k \in \listvar u.\ p \neq \pointGenEntry k$ implies 
$H_{n+1}(p) = v$
since the $\mcasAlg$ steps only change pointers in 
the set $\{ \pointGenEntry k \mid k \in \listvar u \}$. 
We now focus on the case 
\eqref{eqn::appendix::lin::mcas-non-writer}.

By Axiom 
\axiomHRef{help-focused::fin-predicate}, we have for some $v' \in \BoolType$,
\begin{align}
\label{eqn::appendix::lin::mcas-post-1}
\begin{cases}
\forall k \in \listvar u.\ \exists z.\ \visObs {\pointerIndx {\pointGenEntry k}} z {\generatorEvent{proc_{n+1}}} \wedge 
\inputVal z {\pointGenEntry k} {\expGenEntry k} & \text{if } v' = true \\
\begin{aligned}
\exists k \in \listvar u,v'' \in \ValType.\ \exists z.\ \visObs {\pointerIndx {\pointGenEntry k}} z {\generatorEvent{proc_{n+1}}} \wedge {}\\ 
\inputVal z {\pointGenEntry k} {v''} \wedge v'' \neq {\expGenEntry k}
\end{aligned} & \text{if } v' = false \\
\end{cases}
\end{align} 

If $v' = true$, by applying \eqref{eqn::appendix::lin::mcas-post-1}
with $m$, it would contradict 
\eqref{eqn::appendix::lin::mcas-non-writer}.
Therefore, $v' = false$. Hence, for some $k \in \listvar u$, $z$ and $v''$,
\begin{align}
\label{eqn::appendix::lin::mcas-post-2}
\visObs {\pointerIndx {\pointGenEntry k}} z {\generatorEvent{proc_{n+1}}} \wedge 
\inputVal z {\pointGenEntry k} {v''} \wedge v'' \neq {\expGenEntry k}
\end{align}

So, by Axiom \axiomHRef{help-focused::helped-are-writers}, 
$z \in \writesAbs {\pointGenEntry k}$. 

We know that $\left[ proc_1 , \ldots , proc_{n} \right]$ is a subpath 
of length $n$ that matches $\leq$, and $z$ must
appear before $\generatorEvent{proc_{n+1}}$ in $\leq$,
since $\leq$ contains $\genVisSymbol$.
So, there is $proc_l$ such that $\generatorEvent{proc_{l}} = z$,
for some $l \natorderEqSymbol n$. 
 
By Axiom 
\axiomHRef{help-focused::helped-are-writers} and
property \eqref{eq::appendix::lin::sequential-recency},
we also have the following facts:
$z \in \writesAbs {\pointGenEntry k}$,
and for all $l \natorderSymbol j \natorderEqSymbol n$,
$\generatorEvent {proc_{j}} \notin \writesAbs {\pointGenEntry k}$.

Therefore, by the inductive hypothesis applied on the subpath $\left[ proc_1 , \ldots , proc_{n} \right]$
and with $proc_l$, we must have $H_n(\pointGenEntry k) = v''$, where $v'' \neq \expGenEntry k$.

Hence, $proc_{n+1}$ must be a failing $\mcasAlg$ step in the operational semantics.
This means that if either $p = \pointGenEntry k$
or not, we will have $H_{n+1}(p) = v$,
since $H_n(p) = v$ and the $n+1$ step does not modify pointers.

\item Case $proc_{n+1}(in_{n+1}) = \mcasAllocAlg(v)$. 
Identical to the case for $\rdcssAllocAlg$ in Lemma \ref{lem::appendix::lin::rdcss-writers-change-heap}.

\item Case $proc_{n+1}(in_{n+1}) = \mcasWriteAlg(r,m)$. 
Identical to the case for $\rdcssWriteAlg$ in Lemma \ref{lem::appendix::lin::rdcss-writers-change-heap}.

\item Case $proc_{n+1}(in_{n+1}) = \rdcssReadAlg(r)$. 
Identical to the case for $\rdcssReadAlg$ in Lemma \ref{lem::appendix::lin::rdcss-writers-change-heap}.
\end{itemize}

\item Case $i = n+1$. So, we have $\generatorEvent{proc_{n+1}} \in 
\writesAbs p$ and
$\inputVal {\generatorEvent{proc_{n+1}}} p v$. 
We do a case analysis on
$proc_{n+1}$.

\begin{itemize}
\item Case $proc_{n+1}(in_{n+1}) = \mcasAlg(\listvar u)$. 

By definitions of $\writesAbs p$
and $\inputValName$, 
we have for some $l \in \listvar u$, $p = \pointGenEntry l$, 
$v = \newGenEntry l$, and
\begin{align*}
\forall k \in \listvar u.\ \exists z.\ \visObs {\pointerIndx {\pointGenEntry k}} z 
{\generatorEvent{proc_{n+1}}} \wedge \inputVal z {\pointGenEntry k} {\expGenEntry k}
\end{align*}

We claim that for every $k \in \listvar u$,  
$H_n(\pointGenEntry k) = \expGenEntry k$. 

Let $k \in \listvar u$.

By a similar argument as the case for $i \natorderSymbol n + 1$, event $z$ must be one of the 
procedures in the subpath $\left[ proc_1 , \ldots , proc_{n} \right]$.
Therefore, by Axiom 
\axiomHRef{help-focused::helped-are-writers}, property \eqref{eq::appendix::lin::sequential-recency},
and the inductive hypothesis, $H_n(\pointGenEntry k) = \expGenEntry k$ must hold.

This proves the claim.

Therefore, the $proc_{n+1}$ step
must be the successful $\mcasAlg$ step, which means 
$H_{n+1}(\pointGenEntry k) = \newGenEntry k$, for every $k \in \listvar u$.
In particular,
$H_{n+1}(p) = H_{n+1}(\pointGenEntry l) = \newGenEntry l = v$.

\item Case $proc_{n+1}(in_{n+1}) = \mcasAllocAlg(v)$. 
Identical to the case for $\rdcssAllocAlg$ in Lemma \ref{lem::appendix::lin::rdcss-writers-change-heap}.

\item Case $proc_{n+1}(in_{n+1}) = \mcasWriteAlg(q,m)$. 
Identical to the case for $\rdcssWriteAlg$ in Lemma \ref{lem::appendix::lin::rdcss-writers-change-heap}.

\item Case $proc_{n+1}(in_{n+1}) = \mcasReadAlg(q)$. 
Identical to the case for $\rdcssReadAlg$ in Lemma \ref{lem::appendix::lin::rdcss-writers-change-heap}.
\end{itemize}
\end{itemize}
\end{itemize}
\end{prf}

\begin{lem}[Allocs lemma]
\label{lem::appendix::lin::mcas-pointers-are-allocated}
Suppose hypotheses \ref{defn::appendix::lin::hypotheses-for-sequential-soundness-mcas}.
Let $\leq$ be the linear order of Lemma \ref{lem::appendix::lin::existence-basic}.
Suppose $\mathcal P$ is a path of length $1 \natorderEqSymbol n \natorderEqSymbol \vert \closedEvent \vert$ in $Op^M$ that matches
$\leq$,
\[
H_0 = \emptyset \xrightarrow{proc_1(in_1)\ \langle out_1 \rangle} H_1 \xrightarrow{proc_2(in_2)\ \langle out_2 \rangle} 
\ldots \xrightarrow{proc_n(in_n)\ \langle out_n \rangle} H_n 
\]

Then, $p \in dom(H_n)$ if and only if 
there is $i \natorderEqSymbol n$, 
such that $\generatorEvent{proc_i} \in \allocsAbs p$.
\end{lem}

\begin{prf}
Identical to the proof of Lemma
\ref{lem::appendix::lin::rdcss-pointers-are-allocated}, but using $\mcasAllocAlg$ instead of $\rdcssAllocAlg$.
\end{prf}

\begin{lem}[Path Existence]
\label{lem::appendix::lin::mcas-legality-lemma}
Suppose hypotheses \ref{defn::appendix::lin::hypotheses-for-sequential-soundness-mcas}.
Let $\leq$ be the linear order of Lemma \ref{lem::appendix::lin::existence-basic}.
For any $1 \natorderEqSymbol n \natorderEqSymbol \lvert \closedEvent \rvert$, 
there is a path in $Op^M$ of length $n$ that matches $\leq$ and starts from the empty heap.
\end{lem}

\begin{prf}
By induction on $n$.

\begin{itemize}
\item Case $n = 1$. Denote by $x_1$ the first event in $\leq$. 

If $x_1$ is any of 
$\mcasReadAlg$ and $\mcasAlg(\listvar u)$ (since $\listvar u$ is not empty), 
Axiom \axiomHRef{help-focused::fin-predicate}
and the definition of the postcondition predicate force these events
to observe another event in at least one pointer. For example,
$\mcasAlg(\listvar u)$ observes a $z$ in every pointer in the non-empty list $\listvar u$, and 
$\mcasReadAlg(q)$ observes a $z$ in pointer $q$.
This $z$ must be in $\closedEvent$ because $\closedEvent$ is $\genVisSymbol$-downward closed
(Lemma \ref{eq::appendix::lin::committed-downward-closed}).
Therefore, there must exist an event occurring \emph{before}
$x_1$ in $\leq$, which is impossible.

Hence, $x_1$ must be either a $\mcasWriteAlg$ or 
$\mcasAllocAlg$. However, if $x_1$ is   
$\mcasWriteAlg$, Axiom \axiomHRef{help-focused::all-writers-are-willing-helpers}
forces the existence of an alloc $z$ such that $\visSep {\pointerIndx p} z {x_1}$ for some
$p$ (equality cannot hold because $x_1$ is not an alloc).
This $z$ must be in $\closedEvent$ because $\closedEvent$ is $\genVisSymbol$-downward closed
(Lemma \ref{eq::appendix::lin::committed-downward-closed}).
In other words, $z$ must occur \emph{before} $x_1$ in $\leq$, which
is impossible. 

Hence, $x_1$ must be an alloc of the form $\mcasAllocAlg(v)$ for some $v$.
Axiom \axiomHRef{help-focused::fin-predicate}
concludes $\ETimeProp{x_1} \neq \bot$. Define $q \defini \outputProp{x_1}$,
and $H_1 \defini \mapExt {\emptyset} {\mapEntry q v}$.
Then, $\emptyset \xrightarrow{\mcasAllocAlg(v)\ \left\langle q \right\rangle} H_1$ is a path
of length $1$ starting from the empty heap that matches $\leq$.

\item Inductive case. Let $n \natorderEqSymbolRight 1$. Suppose 
$n + 1 \natorderEqSymbol \vert \closedEvent \vert$. 
We need to show that there is a matching path of length $n+1$.
Since $1 \natorderEqSymbol n \natorderEqSymbol \vert \closedEvent \vert$, 
the inductive hypothesis implies that there is a matching path of length $n$,
\begin{align}
\label{eqn::appendix::lin::mcas-legality-path-1}
H_0 = \emptyset \xrightarrow{proc_1(in_1)\ \langle out_1 \rangle} H_1 
\xrightarrow{proc_2(in_2)\ \langle out_2 \rangle} 
\ldots \xrightarrow{proc_n(in_n)\ \langle out_n \rangle} H_n
\end{align}

We need to show that we can extend this path with a matching $n+1$ step 
for the $n+1$ event in $\leq$. Denote the $n+1$ event in $\leq$ as $x_{n+1}$.

We do a case analysis on $x_{n+1}$,

\begin{itemize}

\item Case $x_{n+1} = \mcasAlg(\listvar u)$.

By Axiom 
\axiomHRef{help-focused::fin-predicate}, we have for some $v \in \BoolType$,
\begin{align}
\label{eqn::appendix::lin::mcas-post}
\begin{cases}
\forall k \in \listvar u.\ \exists z.\ \visObs {\pointerIndx {\pointGenEntry k}} z {x_{n+1}} \wedge 
\inputVal z {\pointGenEntry k} {\expGenEntry k} & \text{if } v = true \\
\begin{aligned}
\exists k \in \listvar u,v'\in\ValType.\ \exists z.\ \visObs {\pointerIndx {\pointGenEntry k}} z {x_{n+1}} \wedge {}\\ 
\inputVal z {\pointGenEntry k} {v'} \wedge v' \neq {\expGenEntry k} 
\end{aligned} & \text{if } v = false \\
\end{cases}
\end{align} 

\begin{itemize}
\item Case $v = true$. 

We claim that for any $k \in \listvar u$,
$H_n(\pointGenEntry k) = \expGenEntry k$. 

Let $k \in \listvar u$. By \eqref{eqn::appendix::lin::mcas-post}
there is $z$ such that $\visObs {\pointerIndx {\pointGenEntry k}} z {x_{n+1}}$
and $\inputVal z {\pointGenEntry k} {\expGenEntry k}$.
Since path \eqref{eqn::appendix::lin::mcas-legality-path-1} matches $\leq$
and $z$ occurs before $x_{n+1}$ in $\leq$,
we have
$z = \generatorEvent{proc_i}$ for some $i \natorderEqSymbol n$. 

In addition, by Axiom 
\axiomHRef{help-focused::helped-are-writers},
$\generatorEvent{proc_i} \in \writesAbs {\pointGenEntry k}$.
Also, for every $j \natorderSymbolRight i$, 
$\generatorEvent{proc_j} \notin \writesAbs {\pointGenEntry k}$ 
by property \eqref{eq::appendix::lin::sequential-recency}, since 
$\visObs {\pointerIndx {\pointGenEntry k}} {z} 
{x_{n+1}}$. 
Therefore, by Lemma 
\ref{lem::appendix::lin::mcas-writers-change-heap} with
path \eqref{eqn::appendix::lin::mcas-legality-path-1} and 
$\inputVal z {\pointGenEntry k} {\expGenEntry k}$, 
we have $H_n(\pointGenEntry k) = \expGenEntry k$.

This proves the claim.

Define $H_{n+1} \defini 
\mapExtThree {H_n} {\mapEntry {\pointGenEntry k} {\newGenEntry k}} {k \in \listvar u}$.
Hence, $H_n \xrightarrow{\mcasAlg(\listvar u)\ \left\langle true \right\rangle} H_{n+1}$ is a valid
step corresponding to a successful $\mcasAlg$ step.

In case $x_{n+1} \in \terminatedEvent$, Axiom \axiomHRef{help-focused::fin-predicate}
also concludes that $true = v = \outputProp{x_{n+1}}$, which means that
the step also matches $\leq$ in case $x_{n+1}$ is a terminated event.

\item Case $v = false$. 
By \eqref{eqn::appendix::lin::mcas-post} and Axiom \axiomHRef{help-focused::helped-are-writers}, 
we have $z \in \writesAbs {\pointGenEntry k}$
and $\inputVal z {\pointGenEntry k} {v'}$ for some
$k \in \listvar u$ and $v' \neq \expGenEntry k$.

By a similar argument as in the previous case,
$H_n(\pointGenEntry k) = v'$, where $v' \neq \expGenEntry k$.

Define $H_{n+1} \defini H_n$.
Hence, $H_n \xrightarrow{\mcasAlg(\listvar u)\ \left\langle false \right\rangle} H_{n+1}$ is a valid
step corresponding to a failing $\mcasAlg$ step.

In case $x_{n+1} \in \terminatedEvent$, Axiom \axiomHRef{help-focused::fin-predicate}
also concludes that $false = v = \outputProp{x_{n+1}}$, which means that
the step also matches $\leq$ in case $x_{n+1}$ is a terminated event.

\end{itemize}

\item Case $x_{n+1} = \mcasAllocAlg(v)$. 
Identical to the case for $\rdcssAllocAlg$ in Lemma \ref{lem::appendix::lin::rdcss-legality-lemma}.

\item Case $x_{n+1} = \mcasWriteAlg(q,m)$. 
Identical to the case for $\rdcssWriteAlg$ in Lemma \ref{lem::appendix::lin::rdcss-legality-lemma}.

\item Case $x_{n+1} = \mcasReadAlg(q)$. 
Identical to the case for $\rdcssReadAlg$ in Lemma \ref{lem::appendix::lin::rdcss-legality-lemma}.

\end{itemize}
\end{itemize}
\end{prf}

\begin{lem}[Sequential Soundness]
\label{lem::appendix::lin::linear-order-mcas-sequential-soundness}
Suppose hypotheses \ref{defn::appendix::lin::hypotheses-for-sequential-soundness-mcas}.
The linear order $\leq$ of Lemma \ref{lem::appendix::lin::existence-basic} is sequentially sound.
\end{lem}

\begin{prf}
If $\closedEvent = \emptyset$, then the empty path matches $\leq$. 
If $\closedEvent \neq \emptyset$, then Lemma \ref{lem::appendix::lin::mcas-legality-lemma}
applied with $n = \vert \closedEvent \vert$ ensures the existence of a matching path
for $\leq$.
\end{prf}

\begin{thm}
\label{thm::appendix::lin::mcas-linearizability-from-vis-structure}
Given an implementation of the MCAS, suppose that for any set of
abstract events $\absEvent$ generated from an arbitrary execution
history in the implementation, there are relations
$\visObsSymbol{\pointerIndx p}$,
$\visSepSymbol{\pointerIndx p}$ such that
$\MCASFam(\visObsSymbol {\pointerIndx p},
{\visSepSymbol{\pointerIndx p}})$ is valid. Then, the
implementation is linearizable.
\end{thm}

\begin{prf}
  Let $\absEvent$ be a set of events generated from an arbitrary
  execution history in the implementation. 
  From the hypothesis, $\MCASFam(\visObsSymbol {\pointerIndx p},
  {\visSepSymbol{\pointerIndx p}})$ is valid 
  for some $\visObsSymbol {\pointerIndx p}$ and
  $\visSepSymbol{\pointerIndx p}$.
  Let $\leq$ be the linear order of Lemma
  \ref{lem::appendix::lin::existence-basic}.  Then, we take the general
  visibility relation $\genVisSymbol$ and $\leq$ to be the
  relations required by the definition of linearizability.  
  By construction, $\leq$ respects both $\genVisSymbol$ and
  $\precedesAbsSymbol$. Also,
  $\leq$ is sequentially sound by Lemma
  \ref{lem::appendix::lin::linear-order-mcas-sequential-soundness}.
\end{prf}

\section{Proof of Validity for RDCSS and helping MCAS}
\label{appendix::sect::impl::proof-of-validity}

This section focuses on proving the validity of $\genStructName{\mu}$
($\mu \in \{ \RDCSS, \MCAS \}$) as defined in 
\ref{defn::appendix::lin::rdcss-vis-spec} for $\mu = \RDCSS$, 
and in \ref{defn::appendix::lin::mcas-vis-spec} for $\mu = \MCAS$,
under the assumption that $\genStructName{\mu}$ can be implemented
by spans (i.e., the arrow stating that the span axioms imply the visibility axioms in Figure \ref{fig::proof-diagram}).
Appendix \ref{appendix::sect::impl::RDCSS-and-MCAS-proofs} will then focus on proving that
$\genStructName{\mu}$ is implemented by spans. 

This section makes use of all concepts in Section~\ref{sect::key-concepts-spans} but
augments the notions of span structure and implementation by span
structure to include allocs.

\begin{defn}[Span Structure] 
\label{sect::defn-span-structure}
A \emph{span structure}
$(\spans p, \writesSpans p, \allocsSpans p,
\runFuncSymbol)$ consists
of:
\begin{itemize}
	\item For every pointer $p$, a set $\spans p \subseteq \repEvent \times \repEvent$, called the \emph{spans} accessing $p$.
	\item For every pointer $p$, a set $\writesSpans p \subseteq \spans p$, called the \emph{successful write spans} into $p$.
	\item For every pointer $p$, a set $\allocsSpans p \subseteq \writesSpans p$, called the \emph{alloc spans} creating $p$.
	\item A function $\runFuncSymbol: \absEvent \rightarrow (\mathcal{P}(\bigcup_p \spans p) \times (\ValType \cup \{ \bot \}))$,
	called the \emph{event denotation}, written $\runFunc{x}$ for event $x$. Here, $\mathcal{P}(\bigcup_p \spans p)$ denotes the power set of $\bigcup_p \spans p$. 
\end{itemize}
\end{defn}

Each span in $\spans p$ is a pair of rep events. We denote as
$\firstRepFuncSymbol$ (``first'') and $\lastRepFuncSymbol$ (``last'')
the standard pair projection functions. We write spans of the form
$(a,a)$ as $(a)$.

\begin{defn}[Implementation by a span structure]
	\label{defn::implementation-by-span-structure}
	Given arbitrary relations
	$\visObsSymbol {\pointerIndx p}$,
	$\visSepSymbol{\pointerIndx p}$, we say that
	$\genStructName{}(\visObsSymbol {\pointerIndx p},
	\visSepSymbol{\pointerIndx p}) = (\writesAbs p,
	\allocsAbs p, \postPredSymbol)$ is \emph{implemented by span
		structure}
	$(\spans p, \writesSpans p, \allocsSpans p,
	\runFuncSymbol)$ if all
	the span axioms in Figure \ref{fig:descriptor-lifespans-axioms} are
	satisfied.  We also say that
	$\genStructName{}(\visObsSymbol {\pointerIndx p},
	\visSepSymbol{\pointerIndx p}) = (\writesAbs p,
	\allocsAbs p, \postPredSymbol)$ is \emph{implemented by
		spans} if there exists some span structure that implements it.
\end{defn}

\begin{figure}[t]
	\begin{subfigwrap}{Defined notions for span axioms.}{fig::sub::defined-spans}
    \centering
		\begin{tabular}{l}
			Span returns-before relation \\
			\quad $\precedesSpans b c \defini \linReps {\lastRep b} {\firstRep c}$ \\
		\end{tabular}
		\begin{tabular}{l}
			Set of spans in a denotation \\
			\quad $\hspans{}{x} \defini \projFirst {\runFunc x}$ \\
		\end{tabular}
		\begin{tabular}{l}
			Denotation output \\
			\quad $\outputRunFunc x \defini \projSecond {\runFunc x}$ \\
		\end{tabular}
	\end{subfigwrap}
	
	\begin{subfigwrap}{Key Axioms}{fig::sub::spans-key-axioms}  
    \centering
		 \begin{tabular}{l}
		 	\begin{tabular}{ll}
		 		\axiomDLabel{principle::descriptors-do-not-interfere} Disjointness &     \axiomDLabel{principle::all-descriptors-are-written-before-any-resolution} Bunching \\
		 		\qquad $b, c \in \spans p \implies ({\precedesSpans b c} \vee {\precedesSpans c b} \vee {b = c})$ & 
		 		\qquad $b, c \in \hspans{}{x} \implies \linRepsEq {\firstRep b} {\lastRep c}$ \\
		 	\end{tabular}\\[1em]
		 	\begin{tabular}{l}
		 		\axiomDLabel{principle::finished-operations-have-a-run} Adequacy \\
		 		\qquad $x \in \terminatedEvent \implies \outputRunFunc x = \outputProp x \wedge \hspans{}{x} \neq \emptyset$ \\
		 	\end{tabular}
		 \end{tabular}
	\end{subfigwrap}
	
	\begin{subfigwrap}{Structural Axioms}{fig::sub::spans-structural-axioms}  
        \centering
		\begin{tabular}{c l}  
			\axiomDLabel{principle::descriptor-write-precedes-resolution} & $\linRepsEq {\firstRep b} {\lastRep b}$ \\
			
			\axiomDLabel{principle::runs-are-injective} & $\hspans{}{x} \cap \hspans{}{y} \neq \emptyset \implies x = y$ \\
			
			\axiomDLabel{principle::writer-blocks-belong-to-runs} & $b \in \writesSpans p \implies \exists x.\ b \in \hspans{}{x}$ \\
			
			\axiomDLabel{principle::postcondition-predicate-holds} & $\hspans{}{x} \neq \emptyset \implies 
			\outputRunFunc x \neq \bot \wedge \postPred x {\outputRunFunc x}$ \\

			\axiomDLabel{principle::writers-have-writer-blocks} & $\hspans{}{x} \neq \emptyset \implies (x \in \writesAbs p \Leftrightarrow \hspans{}{x} \cap \writesSpans p \neq \emptyset)$ \\
			
			\axiomDLabel{principle::blocks-contained-in-abstract-time-interval} & (i) $b \in \hspans{}{x} \implies \STimeProp x \natorderEqSymbol \STimeProp {\firstRep b}$ \\
			
			& (ii) $(b \in \hspans{}{x} \wedge x \in \terminatedEvent) \implies {}$ \\ 
			& \qquad $\exists i \in \repsCompl.\ \ETimeProp i \natorderEqSymbol \ETimeProp x \wedge \linRepsEq {\lastRep b} i$ \\
						
            \axiomDLabel{principle::containment-and-uniqueness-of-alloc-blocks} & $b, c \in \allocsSpans p \implies b = c$ \\
						
            \axiomDLabel{principle::every-block-must-have-an-allocated-pointer} & $b \in \spans p \implies \exists c \in \allocsSpans p.\ \precedesSpansEq c b$ \\
            
            \axiomDLabel{principle::allocs-have-alloc-blocks} & $\hspans{}{x} \neq \emptyset \implies (x \in \allocsAbs p \Leftrightarrow \hspans{}{x} \cap \allocsSpans p \neq \emptyset)$ \\
		\end{tabular}
	\end{subfigwrap}
	\caption{Span axioms for span structure $(\spans p, \writesSpans p, \allocsSpans p, \runFuncSymbol)$ implementing $\genStructName{}(\visObsSymbol {\pointerIndx p}, \visSepSymbol{\pointerIndx p}) = (\writesAbs p, \allocsAbs p, \postPredSymbol)$. Variables $x$, $y$ range over $\absEvent$. Variables $b$, $c$ over $\bigcup_p \spans p$. Variable $p$ over $\PtsType$.}
	\label{fig:descriptor-lifespans-axioms}
\end{figure}

All axioms from \axiomDRef{principle::descriptors-do-not-interfere} to \axiomDRef{principle::finished-operations-have-a-run}
have already been explained in Section~\ref{sect::key-concepts-spans}. Here we explain the three last ones involving allocs. \emph{Axiom \axiomDRef{principle::containment-and-uniqueness-of-alloc-blocks}}
states that alloc spans are unique.
\emph{Axiom \axiomDRef{principle::every-block-must-have-an-allocated-pointer}}
states that every span must be preceded by an alloc span that
created the pointer.
\emph{Axiom \axiomDRef{principle::allocs-have-alloc-blocks}} states
that $p$-alloc events must contain a $p$-alloc span in their denotation.

We will prove the validity of $\genStructName{\mu}$
for the two specific separable-before \eqref{eq::overview::helpers-separability}
and observation \eqref{eq::overview::helpers-observation} 
relations of Section \ref{subsubsect::visibility-mcas},
which we define here again.

\begin{defn}[Visibility Relations for Helping]
\label{defn::appendix::valid::visibility-relations}
Given span structure $(\spans p, \writesSpans p, \allocsSpans p, \runFuncSymbol)$,
we define the pointer-indexed separable-before $\visSepSymbol{\pointerIndx p}$ and 
observation $\visObsSymbol{\pointerIndx p}$ relations for helping implementations,
\begin{align}
\label{eq::appendix::valid::helpers-separability-copy}
\visSep{\pointerIndx p} A B & \defini \exists b \in \hspans p A, c \in \hspans p B.\ \precedesSpans b c \\
\label{eq::appendix::valid::helpers-observation-copy}
\visObs {\pointerIndx p} A B & \defini \exists b \in 
\hspans p A, c \in \hspans p B.\  b = \max_{\precedesSpansEqSymbol} \{ d \in \writesSpans p \mid \precedesSpans d c \}
\end{align}
\end{defn}

\begin{defn}[Section Hypotheses]
\label{defn::appendix::valid::hypotheses-for-validity}
To shorten the statements of propositions from \ref{lma::appendix::valid::strict-partial-order-lifespans} 
to \ref{thm::appendix::valid::span-axioms-imply-visibility}, the
following hypotheses will apply,
\begin{itemize}
\item $\absEvent$ is an arbitrary set of events.
\item $\repEvent$ is an arbitrary set of rep events, linearized under a given $\linRepsEqSymbol$.
\item $Z \defini (\spans p, \writesSpans p, \allocsSpans p, \runFuncSymbol)$ 
is an arbitrary span structure.
\item $\genStructName{\mu}(\visObsSymbol{\pointerIndx p}, \visSepSymbol{\pointerIndx p})$
(for $\mu \in \{ \RDCSS, \MCAS \}$)
is implemented by $Z$, where relations $\visObsSymbol{\pointerIndx p}$, $\visSepSymbol{\pointerIndx p}$
are those in Definition \ref{defn::appendix::valid::visibility-relations} and instantiated with
$Z$.
\end{itemize}
\end{defn}

The first lemmas describe basic results whose statements are 
self-explanatory.

\begin{lem}
\label{lma::appendix::valid::strict-partial-order-lifespans}
Suppose hypotheses \ref{defn::appendix::valid::hypotheses-for-validity}.
Relation $\precedesSpansSymbol$ defines a strict partial order on the set
$\bigcup_p \spans p$. 
\end{lem}

\begin{prf}
Each required property follows,
\begin{itemize}
\item Irreflexivity. Suppose for a contradiction that 
$\precedesSpans a a$ for some $a \in \bigcup_p \spans p$.
Hence, $\lastRep a \linRepsSymbol \firstRep a$ by
definition. 
But by Axiom
\axiomDRef{principle::descriptor-write-precedes-resolution}, 
$\linRepsEq {\firstRep a} {\lastRep a}$.
Therefore, $\linReps {\lastRep a} {\lastRep a}$
(Contradiction).

\item Transitivity. Let $\precedesSpans a b$ and 
$\precedesSpans b c$.

We have by definition that 
$\lastRep a \linRepsSymbol \firstRep b$ and 
$\lastRep b \linRepsSymbol \firstRep c$. 
But by Axiom
\axiomDRef{principle::descriptor-write-precedes-resolution}, 
$\linRepsEq {\firstRep b} {\lastRep b}$.
Therefore, $\linReps {\lastRep a} {\firstRep c}$,
which means $\precedesSpans a c$.
\end{itemize}
\end{prf}

\begin{lem}
\label{lma::appendix::valid::partial-order-lifespans}
Suppose hypotheses \ref{defn::appendix::valid::hypotheses-for-validity}.
Relation $\precedesSpansEqSymbol$ defines a partial order on the set
$\bigcup_p \spans p$. In addition, for every $q$, 
relation $\precedesSpansEqSymbol$ defines a linear order on the set
$\spans q$.
\end{lem}

\begin{prf}
By Lemma 
\ref{lma::appendix::valid::strict-partial-order-lifespans},
$\precedesSpansSymbol$ defines a strict partial order. It is a 
standard result that the reflexive closure of a strict partial
order defines a partial order (i.e. a reflexive, transitive,
and antisymmetric relation).

Since for every $q$, $\spans q \subseteq \bigcup_p \spans p$
holds, relation $\precedesSpansEqSymbol$ is also a partial order 
on $\spans q$ for every $q$. 
That $\precedesSpansEqSymbol$ is a linear order
for $\spans q$ is simply Axiom
\axiomDRef{principle::descriptors-do-not-interfere}.
\end{prf}

Therefore, Lemma \ref{lma::appendix::valid::partial-order-lifespans}
implies that it makes sense to take a maximum under $\precedesSpansEqSymbol$
in Definition \ref{defn::appendix::valid::visibility-relations}.

The next lemma states that events related under a visibility relation 
must contain disjoint spans. This is a direct consequence of
the definitions of the relations in \ref{defn::appendix::valid::visibility-relations}.

\begin{lem}
\label{lma::appendix::valid::abs-visibility-implies-block-precedence}
Suppose hypotheses \ref{defn::appendix::valid::hypotheses-for-validity}.
If $\genVis x y$ then $\exists b_x \in 
\hspans{}{x}, b_y \in \hspans{}{y}.\ 
\precedesSpans {b_x} {b_y}$.
\end{lem}

\begin{prf}
The case $\visSep {\pointerIndx p} x y$ (for some $p$) implies
by definition that $\precedesSpans {b_x} {b_y}$ for some $b_x \in 
\hspans{p}{x} \subseteq \hspans{}{x}$ and $b_y \in \hspans{p}{y} \subseteq \hspans{}{y}$.

The case 
$\visObs {\pointerIndx p} x y$ (for some $p$), 
also implies by definition that $\precedesSpans {b_x} {b_y}$ for some $b_x \in 
\hspans{p}{x} \subseteq \hspans{}{x}$ and $b_y \in \hspans{p}{y} \subseteq \hspans{}{y}$.
\end{prf}

The next lemma states that when we apply the transitive closure to $\genVisSymbol$, the property
of Lemma \ref{lma::appendix::valid::abs-visibility-implies-block-precedence}
remains true. This is a direct consequence of the bunching axiom.

\begin{lem}
\label{lma::appendix::valid::vis-trans-implies-having-runs}
Suppose hypotheses \ref{defn::appendix::valid::hypotheses-for-validity}.
If $\genVisTrans x y$, then $\exists b_x \in \hspans{}{x},
b_y \in \hspans{}{y}.\ \precedesSpans {b_x} {b_y}$.
\end{lem}

\begin{prf}
Define the following binary relation on $\absEvent$,
\[
P(w,z) \defini \exists b_w \in \hspans{}{w}, b_z \in 
\hspans{}{z}.\ \precedesSpans {b_w} {b_z}
\]

So, we need to show $\genVisTrans x y \implies P(x,y)$. 
But it suffices to show the following two properties,
\begin{itemize}
\item ${\genVisSymbol} \subseteq P$.
\item $P$ is transitive. 
\end{itemize}
because $\genVisTransSymbol$ is the smallest transitive 
relation containing $\genVisSymbol$.

Let us show the two required properties.
\begin{itemize}
\item ${\genVisSymbol} \subseteq P$. This is Lemma \ref{lma::appendix::valid::abs-visibility-implies-block-precedence}.

\item $P$ is transitive. By hypotheses $P(u,v)$ and $P(v,z)$, 
we have,
\begin{align}
\begin{split}
\exists b_u \in \hspans{}{u}, b_v \in \hspans{}{v}.\ 
\precedesSpans {b_u} {b_v} \\
\exists b_v' \in \hspans{}{v}, b_z \in \hspans{}{z}.\ 
\precedesSpans {b_v'} {b_z}
\end{split}
\end{align}

We need to show $P(u,z)$. By the bunching axiom \axiomDRef{principle::all-descriptors-are-written-before-any-resolution}
on the denotation $\hspans{}{v}$,
$\linRepsEq {\firstRep {b_v}} {\lastRep {b_v'}}$ must hold. Hence,
\[
\lastRep {b_u} \linRepsSymbol \firstRep {b_v} \linRepsEqSymbol \lastRep {b_v'} \linRepsSymbol \firstRep {b_z}
\]
which means $\precedesSpans {b_u} {b_z}$.
\end{itemize}
\end{prf}

A direct consequence of Lemma \ref{lma::appendix::valid::vis-trans-implies-having-runs}
is the following, which states that events chosen for linearization
must have carried out some non-empty execution path.

\begin{lem}
\label{lma::appendix::valid::visible-implies-having-run}
Suppose hypotheses \ref{defn::appendix::valid::hypotheses-for-validity}.
If $x \in \closedEvent$ then $\hspans{}{x} \neq \emptyset$.
\end{lem}

\begin{prf}
By definition of $\closedEvent$, $x \refleTransCl{\genVisSymbol} y$
for some $y \in \terminatedEvent$.
\begin{itemize}
\item Case $x = y$. Hence, $x \in \terminatedEvent$. By Axiom 
\axiomDRef{principle::finished-operations-have-a-run},
$\hspans{}{x} \neq \emptyset$ must hold.

\item Case $x \neq y$. Hence, $x \transCl{\genVisSymbol} y$.
By Lemma \ref{lma::appendix::valid::vis-trans-implies-having-runs},
$\hspans{}{x} \neq \emptyset$ must hold.
\end{itemize}
\end{prf}

The following lemma states that events ordered in real-time 
have all their spans disjoint from each other, i.e., spans 
do not go beyond the events they belong to.

\begin{lem}
\label{lma::appendix::valid::abs-precedence-implies-block-precedence}
Suppose hypotheses \ref{defn::appendix::valid::hypotheses-for-validity}.
If $\precedesAbs x y$ then $\forall b_x \in \hspans{}{x}, b_y 
\in \hspans{}{y}.\ \precedesSpans {b_x} {b_y}$.
\end{lem}

\begin{prf}
Let $b_x \in \hspans{}{x}, b_y \in \hspans{}{y}$. By Axiom 
\axiomDRef{principle::blocks-contained-in-abstract-time-interval},
we have $\STimeProp y \natorderEqSymbol \STimeProp {\firstRep {b_y}}$ and
there is $i$ such that $\linRepsEq {\lastRep {b_x}} i$ and
$\ETimeProp i \natorderEqSymbol \ETimeProp x$.

Hence, by definition of $\precedesAbsSymbol$, we have,
\[
\ETimeProp i \natorderEqSymbol \ETimeProp x 
\natorderSymbol \STimeProp y \natorderEqSymbol 
\STimeProp {\firstRep {b_y}} 
\]

This means that $i$ finishes in real-time before 
$\firstRep {b_y}$ starts. But since $\linRepsEqSymbol$ respects the real time order of rep events
(i.e., $\linRepsEqSymbol$ is a linearization), 
we have $\linReps i {\firstRep {b_y}}$. 

But $\linRepsEq {\lastRep {b_x}} i$, which means
$\linReps {\lastRep {b_x}} {\firstRep {b_y}}$. Therefore,
$\precedesSpans {b_x} {b_y}$.
\end{prf}

We can now prove each visibility axiom.

\begin{lem}
\label{lma::appendix::valid::no-inbetween-vis-axiom-holds}
Suppose hypotheses \ref{defn::appendix::valid::hypotheses-for-validity}.
Axiom \axiomHRef{help-focused::non-helpers} holds.
\end{lem}

\begin{prf}
Let $\visObs {\pointerIndx {p}} x y$ and $z \in \writesAbs {p} \cap \closedEvent$.

By definition of $\visObs {\pointerIndx {p}} x y$
there are 
$b_x \in \hspans{p}{x}$ and $b_y \in \hspans{p}{y}$ such that 
$b_x$ is the maximum satisfying
$b_x \in \writesSpans {p}$ and 
$\precedesSpans {b_x} {b_y}$.

From hypothesis $z \in \closedEvent$ and Lemma \ref{lma::appendix::valid::visible-implies-having-run},
$\hspans{}{z} \neq \emptyset$ holds. So, by 
Axiom \axiomDRef{principle::writers-have-writer-blocks}, 
there is $b_z \in \hspans{}{z} \cap \writesSpans {p}$. 
 
But $b_y \in \spans {p}$ and $b_z \in 
\writesSpans {p} \subseteq \spans {p}$, 
which means that either $\precedesSpans {b_y} {b_z}$ or 
$\precedesSpans {b_z} {b_y}$ or $b_y = b_z$ by Axiom 
\axiomDRef{principle::descriptors-do-not-interfere}.

\begin{itemize}
\item Case $\precedesSpans {b_y} {b_z}$. By definition,
$\visSep {\pointerIndx p} y z$ holds.

\item Case $\precedesSpans {b_z} {b_y}$. 
Again, since $b_x, b_z \in \spans {p}$, we have three subcases by Axiom 
\axiomDRef{principle::descriptors-do-not-interfere}.

\begin{itemize}
\item Case $\precedesSpans {b_x} {b_z}$.
We have 
$b_x \precedesSpansSymbol b_z \precedesSpansSymbol b_y$, which 
is a contradiction, because $b_x$ was the maximum satisfying
the conditions $b_x \in \writesSpans {p}$ and 
$\precedesSpans {b_x} {b_y}$, but now $b_z$ is a more recent
span satisfying the conditions.

\item Case $\precedesSpans {b_z} {b_x}$. By definition,
$\visSep {\pointerIndx p} z x$ holds.

\item Case $b_x = b_z$. We have $\hspans{}{x} \cap \hspans{}{z} 
\neq \emptyset$, which means $x = z$ by Axiom 
\axiomDRef{principle::runs-are-injective}.
\end{itemize}

\item Case $b_y = b_z$. We have $\hspans{}{y} \cap \hspans{}{z} 
\neq \emptyset$, which means $y = z$ by Axiom 
\axiomDRef{principle::runs-are-injective}.
\end{itemize}
\end{prf}

\begin{lem}
Suppose hypotheses \ref{defn::appendix::valid::hypotheses-for-validity}.
Axiom \axiomHRef{help-focused::helped-are-writers} holds.
\end{lem}

\begin{prf}
Let $x,y \in \closedEvent$, and
$\visObs {\pointerIndx {p}} x y$.

By definition of $\visObsSymbol{\pointerIndx p}$, 
there is $b_x \in \hspans{p}{x}$ such that $b_x \in \writesSpans {p}$.
So, $x \in \writesAbs {p}$ by Axiom 
\axiomDRef{principle::writers-have-writer-blocks}.
\end{prf}

\begin{lem}
\label{lma::appendix::valid::acyclic-vis-axiom-holds}
Suppose hypotheses \ref{defn::appendix::valid::hypotheses-for-validity}.
Axiom \axiomHRef{help-focused::vis-acyclic} holds.
\end{lem}

\begin{prf}
Let $\genVisTrans x y$.
For a contradiction, suppose $\precedesAbsEq y x$.

From $\genVisTrans x y$ and Lemma 
\ref{lma::appendix::valid::vis-trans-implies-having-runs}, 
there are $b_x \in \hspans{}{x}$ and 
$b_y \in \hspans{}{y}$ such that 
$\precedesSpans {b_x} {b_y}$.

From $\precedesAbsEq y x$, we have two cases,

\begin{itemize}
\item Case $\precedesAbs y x$. By Lemma
\ref{lma::appendix::valid::abs-precedence-implies-block-precedence},
we also have $\precedesSpans {b_y} {b_x}$.
Hence, $b_y \precedesSpansSymbol {b_x} \precedesSpansSymbol b_y$
(Contradiction, since $\precedesSpansSymbol$ is irreflexive by Lemma \ref{lma::appendix::valid::strict-partial-order-lifespans}).

\item Case $y = x$. Hence, $b_x, b_y \in \hspans{}{x}$.

By Axiom 
\axiomDRef{principle::all-descriptors-are-written-before-any-resolution},
we also have $\firstRep {b_y} \linRepsEqSymbol \lastRep {b_x}$.
This means, $\lastRep {b_x} \linRepsSymbol \firstRep {b_y} 
\linRepsEqSymbol \lastRep {b_x}$, since $\precedesSpans {b_x} {b_y}$.
Hence, $\lastRep {b_x} \linRepsSymbol \lastRep {b_x}$ 
(Contradiction).
\end{itemize}
\end{prf}

\begin{lem}
	Suppose hypotheses \ref{defn::appendix::valid::hypotheses-for-validity}.
	Axiom \axiomHRef{help-focused::fin-predicate} holds.
\end{lem}

\begin{prf}
	Let $x \in \closedEvent$. By Lemma \ref{lma::appendix::valid::visible-implies-having-run},
	$\hspans{}{x} \neq \emptyset$ holds. Hence, by Axiom \axiomDRef{principle::postcondition-predicate-holds},
	$\outputRunFunc x \neq \bot$ and $\postPred x {\outputRunFunc x}$ hold. 
	In addition, if $x \in \terminatedEvent$, 
	then Axiom \axiomDRef{principle::finished-operations-have-a-run} implies
	$\outputRunFunc x = \outputProp x$.
\end{prf}

\begin{lem}
	Suppose hypotheses \ref{defn::appendix::valid::hypotheses-for-validity}.
	Axiom \axiomHRef{help-focused::allocs} holds.
\end{lem}

\begin{prf}
	Let $x,y \in \allocsAbs p \cap \closedEvent$.
	By Lemma \ref{lma::appendix::valid::visible-implies-having-run},
	$\hspans{}{x} \neq \emptyset$ and $\hspans{}{y} \neq \emptyset$ hold.
	
	By Axiom 
	\axiomDRef{principle::allocs-have-alloc-blocks},
	there are $b_x \in \allocsSpans p \cap \hspans{}{x}$ and
	$b_y \in \allocsSpans p \cap \hspans{}{y}$.
	But then $b_x = b_y$ by Axiom 
	\axiomDRef{principle::containment-and-uniqueness-of-alloc-blocks},
	which implies $x = y$ by Axiom
	\axiomDRef{principle::runs-are-injective}.
\end{prf}

\begin{lem}
	\label{lma::appendix::valid::writers-have-allocated-pointers-vis-axiom-holds}
	Suppose hypotheses \ref{defn::appendix::valid::hypotheses-for-validity}.
	Axiom \axiomHRef{help-focused::all-writers-are-willing-helpers} holds.
\end{lem}

\begin{prf}
	Let $x \in \writesAbs {p} \cap \closedEvent$.
	
	By Lemma \ref{lma::appendix::valid::visible-implies-having-run}, 
	$\hspans{}{x} \neq \emptyset$ holds. Hence, by Axiom 
	\axiomDRef{principle::writers-have-writer-blocks}, 
	there is $b_x \in \hspans{}{x} \cap \writesSpans {p}$. 
	But then by Axiom \axiomDRef{principle::every-block-must-have-an-allocated-pointer} 
	there is $b \in \allocsSpans {p}$ 
	such that $\precedesSpansEq b {b_x}$. 
	
	\begin{itemize}
		\item Case $b = b_x$. By Axiom 
		\axiomDRef{principle::allocs-have-alloc-blocks} we have 
		$x \in \allocsAbs {p}$, and trivially 
		$\visSepEq {\pointerIndx p} x x$.
		
		\item Case $\precedesSpans b {b_x}$.
		By Axiom \axiomDRef{principle::writer-blocks-belong-to-runs}, 
		$b \in \hspans{}{z}$ for some $z$. 
		By Axiom \axiomDRef{principle::allocs-have-alloc-blocks} we have 
		$z \in \allocsAbs {p}$.
		But since $\precedesSpans b {b_x}$, we have 
		$\visSep {\pointerIndx p} z x$ by definition.
	\end{itemize}
\end{prf}

\begin{thm}
\label{thm::appendix::valid::span-axioms-imply-visibility}
If hypotheses \ref{defn::appendix::valid::hypotheses-for-validity} hold, then 
$\genStructName{\mu}(\visObsSymbol{\pointerIndx p}, \visSepSymbol{\pointerIndx p})$
is valid.
\end{thm}

\begin{prf}
All visibility axioms hold by Lemmas from \ref{lma::appendix::valid::no-inbetween-vis-axiom-holds} to \ref{lma::appendix::valid::writers-have-allocated-pointers-vis-axiom-holds}.
\end{prf}

\section{Proof of Span Axioms for RDCSS and helping MCAS}
\label{appendix::sect::impl::RDCSS-and-MCAS-proofs}

This section focuses on proving that
$\genStructName{\mu}(\visObsSymbol{\pointerIndx p}, \visSepSymbol{\pointerIndx p})$ is implemented by spans 
(for $\mu \in \{ \RDCSS, \MCAS \}$).
As such, we will define a span structure for RDCSS and MCAS and then
show that this structure implements $\genStructName{\mu}$.
Here,
$\RDCSSFam$ was defined in
\ref{defn::appendix::lin::rdcss-vis-spec} and
$\MCASFam$ in \ref{defn::appendix::lin::mcas-vis-spec}.
Also, we use relations $\visObsSymbol{\pointerIndx p}$ and
$\visSepSymbol{\pointerIndx p}$ 
as defined in \ref{defn::appendix::valid::visibility-relations}

This section uses the notion of span structure (Definition~\ref{sect::defn-span-structure}) 
and implementation of span structure (Definition~\ref{defn::implementation-by-span-structure}). 
See also the concepts in Section \ref{sect::key-concepts-spans}.
 
\subsection{RDCSS}
\label{appendix::sub::sect::impl::RDCSS}

The full pseudocode for RDCSS is shown in
Figure \ref{appendix::alg-RDCSS}. The exportable 
procedures are $\rdcssAlg$, $\rdcssReadAlg$,
$\rdcssCasAlg$, $\rdcssWriteAlg$, $\rdcssReadCtlAlg$, 
$\rdcssCasCtlAlg$, $\rdcssWriteCtlAlg$, and 
$\rdcssAllocAlg$. 

$\ValType$ denotes the set of all
possible input values. $\ValType$ contains 
neither RDCSS descriptors nor pointers storing RDCSS descriptors.

RDCSS requires that every pointer exposed to the clients be classified
as a control or data pointer ($\ControlPtType$ and $\DataPtType$,
respectively). The data pointers may store values in $\ValType$, 
descriptors, and pointers. Control
pointers can only store values in $\ValType$.

No particular implementation is provided for boolean predicate
$\isRdcssDescRepAlg(p)$, but it is assumed that it returns 
true if and only if $p$ is a pointer storing an RDCSS descriptor.
For example,~\cite{Harris} suggests that $\isRdcssDescRepAlg(p)$ 
could be implemented by checking a reserved bit in $p$; this
reserved bit indicates whether or not the pointer stores a descriptor.

The implementation makes the following assumptions:
\begin{itemize}
\item The alloc at line~\ref{appendix::alg-alloc-desc-RDCSS}
creates a data pointer $d$ such that $\isRdcssDescRepAlg(d)$ returns true.
\item No implementation details are provided for the primitive allocs at
lines \ref{appendix::alloc-control-Alloc-RDCSS} and
\ref{appendix::alloc-data-Alloc-RDCSS}, but they are assumed to 
have linearizable implementations.
\item Any input data pointer $p: \DataPtType$ to any
exportable procedure must satisfy $\neg \isRdcssDescRepAlg(p)$.
\item Any RDCSS descriptor $desc$ must satisfy
$\neg \isRdcssDescRepAlg(\pointOne {desc})$ and
$\neg \isRdcssDescRepAlg(\pointTwo {desc})$.
\item Any input pointer to any procedure must have been previously
allocated with an invocation to $\rdcssAllocAlg$.
\item The allocs at either line \ref{appendix::alloc-control-Alloc-RDCSS} 
or line \ref{appendix::alloc-data-Alloc-RDCSS}
return a pointer $p$ such that $\neg \isRdcssDescRepAlg(p)$.
\end{itemize}   

\begin{figure}[t]
\begin{multicols*}{2}
\begin{algorithmic}[1]
\Record{$\rdcssDesc$}
\State $pt_1$ : $\ControlPtType$
\State $pt_2$ : $\DataPtType$
\State $exp_1$, $exp_2$, $new_2$ : $\ValType$
\EndRecord
\Enum{$\ptKind$}
\State $\CONTROL$, $\DATA$
\EndEnum
%
\State 
%
\Proc{$\rdcssAlg$}{$desc: \rdcssDesc$}
\State \label{appendix::alg-alloc-desc-RDCSS} $d \gets \allocRepAlg(desc)$ as $\DataPtType$
\State \returnCmd{$\rdcssLoopAlg(d, desc)$}
\EndProc
%
\State 
%
\Proc{$\rdcssLoopAlg$}{$d: \DataPtType$, $desc: \rdcssDesc$}
\State \label{appendix::alg-CAS-RDCSS} $old \gets \casRepAlg(\pointTwo{desc}, \expTwo{desc}, d)$
\If{$\isRdcssDescRepAlg(old)$} \label{appendix::alg-is-desc-RDCSS}
  \State \label{appendix::alg-help-complete-invoke-RDCSS} $\completeAlg(old)$
  \State \returnCmd{$\rdcssLoopAlg(d, desc)$} \label{appendix::rdcss-try-again-RDCSS}
\Else
  \If {$old = desc.exp_2$} \label{appendix::alg-is-cas-successful-RDCSS}
  	\State \label{appendix::alg-self-complete-invoke-RDCSS} $\completeAlg(d)$
  \EndIf  
  \State \label{appendix::alg-return-old-value-RDCSS} \returnCmd{$old$}
\EndIf
\EndProc
%
\State 
%
\Proc{$\completeAlg$}{$d: \DataPtType$}
\State \label{appendix::alg-desc-read-RDCSS} $desc \gets \readRepAlg d$
\State \label{appendix::alg-pt1-read-RDCSS} $x \gets \readRepAlg \pointOne{desc}$
\If {$x = \expOne{desc}$}
  \State \label{appendix::alg-pt2-write-success-RDCSS} $\casRepAlg(\pointTwo{desc}, d, \newTwo{desc})$
\Else
  \State \label{appendix::alg-pt2-write-fail-RDCSS} $\casRepAlg(\pointTwo{desc}, d, \expTwo{desc})$
\EndIf
\EndProc
%
\State 
%
\Proc{$\rdcssReadAlg$}{$pt : \DataPtType$}
\State \label{appendix::alg-access-Read-RDCSS} $old \gets \readRepAlg pt$
\If {$\isRdcssDescRepAlg(old)$} \label{appendix::alg-is-desc-Read-RDCSS}
  \State \label{appendix::alg-help-complete-invoke-Read-RDCSS} $\completeAlg(old)$
  \State \returnCmd{$\rdcssReadAlg(pt)$}
\Else 
  \State \label{appendix::alg-return-old-value-Read-RDCSS} \returnCmd{$old$}
\EndIf
\EndProc
%
\columnbreak
%
\Proc{$\rdcssCasAlg$}{$pt : \DataPtType$, $exp$, $new: \ValType$}
\State \label{appendix::alg-access-CAS-RDCSS} $old \gets \casRepAlg(pt, exp, new)$
\If {$\isRdcssDescRepAlg(old)$} \label{appendix::alg-is-desc-CAS-RDCSS}
  \State \label{appendix::alg-help-complete-invoke-CAS-RDCSS} $\completeAlg(old)$
  \State \returnCmd{$\rdcssCasAlg(pt, exp, new)$}
\Else
  \State \label{appendix::alg-return-old-value-CAS-RDCSS} \returnCmd{$old$}
\EndIf
\EndProc
%
\State 
%
\Proc{$\rdcssWriteAlg$}{$pt : \DataPtType$, $v: \ValType$}
\State \label{appendix::alg-access-Write-RDCSS} $old \gets \readRepAlg pt$
\If {$\isRdcssDescRepAlg(old)$} \label{appendix::alg-is-desc-Write-RDCSS}
  \State \label{appendix::alg-help-complete-invoke-Write-RDCSS} $\completeAlg(old)$
  \State $\rdcssWriteAlg(pt, v)$
\Else 
  \State \label{appendix::alg-attempt-write-Write-RDCSS} $x \gets \casRepAlg(pt, old, v)$
  \If {$x \neq old$}
    \State $\rdcssWriteAlg(pt, v)$
  \EndIf
\EndIf
\EndProc
%
\State 
%
\Proc{$\rdcssReadCtlAlg$}{$pt : \ControlPtType$}
\State \label{appendix::control-Read-RDCSS} \returnCmd{$\readRepAlg pt$}
\EndProc
%
\State 
%
\Proc{$\rdcssCasCtlAlg$}{$pt : \ControlPtType$, $exp$, $new: \ValType$}
\State \label{appendix::control-CAS-RDCSS} \returnCmd{$\casRepAlg(pt, exp, new)$}
\EndProc
%
\State 
%
\Proc{$\rdcssWriteCtlAlg$}{$pt : \ControlPtType$, $v: \ValType$}
\State \label{appendix::control-Write-RDCSS} $pt\ \writeRepAlg\ v$
\EndProc
%
\State 
%
\Proc{$\rdcssAllocAlg$}{$v: \ValType$, $k : \ptKind$}
\If {$k = \CONTROL$}
  \State \label{appendix::alloc-control-Alloc-RDCSS} \returnCmd{$\allocRepAlg(v)$} as $\ControlPtType$
\Else 
  \State \label{appendix::alloc-data-Alloc-RDCSS} \returnCmd{$\allocRepAlg(v)$} as $\DataPtType$
\EndIf
\EndProc
\end{algorithmic}
\end{multicols*}
\caption{RDCSS implementation.}
\label{appendix::alg-RDCSS}
\end{figure}

We now define the span structure for RDCSS.

\begin{defn}[Span Structure for RDCSS]
We denote the structure by $\genSpanStructName{\RDCSS}$.

A span is either a 2-tuple of the form $(a,b)$ or
a 1-tuple of the form $(a)$, where $a,b$ are rep events in the execution history.

For each pointer $p$, set $\spans p$ is defined by the
following list of spans containing rep events,
\begin{enumerate}
\item Any $(b)$ such that, 

\begin{itemize}
\item $p$ is a $\DATA$ pointer
\item $\lineProp b = \ref{appendix::alg-CAS-RDCSS}$
\item $\neg \isRdcssDescRepAlg(\outputProp b)$
\item $b$ is a failed $\casRepAlg$ reading pointer $p$
\end{itemize}

\item Any $(b,c)$ such that, 

\begin{itemize}
\item $p$ is a $\DATA$ pointer
\item $\lineProp b = \ref{appendix::alg-CAS-RDCSS}$
\item $\lineProp c = \ref{appendix::alg-pt2-write-success-RDCSS}$
\item $\neg \isRdcssDescRepAlg(\outputProp b)$
\item $b$ executes before $c$ and there is no other $p$-write rep event between $b$ and $c$.
\item $b$ and $c$ are successful $\casRepAlg$es having the form 
$b = \casRepAlg(p,\_,n)$ and $c = \casRepAlg(p,n,\_)$ for some $n$
such that $\isRdcssDescRepAlg(n)$
\end{itemize}

\item Any $(b,c)$ such that,

\begin{itemize}
\item $p$ is a $\DATA$ pointer
\item $\lineProp b = \ref{appendix::alg-CAS-RDCSS}$
\item $\lineProp c = \ref{appendix::alg-pt2-write-fail-RDCSS}$
\item $\neg \isRdcssDescRepAlg(\outputProp b)$
\item $b$ executes before $c$ and there is no other $p$-write rep event between $b$ and $c$.
\item $b$ and $c$ are successful $\casRepAlg$es having the form 
$b = \casRepAlg(p,e,n)$ and $c = \casRepAlg(p,n,e)$ for some $e,n$
such that $\isRdcssDescRepAlg(n)$
\end{itemize}

\item Any $(b)$ such that,

\begin{itemize}
\item $p$ is a $\CONTROL$ pointer
\item $\lineProp b = \ref{appendix::alg-pt1-read-RDCSS}$
\item $b$ reads pointer $p$
\end{itemize}

\item Any $(b)$ such that,

\begin{itemize}
\item $p$ is a $\DATA$ pointer
\item $\lineProp b = \ref{appendix::alg-access-Read-RDCSS}$
\item $\neg \isRdcssDescRepAlg(\outputProp b)$
\item $b$ reads pointer $p$
\end{itemize}

\item Any $(b)$ such that,

\begin{itemize}
\item $p$ is a $\DATA$ pointer
\item $\lineProp b = \ref{appendix::alg-access-CAS-RDCSS}$
\item $\neg \isRdcssDescRepAlg(\outputProp b)$
\item $b$ accesses pointer $p$
\end{itemize}

\item Any $(b)$ such that,

\begin{itemize}
\item $p$ is a $\DATA$ pointer
\item $\lineProp b = \ref{appendix::alg-attempt-write-Write-RDCSS}$
\item $\neg \isRdcssDescRepAlg(\outputProp b)$
\item $b$ is a successful $\casRepAlg$ writing into pointer $p$
\end{itemize}

\item Any $(b)$ such that,

\begin{itemize}
\item $p$ is a $\CONTROL$ pointer
\item $\lineProp b = \ref{appendix::control-Read-RDCSS}$
\item $b$ reads pointer $p$
\end{itemize}

\item Any $(b)$ such that,

\begin{itemize}
\item $p$ is a $\CONTROL$ pointer
\item $\lineProp b = \ref{appendix::control-CAS-RDCSS}$
\item $b$ accesses pointer $p$
\end{itemize}

\item Any $(b)$ such that, 

\begin{itemize}
\item $p$ is a $\CONTROL$ pointer
\item $\lineProp b = \ref{appendix::control-Write-RDCSS}$
\item $b$ writes at pointer $p$
\end{itemize}

\item Any $(b)$ such that $\lineProp b = \ref{appendix::alloc-control-Alloc-RDCSS}$ and
$b$ has as output $\CONTROL$ pointer $p$.
\item Any $(b)$ such that $\lineProp b = \ref{appendix::alloc-data-Alloc-RDCSS}$ and
$b$ has as output $\DATA$ pointer $p$.
\end{enumerate}

With this, the set of writer spans $\writesAbs p$ can be defined as,
\begin{align*}
\writesSpans p & \defini \{ (b) \in \spans p \mid b \text{ writes or allocs pointer } p \} \cup 
\{ (b,c) \in \spans p \mid \lineProp c = \ref{appendix::alg-pt2-write-success-RDCSS} \} 
\end{align*}

The set of alloc spans $\allocsAbs p$ as follows,
\begin{align*}
\allocsSpans p & \defini \{ (b) \in \writesSpans p \mid b \text{ allocs pointer } p \wedge 
(\lineProp b = \ref{appendix::alloc-control-Alloc-RDCSS} \vee \lineProp p = \ref{appendix::alloc-data-Alloc-RDCSS}) \}
\end{align*}

We now define the denotation $\runFunc x$ by cases on 
event $x$,
\begin{enumerate}

\item Case $x = \rdcssAlg(desc)$.

\begin{enumerate}

\item If there are $(c) \in \spans {\pointTwo {desc}}$, and rep event $i$ such that,

\begin{itemize}
\item $i = \allocRepAlg(desc)$ with code line \ref{appendix::alg-alloc-desc-RDCSS},
\item If $T$ is the thread that invoked $x$, then $T$ executes $i$ 
within the invocation of $x$,
\item $c = \casRepAlg(\pointTwo{desc}, \expTwo{desc}, \outputProp i)$ with code line \ref{appendix::alg-CAS-RDCSS},
\item $\outputProp c \neq \expTwo{desc}$,
\end{itemize}
then
$\runFunc x \defini (\{ (c) \},\ \outputProp c)$.

\item If there are $(r) \in \spans {\pointOne {desc}}$,
$(c_1,c_2) \in \writesSpans {\pointTwo {desc}}$, and rep event $i$ such that,

\begin{itemize}
\item $i = \allocRepAlg(desc)$ with code line \ref{appendix::alg-alloc-desc-RDCSS},
\item If $T$ is the thread that invoked $x$, then $T$ executes $i$ 
within the invocation of $x$,
\item $r = {\readRepAlg \pointOne {desc}}$ with code line \ref{appendix::alg-pt1-read-RDCSS},
\item $c_1 = \casRepAlg(\pointTwo{desc}, \expTwo{desc}, \outputProp i)$ with code line \ref{appendix::alg-CAS-RDCSS},
\item $c_2 = \casRepAlg(\pointTwo{desc}, \outputProp i, \newTwo{desc})$ with code line \ref{appendix::alg-pt2-write-success-RDCSS},
\item If $T$ is the thread that executed $c_2$, then $r$ is the last read
carried out by $T$ before executing $c_2$,
\end{itemize}
then 
$\runFunc x \defini (\{ (r),\ (c_1,c_2) \},\ \expTwo {desc})$.

\item If there are $(r) \in \spans {\pointOne {desc}}$,
$(c_1,c_2) \in \spans {\pointTwo {desc}}$, and rep event $i$ such that,

\begin{itemize}
\item $i = \allocRepAlg(desc)$ with code line \ref{appendix::alg-alloc-desc-RDCSS},
\item If $T$ is the thread that invoked $x$, then $T$ executes $i$ 
within the invocation of $x$,
\item $r = {\readRepAlg \pointOne {desc}}$ with code line \ref{appendix::alg-pt1-read-RDCSS},
\item $c_1 = \casRepAlg(\pointTwo{desc}, \expTwo{desc}, \outputProp i)$ with code line \ref{appendix::alg-CAS-RDCSS},
\item $c_2 = \casRepAlg(\pointTwo{desc}, \outputProp i, \expTwo{desc})$ with code line \ref{appendix::alg-pt2-write-fail-RDCSS},
\item If $T$ is the thread that executed $c_2$, then $r$ is the last read
carried out by $T$ before executing $c_2$,
\end{itemize} 
then 
$\runFunc x \defini (\{ (r),\ (c_1,c_2) \},\ \expTwo {desc})$.

\item Otherwise, $\runFunc x \defini (\emptyset, \bot)$.
\end{enumerate}

\item Case $x = \rdcssReadAlg(p)$.
\begin{enumerate}
\item If there is $(r) \in \spans p$ such that,

\begin{itemize}
\item If $T$ is the thread that invoked $x$, then $T$ executes $r$ 
within the invocation of $x$,
\item $r = {\readRepAlg p}$ with code line \ref{appendix::alg-access-Read-RDCSS},
\end{itemize}
then
$\runFunc x \defini (\{ (r) \},\ \outputProp r)$.

\item Otherwise, $\runFunc x \defini (\emptyset, \bot)$.
\end{enumerate}

\item Case $x = \rdcssReadCtlAlg(p)$.
\begin{enumerate}

\item If there is $(r) \in \spans p$ such that,

\begin{itemize}
\item If $T$ is the thread that invoked $x$, then $T$ executes $r$ 
within the invocation of $x$,
\item $r = {\readRepAlg p}$ with code line \ref{appendix::control-Read-RDCSS},
\end{itemize}
then
$\runFunc x \defini (\{ (r) \},\ \outputProp r)$.

\item Otherwise, $\runFunc x \defini (\emptyset, \bot)$.
\end{enumerate}

\item Case $x = \rdcssCasAlg(p,e,n)$.
\begin{enumerate}
\item If there is $(c) \in \spans p$ such that,

\begin{itemize}
\item If $T$ is the thread that invoked $x$, then $T$ executes $c$ 
within the invocation of $x$,
\item $c = \casRepAlg(p,e,n)$ with code line \ref{appendix::alg-access-CAS-RDCSS},
\end{itemize}
then
$\runFunc x \defini (\{ (c) \},\ \outputProp c)$.

\item Otherwise, $\runFunc x \defini (\emptyset, \bot)$.

\end{enumerate}

\item Case $x = \rdcssCasCtlAlg(p,e,n)$.

\begin{enumerate}
\item If there is $(c) \in \spans p$ such that,

\begin{itemize}
\item If $T$ is the thread that invoked $x$, then $T$ executes $c$ 
within the invocation of $x$,
\item $c = \casRepAlg(p,e,n)$ with code line \ref{appendix::control-CAS-RDCSS},
\end{itemize}
then
$\runFunc x \defini (\{ (c) \},\ \outputProp c)$.

\item Otherwise, $\runFunc x \defini (\emptyset, \bot)$.
\end{enumerate}

\item Case $x = \rdcssWriteAlg(p,v)$.
\begin{enumerate}
\item If there are $(c) \in \writesSpans p$ and rep event $r$ such that,

\begin{itemize}
\item If $T$ is the thread that invoked $x$, then $T$ executes $c$ 
within the invocation of $x$,
\item $r = {\readRepAlg p}$ with code line \ref{appendix::alg-access-Write-RDCSS},
\item $c = \casRepAlg(p, \outputProp r, v)$ with code line \ref{appendix::alg-attempt-write-Write-RDCSS},
\item If $T$ is the thread that invoked $x$, then $r$ is the last read
carried out by $T$ before the execution of $c$,
\end{itemize}
then
$\runFunc x \defini (\{ (c) \},\ \unitValue)$.

\item Otherwise, $\runFunc x \defini (\emptyset, \bot)$.
\end{enumerate}

\item Case $x = \rdcssWriteCtlAlg(p,v)$.
\begin{enumerate}
\item If there is $(c) \in \writesSpans p$ such that,

\begin{itemize}
\item If $T$ is the thread that invoked $x$, then $T$ executes $c$ 
within the invocation of $x$,
\item $c = {(p\ \writeRepAlg\ v)}$ with code line \ref{appendix::control-Write-RDCSS},
\end{itemize}
then
$\runFunc x \defini (\{ (c) \},\ \unitValue)$.

\item Otherwise, $\runFunc x \defini (\emptyset, \bot)$.
\end{enumerate}

\item Case $x = \rdcssAllocAlg(v,k)$.
\begin{enumerate}
\item If $\outputProp x \neq \bot$, $\ETimeProp x \neq \bot$, $k = \CONTROL$ and
there is $(i) \in \writesSpans {\outputProp x}$ such that,

\begin{itemize}
\item If $T$ is the thread that invoked $x$, then $T$ executes $i$ 
within the invocation of $x$,
\item $i = \allocRepAlg(v)$ with code line \ref{appendix::alloc-control-Alloc-RDCSS},
\end{itemize}
then
$\runFunc x \defini (\{ (i) \},\ \outputProp i)$.

\item If $\outputProp x \neq \bot$, $\ETimeProp x \neq \bot$, $k = \DATA$ and
there is $(i) \in \writesSpans {\outputProp x}$ such that,

\begin{itemize}
\item If $T$ is the thread that invoked $x$, then $T$ executes $i$ 
within the invocation of $x$,
\item $i = \allocRepAlg(v)$ with code line \ref{appendix::alloc-data-Alloc-RDCSS},
\end{itemize}
then
$\runFunc x \defini (\{ (i) \},\ \outputProp i)$.

\item Otherwise, $\runFunc x \defini (\emptyset, \bot)$.
\end{enumerate}

\end{enumerate}

\end{defn}

The list defining $\spans p$ could contain further 
single-instruction spans (for example, by including
code lines \ref{appendix::alg-alloc-desc-RDCSS},
\ref{appendix::alg-desc-read-RDCSS}, and
\ref{appendix::alg-access-Write-RDCSS}), but the listed 
ones define the minimal set that will allow us to prove 
the span axioms. 
 
Notice that the only spans in $\spans p$ with 
more than one instruction are those of the form
$(b,c)$, where $b$ is the instruction that writes the 
descriptor and $c$ is the instruction that resolves the 
descriptor.

Also, notice that the only spans of the form $(b,c)$ 
that are considered to be writer spans in set 
$\writesSpans p$ are those where $c$ successfully
updates the pointer to its new value at Line
\ref{appendix::alg-pt2-write-success-RDCSS}. In
other words, spans that ``undo'' the descriptor
by successfully executing the CAS at
Line \ref{appendix::alg-pt2-write-fail-RDCSS}
are not considered to be writer spans, since they
belong to a failed RDCSS.

We argue that $\runFunc x$ is well-defined.
For that matter, we need to show that the 
function's conditions pick spans uniquely.

First, let us focus on the $\rdcssAlg$ cases. Notice that whenever
an $\rdcssAlg(desc)$ invocation starts, a fresh pointer
$d$ is created at Line \ref{appendix::alg-alloc-desc-RDCSS} 
to store the descriptor $desc$.
This pointer $d$ will serve as unique identifier for 
the invocation, because the invoking thread executes
Line \ref{appendix::alg-alloc-desc-RDCSS} only once.

The invoking thread then enters the $\rdcssLoopAlg$
procedure and tries to write $d$ into pointer 
$\pointTwo {desc}$ by attempting the CAS at Line
\ref{appendix::alg-CAS-RDCSS}. 

If the CAS fails and its output $v$ satisfies 
$\neg \isRdcssDescRepAlg(v)$, then the invoking 
thread will finish the invocation by returning $v$. 
Notice that once these
two conditions are satisfied (i.e. CAS fails and 
$\neg \isRdcssDescRepAlg(v)$ holds), there will
be no more attempts of executing Line
\ref{appendix::alg-CAS-RDCSS}.
In other words, there is at most one instruction 
of the form 
$\casRepAlg(\pointTwo {desc}, \expTwo {desc}, d)$ 
at line code \ref{appendix::alg-CAS-RDCSS}, such that
it is a failing CAS and its output does not
satisfy $\isRdcssDescRepAlg$. 
Therefore, if there is at least one such instruction,
it will be unique. This justifies case 1.(a) of
$\runFunc x$ (also, by case (1) in the definition of set 
$\spans {\pointTwo {desc}}$).

On the contrary, if the instruction 
$\casRepAlg(\pointTwo {desc}, \expTwo {desc}, d)$
at Line \ref{appendix::alg-CAS-RDCSS} succeeds,
then this CAS with those parameters cannot succeed again: the  
thread will invoke the $\completeAlg$ procedure
afterwards and finish by returning $\expTwo {desc}$, 
and other threads cannot execute Line
\ref{appendix::alg-CAS-RDCSS} with parameter $d$
because those threads must have executed Line
\ref{appendix::alg-alloc-desc-RDCSS} to
create a fresh pointer to pass to the CAS
(i.e. their CAS will have the form
$\casRepAlg(\_, \_, d')$, where $d' \neq d$).

Therefore, since 
$\casRepAlg(\pointTwo {desc}, \expTwo {desc}, d)$
at Line \ref{appendix::alg-CAS-RDCSS} can succeed 
at most once, either of
$\casRepAlg(\pointTwo {desc}, d, \newTwo {desc})$,
$\casRepAlg(\pointTwo {desc}, d, \expTwo {desc})$
at Lines \ref{appendix::alg-pt2-write-success-RDCSS}
and \ref{appendix::alg-pt2-write-fail-RDCSS} can
succeed at most once because
these CASes expect the value $d$ to be present 
in the pointer and $d$ can only be written by the 
CAS at Line \ref{appendix::alg-CAS-RDCSS}.

This justifies cases 1.(b) and 1.(c) of
$\runFunc x$. Notice that the read $r$
in those cases is uniquely chosen because it is
required to be the \emph{last} read 
before $c_2$ such that $r$ was executed by the thread that 
successfully executed $c_2$.

For the cases $x = \rdcssReadAlg(p)$ and
$x = \rdcssCasAlg(p,e,n)$,
their condition is choosing instructions at Lines
\ref{appendix::alg-access-Read-RDCSS} and
\ref{appendix::alg-access-CAS-RDCSS}
such that their output do not satisfy 
$\isRdcssDescRepAlg$ (by definition of set $\spans p$).
But this condition can happen at most once during the
invocation of $x$, 
because the invoking thread finishes $x$ once
the condition is satisfied.
 
The cases for $x = \rdcssWriteAlg(p,v)$ follow a 
similar strategy. While the cases for 
$x = \rdcssAllocAlg(v,k)$ are trivial, since the
allocs execute at most once within the invocation of
$x$.

Finally, the cases for $x = \rdcssReadCtlAlg$, $\rdcssCasCtlAlg$,
and $\rdcssWriteCtlAlg$ are trivial, since
their code execute at most once during their invocation.

We are ready to prove that 
$\genStructName{\RDCSS}(\visObsSymbol{\pointerIndx p}, \visSepSymbol{\pointerIndx p})$ is implemented by span
structure $\genSpanStructName{\RDCSS}$, i.e., we are going to prove that
the span axioms of Figure \ref{fig:descriptor-lifespans-axioms} are satisfied. We require a couple of lemmas.

\begin{lem}
\label{lem::appendix::impl::rdcss::unique-alloc}
If $i_1$, $i_2$ are allocs at either 
Line \ref{appendix::alloc-control-Alloc-RDCSS} or Line
\ref{appendix::alloc-data-Alloc-RDCSS} 
such that they allocate the same pointer, then $i_1 = i_2$.
\end{lem}

\begin{prf}
Allocs create fresh pointers, and pointers are not deallocated.
Therefore, if $\linReps {i_1} {i_2}$, then $i_2$ must create
a pointer that is different from the one created by $i_1$ (Contradicting 
the hypothesis).
Similarly, if $\linReps {i_2} {i_1}$.
Hence, $i_1 = i_2$.
\end{prf}

\begin{lem}
\label{lem::appendix::impl::rdcss::writers-belong-to-writer-procs}
If $b \in \writesSpans p$, 
then there is $x$ such that $b \in \hspans{}{x}$ and $\inputVal x p v$,
where $v$ is the value written by $\lastRep b$.
\end{lem}

\begin{prf}
Each span in $\writesSpans p$ occurs in one of the cases in
the definition of $\runFunc x$. 

For example, consider the span of the form $(b,c)$ at entry 2 in the definition of $\spans p$.
Since $(b,c)$ contains the successful CASes at lines \ref{appendix::alg-CAS-RDCSS}
and \ref{appendix::alg-pt2-write-success-RDCSS}, those CASes must have the 
form $\casRepAlg(\pointTwo {desc},\expTwo {desc},d)$ and 
$\casRepAlg(\pointTwo {desc},d,\newTwo {desc})$, for some $desc$, $d$, because
the CASes are instances of the code. In addition, $d$ must have been generated 
by executing line \ref{appendix::alg-alloc-desc-RDCSS} by some invocation
of procedure $\rdcssAlg(desc)$. Even more, the thread $T$ that executed line
\ref{appendix::alg-pt2-write-success-RDCSS} must have executed the read
at line \ref{appendix::alg-pt1-read-RDCSS}, and this read must have occurred
after $d$ was written at Line \ref{appendix::alg-CAS-RDCSS}, otherwise, $T$
would not be helping $d$ at the $\completeAlg$ procedure.

All the above satisfy case 1.(b) in the definition of $\runFunc x$.
Notice that $\lastRep c$ writes value $\newTwo {desc}$ at $\pointTwo {desc}$, and
$\inputVal x {\pointTwo {desc}} {\newTwo {desc}}$ holds by
definition of $\inputValName$ for the $\rdcssAlg$ case.

The rest of spans in $\writesSpans p$ are even simpler, because 
they are 1-tuple spans of the form $(b)$, and $b$ is executed by the same thread
that invoked the exportable procedure that $b$ is part of.
\end{prf}

\begin{lem}
\label{lem::appendix::impl::rdcss::inputs-are-written-by-lifespans}
Let $b \in \hspans{}{x} \cap \writesSpans p$. If 
$\inputVal x p v$, then $\lastRep b$ writes value
$v$ into $p$.
\end{lem}

\begin{prf}
By simple inspection on the cases for
$\runFunc x$ and $\inputValName$.
\end{prf}

\begin{lem}
\label{lem::appendix::impl::rdcss::visibility-lemma}
Suppose $x$ is one of $\rdcssAlg$, $\rdcssCasAlg$, $\rdcssReadAlg$, $\rdcssCasCtlAlg$,
or $\rdcssReadCtlAlg$. Suppose $p \in \PtsType$ and $v \in \ValType$. Then,
the following statements are equivalent,
\begin{itemize}
\item There is $y$ such that $\visObs {\pointerIndx p} y x$
and $\inputVal y p v$.
\item There is $b \in \hspans{p}{x}$ such that $\firstRep b$ reads value $v$.
\end{itemize}
\end{lem}

\begin{prf}
We prove each direction.

$\Longrightarrow$. By definition of $\visObs {\pointerIndx p} y x$, 
there are $b_y \in \hspans{p}{y}$,
$b_x \in \hspans{p}{x}$ such that $b_y$ is the most recent span under $\precedesSpansEqSymbol$ 
satisfying $b_y \in \writesSpans p$ and $\precedesSpans {b_y} {b_x}$.
By Lemma \ref{lem::appendix::impl::rdcss::inputs-are-written-by-lifespans}, 
$\lastRep {b_y}$ writes value $v$ into $p$.

We argue that any $p$-mutating rep event in between $\lastRep {b_y}$ and $\firstRep {b_x}$ must eventually
restore the value to $v$ before $\firstRep {b_x}$ reads $p$.

If the $p$-write rep events correspond to lines \ref{appendix::alg-access-CAS-RDCSS},
\ref{appendix::alg-attempt-write-Write-RDCSS},
\ref{appendix::control-CAS-RDCSS}, 
\ref{appendix::control-Write-RDCSS}, they would 
belong to a 1-span, contradicting that there are no $p$-writing spans in between 
$b_y$ and $b_x$. So, let us focus on rep events at lines \ref{appendix::alg-CAS-RDCSS},
\ref{appendix::alg-pt2-write-success-RDCSS}, and \ref{appendix::alg-pt2-write-fail-RDCSS}.
If the CAS at line \ref{appendix::alg-CAS-RDCSS} succeeds, no other $p$-write rep event can occur
up to the point where the descriptor is resolved, and any read would discover the descriptor up
to the point where the descriptor is resolved. Therefore $\firstRep {b_x}$
must occur after the descriptor is resolved by either line 
\ref{appendix::alg-pt2-write-success-RDCSS} or \ref{appendix::alg-pt2-write-fail-RDCSS}.
If resolution happens at line \ref{appendix::alg-pt2-write-success-RDCSS}, we would have
a $p$-write span in between $b_y$ and $b_x$ which is impossible by hypothesis. Therefore,
resolution must happen at line \ref{appendix::alg-pt2-write-fail-RDCSS}. 
But line \ref{appendix::alg-pt2-write-fail-RDCSS} undoes or restores to whatever value was in $p$
before line \ref{appendix::alg-CAS-RDCSS} wrote the descriptor.
This means that if $v$ was in $p$ before line \ref{appendix::alg-CAS-RDCSS} executed,
$v$ will be in $p$ after line \ref{appendix::alg-pt2-write-fail-RDCSS} executes. 
Therefore $v$ is preserved and $\firstRep{b_x}$ reads $v$.

$\Longleftarrow$. Let $b_x \in \hspans{p}{x}$ 
such that $\firstRep {b_x}$ reads value $v \in \ValType$. 
Therefore, there is a most recent rep event $w$ that wrote $v$ into $p$.
If $w$ corresponds to line \ref{appendix::alg-pt2-write-fail-RDCSS} 
then $w$ is the last rep event in a span that failed 
to modify the pointer, which means that the pointer is getting restored 
to the value it had before the descriptor was written at line \ref{appendix::alg-CAS-RDCSS}.
If this is the case, we just keep searching for a $p$-write rep event of $v$ 
that does not correspond to line \ref{appendix::alg-pt2-write-fail-RDCSS}.
This rep event exists because there is an alloc of $p$ as first rep event for $p$.
If $w'$ is the chosen $p$-write event, we will have $w' = \lastRep b$ for some span $b$ in $\writesSpans p$,
since the set $\spans p$ covers all possible places where a write of a $\ValType$
value could occur in the code. Span $b$ will be the most recent $p$-write span 
before $b_x$, since all the more recent spans we found correspond to failing spans resolving
at line \ref{appendix::alg-pt2-write-fail-RDCSS}.
By Lemma \ref{lem::appendix::impl::rdcss::writers-belong-to-writer-procs},
there is $y$ such that $b \in \hspans{}{y}$ and $\inputVal y p v$.
Therefore, $\visObs {\pointerIndx p} y x$
by definition.
\end{prf}

\begin{lem}
\label{lem::appendix::impl::rdcss::non-interference-of-lifespans}
Axiom \axiomDRef{principle::descriptors-do-not-interfere} holds.
\end{lem}

\begin{prf}
Since this is trivial when $b_1$ and $b_2$ are 1-tuple spans, 
we only need to consider the cases 
when $b_1$ is a 1-span and $b_2$ a 2-tuple span or when both
$b_1$ and $b_2$ are 2-tuple spans.

\begin{itemize}
\item Case $b_1 = (c)$ is a 1-tuple span and $b_2 = (e,g)$ a 2-tuple span.

Since $b_2$ is a 2-tuple, by definition of $\spans p$, pointer $p$ is of type
$\DATA$. Since every procedure invocation requires that input pointers be previously 
allocated, there must be some $\DATA$ rep alloc $i$ such that
$\linReps i e$. We also know that there cannot be a $p$-writer rep in 
between $e$ and $g$, because once a descriptor is written by $e$, no other write can
occur until the descriptor is removed by $g$.

If $c$ is an alloc, by Lemma \ref{lem::appendix::impl::rdcss::unique-alloc},
$c = i$ and $\precedesSpans {b_1} {b_2}$ follows. So, we can assume that $c$ is not an alloc. 

By going through all the $\DATA$ pointer cases for $(c) \in \spans p$,
we see that all the cases imply $\neg \isRdcssDescRepAlg(\outputProp c)$,
i.e., $c$ does not read a descriptor.

Therefore, $c$ cannot occur in between $e$ and $g$, because $e$ writes a descriptor. 
Also, the cases $c = e$ or $c = g$ are excluded because $\spans p$
is defined so that 1-spans and 2-tuple spans do not share rep events.
 
\item Case $b_1 = (c,d)$ is a 2-tuple span and $b_2 = (e,g)$ a 2-tuple span.

By definition of $\spans p$, we know there is no writer into $p$
in between $c$ and $d$ and in between $e$ and $g$. But $c$, $d$, $e$, $g$
are writers into $p$ by definition of $\spans p$. 
Therefore, we must have $\linReps e c$, or 
$e = c$, or $\linReps d e$ (the case $d = e$ is impossible, because
$d$ and $e$ correspond to different code lines: $e$ writes a descriptor, while
$d$ removes it). 

\begin{itemize}
\item Case $\linReps d e$. Hence $\precedesSpans {b_1} {b_2}$.

\item Case $\linReps e c$. We have either $\linReps g c$, or 
$g = d$, or $\linReps d g$ (the case $g = c$ is impossible,
because $g$ removes a descriptor, while $c$ writes one).

The cases $g = d$ and $\linReps d g$ would lead to 
$e \linRepsSymbol c \linRepsSymbol g$ (Contradiction).
Therefore, $\linReps g c$ and $\precedesSpans {b_2} {b_1}$ follows.

\item Case $e = c$. We have either  
$g = d$ or $\linReps d g$ (the case $\linReps g c$ is
impossible because $g$ occurs after $e = c$).

But the case $\linReps d g$ leads to 
$e \linRepsSymbol d \linRepsSymbol g$ (Contradiction).
Therefore $g = d$ also holds and $b_1 = b_2$ follows.
\end{itemize}
\end{itemize}
\end{prf}

\begin{lem}
Axiom \axiomDRef{principle::all-descriptors-are-written-before-any-resolution} holds.
\end{lem}

\begin{prf}
This is trivial when $x$ has a denotation with only one span. 
So, it is enough to check the two cases for $x = \rdcssAlg(desc)$ when
the denotation is of the form $\hspans{}{x} = \{ (c_1,c_2),\ (r) \}$. 

But in both cases we have $c_1 \linRepsEqSymbol r \linRepsEqSymbol c_2$
because the control read $r$ and descriptor removal $c_2$ are executed by the same thread
inside the $\completeAlg$ procedure, in that order.
Additionally, $r$ must occur after $c_1$ because $r$ occurs inside the 
$\completeAlg$ procedure, which can only be invoked after the descriptor written
by $c_1$ is discovered.
Therefore, it does not matter how we choose 
$b_1, b_2 \in \hspans{}{x}$, we will always have 
$\linRepsEq {\firstRep {b_1}} {\lastRep {b_2}}$.
\end{prf}

\begin{lem}
	\label{lem::appendix::impl::rdcss::finished-are-non-empty-axiom-holds}
	Axioms
	\axiomDRef{principle::finished-operations-have-a-run} holds.
\end{lem}

\begin{prf}
	We prove each procedure in turn.
	\begin{itemize}
		\item Case $\rdcssAlg(desc)$. Suppose that $\rdcssAlg(desc)$ finished.
		Then, the invoking thread $T$ must have created a pointer $d$ storing $desc$
		at line~\ref{appendix::alg-alloc-desc-RDCSS}, and finished the invocation of $\rdcssLoopAlg(d, desc)$.  
		Since $\rdcssLoopAlg$ is a recursive method and we know that $T$
		finished executing it, this means that $T$ must have reached an
		invocation of $\rdcssLoopAlg$ that did not recurse, i.e., an invocation
		where lines \ref{appendix::alg-help-complete-invoke-RDCSS}-\ref{appendix::rdcss-try-again-RDCSS} were not executed. So, we can assume we are on
		such execution. Since $T$ reached line~20, it means that the CAS at
		line~\ref{appendix::alg-CAS-RDCSS} must have returned a non-descriptor value $v$ (otherwise, $T$
		would have enter the conditional at line~\ref{appendix::alg-is-desc-RDCSS}, executing lines  \ref{appendix::alg-help-complete-invoke-RDCSS}-\ref{appendix::rdcss-try-again-RDCSS}).
		
		\begin{itemize}
			\item Case $v \neq \expTwo{desc}$. Then the CAS at line~\ref{appendix::alg-CAS-RDCSS} failed, $T$ did not
			enter the conditional at line~\ref{appendix::alg-is-desc-RDCSS} and returned $v$. This corresponds to
			(1).(a) in the definition of $\runFuncSymbol$ with output whatever line~\ref{appendix::alg-CAS-RDCSS}
			returned (which is equal to $v$) and the denotation is not empty.
			
			\item Case $v = \expTwo{desc}$. Then the CAS at line~\ref{appendix::alg-CAS-RDCSS} wrote the descriptor
			$d$, and $T$ terminated $\completeAlg(d)$ at line~19 before returning $v$ at
			line~20. Since $T$ terminated $\completeAlg(d)$, $T$ must have reached
			either line \ref{appendix::alg-pt2-write-success-RDCSS} or \ref{appendix::alg-pt2-write-fail-RDCSS}.
			
			\begin{itemize}
				\item Subcase: $T$ reached line 26. Let us suppose that the CAS at line
				26 succeeded for $T$ (so that the descriptor was removed and
				replaced with the new value). The read at line 24 must have
				returned $\expOne{desc}$. Since $T$ executed both lines 24 and 26, this
				corresponds to (1).(b) in the definition of $\runFuncSymbol$, with output
				$\expTwo{desc}= v$, and the denotation is not empty.  Let us now suppose that the
				CAS at line 26 did not succeed for $T$. This means that some other
				thread $T'$ has already removed the descriptor, which means that
				$T'$ successfully executed either line 26 or 28. If $T'$ reached
				line 26, then $T'$ must have read $\expOne{desc}$ at line 24 and once
				again this corresponds to (1).(b) in the definition of $\runFuncSymbol$, with
				output $\expTwo{desc} = v$ and the denotation is not empty. If $T'$ reached line 28,
				then $T'$ must have read a value different from $\expOne{desc}$ at line
				24, in which case, (1).(c) applies in the definition of $\runFuncSymbol$, with
				output $\expTwo{desc} = v$ and the denotation is not empty.
				
				\item Subcase: $T$ reached line 28. Let us suppose that the CAS at line
				28 succeeded for $T$ (so that the descriptor was removed and the
				pointer restored to the expected value). The read at line 24
				must have returned a value different from $\expOne{desc}$. Since $T$
				executed both lines 24 and 28, this corresponds to (1).(c) in the definition of $\runFuncSymbol$, 
				with output $\expTwo{desc} = v$, and the denotation is not
				empty.  Let us now suppose that the CAS at line 28 did not
				succeed for $T$. This means that some other thread $T'$ has already
				removed the descriptor, which means that $T'$ successfully
				executed either line 26 or 28. If $T'$ reached line 26, then $T'$
				must have read $\expOne{desc}$ at line 24 and this corresponds to
				(1).(b) in the definition of $\runFuncSymbol$, with output $\expTwo{desc} = v$ 
				and the denotation is
				not empty. If $T'$ reached line 28, then $T'$ must have read a value
				different from $\expOne{desc}$ at line 24, in which case, (1).(c)
				applies in the definition of $\runFuncSymbol$, with output $\expTwo{desc} = v$ and 
				the denotation is
				not empty.
				
			\end{itemize}
		\end{itemize}
		
		\item Case $\rdcssReadAlg(p)$. Suppose that $\rdcssReadAlg(p)$ finished.
		Let $T$ be the invoking thread. Since $\rdcssReadAlg$ is a recursive method and
		we know that $T$ finished executing it, this means that $T$ must have
		reached an invocation of $\rdcssReadAlg$ that did not recurse, i.e., an
		invocation where lines 33-34 were not executed. So, we can assume we
		are on such execution. Since $T$ reached line 36, the read at line 31
		must have returned a non-descriptor value $v$. This corresponds to
		(2).(a) in the definition of $\runFuncSymbol$, with denotation output
		whatever line 10 produced (which is equal to $v$) and it is not empty.

		\item Case $\rdcssReadCtlAlg(p)$. Suppose that $\rdcssReadCtlAlg(p)$ finished.  The invoking thread must have executed line 56. This corresponds to
		case (3).(a) in the definition of $\runFuncSymbol$, with denotation output
		whatever line 56 produced.
		
		\item Case $\rdcssCasAlg(p,e,n)$. Suppose that $\rdcssCasAlg(p,e,n)$ finished.
		Let $T$ be the invoking thread. Since rCAS is a recursive method and
		we know that $T$ finished executing it, this means that $T$ must have
		reached an invocation that did not recurse, i.e., an invocation
		where lines 40-41 were not executed. So, we can assume we are on
		such execution. Since $T$ reached line 36, the CAS at line 38 must
		have returned a non-descriptor value $v$. This corresponds to (4).(a)
		in the definition of $\runFuncSymbol$, with denotation output whatever line
		38 produced (which is equal to $v$) and it is not empty.

		\item Case $\rdcssCasCtlAlg(p,e,n)$. Suppose that $\rdcssCasCtlAlg(p,e,n)$ finished.  The invoking thread must have executed line 59. This corresponds to
		case (5).(a) in the definition of $\runFuncSymbol$, with denotation output
		whatever line 59 produced.

		\item Case $\rdcssWriteAlg(p,v)$. Suppose that $\rdcssWriteAlg(p,v)$ finished.
		Let $T$ be the invoking thread. Since $\rdcssWriteAlg$ is a recursive method and
		we know that $T$ finished executing it, this means that $T$ must have
		reached an invocation that did not recurse, i.e., an invocation
		where lines 48-49,53 were not executed. So, we can assume we are on
		such execution. Since $T$ must have returned after reaching line 54
		without entering the conditional at line 52, the CAS at line 51 must have
		returned $old$ (meaning that the CAS succeeded), and at the same time,
		the read at line 46 must have returned $old$ and $old$ is a
		non-descriptor value. Since the CAS at line 51 succeeded, this
		corresponds to (6).(a) in the definition of $\runFuncSymbol$, with
		denotation output $tt$ and the denotation is not empty.

		\item Case $\rdcssWriteCtlAlg(p,v)$. Suppose that $\rdcssWriteCtlAlg(p,v)$ finished.  The invoking thread must have executed line 62. This corresponds to
		case (7).(a) in the definition of $\runFuncSymbol$, with denotation output $tt$
		and the denotation is not empty.

		\item Case $\rdcssAllocAlg(v,k)$. Suppose that $\rdcssAllocAlg(v,k)$ finished.
		Let $T$ be the invoking thread. If $k = \ControlPtType$, then $T$ must have executed
		line 66 producing some control pointer $p$ as output. This corresponds
		to (8).(a) in the definition of $\runFuncSymbol$ with output
		whatever line 66 produced and the denotation is not empty. If $k = \DataPtType$, then $T$
		must have executed line 68 producing some data pointer $p$ as
		output. This corresponds to (8).(b) in the definition of $\runFuncSymbol$ 
		with output whatever line 68 produced and the denotation is not empty.
	\end{itemize}
\end{prf}

\begin{lem}
\label{lem::appendix::impl::rdcss::lifespans-lifestages}
Axiom \axiomDRef{principle::descriptor-write-precedes-resolution} holds.
\end{lem}

\begin{prf}
Trivial for 1-spans. For 2-tuple spans $(a,b)$, the code can only remove descriptors (rep event $b$)
only if it was previously written at line \ref{appendix::alg-CAS-RDCSS} (rep event $a$).
\end{prf}

\begin{lem}
Axiom
\axiomDRef{principle::runs-are-injective} holds.
\end{lem}

\begin{prf}
First, let us focus when neither $x$ nor $y$ are $\rdcssAlg$ events.

If $b \in \hspans{}{x} \cap \hspans{}{y}$, then (by going through all the 
non-$\rdcssAlg$ cases in $\runFuncSymbol$),
$b = (i_1)$ for some $i_1$ that is executed by the thread that invoked $x$ and 
$b = (i_2)$ for some $i_2$ that is executed by the thread that invoked $y$.
In addition, $i_1$ occurs within the invocation of $x$ and $i_2$ within
the invocation of $y$. Therefore, $i_1 = i_2$, which means that $x$ and $y$ 
are invoked by the same thread.

If $x \neq y$, then $x$ and $y$ cannot overlap in real-time, because they 
are invoked by the same thread. But this contradicts that $i_1$ occurs within
the invocation of both $x$ and $y$. Therefore, $x = y$.

Now, let us check the case when either $x$ is an $\rdcssAlg$ event or $y$ is.
Say, $x$ is an $\rdcssAlg$ event. 

It must be the case that either $\firstRep b$ has code line 
\ref{appendix::alg-CAS-RDCSS} or \ref{appendix::alg-pt1-read-RDCSS}.

If $\firstRep b$ has code line \ref{appendix::alg-CAS-RDCSS}, then
it has the form $\casRepAlg(\_,\_, d)$ where $d$ is the unique
identifier for the invocation $x$ generated at line 
\ref{appendix::alg-alloc-desc-RDCSS} (see discussion that the denotation
is well-defined). Since $b$ also occurs in the denotation of $y$ and $d$ is unique
per-invocation, it must be the case that $x = y$.

If $\firstRep b$ has code line \ref{appendix::alg-pt1-read-RDCSS},
then by definition of $\hspans{}{x}$, $\firstRep b$ is the last
read carried out by some thread $T$ before it successfully executed 
$\casRepAlg(\_,d, \_)$ 
(line \ref{appendix::alg-pt2-write-success-RDCSS} or
line \ref{appendix::alg-pt2-write-fail-RDCSS}),
where $d$ is the unique id of invocation $x$.
Since $b$ also occurs in the denotation of $y$, event $y$ must
also be an $\rdcssAlg$, and $\firstRep b$ is the last
read carried out by the same thread $T$ (it must be the same thread because no two threads
can execute the \emph{same} rep event $\firstRep b$) before it successfully executed 
$\casRepAlg(\_,d', \_)$,
where $d'$ is the unique id of invocation $y$. 

Therefore, $\casRepAlg(\_,d, \_)$ and $\casRepAlg(\_,d', \_)$ 
must be the same rep event (since they are both the most immediate CAS executed
by $T$ after $\firstRep b$).
This means that $d = d'$, which implies $x = y$ since ids are unique per invocation.
\end{prf}

\begin{lem}
Axiom
\axiomDRef{principle::writer-blocks-belong-to-runs} holds.
\end{lem}

\begin{prf}
Directly from Lemma \ref{lem::appendix::impl::rdcss::writers-belong-to-writer-procs}.
\end{prf}

\begin{lem}
Axiom
\axiomDRef{principle::postcondition-predicate-holds} holds.
\end{lem}

\begin{prf}
Suppose $\hspans{}{x} \neq \emptyset$. We see from definition of $\runFunc x$ that all non-empty cases have $\outputRunFunc x \neq \bot$.
To prove $\postPred x {\outputRunFunc x}$, we do a case analysis on $x$.

\begin{itemize}
\item Case $x = \rdcssAlg(d)$. We want to prove,
\begin{align*}
\exists z_2.\ \visObs {\pointerIndx {\pointTwo d}} {z_2} x \wedge \inputVal {z_2} {\pointTwo d} {\outputRunFunc x} \wedge {} \\
(\outputRunFunc x = \expTwo d \rightarrow \exists z_1.\ \visObs {\pointerIndx {\pointOne d}} {z_1} x)
\end{align*}

All the three cases in the definition of $\runFunc x$
have a span $b$ such that $\firstRep b$ is the CAS at line \ref{appendix::alg-CAS-RDCSS}.
In all the three cases, the value read by this CAS serves as denotation output, even when 
the CAS succeeds, in which case the CAS read the value $\expTwo{d}$.
Therefore, by Lemma \ref{lem::appendix::impl::rdcss::visibility-lemma},
there is $z_2$ such that $\visObs {\pointerIndx {\pointTwo d}} {z_2} x$
and $\inputVal{z_2} {\pointTwo d} {\outputRunFunc x}$.

Suppose $\outputRunFunc x = \expTwo d$. Hence, the CAS at line \ref{appendix::alg-CAS-RDCSS} 
must have succeeded, and the only two applicable cases in the definition of
$\runFunc x$ have a span $r$ such that $\firstRep r$ reads some value
$v'$ in pointer $\pointOne d$ at Line \ref{appendix::alg-pt1-read-RDCSS}.
Therefore, by Lemma \ref{lem::appendix::impl::rdcss::visibility-lemma},
there is $z_1$ such that $\visObs {\pointerIndx {\pointOne d}} {z_1} x$
and $\inputVal {z_2} {\pointOne d} {v'}$.

\item Case $x = \rdcssAllocAlg(v,k)$. We want to prove $\outputRunFunc x = \outputProp x$
and $\ETimeProp x \neq \bot$. But this follows trivially from the definition of $\runFunc x$.

\item Case $x = \rdcssCasAlg(q,e,n)$. We want to prove,
\begin{align*}
\exists z.\ \visObs {\pointerIndx q} z x \wedge \inputVal z q {\outputRunFunc x}
\end{align*}

The only case in the definition of $\runFunc x$
has a span $b$ such that $\firstRep b$ reads value $\outputRunFunc x$ 
in pointer $q$ at line \ref{appendix::alg-access-CAS-RDCSS}.
Therefore, by Lemma \ref{lem::appendix::impl::rdcss::visibility-lemma},
there is $z$ such that $\visObs {\pointerIndx q} {z} x$
and $\inputVal z q {\outputRunFunc x}$.

\item Case $x = \rdcssWriteAlg(q,v)$. We want to prove $\outputRunFunc x = \unitValue$
but this follows trivially from the definition of $\runFunc x$.

\item Case $x = \rdcssReadAlg(q)$. We want to prove,
\begin{align*}
\exists z.\ \visObs {\pointerIndx q} z x \wedge \inputVal z q {\outputRunFunc x}
\end{align*}

The only case in the definition of $\runFunc x$
has a span $b$ such that $\firstRep b$ reads value $\outputRunFunc x$ 
in pointer $q$ at line \ref{appendix::alg-access-Read-RDCSS}.
Therefore, by Lemma \ref{lem::appendix::impl::rdcss::visibility-lemma},
there is $z$ such that $\visObs {\pointerIndx q} {z} x$
and $\inputVal z q {\outputRunFunc x}$.

\item The cases for $\rdcssCasCtlAlg$, $\rdcssWriteCtlAlg$, and $\rdcssReadCtlAlg$ are similar
to the cases for $\rdcssCasAlg$, $\rdcssWriteAlg$, and $\rdcssReadAlg$, respectively.
\end{itemize}
\end{prf}

\begin{lem}
Axiom
\axiomDRef{principle::writers-have-writer-blocks} holds.
\end{lem}

\begin{prf}
Suppose $\hspans{}{x} \neq \emptyset$. We do a case analysis on $x$.
\begin{itemize}
\item Case $x = \rdcssAlg(d)$.

$\Longrightarrow$. By definition of $\writesAbs p$, we have $p = \pointTwo d$ and,
\begin{align}
\label{eqn::appendix::impl::rdcss::writers-have-writer-blocks-1}
\begin{split}
\exists z_1, z_2.\ \visObs {\pointerIndx {\pointOne d}} {z_1} x \wedge \visObs {\pointerIndx {\pointTwo d}} {z_2} x \wedge {} \\
\inputVal {z_1} {\pointOne d} {\expOne d} \wedge \inputVal {z_2} {\pointTwo d} {\expTwo d}
\end{split}
\end{align}
By Lemma \ref{lem::appendix::impl::rdcss::visibility-lemma}, there are
$b_1 \in \hspans {\pointOne d} x$, 
$b_2 \in \hspans {\pointTwo d} x$ such that
$\firstRep {b_1}$ reads value $\expOne d$ and 
$\firstRep {b_2}$ reads value $\expTwo d$.

By definition of $\runFuncSymbol$, $b_2$ must be a 2-tuple span where $\firstRep {b_2}$ is the successful CAS 
at line \ref{appendix::alg-CAS-RDCSS} and $b_1$ must be a 1-span where $\firstRep {b_1}$ is
the read at line \ref{appendix::alg-pt1-read-RDCSS}. Since 
$\firstRep {b_1}$ reads the expected value, the thread must have executed the successful CAS at line 
\ref{appendix::alg-pt2-write-success-RDCSS} at rep event $\lastRep {b_2}$.

Therefore, $b_2 \in \hspans{}{x} \cap \writesSpans {\pointTwo d}$.

$\Longleftarrow$. Suppose $b_2 \in \hspans{}{x} \cap \writesSpans {p}$. 
By the definition of $\runFuncSymbol$, $b_2$ must be one of the two kinds of 2-tuple spans. 
In both cases of the 2-tuple spans,
$\firstRep {b_2}$ reads value $\expTwo d$ since the
CAS at line \ref{appendix::alg-CAS-RDCSS} succeeded.
But by definition of $\writesSpans {p}$, $\lastRep {b_2}$ must 
be the successful CAS at line \ref{appendix::alg-pt2-write-success-RDCSS}.
Then, $p = \pointTwo d$
since $\lastRep {b_2}$ writes into $p$ (by definition of $\writesSpans {p}$)
and into $\pointTwo d$ (by definition of $\runFuncSymbol$).

Also, by definition of $\runFuncSymbol$, there is $\firstRep {b_1}$ in the denotation that reads value $\expOne d$
at pointer $\pointOne d$
because the CAS at line \ref{appendix::alg-pt2-write-success-RDCSS} succeeded,
which means that the thread must have read the expected value at line \ref{appendix::alg-pt1-read-RDCSS}.

Hence, \eqref{eqn::appendix::impl::rdcss::writers-have-writer-blocks-1}
follows by Lemma \ref{lem::appendix::impl::rdcss::visibility-lemma}.

\item Case $x = \rdcssAllocAlg(v,k)$.

$\Longrightarrow$. Since $\hspans{}{x} \neq \emptyset$, by definition of $\runFuncSymbol$, 
$\ETimeProp x \neq \bot$, $\outputProp x \neq \bot$, and there is
a span in $\hspans{}{x} \cap \writesSpans {\outputProp x}$.
But since $x \in \writesAbs p$, we have $p = \outputProp x$.

$\Longleftarrow$. Let $b \in \hspans{}{x} \cap \writesSpans {p}$.
By definition of $\runFuncSymbol$, $\ETimeProp x \neq \bot$, $\outputProp x \neq \bot$,
$b = (i)$ for some alloc $i$ at either line 
\ref{appendix::alloc-control-Alloc-RDCSS} or line \ref{appendix::alloc-data-Alloc-RDCSS}, and $(i) \in \writesSpans{\outputProp x}$.

But from $(i) \in \writesSpans{p}$ we also know that $i$ allocates $p$,
and since allocations are unique, $p = \outputProp x$. Hence, $x \in \writesAbs p$.

\item Case $x = \rdcssCasAlg(q,e,n)$.

$\Longrightarrow$. By definition of $\writesAbs p$, we have $p = q$ and,
\begin{align}
\label{eqn::appendix::impl::rdcss::writers-have-writer-blocks-2}
\exists z.\ \visObs {\pointerIndx p} z x \wedge \inputVal z p e
\end{align}
By Lemma \ref{lem::appendix::impl::rdcss::visibility-lemma}, there is
$b \in \hspans{p}{x}$ such that
$\firstRep {b}$ reads value $e$.

By definition of $\runFuncSymbol$, $\firstRep {b}$ must be the successful CAS 
at line \ref{appendix::alg-access-CAS-RDCSS}.

Therefore, $b \in \hspans{}{x} \cap \writesSpans {p}$.

$\Longleftarrow$. Suppose $b \in \hspans{}{x} \cap \writesSpans {p}$. Then, $p = q$
by definition of $\runFuncSymbol$, since the same rep event writes into
$p$ (by definition of $\writesSpans {p}$) and into $q$ (by definition of $\runFuncSymbol$). 
Also, $\firstRep {b}$ reads value $e$ since the
CAS at line \ref{appendix::alg-access-CAS-RDCSS} must succeed because $b \in \writesSpans {p}$.

Hence, \eqref{eqn::appendix::impl::rdcss::writers-have-writer-blocks-2}
follows by Lemma \ref{lem::appendix::impl::rdcss::visibility-lemma}.

\item Case $x = \rdcssWriteAlg(q,v)$.

$\Longrightarrow$. Since $\hspans{}{x} \neq \emptyset$, there is
a span in $\hspans{}{x} \cap \writesSpans {q}$ because all
spans in the denotation write at $q$. But $p = q$ follows from 
$x \in \writesAbs p$.

$\Longleftarrow$. Suppose $b \in \hspans{}{x} \cap \writesSpans {p}$. Then, $p = q$
since the same rep event writes into $p$ (by definition of $\writesSpans {p}$)
and into $q$ (by definition of $\runFuncSymbol$), which means $x \in \writesAbs p$.

\item Case $x = \rdcssReadAlg(q)$. Trivial since $x$ is neither a successful writer
nor it has writer spans in its denotation.

\item The cases for $\rdcssCasCtlAlg$, $\rdcssWriteCtlAlg$, and $\rdcssReadCtlAlg$ are similar
to the cases for $\rdcssCasAlg$, $\rdcssWriteAlg$, and $\rdcssReadAlg$, respectively.
\end{itemize}
\end{prf}

\begin{lem}
Axiom
\axiomDRef{principle::blocks-contained-in-abstract-time-interval} holds.
\end{lem}

\begin{prf}
We prove each item.

\begin{itemize}
\item (i).
Let us focus on all denotations in $\runFunc x$ having
the form $\{ (b) \}$. 
In these denotations, $b$ is a rep event
that is invoked by the same thread that invoked $x$. Hence, $b$ starts after $x$
started.

Now, for the cases producing denotations of the form $\{ (c_1,c_2),\ (r) \}$. $c_1$ can only
execute after $x$ was started because line \ref{appendix::alg-CAS-RDCSS} executes
after line \ref{appendix::alg-alloc-desc-RDCSS} is executed by the thread invoking
$x$. Also, $r$ can only execute after $x$ started because threads enter
the $\completeAlg$ procedure only after the descriptor identifier has been written
at Line \ref{appendix::alg-CAS-RDCSS} by $c_1$.

\item (ii).
Let us focus on all denotations in $\runFunc x$ having
the form $\{ (b) \}$.
In these denotations, $b$ is a rep event
that is invoked by the same thread that invoked $x$. Hence, $b$ finishes before $x$
finishes (hence, in this cases, choose $i \defini b$).

Now, for the cases producing denotations of the form $\{ (c_1,c_2),\ (r) \}$. Let $T$ be the thread
that invoked $x$ and $c_1$. At $c_1$, $T$ wrote the descriptor identifier at Line 
\ref{appendix::alg-CAS-RDCSS} and now it is about to enter the $\completeAlg$
procedure. 

If the thread that executes $c_2$ and $r$ is also $T$, then choose $i \defini c_2$ for any
of the spans $(c_1,c_2)$ and $(r)$, since $r$ must have finished before $c_2$ started.

If the thread that executes $c_2$ and $r$ is another thread $T'$, then $T'$
successfully executed one of the CASes at Lines
\ref{appendix::alg-pt2-write-success-RDCSS} and \ref{appendix::alg-pt2-write-fail-RDCSS} 
together with the read at line \ref{appendix::alg-pt1-read-RDCSS}.
Therefore, once $T$ enters the $\completeAlg$ procedure, $T$ will attempt one of the CASes and fail. 
Hence, $T$ will discover that the descriptor has been resolved already, and then it will finish $x$. 
In this case, for any of the spans, choose $i$ to be the failed CAS executed by $T$. 
Notice that we cannot choose 
$i$ to be $c_2$, because $c_2$ might not have finished by the time $T$ finished the invocation of $x$
(i.e. $T'$ carried out the effect of $c_2$ but $T'$ might not have finished $c_2$ yet).
\end{itemize}
\end{prf}

\begin{lem}
	Axiom \axiomDRef{principle::containment-and-uniqueness-of-alloc-blocks} holds.
\end{lem}

\begin{prf}
	If $b_1, b_2 \in \allocsSpans p$,
	then by definition of $\allocsSpans p$,
	$b_1$ and $b_2$ are 1-tuples allocating $p$ by executing either 
	Line \ref{appendix::alloc-control-Alloc-RDCSS} or line 
	\ref{appendix::alloc-data-Alloc-RDCSS}.
	But by Lemma 
	\ref{lem::appendix::impl::rdcss::unique-alloc},
	$b_1 = b_2$.
\end{prf}

\begin{lem}
	Axiom
	\axiomDRef{principle::every-block-must-have-an-allocated-pointer} holds.
\end{lem}

\begin{prf}
	Let $b \in \spans p$. 
	Each procedure requires that its input pointers be created by a previous alloc invocation.
	Therefore, there is some rep alloc $i$ that allocates $p$ and executes either 
	line \ref{appendix::alloc-control-Alloc-RDCSS} or line 
	\ref{appendix::alloc-data-Alloc-RDCSS}. So that  
	$i$ executes before $\firstRep b$ (in case $b$ does not contain an alloc), or $b = (i)$.
	
	Now, any rep event which is an instance of lines \ref{appendix::alloc-control-Alloc-RDCSS} and line 
	\ref{appendix::alloc-data-Alloc-RDCSS} belongs to some span by definition of $\allocsSpans p$.
	Therefore, $(i) \in \allocsSpans p$. 
	
	So, if $i$ executes before $\firstRep b$, then $\precedesSpans {(i)} b$, and if 
	$b = (i)$, then $\precedesSpansEq {(i)} b$.
\end{prf}

\begin{lem}
	\label{lem::appendix::impl::rdcss::allocs-have-alloc-spans}
	Axioms
	\axiomDRef{principle::allocs-have-alloc-blocks} holds.
\end{lem}

\begin{prf}
	We prove each direction.
	
	$\Longrightarrow$. Since $\hspans{}{x} \neq \emptyset$ and $x \in \allocsAbs p$ then $x$ is an alloc and 
	$\ETimeProp x \neq \bot$, $\outputProp x \neq \bot$, and $p = \outputProp x$. But by definition of $\runFuncSymbol$, 
	there is $(i) \in \writesSpans{\outputProp x}$, where $i$ is either line
	\ref{appendix::alloc-control-Alloc-RDCSS} or line \ref{appendix::alloc-data-Alloc-RDCSS}.
	Therefore, $(i) \in \hspans{}{x} \cap \allocsSpans p$ by definition.
	
	$\Longleftarrow$. Let $b \in \hspans{}{x} \cap \allocsSpans p$. 
	By definition of $\allocsSpans p$, $\firstRep b$ allocates $p$ and it is either
	line \ref{appendix::alloc-control-Alloc-RDCSS} or line \ref{appendix::alloc-data-Alloc-RDCSS}.
	But the only case applicable in the definition of $\runFuncSymbol$ is when $x$ is an alloc, which means 
	$\ETimeProp x \neq \bot$, $\outputProp x \neq \bot$ and $b \in \writesSpans{\outputProp x}$. 
	Also, since rep allocs are unique $p = \outputProp x$, which
	means $x \in \allocsAbs p$.
\end{prf}

\begin{thm}
\label{thm::appendix::impl::span-axioms-satisfied-rdcss}
$\genStructName{\RDCSS}(\visObsSymbol {\pointerIndx p}, \visSepSymbol{\pointerIndx p})$ 
is implemented by
span structure $\genSpanStructName{\RDCSS}$.
\end{thm}

\begin{prf}
All span axioms hold from Lemma \ref{lem::appendix::impl::rdcss::non-interference-of-lifespans} to 
Lemma \ref{lem::appendix::impl::rdcss::allocs-have-alloc-spans}.
\end{prf}

\begin{thm}
\label{thm::appendix::impl::rdcss-is-linearizable}
The RDCSS implementation of Figure \ref{appendix::alg-RDCSS} is linearizable.
\end{thm}

\begin{prf}
By Theorem \ref{thm::appendix::lin::rdcss-linearizability-from-vis-structure}, it
  suffices to show that
  $\genStructName{\RDCSS}(\visObsSymbol {\pointerIndx p},
  \visSepSymbol{\pointerIndx p})$ is valid.  But by
  Theorem \ref{thm::appendix::valid::span-axioms-imply-visibility}, it suffices that
  $\genStructName{\RDCSS}(\visObsSymbol {\pointerIndx p},
  \visSepSymbol{\pointerIndx p})$ is implemented by span
  structure $\genSpanStructName{\RDCSS}$. This is given by Theorem
  \ref{thm::appendix::impl::span-axioms-satisfied-rdcss}.
\end{prf}

\subsection{MCAS}
\label{appendix::sub::sect::impl::MCAS}

The full pseudocode for MCAS is shown in
Figure \ref{appendix::alg-MCAS}. MCAS uses the exportable
procedures of RDCSS as primitives. 
The exportable procedures for MCAS are $\mcasAlg$, $\mcasReadAlg$,
$\mcasWriteAlg$, and $\mcasAllocAlg$. 

$\ValType$ denotes the set of all
possible input values. $\ValType$ contains 
neither MCAS descriptors nor pointers storing MCAS descriptors.

No particular implementation is provided for boolean predicate
$\isMcasDescRepAlg(p)$, but it is assumed that it returns 
true if and only if $p$ is a pointer storing an MCAS descriptor.
For example,~\cite{Harris} suggests that $\isMcasDescRepAlg(p)$ 
could be implemented by checking a reserved bit in $p$; this
reserved bit indicates whether or not the pointer stores a descriptor.

The implementation makes the following assumptions:
\begin{itemize}
\item The list of entries given as input to procedure $\mcasAlg$
is not empty.
\item The alloc at line~\ref{appendix::alloc-desc-MCAS}
creates a pointer $d$ such that $\isMcasDescRepAlg(d)$ returns true.
\item Any input pointer $p: \PtsType$ to any
exportable procedure must satisfy $\neg \isMcasDescRepAlg(p)$.
\item Any input pointer to any procedure must have been previously
allocated with an invocation to $\mcasAllocAlg$.
\item The alloc at line \ref{appendix::alloc-data-Alloc-MCAS} 
returns a pointer $p$ such that $\neg \isMcasDescRepAlg(p)$.
\end{itemize}   

\begin{figure}[t]
\begin{multicols*}{2}

\begin{algorithmic}[1]
\Record{$\textsc{update\_entry}$} 
	\State $\textit{pt}$ : $\DataPtType$
	\State $\textit{exp}$, $\textit{new}$ : $\ValType$
\EndRecord
\EnumLine{$\textsc{status}$}{$\UNDECIDED$, $\SUCCEEDED$, $\FAILED$}
\EndEnumLine
\Record{$\mcasDesc$}
\State $\textit{status}$ : $\ControlPtType$ $\textsc{status}$
\State $\textit{entries}$ : $\textsc{list}\,\textsc{update\_entry}$
\EndRecord
\State 
%
\Proc{$\mcasReadAlg$}{$pt: \DataPtType$}
\State \label{appendix::alg-access-Read-MCAS} $old \gets \rdcssReadAlg(pt)$
\If {$\isMcasDescRepAlg(old)$} \label{appendix::alg-is-desc-Read-MCAS}
  \State \label{appendix::alg-help-complete-invoke-Read-MCAS} $\mcasHelpAlg(old)$
  \State \returnCmd $\mcasReadAlg(pt)$ \label{appendix::try-again-Read-MCAS}
\Else 
  \State \label{appendix::alg-return-old-value-Read-MCAS} \returnCmd $old$
\EndIf
\EndProc
%
\State 
%
\Proc{$\mcasWriteAlg$}{$pt: \DataPtType$, $v: \ValType$}
\State \label{appendix::alg-access-Write-MCAS} $old \gets \rdcssReadAlg(pt)$
\If {$\isMcasDescRepAlg(old)$} \label{appendix::alg-is-desc-Write-MCAS}
  \State \label{appendix::alg-help-complete-invoke-Write-MCAS} $\mcasHelpAlg(old)$
  \State $\mcasWriteAlg(pt, v)$ \label{appendix::try-again-Write-One-MCAS}
\Else 
  \State \label{appendix::alg-attempt-write-Write-MCAS} $x \gets \rdcssCasAlg(pt, old, v)$
  \If {$x \neq old$} \label{appendix::last-if-in-Write-MCAS}
    \State $\mcasWriteAlg(pt, v)$ \label{appendix::try-again-Write-Two-MCAS}
  \EndIf
\EndIf
\EndProc
%
\State \label{appendix::implicit-return-for-Write-MCAS}
%
\Proc{$\mcasAllocAlg$}{$v: \ValType$}
\State \label{appendix::alloc-data-Alloc-MCAS} \returnCmd $\rdcssAllocAlg(v,\DATA)$
\EndProc
%
\columnbreak
%
\Proc{$\mcasAlg$}{$\listvar u : \textsc{list}\,\textsc{update\_entry}$}
\State \label{appendix::alloc-status-MCAS} $s \gets \rdcssAllocAlg(\UNDECIDED, \CONTROL)$
\State \label{appendix::create-mcas-desc-MCAS} $desc \gets \mcasDesc(s, \listvar u)$
\State \label{appendix::alloc-desc-MCAS} $d \gets \rdcssAllocAlg(desc, \DATA)$
\State \returnCmd $\mcasHelpAlg(d)$ \label{appendix::invoke-auxiliary-mcas-MCAS}
\EndProc
%
\State 
%
\Proc{$\mcasHelpAlg$}{$d: \DataPtType$}
\State \label{appendix::read-desc-MCAS} $desc \gets \rdcssReadAlg(d)$
\State \label{appendix::read-phase1-status-MCAS} $phase1 \gets \rdcssReadCtlAlg(\statusProp{desc})$
\If{$phase1 = \UNDECIDED$} \label{appendix::is-phase1-still-undecided}
	\State \label{appendix::write-all-descs-MCAS} $s \gets \writeAllDescsAlg(d,desc)$
	\State \label{appendix::resolve-status-MCAS} $\rdcssCasCtlAlg(\statusProp{desc}, \UNDECIDED, s)$
\EndIf
\State \label{appendix::read-phase2-status-MCAS} $phase2 \gets \rdcssReadCtlAlg(\statusProp{desc})$
\State $r \gets (phase2 = \SUCCEEDED)$
\ForEach{$e$}{$\entriesProp {desc}$} \label{appendix::remove-all-descs-loop-MCAS}
	\State \label{appendix::remove-all-descs-MCAS} $\rdcssCasAlg(\pointGen e, d, r\ ?\ \newGen e : \expGen e)$
\EndForEach \label{appendix::end-phase2-MCAS}
\State \returnCmd $r$ \label{appendix::return-boolean-MCAS}
\EndProc
%
\State 
%
\Proc{$\writeAllDescsAlg$}{$d: \DataPtType, desc: \mcasDesc$}
\ForEach{$e$}{$\entriesProp {desc}$}
	\State \label{appendix::create-rdcss-desc-MCAS} $rD \gets \rdcssDesc(\statusProp {desc},$ 
	\State $\phantom{rD \gets \quad}\pointGen e, \UNDECIDED, \expGen e, d)$
	\State \label{appendix::invoke-rdcss-in-MCAS} $old \gets \rdcssAlg(rD)$
	\If{$\isMcasDescRepAlg(old)$} \label{appendix::alg-is-desc-MCAS-MCAS}
	  \If{$old \neq d$} \label{appendix::is-my-desc-MCAS}
	     \State \label{appendix::alg-help-complete-invoke-MCAS-MCAS} $\mcasHelpAlg(old)$
	     \State \returnCmd $\writeAllDescsAlg(d, desc)$ \label{appendix::try-again-writeall-MCAS}
	  \EndIf
	\ElsIf{$old \neq \expGen e$} \label{appendix::rdcss-failed-MCAS}
	  \State \returnCmd $\FAILED$ \label{appendix::return-failed-writeall-MCAS}
    \EndIf
\EndForEach
\State \returnCmd $\SUCCEEDED$
\EndProc
\end{algorithmic}

\end{multicols*}
\caption{MCAS implementation. It uses as primitives the exportable procedures in RDCSS.}
\label{appendix::alg-MCAS}
\end{figure}

We now define the span structure for MCAS.

\begin{defn}[Span Structure for Helping MCAS]
\label{defn::appendix::impl::span-struture-mcas}
We denote the structure by $\genSpanStructName{\MCAS}$.

A span is either a 2-tuple of the form $(a,b)$ or
a 1-tuple of the form $(a)$, where $a,b$ are rep events in the execution history.

For each pointer $p$, set $\spans p$ is defined by the
following list of spans containing rep events,
\begin{enumerate}
\item Any $(b)$ such that, 

\begin{itemize}
\item $\lineProp b = \ref{appendix::invoke-rdcss-in-MCAS}$
\item $\neg \isMcasDescRepAlg(\outputProp b)$
\item $b = \rdcssAlg(desc)$ for some $exp$ and $desc = \rdcssDesc(\_, p, \UNDECIDED, exp, \_)$
such that $\outputProp b \neq exp$
\end{itemize}

\item Any $(b,c)$ such that, 

\begin{itemize}
\item $\lineProp b = \ref{appendix::invoke-rdcss-in-MCAS}$
\item $\lineProp c = \ref{appendix::resolve-status-MCAS}$
\item $\neg \isMcasDescRepAlg(\outputProp b)$
\item $b$ executes before $c$ and there is no other $p$-write rep event between $b$ and $c$.
\item \begin{sloppypar}
$b = \rdcssAlg(desc)$ and $c = \rdcssCasAlg(s, \UNDECIDED, \SUCCEEDED)$, 
for some $s$, $exp$, $d$, and $desc = \rdcssDesc(s, p, \UNDECIDED, exp, d)$ such that
$c$ is successful, $\outputProp b = exp$, and $\isMcasDescRepAlg(d)$.
\end{sloppypar}
\end{itemize}

\item Any $(b,c)$ such that,

\begin{itemize}
\item $\lineProp b = \ref{appendix::invoke-rdcss-in-MCAS}$
\item $\lineProp c = \ref{appendix::resolve-status-MCAS}$
\item $\neg \isMcasDescRepAlg(\outputProp b)$
\item $b$ executes before $c$ and there is no other $p$-write rep event between $b$ and $c$.
\item \begin{sloppypar}
$b = \rdcssAlg(desc)$ and $c = \rdcssCasAlg(s, \UNDECIDED, \FAILED)$, 
for some $s$, $exp$, $d$, and $desc = \rdcssDesc(s, p, \UNDECIDED, exp, d)$ such that
$c$ is successful, $\outputProp b = exp$, and $\isMcasDescRepAlg(d)$.
\end{sloppypar}
\end{itemize}

\item Any $(b)$ such that,

\begin{itemize}
\item $\lineProp b = \ref{appendix::alg-access-Read-MCAS}$
\item $\neg \isMcasDescRepAlg(\outputProp b)$
\item $b$ reads pointer $p$
\end{itemize}

\item Any $(b)$ such that,

\begin{itemize}
\item $\lineProp b = \ref{appendix::alg-attempt-write-Write-MCAS}$
\item $\neg \isMcasDescRepAlg(\outputProp b)$
\item $b$ is a successful $\rdcssCasAlg$ writing into pointer $p$
\end{itemize}

\item Any $(b)$ such that $\lineProp b = \ref{appendix::alloc-data-Alloc-MCAS}$ and
$b$ has as output pointer $p$.
\end{enumerate}

With this, the set of writer spans $\writesAbs p$ can be defined as,
\begin{align*}
\writesSpans p & \defini \{ (b) \in \spans p \mid b \text{ writes or allocs pointer } p \} \cup 
\{ (b,c) \in \spans p \mid c = \rdcssCasAlg(\_, \UNDECIDED, \SUCCEEDED) \} 
\end{align*}

The set of alloc spans $\allocsAbs p$ as follows,
\begin{align*}
\allocsSpans p & \defini \{ (b) \in \writesSpans p \mid b \text{ allocs pointer } p \wedge 
\lineProp b = \ref{appendix::alloc-data-Alloc-MCAS} \}
\end{align*}

We now define the denotation $\runFunc x$ by cases on 
event $x$,
\begin{enumerate}

\item Case $x = \mcasAlg(\listvar u)$.

\begin{enumerate}

\item If there are rep events $i_a$, $i_s$, $r$, and for every $j \in \listvar u$, there is $(a_j,r) \in \writesSpans {\pointGenEntry j}$,
such that,

\begin{itemize}
\item $i_a = \rdcssAllocAlg(desc, \DATA)$ with code line \ref{appendix::alloc-desc-MCAS},
where $\entriesProp {desc} = \listvar u$.
\item $i_s = \rdcssAllocAlg(\UNDECIDED, \CONTROL)$ with code line \ref{appendix::alloc-status-MCAS},
\item If $T$ is the thread that invoked $x$, then $T$ executes $i_a$ and $i_s$ within the invocation of $x$,
\item $a_j = \rdcssAlg(rD)$, where $rD = \rdcssDesc(\outputProp {i_s},\ \pointGenEntry j,\ \UNDECIDED,\ \expGenEntry j,\ \outputProp {i_a})$ with code 
line \ref{appendix::invoke-rdcss-in-MCAS},
\item $r = \rdcssCasAlg(\outputProp {i_s}, \UNDECIDED, \SUCCEEDED)$ with code 
line \ref{appendix::resolve-status-MCAS},
\item $a_j$ is the most recent such $\rdcssAlg$ call for pointer $\pointGenEntry j$ before $r$.
\end{itemize}
then 
$\runFunc x \defini (\{ (a_j,r) \mid j \in \listvar u \},\ true)$.

\item If there are rep events $i_a$, $i_s$, $r$ and for some $j \in \listvar u$, there is $(c_j) \in \spans {\pointGenEntry j}$,
such that for every $k < j$, there are $(a_k,r) \in \spans {\pointGenEntry k}$, such that,

\begin{itemize}
\item $i_a = \rdcssAllocAlg(desc, \DATA)$ with code line \ref{appendix::alloc-desc-MCAS},
where $\entriesProp {desc} = \listvar u$.
\item $i_s = \rdcssAllocAlg(\UNDECIDED, \CONTROL)$ with code line \ref{appendix::alloc-status-MCAS},
\item If $T$ is the thread that invoked $x$, then $T$ executes $i_a$ and $i_s$ within the invocation of $x$,
\item $a_k = \rdcssAlg(rD)$, where $rD = \rdcssDesc(\outputProp {i_s},\ \pointGenEntry k,\ \UNDECIDED,\ \expGenEntry k,\ \outputProp {i_a})$ with code 
line \ref{appendix::invoke-rdcss-in-MCAS}, and $\outputProp{a_k} = \expGenEntry k$.
\item $r = \rdcssCasAlg(\outputProp {i_s}, \UNDECIDED, \FAILED)$ with code 
line \ref{appendix::resolve-status-MCAS},
\item $a_k$ is the most recent such $\rdcssAlg$ call for pointer $\pointGenEntry k$ before $r$.
\item $c_j = \rdcssAlg(rD)$, where $rD = \rdcssDesc(\outputProp {i_s},\ \pointGenEntry j,\ \UNDECIDED,\ \expGenEntry j,\ \outputProp {i_a})$ with code 
line \ref{appendix::invoke-rdcss-in-MCAS}, and $\outputProp{c_j} \neq \expGenEntry j$, and $\neg \isMcasDescRepAlg(\outputProp {c_j})$.
\item $c_j$ was executed by the same thread that executed $r$, and $c_j$ is the most recent such $\rdcssAlg$ call for pointer $\pointGenEntry j$ before $r$,
\end{itemize}
then 
$\runFunc x \defini (\{ (a_k,r),\ (c_j) \mid k \in \listvar u \wedge k < j \},\ false)$.

\item Otherwise, $\runFunc x \defini (\emptyset, \bot)$.
\end{enumerate}

\item Case $x = \mcasReadAlg(p)$.
\begin{enumerate}
\item If there is $(r) \in \spans p$ such that,

\begin{itemize}
\item If $T$ is the thread that invoked $x$, then $T$ executes $r$ 
within the invocation of $x$,
\item $r = {\rdcssReadAlg(p)}$ with code line \ref{appendix::alg-access-Read-MCAS},
\end{itemize}
then
$\runFunc x \defini (\{ (r) \},\ \outputProp r)$.

\item Otherwise, $\runFunc x \defini (\emptyset, \bot)$.
\end{enumerate}

\item Case $x = \mcasWriteAlg(p,v)$.
\begin{enumerate}
\item If there are $(c) \in \writesSpans p$ and rep event $r$ such that,

\begin{itemize}
\item If $T$ is the thread that invoked $x$, then $T$ executes $c$ 
within the invocation of $x$,
\item $r = {\rdcssReadAlg(p)}$ with code line \ref{appendix::alg-access-Write-MCAS},
\item $c = \rdcssCasAlg(p, \outputProp r, v)$ with code line \ref{appendix::alg-attempt-write-Write-MCAS},
\item If $T$ is the thread that invoked $x$, then $r$ is the last read
carried out by $T$ before the execution of $c$,
\end{itemize}
then
$\runFunc x \defini (\{ (c) \},\ \unitValue)$.

\item Otherwise, $\runFunc x \defini (\emptyset, \bot)$.
\end{enumerate}

\item Case $x = \mcasAllocAlg(v)$.
\begin{enumerate}
\item If $\outputProp x \neq \bot$, $\ETimeProp x \neq \bot$, and
there is $(i) \in \writesSpans {\outputProp x}$ such that,

\begin{itemize}
\item If $T$ is the thread that invoked $x$, then $T$ executes $i$ 
within the invocation of $x$,
\item $i = \rdcssAllocAlg(v,\DATA)$ with code line \ref{appendix::alloc-data-Alloc-MCAS},
\end{itemize}
then
$\runFunc x \defini (\{ (i) \},\ \outputProp i)$.

\item Otherwise, $\runFunc x \defini (\emptyset, \bot)$.
\end{enumerate}

\end{enumerate}
\end{defn}

We argue that $\runFunc x$ is well-defined.
For that matter, we need to show that the 
function's conditions pick spans uniquely.

First, let us focus on the $\mcasAlg$ cases. Notice that whenever
an $\mcasAlg(\listvar u)$ invocation starts, a fresh pointer
$d$ is created at Line \ref{appendix::alloc-desc-MCAS} 
to store the descriptor $desc$.
This pointer $d$ will serve as unique identifier for 
the invocation, because the invoking thread executes
line \ref{appendix::alloc-desc-MCAS} only once.

The invoking thread then enters the $\mcasHelpAlg$
procedure and tries to write $d$ into every pointer in $\listvar u$ 
by executing $\writeAllDescsAlg$. 
Procedure $\writeAllDescsAlg$ will try every pointer in the order of $\listvar u$.

If for some thread $T$ and
$j \in \listvar u$, the $\rdcssAlg$ at line \ref{appendix::invoke-rdcss-in-MCAS} 
returns a non-expected value and the output is not
a descriptor, and then later $T$ manages to set the descriptor status
to $\FAILED$ at line \ref{appendix::resolve-status-MCAS}, then this will correspond to case
(1).(b) in the definition of $\runFuncSymbol$. Notice
that for every $k < j$, the descriptor must have been written
at line \ref{appendix::invoke-rdcss-in-MCAS}, because pointers were tried in the order of $\listvar u$
before it failed for $j$. Also, any $\rdcssAlg$ at line \ref{appendix::invoke-rdcss-in-MCAS}
that returns the expected value and
executes before the descriptor status is set, must
have succeeded because the descriptor status must be $\UNDECIDED$ 
before the descriptor status is changed from $\UNDECIDED$ 
to $\SUCCEEDED$ or $\FAILED$ at line \ref{appendix::resolve-status-MCAS}.
Also, notice that for case (1).(b) in the definition of $\runFuncSymbol$, 
rep events $i_s$, $i_a$, and $r$ are chosen
uniquely because they execute at most once 
(for rep event $r$, this is so
because the descriptor status starts from $\UNDECIDED$ at line \ref{appendix::alloc-status-MCAS} and
the status can only change from 
$\UNDECIDED$ into either $\SUCCEEDED$ or $\FAILED$ at line \ref{appendix::resolve-status-MCAS}),
while the $a_k$'s and the $a_j$ are chosen uniquely
because they are the most recent $\rdcssAlg$ rep events 
of the appropriate kind before $r$.

If on the other hand, for every $j \in \listvar u$,
the $\rdcssAlg$ at line \ref{appendix::invoke-rdcss-in-MCAS} returns the expected value and later 
some thread manages to set the descriptor status
to $\SUCCEEDED$ at line \ref{appendix::resolve-status-MCAS}, then this will correspond to case
(1).(a) in the definition of $\runFuncSymbol$.
The reasoning for the unique choosing of the rep events is identical
as in the (1).(b) case.

For the cases $x = \mcasReadAlg(p)$ and
$x = \mcasWriteAlg(p,v)$,
their condition is choosing instructions at Lines
\ref{appendix::alg-access-Read-MCAS} and
\ref{appendix::alg-attempt-write-Write-MCAS}
such that their output do not satisfy 
$\isMcasDescRepAlg$ (by definition of set $\spans p$).
But this condition can happen at most once during the
invocation of $x$, 
because the invoking thread finishes $x$ once
the condition is satisfied.
 
While the case for 
$x = \mcasAllocAlg(v)$ is trivial, since the
rep alloc executes at most once within the invocation of
$x$.

We are ready to prove that 
$\genStructName{\MCAS}(\visObsSymbol{\pointerIndx p}, \visSepSymbol{\pointerIndx p})$ is implemented by span
structure $\genSpanStructName{\MCAS}$, i.e., we are going to prove that
the span axioms of Figure \ref{fig:descriptor-lifespans-axioms} are satisfied. We require a couple of lemmas.

\begin{lem}
\label{lem::appendix::impl::unique-alloc-mcas}
If $i_1$, $i_2$ are allocs at line \ref{appendix::alloc-data-Alloc-MCAS} 
such that they allocate the same pointer, then $i_1 = i_2$.
\end{lem}

\begin{prf}
Allocs create fresh pointers, and pointers are not deallocated.
Therefore, if $\linReps {i_1} {i_2}$, then $i_2$ must create
a pointer that is different from the one created by $i_1$ (Contradicting 
the hypothesis).
Similarly, if $\linReps {i_2} {i_1}$.
Hence, $i_1 = i_2$.
\end{prf}

For the next lemma we make the following definition,

\begin{defn}
We say that $v \in \ValType$ is the value that $p$ is bound to by $(a,\_) \in \writesSpans p$ 
if there are $rD$, $desc$, $j$ such that,
\begin{itemize}
\item $a = \rdcssAlg(rD)$,
\item $\newTwo {rD}$ was allocated by rep event $\rdcssAllocAlg(desc, \DATA)$ at line
\ref{appendix::alloc-desc-MCAS},
\item $j \in \entriesProp {desc}$,
\item $p = \pointTwo{rD} = \pointGenEntry j$,
\item $v = \newGenEntry j$.
\end{itemize}
Alternatively, one can think of $v$ as the new value that 
line \ref{appendix::remove-all-descs-MCAS} will write into $p$
because span $(a,\_)$ has already resolved the descriptor status to $\SUCCEEDED$.
\end{defn}

\begin{lem}
\label{lem::appendix::impl::writers-belong-to-writer-procs-mcas}
If $b \in \writesSpans p$, 
then there is $x$ such that $b \in \hspans{}{x}$ and $\inputVal x p v$,
where $v$ is the value determined by one of the following cases:
\begin{itemize}
\item If $b$ is of the form $(i)$ for some rep $i$, then $v$ is the value written by $i$.
\item If $b$ is of the form $(a,\_)$ for some rep $a$, then
$v$ is the value that $p$ is bound to by $b$.
\end{itemize}
\end{lem}

\begin{prf}
Each span in $\writesSpans p$ occurs in one of the cases in
the definition of $\runFunc x$. 

For example, consider the span of the form $(a,e)$ at entry 2 in the definition of $\spans p$.
Reps $a$ and $e$ being instances of the code, must have the forms $a = \rdcssAlg(rD)$,  
$e = \rdcssCasAlg(\statusProp{desc}, \UNDECIDED, \SUCCEEDED)$
for some $rD$ and $desc$ such that $\newTwo {rD}$ was allocated by some rep
event $c = \rdcssAllocAlg(desc, \DATA)$ at line
\ref{appendix::alloc-desc-MCAS} such that $\newTwo {rD} = \outputProp c$.
In this case, $p = \pointTwo{rD} = \pointGenEntry j$ and $\outputProp a = \expTwo{rD} = \expGenEntry j$
for some $j \in \entriesProp {desc}$.

In other words, the status pointer was set to $\SUCCEEDED$ by some thread $T$, which means that
$T$ must have tried to write the descriptor at every input pointer and $T$ either succeeded
in doing so or found out some other thread(s) did it already (line \ref{appendix::write-all-descs-MCAS}).
Since $a$ writes into pointer $\pointGenEntry j$, and all the entries in 
$\entriesProp {desc}$ wrote the descriptor in their respective pointers,
$(a,b)$ will be part of the successful case (1).(a) in the definition of $\runFunc x$ for
$x = \mcasAlg$.
Also, notice that by definition, $\inputVal x {\pointGenEntry j} {\newGenEntry j}$.

The rest of spans in $\writesSpans p$ are even simpler, because 
they are 1-tuple spans of the form $(b)$, and $b$ is executed by the same thread
that invoked the exportable procedure that $b$ is part of.
\end{prf}

\begin{lem}
\label{lem::appendix::impl::visibility-lemma-mcas}
Suppose $x$ is one of $\mcasAlg$, or $\mcasReadAlg$. Suppose $p \in \PtsType$ and $v \in \ValType$. Then,
the following statements are equivalent,
\begin{itemize}
\item There is $y$ such that $\visObs {\pointerIndx p} y x$
and $\inputVal y p v$.
\item There is $b \in \hspans{p}{x}$ such that $\firstRep b$ reads value $v$.
\end{itemize}
\end{lem}

\begin{prf}
We prove each direction.

$\Longrightarrow$. By definition of $\visObs {\pointerIndx p} y x$, 
there are $b_y \in \hspans{p}{y}$,
$b_x \in \hspans{p}{x}$ such that $b_y$ is the most recent span under $\precedesSpansEqSymbol$ 
satisfying $b_y \in \writesSpans p$ and $\precedesSpans {b_y} {b_x}$.

The possible spans for $b_x$ in the denotation of $x = \mcasAlg$ or $x = \mcasReadAlg$, they all
read a non-descriptor value in $\firstRep {b_x}$. This means that if $b_y$ is of the 
form $(a,b)$, line \ref{appendix::remove-all-descs-MCAS} must have replaced the descriptor with a $\ValType$ value after $b$ executed, otherwise
the descriptor would still be in the pointer and $\firstRep {b_x}$ would read a descriptor.
Additionally, line \ref{appendix::remove-all-descs-MCAS} must have written $v$, since $v$ is the value that $y$ is going to write into $p$ 
in case it succeeds (by definition of $\inputValName$), which it does, since $b_y = (a,b)$ is a successful $p$-write span,
which means it must belong to the (1).(a) case in the definition of $\runFuncSymbol$ (see proof of Lemma 
\ref{lem::appendix::impl::writers-belong-to-writer-procs-mcas}).
If $b_y$ has the form $(i)$, then $i$ must write $v$ because $i$ is a successful write, 
which means that $(i)$ must belong to the (3).(a) or (4).(a) cases in the definition of $\runFuncSymbol$.
Therefore, we can assume that there is a rep event $i$ that writes $v$, such that 
$\lastRep{b_y} \linRepsEqSymbol i \linRepsSymbol \firstRep {b_x}$ and 
that there is no rep $p$-write in between $\lastRep {b_y}$ and $i$.

We argue that any $p$-mutating rep event in between $i$ and $\firstRep {b_x}$ must eventually
restore the value to $v$ before $\firstRep {b_x}$ reads $p$.

If the $p$-write rep event corresponds to line \ref{appendix::alg-attempt-write-Write-MCAS}, it would 
be a 1-span, contradicting that there are no $p$-writing spans in between 
$b_y$ and $b_x$. So, let us focus on rep events at lines \ref{appendix::resolve-status-MCAS},
\ref{appendix::remove-all-descs-MCAS}, and \ref{appendix::invoke-rdcss-in-MCAS}.

If the $\rdcssAlg$ at line \ref{appendix::invoke-rdcss-in-MCAS} returns the expected value
and executes before the descriptor status is set at line \ref{appendix::resolve-status-MCAS},
then the $\rdcssAlg$ succeeded in writing a descriptor, up to the point where the 
descriptor is removed at line \ref{appendix::remove-all-descs-MCAS}. Therefore 
$\firstRep {b_x}$ must occur after the rep event at line \ref{appendix::remove-all-descs-MCAS}, otherwise
$\firstRep {b_x}$ would read a descriptor. However, the resolution
at line \ref{appendix::resolve-status-MCAS} cannot be $\SUCCEEDED$, since this would 
contradict that there are no $p$-write spans in between $b_x$ and $b_y$. Since
the resolution must be $\FAILED$, the removal of the descriptor at line 
\ref{appendix::remove-all-descs-MCAS} will restore the pointers to the values they had before the
descriptor was written.
In other words, $v$ is preserved and $\firstRep{b_x}$ reads $v$.

Notice that we can ignore the failing cases of $\rdcssAlg$ at line \ref{appendix::invoke-rdcss-in-MCAS}: 
when $\rdcssAlg$ returns a non-expected value
or when $\rdcssAlg$ does return the expected value but executes after the descriptor status is set at line \ref{appendix::resolve-status-MCAS},
since the status is no longer $\UNDECIDED$.

$\Longleftarrow$. Let $b_x \in \hspans{p}{x}$ 
such that $\firstRep {b_x}$ reads value $v \in \ValType$. 
Therefore, there is a most recent rep event $w$ that wrote $v$ into $p$.
If $w$ corresponds to line \ref{appendix::remove-all-descs-MCAS} 
and it is removing a descriptor that was set as $\FAILED$ by a rep event $r$ at line 
\ref{appendix::resolve-status-MCAS}, then $r$ is the last rep event in a span that failed 
to modify the pointer, which means that $w$ is restoring the pointer 
to the value it had before the descriptor was written at line \ref{appendix::invoke-rdcss-in-MCAS}.
If this is the case, we just keep searching for a $p$-write rep event of $v$ 
that does not correspond to line \ref{appendix::remove-all-descs-MCAS} having a descriptor resolution of $\FAILED$
at line \ref{appendix::resolve-status-MCAS}.
This rep event exists because there is an alloc of $p$ as first rep event for $p$.

Suppose $w'$ is the chosen $p$-write event. If $w'$ corresponds to line \ref{appendix::remove-all-descs-MCAS}
then there will be a previous rep event $r'$ resolving the status to $\SUCCEEDED$ at line \ref{appendix::resolve-status-MCAS}.
We will have $r' = \lastRep b$ for some span $b$ in $\writesSpans p$,
since the set $\spans p$ covers all possible places where a successful resolution could occur
in the code. 
If $w'$ does not correspond to line \ref{appendix::remove-all-descs-MCAS}
then $w' = \lastRep b$ for some span $b$ in $\writesSpans p$
since $w'$ can only be either line \ref{appendix::alg-attempt-write-Write-MCAS} or 
line \ref{appendix::alloc-data-Alloc-MCAS}.

Span $b$ will be the most recent $p$-write span 
before $b_x$, since all the more recent spans we found correspond to failing spans.
By Lemma \ref{lem::appendix::impl::writers-belong-to-writer-procs-mcas},
there is $y$ such that $b \in \hspans{}{y}$ and $\inputVal y p v$.
Therefore, $\visObs {\pointerIndx p} y x$
by definition.
\end{prf}

\begin{lem}
\label{lem::appendix::impl::non-interference-of-lifespans-mcas}
Axiom \axiomDRef{principle::descriptors-do-not-interfere} holds.
\end{lem}

\begin{prf}
Since this is trivial when $b_1$ and $b_2$ are 1-tuple spans, 
we only need to consider the cases 
when $b_1$ is a 1-span and $b_2$ a 2-tuple span or when both
$b_1$ and $b_2$ are 2-tuple spans.

\begin{itemize}
\item Case $b_1 = (c)$ is a 1-tuple span and $b_2 = (e,g)$ a 2-tuple span.

Since every procedure invocation requires that input pointers be previously 
allocated, there must be some rep alloc $i$ such that
$\linReps i e$. We also know that there cannot be a $p$-writer rep in 
between $e$ and $g$, because once a descriptor is written by $e$, no other write can
occur up to the point where the descriptor is resolved by $g$ (and even until the 
descriptor is removed at line \ref{appendix::remove-all-descs-MCAS} which is not part of the span).

If $c$ is an alloc, by Lemma \ref{lem::appendix::impl::rdcss::unique-alloc},
$c = i$ and $\precedesSpans {b_1} {b_2}$ follows. So, we can assume that $c$ is not an alloc. 

By going through all the cases for $(c) \in \spans p$,
we see that all the cases imply $\neg \isRdcssDescRepAlg(\outputProp c)$,
i.e., $c$ does not read a descriptor.

Therefore, $c$ cannot occur in between $e$ and $g$, because $e$ writes a descriptor. 
Also, the cases $c = e$ or $c = g$ are excluded because $\spans p$
is defined so that 1-spans and 2-tuple spans do not share rep events.
 
\item Case $b_1 = (c,d)$ is a 2-tuple span and $b_2 = (e,g)$ a 2-tuple span.

By definition of $\spans p$, we know there is no writer into $p$
in between $c$ and $d$ and in between $e$ and $g$. But $c$ and $e$ 
are writers into $p$ by definition of $\spans p$.
Therefore, we must have $\linReps e c$, or 
$e = c$, or $\linReps d e$ (the case $d = e$ is impossible, because
$d$ and $e$ correspond to different code lines: $e$ writes a descriptor, while
$d$ resolves resolves a descriptor). 

\begin{itemize}
\item Case $\linReps d e$. Hence $\precedesSpans {b_1} {b_2}$.

\item Case $\linReps e c$. We have either $\linReps g c$, or 
$c \linRepsSymbol g \linRepsSymbol d$, or
$g = d$, or $\linReps d g$ (the case $g = c$ is impossible,
because $g$ resolves a descriptor, while $c$ writes one).

The cases $g = d$ and $\linReps d g$ would lead to 
$e \linRepsSymbol c \linRepsSymbol g$ (Contradiction).
If $c \linRepsSymbol g \linRepsSymbol d$,
then we also have $e \linRepsSymbol c \linRepsSymbol g$,
since $\linReps e c$ (Contradiction).
Therefore, $\linReps g c$ and $\precedesSpans {b_2} {b_1}$ follows.

\item Case $e = c$. We have either  
$c \linRepsSymbol g \linRepsSymbol d$, or
$g = d$, or $\linReps d g$ (the case $\linReps g c$ is
impossible because $g$ occurs after $e = c$).
Since $e = c$, it means that $e$ and $c$ wrote the same descriptor.

The case $e = c \linRepsSymbol g \linRepsSymbol d$
means that the descriptor status was resolved twice, which is impossible (Contradiction).

The case $\linReps d g$ leads to 
$c = e \linRepsSymbol d \linRepsSymbol g$, which means that
the descriptor was again resolved twice (Contradiction).

Therefore $g = d$ also holds and $b_1 = b_2$ follows.
\end{itemize}
\end{itemize}
\end{prf}

\begin{lem}
\label{lem::appendix::impl::all-descriptors-are-written-before-any-resolution-mcas}
Axiom \axiomDRef{principle::all-descriptors-are-written-before-any-resolution} holds.
\end{lem}

\begin{prf}
This is trivial when $x$ has a denotation with only one span. 
So, it is enough to check the cases for $x = \mcasAlg(\listvar u)$. 

The failing denotation 
has the form $\{ (a_k,r),\ (c_j) \mid k \in \listvar u \wedge k < j \}$ for some $j \in \listvar u$,
where $c_j$ occurs after all the $a_k$'s since $c_j$ is the last $\rdcssAlg$ that fails 
at line \ref{appendix::invoke-rdcss-in-MCAS}. Also, $c_j$ occurs before $r$, because
the failing $\rdcssAlg$ at line \ref{appendix::invoke-rdcss-in-MCAS} precedes the descriptor status change 
at line \ref{appendix::resolve-status-MCAS}.
Therefore $(c_j)$ overlaps with all the $(a_k,r)$, while any two $(a_k,r)$ overlap because 
they share $r$.

The success denotation has the form $\{ (a_j,r) \mid j \in \listvar u \}$. Any two $(a_j,r)$ overlap because
they share $r$.
\end{prf}

\begin{lem}
	\label{lem::appendix::impl::finished-are-non-empty-axiom-holds-mcas}
	Axioms
	\axiomDRef{principle::finished-operations-have-a-run} holds.
\end{lem}

\begin{prf}
	We prove each procedure in turn.
	\begin{itemize}
		\item Case $\mcasAlg(\listvar u)$. 
		Suppose that $\mcasAlg(\listvar u)$ finished.
		Then, the invoking thread $T$ must have created a pointer $d$ to the
		descriptor at line~\ref{appendix::alloc-desc-MCAS}, and finished the invocation of
		$\mcasHelpAlg(d)$ at line~\ref{appendix::invoke-auxiliary-mcas-MCAS}. 
		Thread $T$ must have reached line~\ref{appendix::return-boolean-MCAS}, which returns a
		boolean $r$.
		
		\begin{itemize}
			\item Case $r = true$. $T$ must have read $d$'s status pointer to be $\SUCCEEDED$ 
			at line~\ref{appendix::read-phase2-status-MCAS}. This implies that some thread $T'$ (which could be $T$ as well)
			must have set the status pointer to $\SUCCEEDED$ at line~\ref{appendix::resolve-status-MCAS} and $T'$ must
			have finished the execution of $\writeAllDescsAlg$ with a $\SUCCEEDED$ at line~\ref{appendix::write-all-descs-MCAS}.
			Since $\writeAllDescsAlg$ is a recursive method which only recurses when it needs to try again, and we know that $T'$
			finished executing $\writeAllDescsAlg$, $T'$ must have reached
			an invocation of $\writeAllDescsAlg$ that did not recurse, i.e., an
			invocation of $\writeAllDescsAlg$ where 
			lines \ref{appendix::alg-help-complete-invoke-MCAS-MCAS}-\ref{appendix::try-again-writeall-MCAS} were not executed. So, we
			can assume we are on such execution. 
			
			Since $\writeAllDescsAlg$ returned $\SUCCEEDED$,
			thread $T'$ must have tried all the entries in the loop. 
			For each entry $i \in \listvar{u}$, $T'$ must have tried to write $d$ using the 
			$\rdcssAlg$ at line~\ref{appendix::invoke-rdcss-in-MCAS}. 
			The $\rdcssAlg$ must have returned either $d$ or the expected
			value $\expGenEntry{i}$ (it cannot return a descriptor different
			from $d$, because lines \ref{appendix::alg-help-complete-invoke-MCAS-MCAS}-\ref{appendix::try-again-writeall-MCAS} would be executed). 
			If the returned
			value was the expected value $\expGenEntry{i}$, then the $\rdcssAlg$ succeeded (since we
			already know that the status was set to $\SUCCEEDED$ by $T'$ later and the
			status is $\UNDECIDED$ previous to the change). If the $\rdcssAlg$ at line~\ref{appendix::invoke-rdcss-in-MCAS}
			returned $d$, then some other thread (or $T'$ in a previous recursive
			$\writeAllDescsAlg$ invocation) already wrote the descriptor. In other words,
			in both cases the descriptor was written. Since we know that $d$ was
			written for each entry and its status pointer was set to $\SUCCEEDED$
			later, this corresponds to case (1).(a) in the definition of $\runFunc {\mcasAlg(\listvar u)}$,
			which has as denotation output $true$ and the denotation is not empty.
			
			\item Case $r = false$. 
			Notice that no thread can reach line~\ref{appendix::remove-all-descs-MCAS} unless some thread reaches
			line~\ref{appendix::resolve-status-MCAS} first and changes the status to either 
			$\SUCCEEDED$ or $\FAILED$. The
			reason is that while the status is $\UNDECIDED$, threads starting $\mcasHelpAlg$
			will enter the true case of the conditional at line~\ref{appendix::is-phase1-still-undecided}.
			
			Having said this, $T$ must have read the status pointer of $d$ to be
			different from $\SUCCEEDED$ at line~\ref{appendix::read-phase2-status-MCAS}. But since no thread can reach 
			line~\ref{appendix::remove-all-descs-MCAS} while the status is $\UNDECIDED$, $T$ must have read $\FAILED$ at line~\ref{appendix::read-phase2-status-MCAS}. 
			This implies that some thread $T'$ (which could be $T$ as well)
			must have set the status pointer to $\FAILED$ at line~\ref{appendix::resolve-status-MCAS} and $T'$ must
			have finished the execution of $\writeAllDescsAlg$ with a $\FAILED$ at 
			line~\ref{appendix::write-all-descs-MCAS}. Since $\writeAllDescsAlg$ is a recursive method that recurses only to try again,
			and we know that $T'$ finished executing $\writeAllDescsAlg$, thread $T'$ must have reached
			an invocation of $\writeAllDescsAlg$ that did not recurse, i.e., an
			invocation of $\writeAllDescsAlg$ where lines \ref{appendix::alg-help-complete-invoke-MCAS-MCAS}-\ref{appendix::try-again-writeall-MCAS} were not executed. 
			So, we can assume we are on such execution. Since $\writeAllDescsAlg$ returned $\FAILED$,
			thread $T'$ must have tried a last entry $l$ in the loop for
			which the $\rdcssAlg$ at line~\ref{appendix::invoke-rdcss-in-MCAS} returned a value different from the
			expected value (otherwise $T'$ could not have reached line~\ref{appendix::return-failed-writeall-MCAS}),
			meaning that the $\rdcssAlg$ failed for entry $l$. But this means that
			for all entries $i < l$, the $\rdcssAlg$ must have returned
			either $d$ or the expected value $\expGenEntry i$ (it cannot return
			a descriptor different from $d$, because lines \ref{appendix::alg-help-complete-invoke-MCAS-MCAS}-\ref{appendix::try-again-writeall-MCAS} would be
			executed), otherwise the loop would have been interrupted before
			entry $l$. For entry $i$, if the returned value at line~\ref{appendix::invoke-rdcss-in-MCAS} was the
			expected value, then the $\rdcssAlg$ succeeded (since we already know
			that the status was set to $\FAILED$ by $T'$ later and the status is
			$\UNDECIDED$ previous to the change). For entry $i$, if the $\rdcssAlg$ at 
			line~\ref{appendix::invoke-rdcss-in-MCAS} returned $d$, then some other thread (or $T'$ in a previous
			recursive $\writeAllDescsAlg$ invocation) already wrote the descriptor. In
			other words, in both cases the descriptor was written for every
			entry before $l$. Since we know that $d$ was written for each entry
			before $l$, and the $\rdcssAlg$ at line~\ref{appendix::invoke-rdcss-in-MCAS} failed for $T'$ at entry $l$
			and the status pointer was set to $\FAILED$ by $T'$ later, this
			corresponds to case (1).(b) in the definition of $\runFunc {\mcasAlg(\listvar u)}$, which
			has as denotation output $false$ and the denotation is not empty.
		\end{itemize}
		
		\item Case $\mcasReadAlg(p)$. Suppose that $\mcasReadAlg(p)$ finished. 
		Let $T$ be the invoking thread. Since $\mcasReadAlg$ is a recursive method and
		we know that $T$ finished executing it, this means that $T$ must have
		reached an invocation of $\mcasReadAlg$ that did not recurse, i.e., an
		invocation where lines 
		\ref{appendix::alg-help-complete-invoke-Read-MCAS}-\ref{appendix::try-again-Read-MCAS} were not executed. 
		So, we can assume we
		are on such execution. Since $T$ reached line~\ref{appendix::alg-return-old-value-Read-MCAS}, 
		the read at line~\ref{appendix::alg-access-Read-MCAS}
		must have returned a non-descriptor value $v$. This corresponds to
		(2).(a) in the definition of $\runFuncSymbol$, with denotation output
		whatever line~\ref{appendix::alg-access-Read-MCAS} produced (which is equal to $v$) 
		and the denotation is not empty.

		\item Case $\mcasWriteAlg(p,v)$. Suppose that $\mcasWriteAlg(p,v)$ finished.
		Let $T$ be the invoking thread. Since $\mcasWriteAlg$ is a recursive method and
		we know that $T$ finished executing it, this means that $T$ must have
		reached an invocation of $\mcasWriteAlg$ that did not recurse, i.e., an
		invocation where lines 
		\ref{appendix::alg-help-complete-invoke-Write-MCAS}-\ref{appendix::try-again-Write-One-MCAS},
		\ref{appendix::try-again-Write-Two-MCAS} were not executed. So, we can assume
		we are on such execution. Since $T$ must have returned after reaching
		line~\ref{appendix::implicit-return-for-Write-MCAS} without entering the conditional 
		at line~\ref{appendix::last-if-in-Write-MCAS}, the CAS at line~\ref{appendix::alg-attempt-write-Write-MCAS} must
		have returned $old$ (meaning that the CAS succeeded), and at the same
		time, the read at line~\ref{appendix::alg-access-Write-MCAS} must have returned $old$, and $old$ is a
		non-descriptor value. Since the CAS at line~\ref{appendix::alg-attempt-write-Write-MCAS} succeeded, this
		corresponds to (3).(a) in the definition of $\runFuncSymbol$, with
		denotation output $tt$ and the denotation is not empty.
		
		\item Case $\mcasAllocAlg(v)$. Suppose that $\mcasAllocAlg(v)$ finished.
		Let $T$ be the invoking thread. Then $T$ must have executed line~\ref{appendix::alloc-data-Alloc-MCAS}
		producing some pointer $p$ as output. This corresponds to (4).(a) in
		the definition of $\runFuncSymbol$ with output whatever line~\ref{appendix::alloc-data-Alloc-MCAS}
		produced and the denotation is not empty.
	\end{itemize}
\end{prf}

\begin{lem}
Axiom \axiomDRef{principle::descriptor-write-precedes-resolution} holds.
\end{lem}

\begin{prf}
Trivial for 1-spans. For 2-tuple spans $(a,b)$, the code can only resolve descriptors (rep event $b$)
only if it was previously written at line \ref{appendix::invoke-rdcss-in-MCAS} (rep event $a$).
\end{prf}

\begin{lem}
\label{lem::appendix::impl::runs-are-injective-mcas}.
Axiom
\axiomDRef{principle::runs-are-injective} holds.
\end{lem}

\begin{prf}
First, let us focus when neither $x$ nor $y$ are $\mcasAlg$ events.

If $b \in \hspans{}{x} \cap \hspans{}{y}$, then (by going through all the 
non-$\mcasAlg$ cases in $\runFuncSymbol$),
$b = (i_1)$ for some $i_1$ that is executed by the thread that invoked $x$ and 
$b = (i_2)$ for some $i_2$ that is executed by the thread that invoked $y$.
In addition, $i_1$ occurs within the invocation of $x$ and $i_2$ within
the invocation of $y$. Therefore, $i_1 = i_2$, which means that $x$ and $y$ 
are invoked by the same thread.

If $x \neq y$, then $x$ and $y$ cannot overlap in real-time, because they 
are invoked by the same thread. But this contradicts that $i_1$ occurs within
the invocation of both $x$ and $y$. Therefore, $x = y$.

Now, let us check the case when either $x$ is an $\mcasAlg$ event or $y$ is.
Say, $x$ is an $\mcasAlg$ event. 

All the spans in the cases for $\mcasAlg$ satisfy that $\firstRep b$ has code line 
\ref{appendix::invoke-rdcss-in-MCAS}. This means that $\firstRep b$
has the form $\rdcssAlg(rD)$ such that $\newTwo{rD} = d$, where $d$ is the unique
identifier for the invocation $x$ generated at line 
\ref{appendix::alloc-desc-MCAS} (see discussion that the denotation
is well-defined). Since $b$ also occurs in the denotation of $y$ and $d$ is unique
per-invocation, it must be the case that $x = y$.
\end{prf}

\begin{lem}
Axiom
\axiomDRef{principle::writer-blocks-belong-to-runs} holds.
\end{lem}

\begin{prf}
Directly from Lemma \ref{lem::appendix::impl::writers-belong-to-writer-procs-mcas}.
\end{prf}

\begin{lem}
\label{lem::appendix::impl::postcondition-predicate-holds-mcas}
Axiom
\axiomDRef{principle::postcondition-predicate-holds} holds.
\end{lem}

\begin{prf}
Suppose $\hspans{}{x} \neq \emptyset$. We see from definition of $\runFunc x$ that all non-empty cases have $\outputRunFunc x \neq \bot$.
To prove $\postPred x {\outputRunFunc x}$, we do a case analysis on $x$.

\begin{itemize}
\item Case $x = \mcasAlg(\listvar u)$. We want to prove,
\begin{align*}
\outputRunFunc x \in \BoolType \wedge
\begin{cases}
\forall i \in \listvar u.\ \exists z.\ \visObs {\pointerIndx {\pointGenEntry i}} z {x} \wedge \inputVal z {\pointGenEntry i} {\expGenEntry i} & \text{ if }\outputRunFunc x=true\\
\begin{aligned}
\exists i \in \listvar u,v\in\ValType.\ \exists z.\ \visObs {\pointerIndx {\pointGenEntry i}} z {x} \wedge {} \\
\inputVal z {\pointGenEntry i} v \wedge v \neq {\expGenEntry i} 
\end{aligned} & \text{ if }\outputRunFunc x=false
\end{cases}
\end{align*}

That $\outputRunFunc x \in \BoolType$ follows directly from definition of $\runFuncSymbol$.
We now do a case analysis.
\begin{itemize}
\item Case $\outputRunFunc x = true$.
The success denotation has the form $\{ (a_i,r) \mid i \in \listvar u \}$, where $r$
is the event that sets the descriptor status to $\SUCCEEDED$ and
each $a_i$ writes the descriptor into $\pointGenEntry i$
(which means that each $a_i$ reads the value $\expGenEntry i$).
Therefore, by Lemma \ref{lem::appendix::impl::visibility-lemma-mcas},
for every $i \in \listvar u$,
there is $z$ such that $\visObs {\pointerIndx {\pointGenEntry i}} {z} x$
and $\inputVal{z} {\pointGenEntry i} {\expGenEntry i}$.

\item Case $\outputRunFunc x = false$.
The failing denotation has the form $\{ (a_k,r),\ (c_i) \mid k \in \listvar u \wedge k < i \}$, where $r$
is the event that sets the descriptor status to $\FAILED$ and
each $a_k$ writes the descriptor into $\pointGenEntry k$
(which means that each $a_k$ reads the value $\expGenEntry k$), but
$c_i$ failed to write the descriptor (so $c_i$ read a value different from $\expGenEntry i$).
Therefore, by Lemma \ref{lem::appendix::impl::visibility-lemma-mcas},
there is $z$ such that $\visObs {\pointerIndx {\pointGenEntry i}} {z} x$
and $\inputVal{z} {\pointGenEntry i} {v}$ for some $v \neq \expGenEntry i$.
\end{itemize}

\item Case $x = \mcasAllocAlg(v)$. We want to prove $\outputRunFunc x = \outputProp x$
and $\ETimeProp x \neq \bot$. But this follows trivially from the definition of $\runFunc x$.

\item Case $x = \mcasWriteAlg(q,v)$. We want to prove $\outputRunFunc x = \unitValue$
but this follows trivially from the definition of $\runFunc x$.

\item Case $x = \rdcssReadAlg(q)$. We want to prove,
\begin{align*}
\exists z.\ \visObs {\pointerIndx q} z x \wedge \inputVal z q {\outputRunFunc x}
\end{align*}

The only case in the definition of $\runFunc x$
has a span $b$ such that $\firstRep b$ reads value $\outputRunFunc x$ 
in pointer $q$ at line \ref{appendix::alg-access-Read-MCAS}.
Therefore, by Lemma \ref{lem::appendix::impl::visibility-lemma-mcas},
there is $z$ such that $\visObs {\pointerIndx q} {z} x$
and $\inputVal z q {\outputRunFunc x}$.
\end{itemize}
\end{prf}

\begin{lem}
\label{lem::appendix::impl::writers-have-writer-blocks-mcas}
Axiom
\axiomDRef{principle::writers-have-writer-blocks} holds.
\end{lem}

\begin{prf}
Suppose $\hspans{}{x} \neq \emptyset$. We do a case analysis on $x$.
\begin{itemize}
\item Case $x = \mcasAlg(\listvar u)$.

$\Longrightarrow$. By definition of $\writesAbs p$, we have $p = \pointGenEntry j$ for some $j \in \listvar u$ and,
\begin{align}
\label{eqn::appendix::impl::writers-have-writer-blocks-1-mcas}
\forall i \in \listvar u.\ \exists z.\ \visObs {\pointerIndx {\pointGenEntry i}} z {x} \wedge \inputVal z {\pointGenEntry i} {\expGenEntry i}
\end{align}
By Lemma \ref{lem::appendix::impl::visibility-lemma-mcas}, for every $i \in \listvar u$, there are
$b_i \in \hspans {\pointGenEntry i} x$ such that
$\firstRep {b_i}$ reads value $\expGenEntry i$.

The only case in the denotation that matches this conditions is (1).(a), i.e.,
the successful denotation. In particular, for $j \in \listvar u$,
$b_j \in \hspans{}{x} \cap \writesSpans {\pointGenEntry j}$.

$\Longleftarrow$. Suppose $b \in \hspans{}{x} \cap \writesSpans {p}$. 
The only applicable case in the denotation is the successful one (1).(a).
Hence, $b$ must be of the form $(a_j,r)$, for some $j \in \listvar u$, where $a_j$ writes the descriptor into
$\pointGenEntry j$. Hence, $p = \pointGenEntry j$.

Also, since the denotation contains all spans of the form $(a_i, r)$ for every $i \in \listvar u$, and each $a_i$
is a successful $\rdcssAlg$ that reads the expected value $\expGenEntry i$,
\eqref{eqn::appendix::impl::writers-have-writer-blocks-1-mcas} follows 
by Lemma \ref{lem::appendix::impl::visibility-lemma-mcas}.

\item Case $x = \mcasAllocAlg(v)$.

$\Longrightarrow$. Since $\hspans{}{x} \neq \emptyset$, by definition of $\runFuncSymbol$, 
$\ETimeProp x \neq \bot$, $\outputProp x \neq \bot$, and there is
a span in $\hspans{}{x} \cap \writesSpans {\outputProp x}$.
But since $x \in \writesAbs p$, we have $p = \outputProp x$.

$\Longleftarrow$. Let $b \in \hspans{}{x} \cap \writesSpans {p}$.
By definition of $\runFuncSymbol$, $\ETimeProp x \neq \bot$, $\outputProp x \neq \bot$,
$b = (i)$ for some alloc $i$ at line 
\ref{appendix::alloc-control-Alloc-RDCSS}, and $(i) \in \writesSpans{\outputProp x}$.

But from $(i) \in \writesSpans{p}$ we also know that $i$ allocates $p$,
and since allocations are unique, $p = \outputProp x$. Hence, $x \in \writesAbs p$.

\item Case $x = \mcasWriteAlg(q,v)$.

$\Longrightarrow$. Since $\hspans{}{x} \neq \emptyset$, there is
a span in $\hspans{}{x} \cap \writesSpans {q}$ because all
spans in the denotation write in $q$. But $p = q$ follows from 
$x \in \writesAbs p$.

$\Longleftarrow$. Suppose $b \in \hspans{}{x} \cap \writesSpans {p}$. Then, $p = q$
since the same rep event writes into $p$ (by definition of $\writesSpans {p}$)
and into $q$ (by definition of $\runFuncSymbol$), which means $x \in \writesAbs p$.

\item Case $x = \mcasReadAlg(q)$. Trivial since $x$ is neither a successful writer
nor it has writer spans in its denotation.
\end{itemize}
\end{prf}

\begin{lem}
\label{lem::appendix::impl::blocks-contained-in-abstract-time-interval-mcas}
Axiom
\axiomDRef{principle::blocks-contained-in-abstract-time-interval} holds.
\end{lem}

\begin{prf}
We prove each item.

\begin{itemize}
\item (i).
Let us focus on all denotations in $\runFunc x$ having
the form $\{ (b) \}$. 
In these denotations, $b$ is a rep event
that is invoked by the same thread that invoked $x$. Hence, $b$ starts after $x$
started.

Now, let us focus on $\mcasAlg(\listvar u)$. First, case (1).(a), which produces a denotation of the form 
$\{ (a_j,r) \mid j \in \listvar u \}$. Rep event $a_j$ can only
execute after $x$ was started because line \ref{appendix::invoke-rdcss-in-MCAS} executes
after the unique id was created by line \ref{appendix::alloc-desc-MCAS}, which is executed 
by the thread invoking $x$. The reasoning for case (1).(b) is similar.

\item (ii).
Let us focus on all denotations in $\runFunc x$ having
the form $\{ (b) \}$.
In these denotations, $b$ is a rep event
that is invoked by the same thread that invoked $x$. Hence, $b$ finishes before $x$
finishes (hence, in this cases, choose $i \defini b$).

Now, let us focus on $\mcasAlg(\listvar u)$. First, case (1).(a), which produces a denotation of the form 
$\{ (a_j,r) \mid j \in \listvar u \}$. Let $T$ be the thread that invoked $x$. Before finishing 
$\mcasAlg$, thread $T$ will have to execute line \ref{appendix::remove-all-descs-MCAS}.
Therefore, we can choose $i$ to be this line, since this line executes before
$x$ finishes. Also, no thread can reach line \ref{appendix::remove-all-descs-MCAS} 
unless the status pointer 
was changed to either $\SUCCEEDED$ or $\FAILED$ at line 
\ref{appendix::resolve-status-MCAS}, i.e., rep event $r$ (notice the \textbf{if} at line
\ref{appendix::is-phase1-still-undecided}). The reasoning for case (1).(b) is similar.
\end{itemize}
\end{prf}

\begin{lem}
	\label{lem::appendix::impl::containment-and-uniqueness-of-alloc-blocks-mcas}
	Axiom \axiomDRef{principle::containment-and-uniqueness-of-alloc-blocks} holds.
\end{lem}

\begin{prf}
	If $b_1, b_2 \in \allocsSpans p$,
	then by definition of $\allocsSpans p$,
	$b_1$ and $b_2$ are 1-tuples allocating $p$ by executing 
	line \ref{appendix::alloc-data-Alloc-MCAS}.
	But by Lemma 
	\ref{lem::appendix::impl::unique-alloc-mcas},
	$b_1 = b_2$.
\end{prf}

\begin{lem}
	\label{lem::appendix::impl::every-block-must-have-an-allocated-pointer-mcas}
	Axiom
	\axiomDRef{principle::every-block-must-have-an-allocated-pointer} holds.
\end{lem}

\begin{prf}
	Let $b \in \spans p$. 
	Each procedure requires that its input pointers be created by a previous alloc invocation.
	Therefore, there is some rep alloc $i$ that allocates $p$ and executes 
	line \ref{appendix::alloc-data-Alloc-MCAS}. So that  
	$i$ executes before $\firstRep b$ (in case $b$ does not contain an alloc), or $b = (i)$.
	
	Now, any rep event which is an instance of line \ref{appendix::alloc-data-Alloc-MCAS} 
	belongs to some span by definition of $\allocsSpans p$.
	Therefore, $(i) \in \allocsSpans p$. 
	
	So, if $i$ executes before $\firstRep b$, then $\precedesSpans {(i)} b$, and if 
	$b = (i)$, then $\precedesSpansEq {(i)} b$.
\end{prf}

\begin{lem}
	\label{lem::appendix::impl::allocs-have-alloc-blocks-mcas}
	Axioms
	\axiomDRef{principle::allocs-have-alloc-blocks} holds.
\end{lem}

\begin{prf}
	We prove each direction.
	
	$\Longrightarrow$. Since $\hspans{}{x} \neq \emptyset$ and $x \in \allocsAbs p$ then $x$ is an alloc and 
	$\ETimeProp x \neq \bot$, $\outputProp x \neq \bot$, and $p = \outputProp x$. But by the definition of the denotation, 
	there is $(i) \in \writesSpans{\outputProp x}$, where $i$ is line
	\ref{appendix::alloc-data-Alloc-MCAS}.
	Therefore, $(i) \in \hspans{}{x} \cap \allocsSpans p$ by definition.
	
	$\Longleftarrow$. Let $b \in \hspans{}{x} \cap \allocsSpans p$. 
	By definition of $\allocsSpans p$, $\firstRep b$ allocates $p$ and it is 
	line \ref{appendix::alloc-data-Alloc-MCAS}.
	But the only case applicable in the denotation is when $x$ is an alloc, which means 
	$\ETimeProp x \neq \bot$, $\outputProp x \neq \bot$ and $b \in \writesSpans{\outputProp x}$. 
	Also, since rep allocs are unique $p = \outputProp x$, which
	means $x \in \allocsAbs p$.
\end{prf}

\begin{thm}
\label{thm::appendix::impl::span-axioms-satisfied-mcas}
$\genStructName{\MCAS}(\visObsSymbol {\pointerIndx p}, \visSepSymbol{\pointerIndx p})$ 
is implemented by
span structure $\genSpanStructName{\MCAS}$.
\end{thm}

\begin{prf}
All span axioms hold from Lemma \ref{lem::appendix::impl::non-interference-of-lifespans-mcas} to 
Lemma \ref{lem::appendix::impl::allocs-have-alloc-blocks-mcas}.
\end{prf}

\begin{thm}
\label{thm::appendix::impl::mcas-is-linearizable}
The MCAS implementation of Figure \ref{appendix::alg-MCAS} is linearizable.
\end{thm}

\begin{prf}
By Theorem \ref{thm::appendix::lin::rdcss-linearizability-from-vis-structure}, it
  suffices to show that
  $\genStructName{\MCAS}(\visObsSymbol {\pointerIndx p},
  \visSepSymbol{\pointerIndx p})$ is valid.  But by
  Theorem \ref{thm::appendix::valid::span-axioms-imply-visibility}, it suffices that
  $\genStructName{\MCAS}(\visObsSymbol {\pointerIndx p},
  \visSepSymbol{\pointerIndx p})$ is implemented by span
  structure $\genSpanStructName{\MCAS}$. This is given by Theorem
  \ref{thm::appendix::impl::span-axioms-satisfied-mcas}.
\end{prf}

\section{Proof of Validity and Proof of Opportunism Axioms for MCAS with Opportunistic Readers}
\label{sect::appendix::opportunistic-impl-proof}

\begin{figure}[t]
\centering
\includegraphics[scale=0.5]{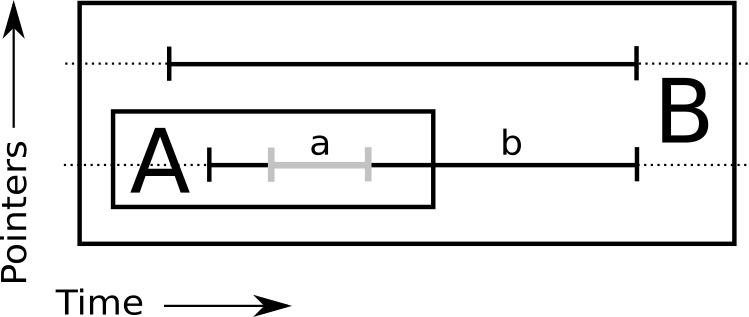}
\caption{Two executions of opportunistic $\mcasReadAlg$ event $A$ and $\mcasAlg$ event $B$.
$A$ executes fully within the time extension of $B$.
Black spans are descriptor spans, while the gray span is an opportunistic
span.
Spans $a$ and $b$ are named for later reference.}
\label{fig::sub::appendix::opor::opport-spans-example}
\end{figure}

The relations in Definition \ref{defn::appendix::valid::visibility-relations} 
suffice to verify RDCSS and MCAS with standard
helping implementations.
However, in the case of MCAS with opportunistic $\mcasReadAlg$, we need
to consider new cases in order to capture the new opportunistic
strategy of the readers. To motivate the new definitions for the visibility relations,
consider Figure \ref{fig::sub::appendix::opor::opport-spans-example}. Black spans in the figure are the standard 
descriptor spans (henceforth, d-spans), which we treated in Section \ref{sub::sect::lifespan-axiomatization} with the name spans; while
gray spans are opportunistic spans (henceforth, o-spans), which are the new kind of spans
we will introduce in this section. 

Figure~\ref{fig::sub::appendix::opor::opport-spans-example} shows a typical interaction
between an opportunistic $\mcasReadAlg$ (event $A$) and an $\mcasAlg$
(event $B$).  Event $A$ executes fully during the execution of $B$.
The gray span $a$ in $A$ overlaps with span $b$ in $B$. Both $a$ and
$b$ access the same pointer.  Span $a$ is painted gray to emphasize
that it is a different kind of span: it does not involve writing a
descriptor, but is, rather, the extension of time between discovering
a descriptor at line~\ref{alg-access-Read-Oport-MCAS} in
Figure~\ref{fig::sub::optimal-mcas-read-impl} and then observing the descriptor status
at line~\ref{alg-Read-status-Oport-MCAS}. For this example,
we will assume that the black spans finish when the descriptor
status is set to $\SUCCEEDED$ at line \ref{resolve-status-MCAS} in Figure
\ref{alg-MCAS}.

Even though spans $a$ and $b$ overlap in Figure
\ref{fig::sub::appendix::opor::opport-spans-example}, there is still a sense in which we
can separate $A$ before $B$. In the figure, $a$ observed the descriptor status 
before the descriptor was resolved. Therefore, line~\ref{alg-Read-status-Oport-MCAS} in
Figure~\ref{fig::sub::optimal-mcas-read-impl} returned $\UNDECIDED$,
which implies that $\mcasReadAlg$ returned the expected value, as if the
descriptor was not present. In other words, $A$ missed the effect
of $B$, and therefore $A$ can be \emph{separated} before $B$. 
Another way in which we can separate $A$ before $B$ is if $A$ attempted 
to discover the descriptor before $B$ wrote its descriptor.
These two modes of separation can be expressed by the relation 
$\visSepIndx {\pointerIndx p} {\opporIndx} A B$
which reads ``$A$ is
opportunistically separable-before $B$ at pointer $p$'', and it is defined as:
\[
\visSepIndx {\pointerIndx p} {\opporIndx} A B \defini \exists x \in \ospans{p} A,
y \in \hspans{}{B} \cap \writesSpans p.\ 
(\linReps {\firstRep x} {\firstRep y} \vee \LastBlockRel {x} {y})
\]
Here, $\ospans{p}{A}$ is
the set of o-spans in $A$ accessing pointer $p$ (i.e., the gray spans in $A$);
$\hspans{}{B} \cap \writesSpans p$ the set of successful $p$-writer d-spans in $B$; 
$\linReps {\firstRep x} {\firstRep y}$
means that the first rep event of span $x$ executes before the first rep event
of span $y$ (i.e., $x$ attempted to discover a descriptor before $y$
wrote a descriptor); and
$\LastBlockRel {x} {y}$ means that
the last rep event executed by span $x$ finished before the last
rep event executed by span $y$ (i.e., $x$ observed the descriptor status
before $y$ resolved it).
The ``gap'' between $A$ and $B$ in this case
is not a gap in real time as it was in the helpers case, but is
rather \emph{logical}.

Analogous to how we introduced an observation relation paired with the 
separable-before relation in Section \ref{subsubsect::visibility-mcas}, 
we next introduce an observation relation for
$\visSepIndxSymbol {\pointerIndx p} {\opporIndx}$.

In Figure \ref{fig::sub::appendix::opor::opport-spans-example}, $A$ missed the effect of
$B$ because $a$ observed the descriptor status before $b$ resolved
the descriptor to $\SUCCEEDED$.
Therefore, if $A$ observes the descriptor status \emph{after} the
descriptor's resolution by $b$, $A$ will not miss the effect of $B$
(indeed, $A$ returns the
new value stored in the descriptor when line
\ref{alg-Read-status-Oport-MCAS} in Figure
\ref{fig::sub::optimal-mcas-read-impl} returns $\SUCCEEDED$, as if the
write to the pointer already occurred). This suggests that we should
define that $A$ observes $B$ at $p$, if $B$ executed the most recent
successful $p$-writer span that resolved the descriptor before $A$ 
read the status (and also, if $B$ wrote the descriptor before $A$ attempted to
discover it, otherwise $A$ would not discover a descriptor). We can express
this by the relation $\visObsIndx {\pointerIndx p} {\opporIndx} B A$
which reads ``$B$ is
opportunistically observed by $A$ at pointer $p$'', and it is defined as:
\[
\visObsIndx {\pointerIndx p} {\opporIndx} B A \defini \exists x \in 
\dspans{p}{B}, y \in \ospans{p}{A}.\  x = \max_{\precedesSpansEqSymbol} \{ z \in \writesSpans p \mid 
\linReps {\firstRep z} {\firstRep y} \wedge \LastBlockRel z y \}
\]
Here, $\dspans{p}{B}$ is the set of $p$-accessing d-spans in $B$ and 
$\writesSpans p$ is the set of $p$-writing d-spans. Notice that
$\linReps {\firstRep z} {\firstRep y}$ captures that $z$ wrote the descriptor
before $y$ attempted to discover it, and
$\LastBlockRel z y$ captures that $z$ resolved before $y$ read the status.
Also, $\precedesSpansEqSymbol$ is the span returns-before relation of Section \ref{sub::sect::lifespan-axiomatization},
which still linearizes $p$-accessing descriptor spans, and hence, taking the maximum
under $\precedesSpansEqSymbol$ makes sense.

Since descriptor spans are still present, we need to keep the visibility relations
we used for the standard implementations in Definition \ref{defn::appendix::valid::visibility-relations}, 
we just tag the relations with a superscript
to indicate that these apply over descriptor spans,
\begin{align*}
\visSepIndx{\pointerIndx p}{\descIndx} A B & \defini \exists x \in \dspans{p}{A}, y \in \dspans{p}{B}.\ \precedesSpans x y \\
\visObsIndx{\pointerIndx p}{\descIndx} A B & \defini \exists x \in \dspans{p}{A}, y \in \dspans{p}{B}.\  x = \max_{\precedesSpansEqSymbol} \{ z \in \writesSpans p \mid \precedesSpans z y \}
\end{align*}

Finally, our visibility relations will be the union of the opportunistic and standard components,
\begin{align*}
\visSepSymbol {\pointerIndx p} & \defini {\visSepIndxSymbol {\pointerIndx p} {\opporIndx}} \cup {\visSepIndxSymbol {\pointerIndx p} {\descIndx}} \\
\visObsSymbol {\pointerIndx p} & \defini {\visObsIndxSymbol {\pointerIndx p} {\opporIndx}} \cup {\visObsIndxSymbol {\pointerIndx p} {\descIndx}}
\end{align*}

As we did with spans in Section \ref{sub::sect::lifespan-axiomatization},
we will introduce an abstract notion of opportunistic
span via opportunistic structures. 
We will state axioms that opportunistic structures need to satisfy, which we call opportunism axioms
(Section \ref{sub::sect::appendix::opor::key-concepts}).
The opportunism axioms are a superset of the span axioms of Section \ref{sub::sect::lifespan-axiomatization}, extending 
the invariants of descriptor spans with invariants about opportunistic spans.
The opportunistic structures and opportunism axioms will allow us to split the validity proof 
of $\MCASFam(\visObsSymbol {\pointerIndx p}, \visSepSymbol {\pointerIndx p})$ into two parts.
The first part of the proof concludes the validity of $\MCASFam(\visObsSymbol {\pointerIndx p}, \visSepSymbol {\pointerIndx p})$
from the assumption that some opportunistic structure satisfies the opportunism axioms (Section \ref{sub::sect::appendix::opor::proof-of-validity}).
The second part of the proof defines an opportunistic structure for opportunistic MCAS and shows that 
this structure satisfies the opportunism axioms (Section \ref{sub::sect::appendix::opor::proof-of-opportunism-axioms}). 
Together with Theorem \ref{thm::appendix::lin::mcas-linearizability-from-vis-structure}, 
we then conclude that opportunistic MCAS is linearizable (Theorem \ref{thm::appendix::opor::mcas-is-linearizable}).

\subsection{Key Concepts}
\label{sub::sect::appendix::opor::key-concepts}

As we did in Section \ref{sub::sect::lifespan-axiomatization}, 
we suppose a linearization $\linRepsEqSymbol$ 
over the set of rep events. We also denote as $\repsCompl$ the domain
of $\linRepsEqSymbol$.

\begin{defn}[Opportunistic Structure]
\label{defn::sub::sect::appendix::opor::opor-structure}
A \emph{opportunistic structure}
$(\descriptorSpans p, \opportunisticSpans p, \writesSpans p, \allocsSpans p,
\runFuncSymbol)$ consists
of:
\begin{itemize}
\item For every pointer $p$, a set $\descriptorSpans p \subseteq \repEvent \times \repEvent$, called the \emph{descriptor spans} accessing $p$.
\item For every pointer $p$, a set $\opportunisticSpans p \subseteq \repEvent \times \repEvent$, called the \emph{opportunistic spans} accessing $p$.
\item For every pointer $p$, a set $\writesSpans p \subseteq \descriptorSpans p$, called the \emph{successful write spans} into $p$.
\item For every pointer $p$, a set $\allocsSpans p \subseteq \writesSpans p$, called the \emph{alloc spans} creating $p$.
\item A function $\runFuncSymbol: \absEvent \rightarrow (\mathcal{P}(\bigcup_p \spans p) \times (\ValType \cup \{ \bot \}))$,
called the \emph{event denotation}, written $\runFunc{x}$ for event $x$. Here, $\mathcal{P}(\bigcup_p \spans p)$ denotes the power set of $\bigcup_p \spans p$.
\end{itemize}
where $\spans p \defini {\descriptorSpans p} \cup {\opportunisticSpans p}$.
\end{defn}

Opportunistic structures are like the span structures of Definition~\ref{sect::defn-span-structure}. 
The only differences are that there is a new set $\opportunisticSpans p$ modeling the opportunistic spans, and
what previously was the set of spans $\spans p$ is now the set $\descriptorSpans p$ of descriptor spans.
The set of spans $\spans p$ now refers to the union of $\descriptorSpans p$ and $\opportunisticSpans p$.
Each span in $\spans p$ is a pair of rep events. We denote as
$\firstRepFuncSymbol$ (``first'') and $\lastRepFuncSymbol$ (``last'')
the standard pair projection functions. We write spans of the form
$(a,a)$ as $(a)$.

For o-span $b$, it is useful to think of $\firstRep b$ as the rep event that discovers the descriptor, and 
$\lastRep b$ as the rep event that read the descriptor status. For d-span $b$, we can still think of
$\firstRep b$ as the rep event that writes the descriptor, and $\lastRep b$ as the rep event that resolves the descriptor. 

\begin{defn}[Notation for Opportunistic Structures]
\begin{sloppypar}
Given opportunistic structure
$(\descriptorSpans p, \opportunisticSpans p, \writesSpans p, \allocsSpans p,
\runFuncSymbol)$,
we make the following definitions.
\end{sloppypar}

\begin{align*}
\descriptorSpans{} & \defini \bigcup_p \descriptorSpans p & \text{(Descriptor spans)} \\
\opportunisticSpans{} & \defini \bigcup_p \opportunisticSpans p & \text{(Oportunistic spans)} \\
\hspans{}{x} & \defini \projFirst {\runFunc x} & \text{(Spans in the denotation of $x$)} \\
\hspans{p}{x} & \defini \hspans{}{x} \cap \spans p & \text{($p$-accessing spans in the denotation of $x$)}\\
\dspans{p}{x} & \defini \hspans{}{x} \cap \descriptorSpans p & \text{($p$-accessing descriptor spans in the denotation of $x$)} \\
\ospans{p}{x} & \defini \hspans{}{x} \cap \opportunisticSpans p & \text{($p$-accessing opportunistic spans in the denotation of $x$)} \\
\dspans{}{x} & \defini \hspans{}{x} \cap \descriptorSpans{} & \text{(Descriptor spans in the denotation of $x$)} \\
\ospans{}{x} & \defini \hspans{}{x} \cap \opportunisticSpans{} & \text{(Oportunistic spans in the denotation of $x$)} \\
\outputRunFunc x & \defini \projSecond {\runFunc x} & \text{(Denotation output of $x$)} \\
\end{align*}
\end{defn}

\begin{defn}[Span Relations]
Given opportunistic structure
$(\descriptorSpans p, \opportunisticSpans p, \writesSpans p, \allocsSpans p,
\runFuncSymbol)$,
we define the following relation over $\bigcup_p \spans p$,
\begin{align*}
\precedesSpans b c & \defini \linReps {\lastRep b} {\firstRep c} \\
\end{align*}
\end{defn}

We now define the notion of implementation by an opportunistic structure,

\begin{defn}[Implementation by an opportunistic structure]
  Given arbitrary relations
  $\visObsSymbol {\pointerIndx p}$,
  $\visSepSymbol{\pointerIndx p}$, we say that
  $\genStructName{}(\visObsSymbol {\pointerIndx p},
  \visSepSymbol{\pointerIndx p}) = (\writesAbs p,
  \allocsAbs p, \postPredSymbol)$ is \emph{implemented by opportunistic
    structure}
  $(\descriptorSpans p, \opportunisticSpans p, \writesSpans p, \allocsSpans p,
\runFuncSymbol)$ if all
  the opportunism axioms in Figure \ref{fig::sub::appendix::opor::opor-axioms} are
  satisfied.  We also say that
  $\genStructName{}(\visObsSymbol {\pointerIndx p},
  \visSepSymbol{\pointerIndx p}) = (\writesAbs p,
  \allocsAbs p, \postPredSymbol)$ is \emph{implemented opportunistically} 
  if there exists some opportunistic structure that implements it.
\end{defn}

\begin{figure}[t]
\begin{subfigwrap}{Defined notions}{fig::sub::appendix::opor::defined-terms}
\centering
\begin{tabular}{l}
Span returns-before relation \\
\quad $\precedesSpans b c \defini \linReps {\lastRep b} {\firstRep c}$ \\
Set of $p$-accessing spans \\
\quad $\spans p \defini {\descriptorSpans p} \cup {\opportunisticSpans p}$ \\
Denotation output \\
\quad $\outputRunFunc x \defini \projSecond {\runFunc x}$ \\
Set of spans in a denotation \\
\quad $\hspans{}{x} \defini \projFirst {\runFunc x}$ \\
\end{tabular}
\quad
\begin{tabular}{l}
Set of $p$-accessing descriptor spans in a denotation \\
\quad $\dspans{p}{x} \defini \hspans{}{x} \cap \descriptorSpans p$ \\
Set of $p$-accessing opportunistic spans in a denotation \\
\quad $\ospans{p}{x} \defini \hspans{}{x} \cap \opportunisticSpans p$ \\
Opportunistic access predicate \\
\quad $\oportunisticPred b x \defini \exists p \in \PtsType, c \in \hspans{}{x} \cap \writesSpans p.$ \\
\quad\phantom{$\oportunisticPred b x \defini$} \quad $b \in \opportunisticSpans p \wedge \firstRep{c} \linRepsEqSymbol \lastRep{b} \linRepsEqSymbol \lastRep{c}$
\end{tabular}
\end{subfigwrap}

\begin{subfigwrap}{Key Axioms}{fig::sub::appendix::opor::main-axioms}
\centering
 \begin{tabular}{l}
   \axiomOLabel{opor::descriptors-do-not-interfere} Disjointness \\
  \qquad $b, c \in \descriptorSpans p \implies ({\precedesSpans b c} \vee {\precedesSpans c b} \vee {b = c})$ \\
    
  \axiomOLabel{opor::all-descriptors-are-written-before-any-resolution} Bunching \\
   \qquad $b, c \in \hspans{}{x} \implies \linRepsEq {\firstRep b} {\lastRep c}$ \\ 
 \end{tabular}
 \begin{tabular}{l}
  \axiomOLabel{opor::opportunistic-access} Opportunism \\
  \qquad $\oportunisticPred b x \implies \forall c \in \hspans{}{x}.\ \LastBlockRel b c$ \\  
  \axiomOLabel{opor::finished-operations-have-a-run} Adequacy \\
  \qquad $x \in \terminatedEvent \implies \outputRunFunc x = \outputProp x \wedge \hspans{}{x} \neq \emptyset$ \\
 \end{tabular}
\end{subfigwrap}

\begin{subfigwrap}{Structural Axioms}{fig::sub::appendix::opor::structural-axioms}  
\centering
 \begin{tabular}{c l}
  \axiomOLabel{opor::descriptor-write-precedes-resolution} & $\linRepsEq {\firstRep b} {\lastRep b}$ \\
    
  \axiomOLabel{opor::descriptor-opor-spans-disjoint} & $\descriptorSpans p \cap \opportunisticSpans p = \emptyset$ \\
  
  \axiomOLabel{opor::spans-access-at-most-one-pointer} & $\spans p \cap \spans q \neq \emptyset \implies p = q$ \\
  
  \axiomOLabel{opor::containment-and-uniqueness-of-alloc-blocks} & $b, c \in \allocsSpans p \implies b = c$ \\
     
  \axiomOLabel{opor::every-block-must-have-an-allocated-pointer} & $b \in \spans p \implies \exists c \in \allocsSpans p.\ \precedesSpansEq c b$ \\
    
  \axiomOLabel{opor::runs-are-injective} & $\hspans{}{x} \cap \hspans{}{y} \neq \emptyset \implies x = y$ \\
  
  \axiomOLabel{opor::writer-blocks-belong-to-runs} & $b \in \writesSpans p \implies \exists x.\ b \in \hspans{}{x}$ \\
  
  \axiomOLabel{opor::postcondition-predicate-holds} & $\hspans{}{x} \neq \emptyset \implies \outputRunFunc x \neq \bot \wedge \postPred x {\outputRunFunc x}$ \\
  
      \axiomOLabel{opor::different-span-kinds-do-not-share-reps} & $\neg \exists b \in \descriptorSpans {p}, c \in \opportunisticSpans{p}.\ \firstRep b = \firstRep c$ \\

    \axiomOLabel{opor::runs-have-at-most-one-type-of-span} & $(\hspans{}{x} \subseteq \bigcup_p \descriptorSpans p) \vee 
  (\hspans{}{x} \subseteq \bigcup_p \opportunisticSpans p)$ \\
  
   \axiomOLabel{opor::opportunism-is-unique} & $(b \in \ospans{p}{x} \wedge c \in \ospans{q}{x}) \implies b = c$ \\
   
  \axiomOLabel{opor::writers-have-writer-blocks} & $\hspans{}{x} \neq \emptyset \implies (x \in \writesAbs p \Leftrightarrow \hspans{}{x} \cap \writesSpans p \neq \emptyset)$ \\

  \axiomOLabel{opor::allocs-have-alloc-blocks} & $\hspans{}{x} \neq \emptyset \implies (x \in \allocsAbs p \Leftrightarrow \hspans{}{x} \cap \allocsSpans p \neq \emptyset)$ \\
  
  \axiomOLabel{opor::blocks-contained-in-abstract-time-interval} & (i) $b \in \hspans{}{x} \implies \STimeProp x \natorderEqSymbol \STimeProp {\firstRep b}$ \\
    
                                                                     & (ii) $(b \in \hspans{}{x} \wedge x \in \terminatedEvent) \implies {}$ \\ 
                                                                     & \qquad $\exists i \in \repsCompl.\ \ETimeProp i \natorderEqSymbol \ETimeProp x \wedge \linRepsEq {\lastRep b} i$ \\
  \end{tabular}
\end{subfigwrap}
  \caption{Opportunism axioms for $(\descriptorSpans p, \opportunisticSpans p, \writesSpans p, \allocsSpans p, \runFuncSymbol)$ implementing $\genStructName{}(\visObsSymbol {\pointerIndx p}, \visSepSymbol{\pointerIndx p}) = (\writesAbs p, \allocsAbs p, \postPredSymbol)$. Variables $x$, $y$ range over $\absEvent$. Variables $b$, $c$ over $\bigcup_p \spans p$. Variables $p$, $q$ over $\PtsType$.}
  \label{fig::sub::appendix::opor::opor-axioms}
\end{figure}

We now explain the axioms. Since the opportunism axioms in Figure \ref{fig::sub::appendix::opor::opor-axioms}
extend the span axioms of Figure \ref{fig:descriptor-lifespans-axioms}, we will explain only the new axioms.

\emph{Axiom \axiomORef{opor::opportunistic-access}} makes use of the opportunistic access predicate
$\oportunisticPred b x$, which states that $b$ is an o-span that read the descriptor status of some descriptor 
written by event $x$, where the status read by $b$ happens after the descriptor was written but before the descriptor is 
successfully resolved,
i.e., $\firstRep{c} \linRepsEqSymbol \lastRep{b} \linRepsEqSymbol \lastRep{c}$. 
If $\oportunisticPred b x$ happens, we say that $b$ opportunistically accesses $x$.
Hence, the axiom states that if $b$ opportunistically accesses $x$, then the status read must
have happened before \emph{any} descriptor resolution carried out by $x$. For MCAS this holds, because 
all spans in an $\mcasAlg$ event simultaneously resolve the descriptor at line \ref{resolve-status-MCAS} in
Figure \ref{alg-MCAS},
which means that any opportunistic access before line \ref{resolve-status-MCAS} will happen 
before the resolution of all the spans in the $\mcasAlg$ event.

\emph{Axiom \axiomORef{opor::descriptor-opor-spans-disjoint}} states that there a no $p$-accessing spans that
are both a d-span and an o-span, i.e., d-spans and o-spans are disjoint. 
\emph{Axiom \axiomORef{opor::spans-access-at-most-one-pointer}} states that spans access a unique pointer. 
\emph{Axiom \axiomORef{opor::different-span-kinds-do-not-share-reps}} states that there is no $p$-accessing 
d-spans and o-spans that share their first rep event. 
\emph{Axiom \axiomORef{opor::runs-have-at-most-one-type-of-span}} states that denotations contain only one kind of span,
i.e., you either help or are opportunistic in the entire execution.
\emph{Axiom \axiomORef{opor::opportunism-is-unique}} states that opportunistic accesses are unique inside a denotation.

We now state the visibility relations we will use. These are just the relations discussed while 
explaining Figure \ref{fig::sub::appendix::opor::opport-spans-example}.

\begin{defn}[Visibility Relations for Opportunism]
\label{defn::appendix::valid::visibility-relations-opor}
\begin{sloppypar}
Given opportunistic structure
$(\descriptorSpans p, \opportunisticSpans p, \writesSpans p, \allocsSpans p,
\runFuncSymbol)$,
we define the pointer-indexed separable-before $\visSepSymbol{\pointerIndx p}$ and 
observation $\visObsSymbol{\pointerIndx p}$ relations for implementations having opportunism,
\end{sloppypar}
\begin{align*}
\visSepSymbol {\pointerIndx p} & \defini {\visSepIndxSymbol {\pointerIndx p} {\opporIndx}} \cup {\visSepIndxSymbol {\pointerIndx p} {\descIndx}} \\
\visObsSymbol {\pointerIndx p} & \defini {\visObsIndxSymbol {\pointerIndx p} {\opporIndx}} \cup {\visObsIndxSymbol {\pointerIndx p} {\descIndx}}
\end{align*}
where,
\begin{align*}
\visSepIndx{\pointerIndx p}{\descIndx} x y & \defini \exists b \in \dspans{p}{x}, c \in \dspans{p}{y}.\ \precedesSpans b c \\
\visObsIndx{\pointerIndx p}{\descIndx} x y & \defini \exists b \in \dspans{p}{x}, c \in \dspans{p}{y}.\  b = \max_{\precedesSpansEqSymbol} \{ d \in \writesSpans p \mid \precedesSpans d c \} \\
\visSepIndx {\pointerIndx p} {\opporIndx} x y & \defini \exists b \in \ospans{p} x,
c \in \hspans{}{y} \cap \writesSpans p.\ (\linReps {\firstRep b} {\firstRep c} \vee \LastBlockRel {b} {c}) \\
\visObsIndx {\pointerIndx p} {\opporIndx} x y & \defini \exists b \in \dspans{p}{x}, c \in 
\ospans{p}{y}.\  b = \max_{\precedesSpansEqSymbol} \{ d \in \writesSpans p \mid \linReps {\firstRep d} {\firstRep c} \wedge \LastBlockRel d c \}
\end{align*}
\end{defn}

\begin{defn}[Visibility relations shorthands]
We also introduce the following shorthands. These are technical definitions that some lemmas will use,
\begin{align*}
\visSepIndxSymbol {} {\opporIndx} & \defini \bigcup_p \visSepIndxSymbol {\pointerIndx p} {\opporIndx} \\
\visObsIndxSymbol {} {\opporIndx} & \defini \bigcup_p \visObsIndxSymbol {\pointerIndx p} {\opporIndx} \\
\visSepIndxSymbol {} {\descIndx} & \defini \bigcup_p \visSepIndxSymbol {\pointerIndx p} {\descIndx} \\
\visObsIndxSymbol {} {\descIndx} & \defini \bigcup_p \visObsIndxSymbol {\pointerIndx p} {\descIndx} \\
\end{align*}
\end{defn}

\subsection{Proof of Validity}
\label{sub::sect::appendix::opor::proof-of-validity}

\begin{defn}[Section Hypotheses]
\label{defn::appendix::opor::hypotheses-for-validity}
To shorten the statements of propositions from \ref{lma::appendix::opor::strict-partial-order-lifespans} 
to \ref{thm::appendix::opor::opor-axioms-imply-visibility}, the
following hypotheses will apply,
\begin{itemize}
\item $\absEvent$ is an arbitrary set of events.
\item $\repEvent$ is an arbitrary set of rep events, linearized under a given $\linRepsEqSymbol$.
\item $Z \defini (\descriptorSpans p, \opportunisticSpans p, \writesSpans p, \allocsSpans p, \runFuncSymbol)$ 
is an arbitrary opportunistic structure.
\item $\MCASFam(\visObsSymbol{\pointerIndx p}, \visSepSymbol{\pointerIndx p})$
is implemented by $Z$, where relations $\visObsSymbol{\pointerIndx p}$, $\visSepSymbol{\pointerIndx p}$
are those in Definition \ref{defn::appendix::valid::visibility-relations-opor} and instantiated with
$Z$.
\end{itemize}
\end{defn}

The first lemmas describe basic results whose statements are 
self-explanatory.

\begin{lem}
\label{lma::appendix::opor::strict-partial-order-lifespans}
Suppose hypotheses \ref{defn::appendix::opor::hypotheses-for-validity}.
Relation $\precedesSpansSymbol$ defines a strict partial order on the set
$\bigcup_p \spans p$. 
\end{lem}

\begin{prf}
Each required property follows,
\begin{itemize}
\item Irreflexivity. Suppose for a contradiction that 
$\precedesSpans a a$ for some $a \in \bigcup_p \spans p$.
Hence, $\lastRep a \linRepsSymbol \firstRep a$ by
definition. 
But by Axiom
\axiomORef{opor::descriptor-write-precedes-resolution}, 
$\linRepsEq {\firstRep a} {\lastRep a}$.
Therefore, $\linReps {\lastRep a} {\lastRep a}$
(Contradiction).

\item Transitivity. Let $\precedesSpans a b$ and 
$\precedesSpans b c$.

We have by definition that 
$\lastRep a \linRepsSymbol \firstRep b$ and 
$\lastRep b \linRepsSymbol \firstRep c$. 
But by Axiom
\axiomORef{opor::descriptor-write-precedes-resolution}, 
$\linRepsEq {\firstRep b} {\lastRep b}$.
Therefore, $\linReps {\lastRep a} {\firstRep c}$,
which means $\precedesSpans a c$.
\end{itemize}
\end{prf}

\begin{lem}
\label{lma::appendix::opor::partial-order-lifespans}
Suppose hypotheses \ref{defn::appendix::opor::hypotheses-for-validity}.
Relation $\precedesSpansEqSymbol$ defines a partial order on the set
$\bigcup_p \spans p$. In addition, for every $q$, 
relation $\precedesSpansEqSymbol$ defines a linear order on the set
$\descriptorSpans q$.
\end{lem}

\begin{prf}
By Lemma 
\ref{lma::appendix::opor::strict-partial-order-lifespans},
$\precedesSpansSymbol$ defines a strict partial order. It is a 
standard result that the reflexive closure of a strict partial
order defines a partial order (i.e. a reflexive, transitive,
and antisymmetric relation).

Since for every $q$, $\descriptorSpans q \subseteq \bigcup_p \spans p$
holds, relation $\precedesSpansEqSymbol$ is also a partial order 
on $\descriptorSpans q$ for every $q$. 
That $\precedesSpansEqSymbol$ is a linear order
for $\descriptorSpans q$ is simply Axiom
\axiomORef{opor::descriptors-do-not-interfere}.
\end{prf}

Therefore, Lemma \ref{lma::appendix::opor::partial-order-lifespans}
implies that it makes sense to take a maximum under $\precedesSpansEqSymbol$
in Definition \ref{defn::appendix::valid::visibility-relations-opor}.

The next lemma states that denotations can have at most one $p$-accessing d-span.

\begin{lem}
\label{lma::appendix::opor::dspans-unique-in-runs}
Suppose hypotheses \ref{defn::appendix::opor::hypotheses-for-validity}.
If $b,c \in \dspans{p}{x}$, then $b = c$.
\end{lem}

\begin{prf}
By Axiom \axiomORef{opor::descriptors-do-not-interfere}, $\precedesSpans b c$ or $\precedesSpans c b$ or
$b = c$.

If $\precedesSpans b c$, since they are in the same denotation, by Axiom \axiomORef{opor::all-descriptors-are-written-before-any-resolution},
we get the contradiction,
\[
\lastRep b \linRepsSymbol \firstRep c \linRepsEqSymbol \lastRep b
\]

Similarly, if $\precedesSpans c b$, by Axiom \axiomORef{opor::all-descriptors-are-written-before-any-resolution},
we get the contradiction,
\[
\lastRep c \linRepsSymbol \firstRep b \linRepsEqSymbol \lastRep c
\]

Hence, $b = c$.
\end{prf}

The next lemma states that there is no span that is both a descriptor span and 
a opportunistic span at the same time. It also states that denotations have only one kind
of span.

\begin{lem}
\label{lma::appendix::opor::descriptor-oportunistic-spans-disjoint}
Suppose hypotheses \ref{defn::appendix::opor::hypotheses-for-validity}. We have the following facts,
\begin{enumerate}
\item $\descriptorSpans{} \cap \opportunisticSpans{} = \emptyset$ holds.
\item For any event $x$, it is not the case that both $\dspans{}{x} \neq \emptyset$ and $\ospans{}{x} \neq \emptyset$
hold simultaneously.
\item If $\ospans{}{x} \neq \emptyset$, then $\hspans{}{x} \subseteq \opportunisticSpans{}$.
\item If $\dspans{}{x} \neq \emptyset$, then $\hspans{}{x} \subseteq \descriptorSpans{}$.
\end{enumerate}
\end{lem}

\begin{prf}
We prove each item as follows,
\begin{enumerate}
\item Suppose for a contradiction that $b \in \descriptorSpans{} \cap \opportunisticSpans{}$.
By definition, $b \in \descriptorSpans{p} \cap \opportunisticSpans{q}$, for some
$p$ and $q$. But by Axiom \axiomORef{opor::spans-access-at-most-one-pointer}, $p = q$ holds, 
which contradicts Axiom \axiomORef{opor::descriptor-opor-spans-disjoint}.

\item Suppose for a contradiction that $b \in \dspans{}{x}$ and $c \in \ospans{}{x}$.
By Axiom \axiomORef{opor::runs-have-at-most-one-type-of-span}, we have the cases,
\begin{itemize}
\item Case $\hspans{}{x} \subseteq \descriptorSpans{}$. Therefore, $c \in \descriptorSpans{}$ also.
But $c \in \descriptorSpans{} \cap \opportunisticSpans{}$ contradicts part (1).

\item Case $\hspans{}{x} \subseteq \opportunisticSpans{}$. Therefore, $b \in \opportunisticSpans{}$ also.
But $b \in \descriptorSpans{} \cap \opportunisticSpans{}$ contradicts part (1).
\end{itemize}

\item Let $b \in \ospans{}{x}$ and $c \in \hspans{}{x}$. We know $c \in \descriptorSpans{} \cup \opportunisticSpans{}$.
If $c \in \descriptorSpans{}$, then we have simultaneously $\dspans{}{x} \neq \emptyset$ and $\ospans{}{x} \neq \emptyset$
which contradicts part (2).

\item Similar to the previous item.
\end{enumerate}
\end{prf}

For the next two lemmas, we will use the following relation,

\begin{defn}
Define,
\[
{\leftslice} \defini {\visSepIndxSymbol {} {\descIndx}} \cup {\visObsIndxSymbol {} {\descIndx}} \cup {({\visObsIndxSymbol {} {\opporIndx}} \circ {\visSepIndxSymbol {} {\opporIndx}})}
\]
where ${\visObsIndxSymbol {} {\opporIndx}} \circ {\visSepIndxSymbol {} {\opporIndx}}$
is the relation composition of $\visObsIndxSymbol {} {\opporIndx}$ and $\visSepIndxSymbol {} {\opporIndx}$.
\end{defn}

The next lemma states that events related under the transitive closure of $\leftslice$ must contain disjoint spans.

\begin{lem}
\label{lma::appendix::opor::abs-visibility-implies-block-precedence}
Suppose hypotheses \ref{defn::appendix::opor::hypotheses-for-validity}.
If $x \transCl{\leftslice} y$ then $\exists b_x \in 
\hspans{}{x}, b_y \in \hspans{}{y}.\ 
\precedesSpans {b_x} {b_y}$.
\end{lem}

\begin{prf}
Define the following binary relation on $\absEvent$,
\[
P(w,z) \defini \exists b_w \in \hspans{}{w}, b_z \in 
\hspans{}{z}.\ \precedesSpans {b_w} {b_z}
\]

So, we need to show $x \transCl{\leftslice} y \implies P(x,y)$. 
But it suffices to show the following two properties,
\begin{itemize}
\item ${\leftslice} \subseteq P$.
\item $P$ is transitive. 
\end{itemize}
because $\transCl{\leftslice}$ is the smallest transitive 
relation containing $\leftslice$.

Let us show the two required properties.
\begin{itemize}
\item ${\leftslice} \subseteq P$. 

The cases for ${\visSepIndxSymbol {} {\descIndx}}$ and ${\visObsIndxSymbol {} {\descIndx}}$ are immediate from
their definition. Let us focus on ${\visObsIndxSymbol {} {\opporIndx}} \circ {\visSepIndxSymbol {} {\opporIndx}}$.

So, let $x \mathrel{({\visObsIndxSymbol {} {\opporIndx}} \circ {\visSepIndxSymbol {} {\opporIndx}})} y$. Hence,
$\visObsIndx {\pointerIndx p} {\opporIndx} x z$ and $\visSepIndx {\pointerIndx q} {\opporIndx} z y$
for some $p$, $q$, and $z$.

By definition, there are $b \in \dspans{p}{x}$, $c \in \ospans{p}{z}$, $d \in \ospans{q}{z}$, $e \in \dspans{q}{y}$ such that 
$\linReps {\lastRep b} {\lastRep c}$ and $\linReps {\firstRep b} {\firstRep c}$ and ($\linReps {\lastRep d} {\lastRep e}$ or $\linReps {\firstRep d} {\firstRep e}$).

By Axiom \axiomORef{opor::opportunism-is-unique}, $c = d$. Hence, $c \in \spans p \cap \spans q$, which implies $p = q$ by 
Axiom \axiomORef{opor::spans-access-at-most-one-pointer}.

Since $b \in \dspans{p}{x}$ and $e \in \dspans{p}{y}$, by Axiom \axiomORef{opor::descriptors-do-not-interfere} there are
three cases,
\begin{itemize}
\item Case $\precedesSpans b e$. This is the required conclusion.
\item Case $\precedesSpans e b$. If $\linReps {\lastRep d} {\lastRep e}$, then together with Axiom \axiomORef{opor::descriptor-write-precedes-resolution}, we get the following contradiction,
\[
\lastRep e \linRepsSymbol \firstRep b \linRepsEqSymbol \lastRep b \linRepsSymbol 
\lastRep c = \lastRep d \linRepsSymbol \lastRep e
\]
If $\linReps {\firstRep d} {\firstRep e}$, then together with Axiom \axiomORef{opor::descriptor-write-precedes-resolution}, we get the following contradiction,
\[
\firstRep d \linRepsSymbol \firstRep e \linRepsEqSymbol \lastRep e \linRepsSymbol \firstRep b \linRepsSymbol \firstRep c = \firstRep d
\]
\item Case $e = b$. If $\linReps {\lastRep d} {\lastRep e}$, we get the following contradiction,
\[
\lastRep e = \lastRep b \linRepsSymbol 
\lastRep c = \lastRep d \linRepsSymbol \lastRep e
\]
If $\linReps {\firstRep d} {\firstRep e}$, we get the following contradiction,
\[
\firstRep e = \firstRep b \linRepsSymbol 
\firstRep c = \firstRep d \linRepsSymbol \firstRep e
\]
\end{itemize}

\item $P$ is transitive. By hypotheses $P(u,v)$ and $P(v,z)$, 
we have,
\begin{align}
\begin{split}
\exists b_u \in \hspans{}{u}, b_v \in \hspans{}{v}.\ 
\precedesSpans {b_u} {b_v} \\
\exists b_v' \in \hspans{}{v}, b_z \in \hspans{}{z}.\ 
\precedesSpans {b_v'} {b_z}
\end{split}
\end{align}

We need to show $P(u,z)$. By the bunching axiom \axiomORef{opor::all-descriptors-are-written-before-any-resolution}
on denotation $\hspans{}{v}$,
$\linRepsEq {\firstRep {b_v}} {\lastRep {b_v'}}$ must hold. Hence,
\[
\lastRep {b_u} \linRepsSymbol \firstRep {b_v} \linRepsEqSymbol \lastRep {b_v'} \linRepsSymbol \firstRep {b_z}
\]
which means $\precedesSpans {b_u} {b_z}$.
\end{itemize}
\end{prf}

Using $\leftslice$, we can now extract some information regarding the transitive closure of $\genVisSymbol$.

\begin{lem}
\label{lma::appendix::opor::vis-trans-implies-two-possibilities}
Suppose hypotheses \ref{defn::appendix::opor::hypotheses-for-validity}.
If $\genVisTrans x y$, then $x \neq y$ and $\exists b_x \in \hspans{}{x}, b_y \in \hspans{}{y}.\ \linReps {\firstRep {b_x}} {\lastRep{b_y}}$.
\end{lem}

\begin{prf}
Notice that by definition of $\genVisSymbol$, we can see it as the union of the four relations,
\begin{align}
\label{eq::appendix::opor::gen-vis-four-components}
{\genVisSymbol} = {\visSepIndxSymbol {} {\descIndx}} \cup {\visObsIndxSymbol {} {\descIndx}} \cup {\visSepIndxSymbol {} {\opporIndx}} \cup {\visObsIndxSymbol {} {\opporIndx}}
\end{align}
Hence, from $\genVisTrans x y$, we have a sequence of length $n \natorderEqSymbolRight 1$,
\[
a_0 R_1 a_1 R_2 \ldots R_{n-1} a_{n-1} R_n a_{n}
\]
where each $R_i$ is one of the four relations in \eqref{eq::appendix::opor::gen-vis-four-components},
and $a_0 = x$ and $a_n = y$.

\begin{claim}
Let $1 \natorderEqSymbol i \natorderSymbol n$. If $R_{i+1} = {\visSepIndxSymbol {} {\opporIndx}}$, then $R_i = {\visObsIndxSymbol {} {\opporIndx}}$.
\end{claim}

We now prove the claim. Suppose $\visSepIndx {\pointerIndx p} {\opporIndx} {a_i} {a_{i+1}}$ holds for some $p$.
We do a case analysis on $R_i$ and show that all the cases different from ${\visObsIndxSymbol {} {\opporIndx}}$ lead
to a contradiction.
\begin{itemize}
\item $R_i = {\visSepIndxSymbol {} {\descIndx}}$ or $R_i = {\visObsIndxSymbol {} {\descIndx}}$. Hence, by definition, 
$\dspans{}{a_i} \neq \emptyset$ and $\ospans{}{a_i} \neq \emptyset$, which contradicts Lemma \ref{lma::appendix::opor::descriptor-oportunistic-spans-disjoint}.
\item $R_i = {\visSepIndxSymbol {} {\opporIndx}}$. Again, by definition, 
$\dspans{}{a_i} \neq \emptyset$ and $\ospans{}{a_i} \neq \emptyset$, which contradicts Lemma \ref{lma::appendix::opor::descriptor-oportunistic-spans-disjoint}.
\end{itemize}
This proves the claim.

\begin{claim}
Let $1 \natorderEqSymbol i \natorderSymbol n$. If $R_{i} = {\visObsIndxSymbol {} {\opporIndx}}$, then $R_{i+1} = {\visSepIndxSymbol {} {\opporIndx}}$.
\end{claim}

We now prove the claim. Suppose $\visObsIndx {\pointerIndx p} {\opporIndx} {a_{i-1}} {a_{i}}$ holds for some $p$.
We do a case analysis on $R_{i+1}$ and show that all the cases different from ${\visSepIndxSymbol {} {\opporIndx}}$ lead
to a contradiction.
\begin{itemize}
\item $R_{i+1} = {\visSepIndxSymbol {} {\descIndx}}$ or $R_{i+1} = {\visObsIndxSymbol {} {\descIndx}}$. Hence, by definition, 
$\ospans{}{a_i} \neq \emptyset$ and $\dspans{}{a_i} \neq \emptyset$, which contradicts Lemma \ref{lma::appendix::opor::descriptor-oportunistic-spans-disjoint}.
\item $R_{i+1} = {\visObsIndxSymbol {} {\opporIndx}}$. Again, by definition, 
$\ospans{}{a_i} \neq \emptyset$ and $\dspans{}{a_i} \neq \emptyset$, which contradicts Lemma \ref{lma::appendix::opor::descriptor-oportunistic-spans-disjoint}.
\end{itemize}
This proves the claim.

By the above claims, relations ${\visObsIndxSymbol {} {\opporIndx}}$ and ${\visSepIndxSymbol {} {\opporIndx}}$ always 
occur contiguously in the sequence, first ${\visObsIndxSymbol {} {\opporIndx}}$ followed by ${\visSepIndxSymbol {} {\opporIndx}}$. 
The only places where this could be violated is at the extremes of the sequence, i.e., at $R_1$ and $R_n$, because at $R_1$, a ${\visSepIndxSymbol {} {\opporIndx}}$ could appear without a preceding ${\visObsIndxSymbol {} {\opporIndx}}$, and at $R_n$, a ${\visObsIndxSymbol {} {\opporIndx}}$ could appear without a succeeding ${\visSepIndxSymbol {} {\opporIndx}}$.

First, we case analyze when $n = 1$.
\begin{itemize}
\item Case $\visObsIndx {\pointerIndx p} {\descIndx} {a_0} {a_1}$ or 
$\visSepIndx {\pointerIndx p} {\descIndx} {a_0} {a_1}$ (for some $p$). 
By definition, there are $b \in \hspans{p}{a_0}$ and $c \in \hspans{p}{a_1}$ such that $\precedesSpans b c$.

Therefore, by Axiom \axiomORef{opor::descriptor-write-precedes-resolution},
\[
\firstRep b \linRepsEqSymbol \lastRep b \linRepsSymbol \firstRep c \linRepsEqSymbol \lastRep c
\]

Suppose for a contradiction that $a_0 = a_1$. Then, by Axiom \axiomORef{opor::all-descriptors-are-written-before-any-resolution}, 
we get the contradiction,
\[
\lastRep b \linRepsSymbol \firstRep c \linRepsEqSymbol \lastRep b
\]
Hence, $a_0 \neq a_1$.

\item Case $\visObsIndx {\pointerIndx p} {\opporIndx} {a_0} {a_1}$ (for some $p$).
By definition, there are $b \in \dspans{p}{a_0}$ and $c \in \ospans{p}{a_1}$ such that $\linReps {\firstRep b} {\firstRep c}$ 
and $\linReps {\lastRep b} {\lastRep c}$.
 
Therefore, by Axiom \axiomORef{opor::descriptor-write-precedes-resolution},
\[
\firstRep b \linRepsSymbol \firstRep c \linRepsEqSymbol \lastRep c
\]

Suppose for a contradiction that $a_0 = a_1$. But then, 
$\ospans{}{a_0} \neq \emptyset$ and $\dspans{}{a_0} \neq \emptyset$, which 
contradicts Lemma \ref{lma::appendix::opor::descriptor-oportunistic-spans-disjoint}.

Hence, $a_0 \neq a_1$.

\item Case $\visSepIndx {\pointerIndx p} {\opporIndx} {a_0} {a_1}$ (for some $p$).
By definition, there are $b \in \ospans{p}{a_0}$ and $c \in \dspans{p}{a_1}$ such that ($\linReps {\firstRep b} {\firstRep c}$ 
or $\linReps {\lastRep b} {\lastRep c}$).

If $\linReps {\firstRep b} {\firstRep c}$, then by Axiom \axiomORef{opor::descriptor-write-precedes-resolution},
\[
\firstRep b \linRepsSymbol \firstRep c \linRepsEqSymbol \lastRep c
\]

Similarly, if $\linReps {\lastRep b} {\lastRep c}$, then by Axiom \axiomORef{opor::descriptor-write-precedes-resolution},
\[
\firstRep b \linRepsEqSymbol \lastRep b \linRepsSymbol \lastRep c
\]

Suppose for a contradiction that $a_0 = a_1$. But then, 
$\ospans{}{a_0} \neq \emptyset$ and $\dspans{}{a_0} \neq \emptyset$, which 
contradicts Lemma \ref{lma::appendix::opor::descriptor-oportunistic-spans-disjoint}.

Hence, $a_0 \neq a_1$.
\end{itemize}

Now, let us focus when $n \natorderEqSymbolRight 2$. We consider 4 cases on the extremes of the sequence.

\begin{itemize}
\item Case $R_1 \neq {\visSepIndxSymbol {} {\opporIndx}}$ and $R_n \neq {\visObsIndxSymbol {} {\opporIndx}}$.
Then $a_0 \transCl{\leftslice} a_n$, since every occurrence of ${\visObsIndxSymbol {} {\opporIndx}}$ is followed by ${\visSepIndxSymbol {} {\opporIndx}}$. Therefore, by Lemma \ref{lma::appendix::opor::abs-visibility-implies-block-precedence}, there are $b \in \hspans{}{a_0}$ and $c \in \hspans{}{a_{n}}$ such that $\precedesSpans b c$.

Hence, by Axiom \axiomORef{opor::descriptor-write-precedes-resolution},
\[
\firstRep b \linRepsEqSymbol \lastRep b \linRepsSymbol \firstRep c \linRepsEqSymbol \lastRep c
\]

Suppose for a contradiction that $a_0 = a_n$. Then, by Axiom \axiomORef{opor::all-descriptors-are-written-before-any-resolution}, 
we get the contradiction,
\[
\lastRep b \linRepsSymbol \firstRep c \linRepsEqSymbol \lastRep b
\]
Hence, $a_0 \neq a_n$.

\item Case $R_1 \neq {\visSepIndxSymbol {} {\opporIndx}}$ and $R_n = {\visObsIndxSymbol {} {\opporIndx}}$. 
Then $a_0 \transCl{\leftslice} a_{n-1}$ because $R_{n-1}$ cannot be ${\visObsIndxSymbol {} {\opporIndx}}$ (otherwise 
$R_{n}$ would be ${\visSepIndxSymbol {} {\opporIndx}}$).

By Lemma \ref{lma::appendix::opor::abs-visibility-implies-block-precedence}, there are $b \in \hspans{}{a_0}$ and $c \in \hspans{}{a_{n-1}}$
such that $\precedesSpans b c$. From $\visObsIndx{\pointerIndx p}{\opporIndx} {a_{n-1}} {a_n}$ (for some $p$), we have
$d \in \dspans{p}{a_{n-1}}$ and $e \in \ospans{p}{a_n}$ such that $\firstRep d \linRepsSymbol \firstRep e$ and $\LastBlockRel d e$. 
Together with Axioms \axiomORef{opor::descriptor-write-precedes-resolution} and \axiomORef{opor::all-descriptors-are-written-before-any-resolution},
\[
\firstRep b \linRepsEqSymbol \lastRep b \linRepsSymbol \firstRep c \linRepsEqSymbol \lastRep d \linRepsSymbol \lastRep e
\]

Suppose for a contradiction that $a_0 = a_n$. Then, by Lemma \ref{lma::appendix::opor::descriptor-oportunistic-spans-disjoint}, 
$b \in \ospans{}{a_n}$, which means $b = e$ by Axiom \axiomORef{opor::opportunism-is-unique}. 
But then, by Axiom \axiomORef{opor::all-descriptors-are-written-before-any-resolution}, 
we get the contradiction,
\[
\lastRep d \linRepsSymbol \lastRep e = \lastRep b \linRepsSymbol \firstRep c \linRepsEqSymbol \lastRep d
\]
Hence, $a_0 \neq a_n$.

\item Case $R_1 = {\visSepIndxSymbol {} {\opporIndx}}$ and $R_n \neq {\visObsIndxSymbol {} {\opporIndx}}$. 
Then $a_1 \transCl{\leftslice} a_n$ because $R_2$ cannot be ${\visSepIndxSymbol {} {\opporIndx}}$ (otherwise 
$R_{1}$ would be ${\visObsIndxSymbol {} {\opporIndx}}$).

By Lemma \ref{lma::appendix::opor::abs-visibility-implies-block-precedence}, there are $b \in \hspans{}{a_1}$ and $c \in \hspans{}{a_{n}}$
such that $\precedesSpans b c$. From $\visSepIndx{\pointerIndx p}{\opporIndx} {a_{0}} {a_1}$ (for some $p$), we have
$d \in \ospans{p}{a_{0}}$ and $e \in \hspans{}{a_1} \cap \writesSpans p$ such that ($\LastBlockRel d e$ or $\linReps {\firstRep d}{\firstRep e})$.

\begin{itemize}
\item Case $\linReps {\firstRep d}{\firstRep e}$. Then we get, together with Axioms 
\axiomORef{opor::descriptor-write-precedes-resolution} and \axiomORef{opor::all-descriptors-are-written-before-any-resolution},
\[
\firstRep d \linRepsSymbol \firstRep e \linRepsEqSymbol \lastRep b \linRepsSymbol \firstRep c \linRepsEqSymbol \lastRep c
\]

Suppose for a contradiction that $a_0 = a_n$. Then, by Lemma \ref{lma::appendix::opor::descriptor-oportunistic-spans-disjoint}, 
$c \in \ospans{}{a_0}$, which means $c = d$ by Axiom \axiomORef{opor::opportunism-is-unique}. 
But then, by Axiom \axiomORef{opor::all-descriptors-are-written-before-any-resolution}, 
we get the contradiction,
\[
\firstRep d \linRepsSymbol \firstRep e \linRepsEqSymbol \lastRep b \linRepsSymbol \firstRep c = \firstRep d
\]
Hence, $a_0 \neq a_n$.

\item Case $\LastBlockRel d e$. Since rep events are totally ordered, $\lastRep d \linRepsSymbol \firstRep e$ or $\firstRep e \linRepsEqSymbol \lastRep d$.
\begin{itemize}
\item Case $\lastRep d \linRepsSymbol \firstRep e$. Together with Axioms 
\axiomORef{opor::descriptor-write-precedes-resolution} and \axiomORef{opor::all-descriptors-are-written-before-any-resolution},
\[
\firstRep d \linRepsEqSymbol \lastRep d \linRepsSymbol \firstRep e \linRepsEqSymbol \lastRep b \linRepsSymbol \firstRep c \linRepsEqSymbol \lastRep c
\]

Suppose for a contradiction that $a_0 = a_n$. Then, by Axiom \axiomORef{opor::all-descriptors-are-written-before-any-resolution},
we get the contradiction,
\[
\lastRep d \linRepsSymbol \firstRep e \linRepsEqSymbol \lastRep b \linRepsSymbol \firstRep c \linRepsEqSymbol \lastRep d
\]
Hence, $a_0 \neq a_n$.

\item Case $\firstRep e \linRepsEqSymbol \lastRep d$. Hence, $\firstRep e \linRepsEqSymbol \lastRep d \linRepsSymbol \lastRep e$,
which means $\oportunisticPred d {a_1}$ must be true. So, by Axiom \axiomORef{opor::opportunistic-access}, $\LastBlockRel d b$ holds. Which means,
by Axiom \axiomORef{opor::descriptor-write-precedes-resolution},
\[
\firstRep d \linRepsEqSymbol \lastRep d \linRepsSymbol \lastRep b \linRepsSymbol \firstRep c \linRepsEqSymbol \lastRep c
\]

Suppose for a contradiction that $a_0 = a_n$. Then, by Axiom \axiomORef{opor::all-descriptors-are-written-before-any-resolution},
we get the contradiction,
\[
\lastRep d \linRepsSymbol \lastRep b \linRepsSymbol \firstRep c \linRepsEqSymbol \lastRep d
\]
Hence, $a_0 \neq a_n$.

\end{itemize}
\end{itemize}

\item Case $R_1 = {\visSepIndxSymbol {} {\opporIndx}}$ and $R_n = {\visObsIndxSymbol {} {\opporIndx}}$.
Since this time we are removing $R_1$ and $R_n$ from the sequence, we need to take into account the special case when the sequence from
$R_2$ to $R_{n-1}$ is empty (i.e., $n = 2$).

\begin{itemize}
\item Case: The sequence from $R_2$ to $R_{n-1}$ is empty (equivalently, $n = 2$). 
From $\visSepIndx{\pointerIndx p}{\opporIndx} {a_{0}} {a_1}$
and $\visObsIndx{\pointerIndx q}{\opporIndx} {a_{1}} {a_2}$ (for some $p$ and $q$), there are $d \in \ospans{p}{a_{0}}$ and $e \in \hspans{}{a_1} \cap \writesSpans p$ such that ($\LastBlockRel d e$ or $\linReps {\firstRep d}{\firstRep e}$); and $b \in \hspans{}{a_1} \cap \writesSpans q$ and $c \in \ospans{q}{a_{2}}$
such that $\LastBlockRel b c$ and $\linReps {\firstRep b} {\firstRep c}$.

Since rep events are totally ordered, $\lastRep d \linRepsSymbol \firstRep e$ or $\firstRep e \linRepsEqSymbol \lastRep d$.
\begin{itemize}
\item Case $\lastRep d \linRepsSymbol \firstRep e$. Together with Axioms 
\axiomORef{opor::descriptor-write-precedes-resolution} and \axiomORef{opor::all-descriptors-are-written-before-any-resolution},
\[
\firstRep d \linRepsEqSymbol \lastRep d \linRepsSymbol \firstRep e \linRepsEqSymbol \lastRep b \linRepsSymbol \lastRep c
\]

Suppose for a contradiction that $a_0 = a_2$. Then, by Axiom \axiomORef{opor::opportunism-is-unique}, $c = d$. Hence, by Axiom \axiomORef{opor::all-descriptors-are-written-before-any-resolution},
we get the contradiction,
\[
\lastRep d \linRepsSymbol \firstRep e \linRepsEqSymbol \lastRep b \linRepsSymbol \lastRep c = \lastRep d
\]
Hence, $a_0 \neq a_2$.

\item Case $\firstRep e \linRepsEqSymbol \lastRep d$. We have two subcases,
\begin{itemize}
\item Case $\LastBlockRel d e$. Hence, $\firstRep e \linRepsEqSymbol \lastRep d \linRepsSymbol \lastRep e$,
which means $\oportunisticPred d {a_1}$ must be true. So, by Axiom \axiomORef{opor::opportunistic-access}, $\LastBlockRel d b$ holds. Which means,
by Axiom \axiomORef{opor::descriptor-write-precedes-resolution},
\[
\firstRep d \linRepsEqSymbol \lastRep d \linRepsSymbol \lastRep b \linRepsSymbol \lastRep c
\]

Suppose for a contradiction that $a_0 = a_2$. Then, by Axiom \axiomORef{opor::opportunism-is-unique}, $c = d$. 
We get the contradiction,
\[
\lastRep d \linRepsSymbol \lastRep b \linRepsSymbol \lastRep c = \lastRep d
\]
Hence, $a_0 \neq a_2$.

\item Case $\linReps {\firstRep d}{\firstRep e}$. Hence, together with Axiom \axiomORef{opor::all-descriptors-are-written-before-any-resolution},
\[
\firstRep d \linRepsSymbol \firstRep e \linRepsEqSymbol \lastRep b \linRepsSymbol \lastRep c
\]
Suppose for a contradiction that $a_0 = a_2$. Then, by Axiom \axiomORef{opor::opportunism-is-unique}, $c = d$. Hence, $c \in \spans p \cap \spans q$, which implies $p = q$ by Axiom \axiomORef{opor::spans-access-at-most-one-pointer}. But since $b,e \in \hspans{}{a_1} \cap \writesSpans p$, 
by Lemma \ref{lma::appendix::opor::dspans-unique-in-runs}, $b = e$.
Therefore, we get the following contradiction,
\[
\firstRep b \linRepsSymbol \firstRep c = \firstRep d \linRepsSymbol \firstRep e = \firstRep b
\]
Hence, $a_0 \neq a_2$.
\end{itemize}
\end{itemize}

\item Case: The sequence from $R_2$ to $R_{n-1}$ is not empty (equivalently, $n \natorderEqSymbolRight 3$).
Then $a_1 \transCl{\leftslice} a_{n-1}$ because $R_2$ cannot be ${\visSepIndxSymbol {} {\opporIndx}}$ (otherwise 
$R_{1}$ would be ${\visObsIndxSymbol {} {\opporIndx}}$) and $R_{n-1}$ cannot be ${\visObsIndxSymbol {} {\opporIndx}}$ (otherwise 
$R_{n}$ would be ${\visSepIndxSymbol {} {\opporIndx}}$).

By Lemma \ref{lma::appendix::opor::abs-visibility-implies-block-precedence}, there are $b \in \hspans{}{a_1}$ and $c \in \hspans{}{a_{n-1}}$
such that $\precedesSpans b c$. From $\visSepIndx{\pointerIndx p}{\opporIndx} {a_{0}} {a_1}$
and $\visObsIndx{\pointerIndx q}{\opporIndx} {a_{n-1}} {a_n}$ (for some $p$ and $q$), there are $d \in \ospans{p}{a_{0}}$ and $e \in \hspans{}{a_1} \cap \writesSpans p$ such that ($\LastBlockRel d e$ or $\linReps {\firstRep d}{\firstRep e}$); and $g \in \hspans{}{a_{n-1}} \cap \writesSpans q$ and $h \in \ospans{q}{a_{n}}$ such that $\LastBlockRel g h$ and $\linReps {\firstRep g} {\firstRep h}$.

Since rep events are totally ordered, $\lastRep d \linRepsSymbol \firstRep e$ or $\firstRep e \linRepsEqSymbol \lastRep d$.
\begin{itemize}
\item Case $\lastRep d \linRepsSymbol \firstRep e$. Together with Axioms 
\axiomORef{opor::descriptor-write-precedes-resolution} and \axiomORef{opor::all-descriptors-are-written-before-any-resolution},
\[
\firstRep d \linRepsEqSymbol \lastRep d \linRepsSymbol \firstRep e \linRepsEqSymbol \lastRep b \linRepsSymbol \firstRep c \linRepsEqSymbol \lastRep g \linRepsSymbol \lastRep h
\]

Suppose for a contradiction that $a_0 = a_n$. Then, $h = d$ by Axiom \axiomORef{opor::opportunism-is-unique}. Hence, with Axiom 
\axiomORef{opor::all-descriptors-are-written-before-any-resolution},
we get the contradiction,
\[
\lastRep d \linRepsSymbol \firstRep e \linRepsEqSymbol \lastRep b \linRepsSymbol \firstRep c \linRepsEqSymbol \lastRep g \linRepsSymbol \lastRep h = \lastRep d
\]
Hence, $a_0 \neq a_n$.

\item Case $\firstRep e \linRepsEqSymbol \lastRep d$. We have two subcases,
\begin{itemize}
\item Case $\LastBlockRel d e$.
Hence, $\firstRep e \linRepsEqSymbol \lastRep d \linRepsSymbol \lastRep e$,
which means $\oportunisticPred d {a_1}$ must be true. So, by Axiom \axiomORef{opor::opportunistic-access}, $\LastBlockRel d b$ holds. Which means,
by Axioms \axiomORef{opor::descriptor-write-precedes-resolution} and \axiomORef{opor::all-descriptors-are-written-before-any-resolution},
\[
\firstRep d \linRepsEqSymbol \lastRep d \linRepsSymbol \lastRep b \linRepsSymbol \firstRep c \linRepsEqSymbol \lastRep g \linRepsSymbol \lastRep h
\]

Suppose for a contradiction that $a_0 = a_n$. Then, $h = d$ by Axiom \axiomORef{opor::opportunism-is-unique}. Hence, with Axiom 
\axiomORef{opor::all-descriptors-are-written-before-any-resolution},
we get the contradiction,
\[
\lastRep d \linRepsSymbol \lastRep b \linRepsSymbol \firstRep c \linRepsEqSymbol \lastRep g \linRepsSymbol \lastRep h = \lastRep d
\]

Hence, $a_0 \neq a_n$.

\item Case $\linReps {\firstRep d}{\firstRep e}$. Hence, together with Axiom \axiomORef{opor::all-descriptors-are-written-before-any-resolution},
\[
\firstRep d \linRepsSymbol \firstRep e \linRepsEqSymbol \lastRep b \linRepsSymbol \firstRep c \linRepsEqSymbol \lastRep g \linRepsSymbol \lastRep h
\]
Suppose for a contradiction that $a_0 = a_n$. Then, by Axiom \axiomORef{opor::opportunism-is-unique}, $h = d$. Hence, $h \in \spans p \cap \spans q$, which implies $p = q$ by Axiom \axiomORef{opor::spans-access-at-most-one-pointer}. Since $g,e \in \writesSpans p$, we have three further subcases 
by Axiom \axiomORef{opor::descriptors-do-not-interfere}, the three of them leading to a contradiction,

If $\precedesSpans e g$, then by Axiom \axiomORef{opor::descriptor-write-precedes-resolution}, we have the contradiction, 
\[
\firstRep g \linRepsSymbol \firstRep h = \firstRep d \linRepsSymbol \firstRep e \linRepsEqSymbol \lastRep e \linRepsSymbol \firstRep g
\]

If $\precedesSpans g e$, then by Axiom \axiomORef{opor::all-descriptors-are-written-before-any-resolution}, we have the contradiction, 
\[
\firstRep e \linRepsEqSymbol \lastRep b \linRepsSymbol \firstRep c \linRepsEqSymbol \lastRep g \linRepsSymbol \firstRep e
\]

If $g = e$, then we have the contradiction,
\[
\firstRep e = \firstRep g \linRepsSymbol \firstRep h = \firstRep d \linRepsSymbol \firstRep e
\]

Hence, $a_0 \neq a_n$.
\end{itemize}
\end{itemize}
\end{itemize}
\end{itemize}
\end{prf}

A direct consequence of Lemma \ref{lma::appendix::opor::vis-trans-implies-two-possibilities}
is the following, which states that events chosen for linearization
must have carried out some non-empty execution path.

\begin{lem}
\label{lma::appendix::opor::visible-implies-having-run}
Suppose hypotheses \ref{defn::appendix::opor::hypotheses-for-validity}.
If $x \in \closedEvent$ then $\hspans{}{x} \neq \emptyset$.
\end{lem} 

\begin{prf}
By definition of $\closedEvent$, $x \refleTransCl{\genVisSymbol} y$
for some $y \in \terminatedEvent$.
\begin{itemize}
\item Case $x = y$. Hence, $x \in \terminatedEvent$. By Axiom 
\axiomORef{opor::finished-operations-have-a-run},
$\hspans{}{x} \neq \emptyset$ must hold.

\item Case $x \neq y$. Hence, $x \transCl{\genVisSymbol} y$.
By Lemma \ref{lma::appendix::opor::vis-trans-implies-two-possibilities},
$\hspans{}{x} \neq \emptyset$ must hold.
\end{itemize}
\end{prf}

The following lemma states that events ordered in real-time 
have all their spans disjoint from each other, i.e., spans 
do not go beyond the events they belong to.

\begin{lem}
\label{lma::appendix::opor::abs-precedence-implies-block-precedence}
Suppose hypotheses \ref{defn::appendix::opor::hypotheses-for-validity}.
If $\precedesAbs x y$ then $\forall b_x \in \hspans{}{x}, b_y 
\in \hspans{}{y}.\ \precedesSpans {b_x} {b_y}$.
\end{lem}

\begin{prf}
Let $b_x \in \hspans{}{x}, b_y \in \hspans{}{y}$. By Axiom 
\axiomORef{opor::blocks-contained-in-abstract-time-interval},
we have $\STimeProp y \natorderEqSymbol \STimeProp {\firstRep {b_y}}$ and
there is $i$ such that $\linRepsEq {\lastRep {b_x}} i$ and
$\ETimeProp i \natorderEqSymbol \ETimeProp x$.

Hence, by definition of $\precedesAbsSymbol$, we have,
\[
\ETimeProp i \natorderEqSymbol \ETimeProp x 
\natorderSymbol \STimeProp y \natorderEqSymbol 
\STimeProp {\firstRep {b_y}} 
\]

This means that $i$ finishes in real-time before 
$\firstRep {b_y}$ starts. But since $\linRepsEqSymbol$ respects the real time order of rep events
(i.e., $\linRepsEqSymbol$ is a linearization), 
we have $\linReps i {\firstRep {b_y}}$. 

But $\linRepsEq {\lastRep {b_x}} i$, which means
$\linReps {\lastRep {b_x}} {\firstRep {b_y}}$. Therefore,
$\precedesSpans {b_x} {b_y}$.
\end{prf} 

We can now prove each visibility axiom.

\begin{lem}
\label{lma::appendix::opor::no-inbetween-vis-axiom-holds}
Suppose hypotheses \ref{defn::appendix::opor::hypotheses-for-validity}.
Axiom \axiomHRef{help-focused::non-helpers} holds.
\end{lem}

\begin{prf}
Let $\visObs {\pointerIndx {p}} x y$ and $z \in \writesAbs {p} \cap \closedEvent$.
By definition of $\visObs {\pointerIndx {p}} x y$,
we need to consider two cases,
\begin{itemize}
\item Case $\visObsIndx{\pointerIndx p}{\descIndx} x y$.
By definition, there are 
$b_x \in \dspans{p}{x}$ and $b_y \in \dspans{p}{y}$ such that 
$b_x$ is the maximum satisfying
$b_x \in \writesSpans {p}$ and 
$\precedesSpans {b_x} {b_y}$.

From hypothesis $z \in \closedEvent$ and Lemma \ref{lma::appendix::opor::visible-implies-having-run},
$\hspans{}{z} \neq \emptyset$ holds. So, by 
Axiom \axiomORef{opor::writers-have-writer-blocks}, 
there is $b_z \in \hspans{}{z} \cap \writesSpans {p}$. 
 
But $b_y \in \descriptorSpans {p}$ and $b_z \in 
\writesSpans {p} \subseteq \descriptorSpans {p}$, 
which means that either $\precedesSpans {b_y} {b_z}$ or 
$\precedesSpans {b_z} {b_y}$ or $b_y = b_z$ by Axiom 
\axiomORef{opor::descriptors-do-not-interfere}.

\begin{itemize}
\item Case $\precedesSpans {b_y} {b_z}$. By definition,
$\visSepIndx {\pointerIndx p}{\descIndx} y z$ holds
(So, $\visSep {\pointerIndx p} y z$).

\item Case $\precedesSpans {b_z} {b_y}$. 
Again, since $b_x, b_z \in \descriptorSpans {p}$, we have three subcases by Axiom 
\axiomORef{opor::descriptors-do-not-interfere}.

\begin{itemize}
\item Case $\precedesSpans {b_x} {b_z}$.
We have 
$b_x \precedesSpansSymbol b_z \precedesSpansSymbol b_y$, which 
is a contradiction, because $b_x$ was the maximum satisfying
the conditions $b_x \in \writesSpans {p}$ and 
$\precedesSpans {b_x} {b_y}$, but now $b_z$ is a more recent
span satisfying the conditions.

\item Case $\precedesSpans {b_z} {b_x}$. By definition,
$\visSepIndx {\pointerIndx p}{\descIndx} z x$ holds
(So, $\visSep {\pointerIndx p} z x$).

\item Case $b_x = b_z$. We have $\hspans{}{x} \cap \hspans{}{z} 
\neq \emptyset$, which means $x = z$ by Axiom 
\axiomORef{opor::runs-are-injective}.
\end{itemize}

\item Case $b_y = b_z$. We have $\hspans{}{y} \cap \hspans{}{z} 
\neq \emptyset$, which means $y = z$ by Axiom 
\axiomORef{opor::runs-are-injective}.
\end{itemize}

\item Case $\visObsIndx{\pointerIndx p}{\opporIndx} x y$.
By definition, there are 
$b_x \in \dspans{p}{x}$ and $b_y \in \ospans{p}{y}$ such that 
$b_x$ is the maximum under $\precedesSpansEqSymbol$ satisfying
$b_x \in \writesSpans {p}$, 
$\LastBlockRel {b_x} {b_y}$, and $\linReps {\firstRep {b_x}} {\firstRep {b_y}}$.

From hypothesis $z \in \closedEvent$ and Lemma \ref{lma::appendix::opor::visible-implies-having-run},
$\hspans{}{z} \neq \emptyset$ holds. So, by 
Axiom \axiomORef{opor::writers-have-writer-blocks}, 
there is $b_z \in \hspans{}{z} \cap \writesSpans {p}$. 

Since $b_x, b_z \in \descriptorSpans {p}$, we have three cases by Axiom 
\axiomORef{opor::descriptors-do-not-interfere}.
\begin{itemize}
\item Case $\precedesSpans {b_x} {b_z}$.
Since rep events are totally ordered, $\linRepsEq {\lastRep {b_y}} {\lastRep {b_z}}$ or
$\linReps {\lastRep {b_z}} {\lastRep {b_y}}$.

\begin{itemize}
\item Case $\linRepsEq {\lastRep {b_y}} {\lastRep {b_z}}$.
Again, we have the following cases by the total order on rep events,
\begin{itemize}
\item Case $\linRepsEq {\firstRep {b_z}} {\lastRep {b_y}}$.
Hence, ${\firstRep {b_z}} \linRepsEqSymbol {\lastRep {b_y}} \linRepsEqSymbol {\lastRep {b_z}}$,
which means that $\oportunisticPred {b_y} z$ holds by definition. 
Therefore, $\LastBlockRel {b_y} {b_z}$ by Axiom
\axiomORef{opor::opportunistic-access}. 
But this means $\visSepIndx {\pointerIndx p}{\opporIndx} y z$ by definition
(So, $\visSep {\pointerIndx p} y z$).

\item Case $\linReps {\lastRep {b_y}} {\firstRep {b_z}}$. By Axiom \axiomORef{opor::descriptor-write-precedes-resolution}, 
$\linRepsEq {\firstRep {b_z}} {\lastRep {b_z}}$.
Hence, $\linReps {\lastRep {b_y}} {\lastRep {b_z}}$,
which means
$\LastBlockRel {b_y} {b_z}$. 
Therefore, $\visSepIndx {\pointerIndx p}{\opporIndx} y z$ holds by definition
(So, $\visSep {\pointerIndx p} y z$).
\end{itemize}

\item Case $\linReps {\lastRep {b_z}} {\lastRep {b_y}}$. 
Since rep events are totally ordered, we have $\linReps {\firstRep {b_z}} {\firstRep {b_y}}$
or $\linReps {\firstRep {b_y}} {\firstRep {b_z}}$ or $\firstRep {b_y} = \firstRep {b_z}$.

\begin{itemize}
\item Case $\linReps {\firstRep {b_z}} {\firstRep {b_y}}$.
We have 
$b_x \precedesSpansSymbol b_z$ and $\lastRep {b_z} \linRepsSymbol \lastRep {b_y}$ by hypothesis, which 
is a contradiction, because $b_x$ was the maximum under $\precedesSpansEqSymbol$ satisfying
the conditions $b_x \in \writesSpans {p}$ and 
$\LastBlockRel {b_x} {b_y}$ and $\linReps {\firstRep {b_x}} {\firstRep {b_y}}$, 
but now $b_z$ is a more recent
span under $\precedesSpansEqSymbol$ satisfying the conditions.

\item Case $\linReps {\firstRep {b_y}} {\firstRep {b_z}}$. By definition,
$\visSepIndx {\pointerIndx p}{\opporIndx} y z$ holds
(So, $\visSep {\pointerIndx p} y z$).

\item Case $\firstRep {b_y} = \firstRep {b_z}$. This contradicts Axiom \axiomORef{opor::different-span-kinds-do-not-share-reps}.
\end{itemize}

\end{itemize}
\item Case $\precedesSpans {b_z} {b_x}$. By definition,
$\visSepIndx {\pointerIndx p}{\descIndx} z x$ holds
(So, $\visSep {\pointerIndx p} z x$).

\item Case $b_x = b_z$. We have $\hspans{}{x} \cap \hspans{}{z} 
\neq \emptyset$, which means $x = z$ by Axiom 
\axiomORef{opor::runs-are-injective}.
\end{itemize}
\end{itemize}
\end{prf}

\begin{lem}
Suppose hypotheses \ref{defn::appendix::opor::hypotheses-for-validity}.
Axiom \axiomHRef{help-focused::helped-are-writers} holds.
\end{lem}

\begin{prf}
Let $x,y \in \closedEvent$, and
$\visObs {\pointerIndx {p}} x y$.

By definition of $\visObsSymbol{\pointerIndx p}$, in either case 
$\visObsIndxSymbol{\pointerIndx p}{\descIndx}$ or 
$\visObsIndxSymbol{\pointerIndx p}{\opporIndx}$,
there is $b_x \in \dspans{p}{x}$ such that $b_x \in \writesSpans {p}$.
So, $x \in \writesAbs {p}$ by Axiom 
\axiomORef{opor::writers-have-writer-blocks}.
\end{prf}

\begin{lem}
\label{lma::appendix::opor::acyclic-vis-axiom-holds}
Suppose hypotheses \ref{defn::appendix::opor::hypotheses-for-validity}.
Axiom \axiomHRef{help-focused::vis-acyclic} holds.
\end{lem}

\begin{prf}
Let $\genVisTrans x y$.
For a contradiction, suppose $\precedesAbsEq y x$.
From $\precedesAbsEq y x$, we have two cases,

\begin{itemize}
\item Case $\precedesAbs y x$. 

By Lemma \ref{lma::appendix::opor::vis-trans-implies-two-possibilities}, we have 
$\linReps {\firstRep {b_x}} {\lastRep{b_y}}$ for some $b_x \in \hspans{}{x}$ and 
$b_y \in \hspans{}{y}$.
But by Lemma
\ref{lma::appendix::opor::abs-precedence-implies-block-precedence},
we also have $\precedesSpans {b_y} {b_x}$.
Hence, we have the contradiction,
\[
\firstRep{b_x} \linRepsSymbol \lastRep{b_y} \linRepsSymbol \firstRep{b_x}
\]

\item Case $y = x$. From $\genVisTrans x y$ and Lemma 
\ref{lma::appendix::opor::vis-trans-implies-two-possibilities}
we have $x \neq y$ (Contradiction).
\end{itemize}
\end{prf}

\begin{lem}
	Suppose hypotheses \ref{defn::appendix::opor::hypotheses-for-validity}.
	Axiom \axiomHRef{help-focused::fin-predicate} holds.
\end{lem}

\begin{prf}
	Let $x \in \closedEvent$. By Lemma \ref{lma::appendix::opor::visible-implies-having-run},
	$\hspans{}{x} \neq \emptyset$ holds. Hence, by Axiom \axiomORef{opor::postcondition-predicate-holds},
	$\outputRunFunc x \neq \bot$ and $\postPred x {\outputRunFunc x}$ hold. 
	In addition, if $x \in \terminatedEvent$, 
	then Axiom \axiomORef{opor::finished-operations-have-a-run} implies
	$\outputRunFunc x = \outputProp x$.
\end{prf}

\begin{lem}
	Suppose hypotheses \ref{defn::appendix::opor::hypotheses-for-validity}.
	Axiom \axiomHRef{help-focused::allocs} holds.
\end{lem}

\begin{prf}
	Let $x,y \in \allocsAbs p \cap \closedEvent$.
	By Lemma \ref{lma::appendix::opor::visible-implies-having-run},
	$\hspans{}{x} \neq \emptyset$ and $\hspans{}{y} \neq \emptyset$ hold.
	
	By Axiom 
	\axiomORef{opor::allocs-have-alloc-blocks},
	there are $b_x \in \allocsSpans p \cap \hspans{}{x}$ and
	$b_y \in \allocsSpans p \cap \hspans{}{y}$.
	But then $b_x = b_y$ by Axiom 
	\axiomORef{opor::containment-and-uniqueness-of-alloc-blocks},
	which implies $x = y$ by Axiom
	\axiomORef{opor::runs-are-injective}.
\end{prf}

\begin{lem}
	\label{lma::appendix::opor::writers-have-allocated-pointers-vis-axiom-holds}
	Suppose hypotheses \ref{defn::appendix::opor::hypotheses-for-validity}.
	Axiom \axiomHRef{help-focused::all-writers-are-willing-helpers} holds.
\end{lem}

\begin{prf}
	Let $x \in \writesAbs {p} \cap \closedEvent$.
	
	By Lemma \ref{lma::appendix::opor::visible-implies-having-run}, 
	$\hspans{}{x} \neq \emptyset$ holds. Hence, by Axiom 
	\axiomORef{opor::writers-have-writer-blocks}, 
	there is $b_x \in \hspans{}{x} \cap \writesSpans {p}$. 
	But then by Axiom \axiomORef{opor::every-block-must-have-an-allocated-pointer} 
	there is $b \in \allocsSpans {p}$ 
	such that $\precedesSpansEq b {b_x}$. 
	
	\begin{itemize}
		\item Case $b = b_x$. By Axiom 
		\axiomORef{opor::allocs-have-alloc-blocks} we have 
		$x \in \allocsAbs {p}$, and trivially 
		$\visSepEq {\pointerIndx p} x x$.
		
		\item Case $\precedesSpans b {b_x}$.
		By Axiom \axiomORef{opor::writer-blocks-belong-to-runs}, 
		$b \in \hspans{}{z}$ for some $z$. 
		By Axiom \axiomORef{opor::allocs-have-alloc-blocks} we have 
		$z \in \allocsAbs {p}$.
		But since $\precedesSpans b {b_x}$, we have 
		$\visSepIndx {\pointerIndx p} {\descIndx} z x$ by definition 
		(So, $\visSep {\pointerIndx p} z x$).
	\end{itemize}
\end{prf}

\begin{thm}
\label{thm::appendix::opor::opor-axioms-imply-visibility}
If hypotheses \ref{defn::appendix::opor::hypotheses-for-validity} hold, then 
$\genStructName{\MCAS}(\visObsSymbol{\pointerIndx p}, \visSepSymbol{\pointerIndx p})$
is valid.
\end{thm}

\begin{prf}
All visibility axioms hold by Lemmas from \ref{lma::appendix::opor::no-inbetween-vis-axiom-holds} to 
\ref{lma::appendix::opor::writers-have-allocated-pointers-vis-axiom-holds}.
\end{prf}

\subsection{Proof of Opportunism Axioms}
\label{sub::sect::appendix::opor::proof-of-opportunism-axioms}

The full pseudocode for MCAS with opportunistic readers (or simply opportunistic MCAS) 
is shown in
Figure \ref{appendix::alg-MCAS-opor}. MCAS uses the exportable
procedures of RDCSS as primitives. 
The exportable procedures for MCAS are $\mcasAlg$, $\mcasReadAlg$,
$\mcasWriteAlg$, and $\mcasAllocAlg$. 

$\ValType$ denotes the set of all
possible input values. $\ValType$ contains 
neither MCAS descriptors nor pointers storing MCAS descriptors.

No particular implementation is provided for boolean predicate
$\isMcasDescRepAlg(p)$, but it is assumed that it returns 
true if and only if $p$ is a pointer storing an MCAS descriptor.
For example,~\cite{Harris} suggests that $\isMcasDescRepAlg(p)$ 
could be implemented by checking a reserved bit in $p$; this
reserved bit indicates whether or not the pointer stores a descriptor.

The implementation makes the following assumptions:
\begin{itemize}
\item The list of entries given as input to procedure $\mcasAlg$
is not empty.
\item The alloc at line~\ref{appendix::alloc-desc-Oport-MCAS}
creates a pointer $d$ such that $\isMcasDescRepAlg(d)$ returns true.
\item Any input pointer $p: \PtsType$ to any
exportable procedure must satisfy $\neg \isMcasDescRepAlg(p)$.
\item Any input pointer to any procedure must have been previously
allocated with an invocation to $\mcasAllocAlg$.
\item The alloc at line \ref{appendix::alloc-data-Alloc-Oport-MCAS} 
returns a pointer $p$ such that $\neg \isMcasDescRepAlg(p)$.
\end{itemize}   

\begin{figure}[t]
\begin{multicols*}{2}

\begin{algorithmic}[1]
\Record{$\textsc{update\_entry}$} 
	\State $\textit{pt}$ : $\DataPtType$
	\State $\textit{exp}$, $\textit{new}$ : $\ValType$
\EndRecord
\EnumLine{$\textsc{status}$}{$\UNDECIDED$, $\SUCCEEDED$, $\FAILED$}
\EndEnumLine
\Record{$\mcasDesc$}
\State $\textit{status}$ : $\ControlPtType$ $\textsc{status}$
\State $\textit{entries}$ : $\textsc{list}\,\textsc{update\_entry}$
\EndRecord
\State 
%
\Proc{$\mcasReadAlg$}{$pt : \DataPtType$}
\State \label{appendix::alg-access-Read-Oport-MCAS} $old \gets \rdcssReadAlg(pt)$
\If {$\isMcasDescRepAlg(old)$} \label{appendix::alg-is-desc-Read-Oport-MCAS}
  \State \label{appendix::alg-Read-Desc-Oport-MCAS} $d \gets \rdcssReadAlg(old)$
  \State \label{appendix::alg-Read-status-Oport-MCAS} $s \gets \rdcssReadCtlAlg(\statusProp{d})$
  \State \label{appendix::alg-Read-Entry-Oport-MCAS} $e \gets$ entry for $pt$ in $\entriesProp{d}$
  \If {$s = \SUCCEEDED$}
     \State \label{appendix::alg-return-new-Oport-MCAS} \returnCmd{$\newGen e$}
  \Else
     \State \label{appendix::alg-return-exp-Oport-MCAS} \returnCmd{$\expGen e$}
  \EndIf
\Else 
  \State \label{appendix::alg-return-old-value-Read-Oport-MCAS} \returnCmd{$old$}
\EndIf
\EndProc
\State 
%
\Proc{$\mcasWriteAlg$}{$pt : \DataPtType$, $v: \ValType$}
\State \label{appendix::alg-access-Write-Oport-MCAS} $old \gets \rdcssReadAlg(pt)$
\If {$\isMcasDescRepAlg(old)$} \label{appendix::alg-is-desc-Write-Oport-MCAS}
  \State \label{appendix::alg-help-complete-invoke-Write-Oport-MCAS} $\mcasHelpAlg(old)$
  \State $\mcasWriteAlg(pt, v)$
\Else 
  \State \label{appendix::alg-attempt-write-Write-Oport-MCAS} $x \gets \rdcssCasAlg(pt, old, v)$
  \If {$x \neq old$}
    \State $\mcasWriteAlg(pt, v)$
  \EndIf
\EndIf
\EndProc
%
\State 
%
\Proc{$\mcasAllocAlg$}{$v: \ValType$}
\State \label{appendix::alloc-data-Alloc-Oport-MCAS} \returnCmd $\rdcssAllocAlg(v,\DATA)$
\EndProc
%
\columnbreak
%
\Proc{$\mcasAlg$}{$\listvar u: \textsc{list}\,\textsc{update\_entry}$}
\State \label{appendix::alloc-status-Oport-MCAS} $s \gets \rdcssAllocAlg(\UNDECIDED, \CONTROL)$
\State \label{appendix::create-mcas-desc-Oport-MCAS} $desc \gets \mcasDesc(s, \listvar u)$
\State \label{appendix::alloc-desc-Oport-MCAS} $d \gets \rdcssAllocAlg(desc, \DATA)$
\State \returnCmd $\mcasHelpAlg(d)$
\EndProc
%
\State 
%
\Proc{$\mcasHelpAlg$}{$d : \DataPtType$}
\State \label{appendix::read-desc-Oport-MCAS} $desc \gets \rdcssReadAlg(d)$
\State \label{appendix::read-phase1-status-Oport-MCAS} $phase1 \gets \rdcssReadCtlAlg(\statusProp{desc})$
\If{$phase1 = \UNDECIDED$} \label{appendix::is-phase1-still-undecided-Oport-MCAS}
	\State \label{appendix::write-all-descs-Oport-MCAS} $s \gets \writeAllDescsAlg(d,desc)$
	\State \label{appendix::resolve-status-Oport-MCAS} $\rdcssCasCtlAlg(\statusProp{desc}, \UNDECIDED, s)$
\EndIf
\State \label{appendix::read-phase2-status-Oport-MCAS} $phase2 \gets \rdcssReadCtlAlg(\statusProp{desc})$
\State $r \gets (phase2 = \SUCCEEDED)$
\ForEach{$e$}{$\entriesProp {desc}$} \label{appendix::remove-all-descs-loop-Oport-MCAS}
	\State \label{appendix::remove-all-descs-Oport-MCAS} $\rdcssCasAlg(\pointGen e, d, r\ ?\ \newGen e : \expGen e)$
\EndForEach \label{appendix::end-phase2-Oport-MCAS}
\State \returnCmd $r$
\EndProc
%
\State 
%
\Proc{$\writeAllDescsAlg$}{$d : \DataPtType, desc: \mcasDesc$}
\ForEach{$e$}{$\entriesProp {desc}$}
	\State \label{appendix::create-rdcss-desc-Oport-MCAS} $rD \gets \rdcssDesc(\statusProp {desc},$ 
	\State $\phantom{rD \gets \quad}\pointGen e, \UNDECIDED, \expGen e, d)$
	\State \label{appendix::invoke-rdcss-in-Oport-MCAS} $old \gets \rdcssAlg(rD)$
	\If{$\isMcasDescRepAlg(old)$} \label{appendix::alg-is-desc-MCAS-Oport-MCAS}
	  \If{$old \neq d$} \label{appendix::is-my-desc-Oport-MCAS}
	     \State \label{appendix::alg-help-complete-invoke-MCAS-Oport-MCAS} $\mcasHelpAlg(old)$
	     \State \returnCmd $\writeAllDescsAlg(d, desc)$
	  \EndIf
	\ElsIf{$old \neq \expGen e$} \label{appendix::rdcss-failed-Oport-MCAS}
	  \State \returnCmd $\FAILED$
    \EndIf
\EndForEach
\State \returnCmd $\SUCCEEDED$
\EndProc
\end{algorithmic}

\end{multicols*}
\caption{Opportunistic MCAS implementation. It uses as primitives the exportable procedures in RDCSS.
Notice the opportunistic implementation of $\mcasReadAlg$. All the other procedures are implemented as
in the helping MCAS implementation of Figure \ref{appendix::alg-MCAS}.}
\label{appendix::alg-MCAS-opor}
\end{figure}

We now define the opportunistic structure for opportunistic MCAS.

\begin{defn}[Opportunistic Structure for Opportunistic MCAS]
\label{defn::appendix::opor::span-struture-mcas}
We denote the structure by $\genOportSpanStructName{\MCAS}$.

A span is either a 2-tuple of the form $(a,b)$ or
a 1-tuple of the form $(a)$, where $a,b$ are rep events in the execution history.

For each pointer $p$, set $\descriptorSpans p$ is defined by the
following list of spans containing rep events,
\begin{enumerate}
\item Any $(b)$ such that, 

\begin{itemize}
\item $\lineProp b = \ref{appendix::invoke-rdcss-in-Oport-MCAS}$
\item $\neg \isMcasDescRepAlg(\outputProp b)$
\item $b = \rdcssAlg(desc)$ for some $exp$ and $desc = \rdcssDesc(\_, p, \UNDECIDED, exp, \_)$
such that $\outputProp b \neq exp$
\end{itemize}

\item Any $(b,c)$ such that, 

\begin{itemize}
\item $\lineProp b = \ref{appendix::invoke-rdcss-in-Oport-MCAS}$
\item $\lineProp c = \ref{appendix::resolve-status-Oport-MCAS}$
\item $\neg \isMcasDescRepAlg(\outputProp b)$
\item $b$ executes before $c$ and there is no other $p$-write rep event between $b$ and $c$.
\item \begin{sloppypar}
$b = \rdcssAlg(desc)$ and $c = \rdcssCasAlg(s, \UNDECIDED, \SUCCEEDED)$, 
for some $s$, $exp$, $d$, and $desc = \rdcssDesc(s, p, \UNDECIDED, exp, d)$ such that
$c$ is successful, $\outputProp b = exp$, and $\isMcasDescRepAlg(d)$.
\end{sloppypar}
\end{itemize}

\item Any $(b,c)$ such that,

\begin{itemize}
\item $\lineProp b = \ref{appendix::invoke-rdcss-in-Oport-MCAS}$
\item $\lineProp c = \ref{appendix::resolve-status-Oport-MCAS}$
\item $\neg \isMcasDescRepAlg(\outputProp b)$
\item $b$ executes before $c$ and there is no other $p$-write rep event between $b$ and $c$.
\item \begin{sloppypar}
$b = \rdcssAlg(desc)$ and $c = \rdcssCasAlg(s, \UNDECIDED, \FAILED)$, 
for some $s$, $exp$, $d$, and $desc = \rdcssDesc(s, p, \UNDECIDED, exp, d)$ such that
$c$ is successful, $\outputProp b = exp$, and $\isMcasDescRepAlg(d)$.
\end{sloppypar}
\end{itemize}

\item Any $(b)$ such that,

\begin{itemize}
\item $\lineProp b = \ref{appendix::alg-access-Read-Oport-MCAS}$
\item $\neg \isMcasDescRepAlg(\outputProp b)$
\item $b$ reads pointer $p$
\end{itemize}

\item Any $(b)$ such that,

\begin{itemize}
\item $\lineProp b = \ref{appendix::alg-attempt-write-Write-Oport-MCAS}$
\item $\neg \isMcasDescRepAlg(\outputProp b)$
\item $b$ is a successful $\rdcssCasAlg$ writing into pointer $p$
\end{itemize}

\item Any $(b)$ such that $\lineProp b = \ref{appendix::alloc-data-Alloc-Oport-MCAS}$ and
$b$ has as output pointer $p$.
\end{enumerate}

For each pointer $p$, set $\opportunisticSpans p$ is defined by the
following list of spans containing rep events,
\begin{enumerate}
\item Any $(b,c)$ such that, 

\begin{itemize}
\item $\lineProp b = \ref{appendix::alg-access-Read-Oport-MCAS}$
\item $\lineProp c = \ref{appendix::alg-Read-status-Oport-MCAS}$
\item $\isMcasDescRepAlg(\outputProp b)$
\item $b$ executes before $c$ and they are executed by the thread that invoked 
the $\mcasReadAlg$ invocation that contains both $b$ and $c$.
\item \begin{sloppypar}
$b = \rdcssReadAlg(p)$ and $c = \rdcssReadCtlAlg(\_)$.
\end{sloppypar}
\end{itemize}
\end{enumerate}

With this, the set of writer spans $\writesAbs p$ can be defined as,
\begin{align*}
\writesSpans p & \defini \{ (b) \in \descriptorSpans p \mid b \text{ writes or allocs pointer } p \} \cup 
\{ (b,c) \in \descriptorSpans p \mid c = \rdcssCasAlg(\_, \UNDECIDED, \SUCCEEDED) \} 
\end{align*}

The set of alloc spans $\allocsAbs p$ as follows,
\begin{align*}
\allocsSpans p & \defini \{ (b) \in \writesSpans p \mid b \text{ allocs pointer } p \wedge 
\lineProp b = \ref{appendix::alloc-data-Alloc-Oport-MCAS} \}
\end{align*}

We now define the denotation $\runFunc x$ by cases on 
event $x$,
\begin{enumerate}

\item Case $x = \mcasAlg(\listvar u)$.

\begin{enumerate}

\item If there are rep events $i_a$, $i_s$, $r$, and for every $j \in \listvar u$, there is $(a_j,r) \in \writesSpans {\pointGenEntry j}$,
such that,

\begin{itemize}
\item $i_a = \rdcssAllocAlg(desc, \DATA)$ with code line \ref{appendix::alloc-desc-Oport-MCAS},
where $\entriesProp {desc} = \listvar u$.
\item $i_s = \rdcssAllocAlg(\UNDECIDED, \CONTROL)$ with code line \ref{appendix::alloc-status-Oport-MCAS},
\item If $T$ is the thread that invoked $x$, then $T$ executes $i_a$ and $i_s$ within the invocation of $x$,
\item $a_j = \rdcssAlg(rD)$, where $rD = \rdcssDesc(\outputProp {i_s},\ \pointGenEntry j,\ \UNDECIDED,\ \expGenEntry j,\ \outputProp {i_a})$ with code 
line \ref{appendix::invoke-rdcss-in-Oport-MCAS},
\item $r = \rdcssCasAlg(\outputProp {i_s}, \UNDECIDED, \SUCCEEDED)$ with code 
line \ref{appendix::resolve-status-Oport-MCAS},
\item $a_j$ is the most recent such $\rdcssAlg$ call for pointer $\pointGenEntry j$ before $r$.
\end{itemize}
then 
$\runFunc x \defini (\{ (a_j,r) \mid j \in \listvar u \},\ true)$.

\item If there are rep events $i_a$, $i_s$, $r$ and for some $j \in \listvar u$, there is $(c_j) \in \descriptorSpans {\pointGenEntry j}$,
such that for every $k < j$, there are $(a_k,r) \in \descriptorSpans {\pointGenEntry k}$, such that,

\begin{itemize}
\item $i_a = \rdcssAllocAlg(desc, \DATA)$ with code line \ref{appendix::alloc-desc-Oport-MCAS},
where $\entriesProp {desc} = \listvar u$.
\item $i_s = \rdcssAllocAlg(\UNDECIDED, \CONTROL)$ with code line \ref{appendix::alloc-status-Oport-MCAS},
\item If $T$ is the thread that invoked $x$, then $T$ executes $i_a$ and $i_s$ within the invocation of $x$,
\item $a_k = \rdcssAlg(rD)$, where $rD = \rdcssDesc(\outputProp {i_s},\ \pointGenEntry k,\ \UNDECIDED,\ \expGenEntry k,\ \outputProp {i_a})$ with code 
line \ref{appendix::invoke-rdcss-in-Oport-MCAS}, and $\outputProp{a_k} = \expGenEntry k$.
\item $r = \rdcssCasAlg(\outputProp {i_s}, \UNDECIDED, \FAILED)$ with code 
line \ref{appendix::resolve-status-Oport-MCAS},
\item $a_k$ is the most recent such $\rdcssAlg$ call for pointer $\pointGenEntry k$ before $r$.
\item $c_j = \rdcssAlg(rD)$, where $rD = \rdcssDesc(\outputProp {i_s},\ \pointGenEntry j,\ \UNDECIDED,\ \expGenEntry j,\ \outputProp {i_a})$ with code 
line \ref{appendix::invoke-rdcss-in-Oport-MCAS}, and $\outputProp{c_j} \neq \expGenEntry j$, and $\neg \isMcasDescRepAlg(\outputProp {c_j})$.
\item $c_j$ was executed by the same thread that executed $r$, and $c_j$ is the most recent such $\rdcssAlg$ call for pointer $\pointGenEntry j$ before $r$,
\end{itemize}
then 
$\runFunc x \defini (\{ (a_k,r),\ (c_j) \mid k \in \listvar u \wedge k < j \},\ false)$.

\item Otherwise, $\runFunc x \defini (\emptyset, \bot)$.
\end{enumerate}

\item Case $x = \mcasReadAlg(p)$.
\begin{enumerate}
\item If there is $(r) \in \descriptorSpans p$ such that,

\begin{itemize}
\item If $T$ is the thread that invoked $x$, then $T$ executes $r$ 
within the invocation of $x$,
\item $r = {\rdcssReadAlg(p)}$ with code line \ref{appendix::alg-access-Read-Oport-MCAS},
\end{itemize}
then
$\runFunc x \defini (\{ (r) \},\ \outputProp r)$.

\item If there is $(a,b) \in \opportunisticSpans p$ such that,

\begin{itemize}
\item If $T$ is the thread that invoked $x$, then $T$ executes $a$ and $b$ 
within the invocation of $x$,
\item $a = {\rdcssReadAlg(p)}$ with code line \ref{appendix::alg-access-Read-Oport-MCAS},
\item $b = {\rdcssReadCtlAlg(\statusProp{desc})}$ with code line \ref{appendix::alg-Read-status-Oport-MCAS},
for some $r$, $desc$, $j \in \entriesProp {desc}$, such that $r = \rdcssReadAlg(\_)$ at line \ref{appendix::alg-Read-Desc-Oport-MCAS}, 
and $desc = \outputProp r$, and $p = \pointGenEntry j$, and $r$ is also executed by $T$ within the invocation of $x$,
\item $\outputProp b = \SUCCEEDED$,
\end{itemize}
then
$\runFunc x \defini (\{ (a,b) \},\ \newGenEntry j)$.

\item If there is $(a,b) \in \opportunisticSpans p$ such that,

\begin{itemize}
\item If $T$ is the thread that invoked $x$, then $T$ executes $a$ and $b$ 
within the invocation of $x$,
\item $a = {\rdcssReadAlg(p)}$ with code line \ref{appendix::alg-access-Read-Oport-MCAS},
\item $b = {\rdcssReadCtlAlg(\statusProp{desc})}$ with code line \ref{appendix::alg-Read-status-Oport-MCAS},
for some $r$, $desc$, $j \in \entriesProp {desc}$, such that $r = \rdcssReadAlg(\_)$ at line \ref{appendix::alg-Read-Desc-Oport-MCAS}, 
and $desc = \outputProp r$, and $p = \pointGenEntry j$, and $r$ is also executed by $T$ within the invocation of $x$,
\item $\outputProp b \neq \SUCCEEDED$,
\end{itemize}
then
$\runFunc x \defini (\{ (a,b) \},\ \expGenEntry j)$.

\item Otherwise, $\runFunc x \defini (\emptyset, \bot)$.
\end{enumerate}

\item Case $x = \mcasWriteAlg(p,v)$.
\begin{enumerate}
\item If there are $(c) \in \writesSpans p$ and rep event $r$ such that,

\begin{itemize}
\item If $T$ is the thread that invoked $x$, then $T$ executes $c$ 
within the invocation of $x$,
\item $r = {\rdcssReadAlg(p)}$ with code line \ref{appendix::alg-access-Write-Oport-MCAS},
\item $c = \rdcssCasAlg(p, \outputProp r, v)$ with code line \ref{appendix::alg-attempt-write-Write-Oport-MCAS},
\item If $T$ is the thread that invoked $x$, then $r$ is the last read
carried out by $T$ before the execution of $c$,
\end{itemize}
then
$\runFunc x \defini (\{ (c) \},\ \unitValue)$.

\item Otherwise, $\runFunc x \defini (\emptyset, \bot)$.
\end{enumerate}

\item Case $x = \mcasAllocAlg(v)$.
\begin{enumerate}
\item If $\outputProp x \neq \bot$, $\ETimeProp x \neq \bot$, and
there is $(i) \in \writesSpans {\outputProp x}$ such that,

\begin{itemize}
\item If $T$ is the thread that invoked $x$, then $T$ executes $i$ 
within the invocation of $x$,
\item $i = \rdcssAllocAlg(v,\DATA)$ with code line \ref{appendix::alloc-data-Alloc-Oport-MCAS},
\end{itemize}
then
$\runFunc x \defini (\{ (i) \},\ \outputProp i)$.

\item Otherwise, $\runFunc x \defini (\emptyset, \bot)$.
\end{enumerate}

\end{enumerate}

\end{defn}

Notice how line \ref{appendix::alg-access-Read-Oport-MCAS} 
appears in both the definition of $\descriptorSpans{p}$
and $\opportunisticSpans p$. When the line returns
a non-descriptor value, it is part of a d-span in 
$\descriptorSpans{p}$, but when the line returns a 
descriptor, it is part of an o-span in $\opportunisticSpans p$.

We argue that $\runFunc x$ is well-defined.
For that matter, we need to show that the 
function's conditions pick spans uniquely.

The $\mcasAlg$, $\mcasWriteAlg$, and $\mcasAllocAlg$ 
cases are identical as in Definition \ref{defn::appendix::impl::span-struture-mcas}
for the helping MCAS, the only difference is that we now use
set $\descriptorSpans p$ instead of $\spans p$.
So, let us focus on $\mcasReadAlg$, which is the only entry
that changed.

Case (2).(a) is choosing a rep event at line
\ref{appendix::alg-access-Read-MCAS} and
such that its output does not satisfy 
$\isMcasDescRepAlg$ (by definition of set $\descriptorSpans p$).
But this condition can happen at most once during the
invocation of $x$, 
because the invoking thread finishes $x$ once
the condition is satisfied.

Cases (2).(b) and (2).(c) are choosing an o-span
containing two rep events that execute exactly once
within the invocation of $x$ 
(since in the opportunistic implementation
there are no loops or recursive calls).

We are ready to prove that 
$\genStructName{\MCAS}(\visObsSymbol{\pointerIndx p}, \visSepSymbol{\pointerIndx p})$ is implemented by opportunistic
structure $\genOportSpanStructName{\MCAS}$, i.e., we are going to prove that
the opportunism axioms of Figure \ref{fig::sub::appendix::opor::opor-axioms} are satisfied. We require a couple of lemmas.

\begin{lem}
\label{lem::appendix::opor::unique-alloc-mcas}
If $i_1$, $i_2$ are allocs at line \ref{appendix::alloc-data-Alloc-Oport-MCAS} 
such that they allocate the same pointer, then $i_1 = i_2$.
\end{lem}

\begin{prf}
Identical to the proof of Lemma \ref{lem::appendix::impl::unique-alloc-mcas}.
\end{prf}

For the next lemma we make the following definition,

\begin{defn}
We say that $v \in \ValType$ is the value that $p$ is bound to by $(a,\_) \in \writesSpans p$ 
if there are $rD$, $desc$, $j$ such that,
\begin{itemize}
\item $a = \rdcssAlg(rD)$,
\item $\newTwo {rD}$ was allocated by rep event $\rdcssAllocAlg(desc, \DATA)$ at line
\ref{appendix::alloc-desc-Oport-MCAS},
\item $j \in \entriesProp {desc}$,
\item $p = \pointTwo{rD} = \pointGenEntry j$,
\item $v = \newGenEntry j$.
\end{itemize}
Alternatively, one can think of $v$ is the new value that 
line \ref{appendix::remove-all-descs-Oport-MCAS} will write into $p$
because span $(a,\_)$ has already resolved the descriptor status to $\SUCCEEDED$.
\end{defn}

\begin{lem}
\label{lem::appendix::opor::writers-belong-to-writer-procs-mcas}
If $b \in \writesSpans p$, 
then there is $x$ such that $b \in \hspans{}{x}$ and $\inputVal x p v$,
where $v$ is the value determined by one of the following cases:
\begin{itemize}
\item If $b$ is of the form $(i)$ for some rep $i$, then $v$ is the value written by $i$.
\item If $b$ is of the form $(a,\_)$ for some rep $a$, then
$v$ is the value that $p$ is bound to by $b$.
\end{itemize}
\end{lem}

\begin{prf}
Identical to the proof of Lemma \ref{lem::appendix::impl::writers-belong-to-writer-procs-mcas}.
\end{prf}

\begin{lem}
\label{lem::appendix::opor::visibility-lemma-mcas}
We have the following propositions.
\begin{enumerate}

\item Suppose $x$ is one of $\mcasAlg$, or $\mcasReadAlg$. Suppose $p \in \PtsType$ and $v \in \ValType$. Then,
the following statements are equivalent,
\begin{itemize}
\item There is $y$ such that $\visObsIndx {\pointerIndx p}{\descIndx} y x$
and $\inputVal y p v$.
\item There is $b \in \dspans{p}{x}$ such that $\firstRep b$ reads value $v$.
\end{itemize}

\item Suppose $x$ is $\mcasReadAlg$. If $(a,b) \in \ospans{p}{x}$ and $\outputProp {b} = \SUCCEEDED$, then 
there is $y$ such that $\visObsIndx {\pointerIndx p}{\opporIndx} y x$
and $\inputVal y p {\newGenEntry j}$ for some $desc$ and $j \in \entriesProp {desc}$ such that 
$p = \pointGenEntry j$ and $desc$ is the descriptor discovered by $a$.

\item Suppose $x$ is $\mcasReadAlg$. If $(a,b) \in \ospans{p}{x}$ and $\outputProp {b} \neq \SUCCEEDED$, then 
there is $y$ such that $\visObsIndx {\pointerIndx p}{\opporIndx} y x$
and $\inputVal y p {\expGenEntry j}$ for some $desc$ and $j \in \entriesProp {desc}$ such that 
$p = \pointGenEntry j$ and $desc$ is the descriptor discovered by $a$.
\end{enumerate}
\end{lem}

\begin{prf}
We prove each proposition.

\begin{enumerate}
\item Identical to the proof of Lemma \ref{lem::appendix::impl::visibility-lemma-mcas}, 
but using $\descriptorSpans p$ instead of $\spans p$.

\item Since $a$ discovers a descriptor $desc$ and $b$ finds the descriptor status 
to have succeeded ($\outputProp {b} = \SUCCEEDED$),
it means that $desc$ must have been written and resolved by a 
span $(g,h)$ such that $g \linRepsSymbol a$ and $h \linRepsSymbol b$. 
Also, $p$ must be one of the input pointers in $\entriesProp {desc}$, i.e., 
$p = \pointGenEntry j$ for some $j \in \entriesProp {desc}$, otherwise,
$a$ would not have found a descriptor in $p$ in the first place.

The successful span $(g,h)$ must be the most recent one satisfying 
$g \linRepsSymbol a$ and $h \linRepsSymbol b$
because if there is a more recent successful writer span, it will have to occur 
after $a$, since the descriptor was still present when $a$ discovered it. 
Notice that $\newGenEntry j$ is the value $p = \pointGenEntry j$ is bound to by $(g,h)$,
since line \ref{appendix::remove-all-descs-Oport-MCAS} will eventually write $\newGenEntry j$ into $p$ as 
$h$ marked the descriptor as
succeeded. The result follows by Lemma \ref{lem::appendix::opor::writers-belong-to-writer-procs-mcas}.

\item Since $a$ discovers a descriptor $desc$, it means that $desc$ must have been written by a 
rep event $g$ such that $g \linRepsSymbol a$. 
Also, $p$ must be one of the input pointers in $\entriesProp {desc}$, i.e., 
$p = \pointGenEntry j$ for some $j \in \entriesProp {desc}$, otherwise,
$a$ would not have found a descriptor in $p$ in the first place.

\begin{itemize}
\item If $\outputProp {b} = \FAILED$, then $g$ must be part of a span $(g,h)$, where $h$
must have resolved the descriptor to $\FAILED$ before $b$
discovered the status, i.e., $(g,h)$ is not a successful span.
Rep event $g$ must have found value $\expGenEntry j$ in $p = \pointGenEntry j$,
otherwise it would not have written the descriptor in the first place (line \ref{appendix::invoke-rdcss-in-Oport-MCAS}).
This means that, before rep event $g$, there must exist a recent successful span $(g',h')$ that 
either wrote value $\expGenEntry j$
or had $\expGenEntry j$ as bound value that later got written by line \ref{appendix::remove-all-descs-Oport-MCAS}.
Since $(g',h')$ finishes before $g$, we have $g' \linRepsSymbol a$ and
$h' \linRepsSymbol b$, and $(g',h')$ is the most recent successful span 
satisfying those conditions.
The result follows by Lemma \ref{lem::appendix::opor::writers-belong-to-writer-procs-mcas}.

\item If $\outputProp {b} = \UNDECIDED$, then $desc$ has not been resolved yet, 
which means that $g$ is still not part of a span (or if $g$ is part of a span, such span ends after $b$).
Rep event $g$ must have found value $\expGenEntry j$ in $p = \pointGenEntry j$,
otherwise it would not have written the descriptor in the first place (line \ref{appendix::invoke-rdcss-in-Oport-MCAS}).
This means that, before rep event $g$, there must exist a recent successful span $(g',h')$ that 
either wrote value $\expGenEntry j$
or had $\expGenEntry j$ as bound value that later got written by line \ref{appendix::remove-all-descs-Oport-MCAS}.
Since $(g',h')$ finishes before $g$, we have $g' \linRepsSymbol a$ and
$h' \linRepsSymbol b$, and $(g',h')$ is the most recent successful span 
satisfying those conditions (in case $g$ is part of a span, such span does not satisfy the condition
of finalizing before $b$ since $desc$ is still $\UNDECIDED$).
The result follows by Lemma \ref{lem::appendix::opor::writers-belong-to-writer-procs-mcas}.
\end{itemize}
\end{enumerate}
\end{prf}

\begin{lem}
\label{lem::appendix::opor::non-interference-of-lifespans-mcas}
Axiom \axiomORef{opor::descriptors-do-not-interfere} holds.
\end{lem}

\begin{prf}
Identical to the proof in Lemma \ref{lem::appendix::impl::non-interference-of-lifespans-mcas},
but using $\descriptorSpans p$ instead of $\spans p$.
\end{prf}

\begin{lem}
\label{lem::appendix::opor::all-descriptors-are-written-before-any-resolution-mcas}
Axiom \axiomORef{opor::all-descriptors-are-written-before-any-resolution} holds.
\end{lem}

\begin{prf}
This is trivial when $x$ has a denotation with only one span. And the cases for
$x = \mcasAlg(\listvar u)$ are identical to the proof of Lemma 
\ref{lem::appendix::impl::all-descriptors-are-written-before-any-resolution-mcas}.
\end{prf}

\begin{lem}
Axiom \axiomORef{opor::opportunistic-access} holds.
\end{lem}

\begin{prf}
Suppose $\oportunisticPred b x$, which means $b \in \opportunisticSpans p$ 
and $\firstRep c \linRepsEqSymbol \lastRep b \linRepsEqSymbol \lastRep c$
for some $p$ and $c \in \hspans{}{x} \cap \writesSpans p$.

$c$ cannot be a 1-span because we would have $\lastRep b = \firstRep c = \lastRep c$, contradicting
that d-spans and o-spans do not share rep events. So, $c$ must have the form $(d,e)$,
where $d$ writes the descriptor $desc$ into $p$ and $e$ sets the descriptor status to $\SUCCEEDED$.
This means that the denotation of $x$ must have the form $\{ (a_j,e) \mid j \in \entriesProp{desc} \}$, since
it is the only case that contains a span like $c = (d,e)$. 
Since $\lastRep b \linRepsEqSymbol \lastRep c = e$, then $\lastRep b \linRepsEqSymbol \lastRep{(a_j,e)} = e$ 
for every $j$, which proves the axiom.
\end{prf}

\begin{lem}
	\label{lem::appendix::opor::finished-are-non-empty-axiom-holds-mcas}
	Axiom
	\axiomORef{opor::finished-operations-have-a-run} holds.
\end{lem}

\begin{prf}
	The cases for $\mcasAlg$, $\mcasWriteAlg$ and $\mcasAllocAlg$ are identical to the cases 
	as in Lemma~\ref{lem::appendix::impl::finished-are-non-empty-axiom-holds-mcas}, since their code is identical. 
	Hence, we only need to check the $\mcasReadAlg$ method.
	
	Suppose that $\mcasReadAlg(p)$ finished.
	Let $T$ be the invoking thread.  
	If $T$ reached line 20, then the read at line 10 returned a
	non-descriptor value which is returned by $\mcasReadAlg$. This corresponds to
	(2).(a) in the definition of $\runFuncSymbol$ with output whatever
	line 10 produced and the denotation is not empty.
	If $T$ reached line 16, then the
	read at line 10 returned a descriptor. The status read at line 13 must
	have returned $\SUCCEEDED$, and the method returned $new_p$ corresponding to
	the new value for pointer $p$ in the descriptor. This corresponds to
	(2).(b) in the definition of $\runFuncSymbol$ with output $new_p$ and
	the denotation is not empty. 
	If $T$ reached line 18, then the read at line 10
	returned a descriptor. The status read at line 13 must have returned
	either $\UNDECIDED$ or $\FAILED$, and the method returned $exp_p$ corresponding to
	the expected value for pointer $p$ in the descriptor. This corresponds
	to (2).(c) in the definition of $\runFuncSymbol$ with output $exp_p$
	and the denotation is not empty.
\end{prf}

\begin{lem}
Axiom \axiomORef{opor::descriptor-write-precedes-resolution} holds.
\end{lem}

\begin{prf}
Trivial for 1-spans. For 2-tuple d-spans $(a,b)$, 
the code can only resolve descriptors (rep event $b$)
if the descriptor was previously written at line \ref{appendix::invoke-rdcss-in-MCAS} 
(rep event $a$).
For 2-tuple o-spans, the code can only look into 
the descriptor status (rep event $b$) if the the read
at line \ref{appendix::alg-access-Read-Oport-MCAS} discovered a descriptor (rep event $a$).
\end{prf}

\begin{lem}
Axiom \axiomORef{opor::descriptor-opor-spans-disjoint} holds.
\end{lem}

\begin{prf}
The only possible cases where there could be an overlap between a d-span and an o-span
is in cases (4) of $\descriptorSpans p$ and (1) of $\opportunisticSpans p$. However,
in (4), the rep event at line \ref{appendix::alg-access-Read-Oport-MCAS} is required to return a 
non-descriptor value, while in (1), the rep event at line \ref{appendix::alg-access-Read-Oport-MCAS}
is required to return a descriptor. Therefore, the spans are different because they contain
different rep events.
\end{prf}

\begin{lem}
Axiom \axiomORef{opor::spans-access-at-most-one-pointer} holds.
\end{lem}

\begin{prf}
All spans $b$ in $\descriptorSpans p$ and $\opportunisticSpans p$ are defined so that
the pointer they access is determined by the pointer that $\firstRep b$ accesses.
All rep events access at most one pointer, with the exception of the $\rdcssAlg$ at line
\ref{appendix::invoke-rdcss-in-Oport-MCAS}. However, looking at cases (1), (2), (3) of 
$\descriptorSpans p$, we see that the spans define $pt_2$ in the descriptor as the
pointer accessed by $\firstRep b$.

So, if a span $b$ accesses pointers $p$ and $q$, then $\firstRep b$ accesses $p$ and $q$, which
means $p = q$.
\end{prf}

\begin{lem}
Axiom \axiomORef{opor::containment-and-uniqueness-of-alloc-blocks} holds.
\end{lem}

\begin{prf}
Identical to the proof of Lemma \ref{lem::appendix::impl::containment-and-uniqueness-of-alloc-blocks-mcas}.
\end{prf}

\begin{lem}
Axiom
\axiomORef{opor::every-block-must-have-an-allocated-pointer} holds.
\end{lem}

\begin{prf}
Identical to the proof of Lemma \ref{lem::appendix::impl::every-block-must-have-an-allocated-pointer-mcas}.
\end{prf}

\begin{lem}
Axiom
\axiomORef{opor::runs-are-injective} holds.
\end{lem}

\begin{prf}
First, when neither $x$ nor $y$ are $\mcasReadAlg$ events, the reasoning 
is identical as in the proof of Lemma \ref{lem::appendix::impl::runs-are-injective-mcas}.
So, we can assume that either $x$ is a $\mcasReadAlg$ event or $y$ is.
Say, $x$ is a $\mcasReadAlg$ event.

All the spans in the cases for $\mcasReadAlg$ satisfy that $\firstRep b$ has code line 
\ref{appendix::alg-access-Read-Oport-MCAS} and it is executed by the same thread
that invoked $x$ and it occurs within the invocation of $x$. 
Since $y$ also includes $b$, event $y$ must also be a $\mcasReadAlg$ 
(other events do not have denotations containing line \ref{appendix::alg-access-Read-Oport-MCAS}),
which means that $x$ and $y$ must be invoked by the same thread and $b$ executes within the 
invocation of both $x$ and $y$.

If $x \neq y$, then $x$ and $y$ cannot overlap in real-time, because they 
are invoked by the same thread. But this contradicts that $b$ occurs within
the invocation of both $x$ and $y$. Therefore, $x = y$.
\end{prf}

\begin{lem}
Axiom
\axiomORef{opor::writer-blocks-belong-to-runs} holds.
\end{lem}

\begin{prf}
Directly from Lemma \ref{lem::appendix::opor::writers-belong-to-writer-procs-mcas}.
\end{prf}

\begin{lem}
Axiom
\axiomORef{opor::postcondition-predicate-holds} holds.
\end{lem}

\begin{prf}
Suppose $\hspans{}{x} \neq \emptyset$. We see from definition of $\runFunc x$ that all non-empty cases have $\outputRunFunc x \neq \bot$.
To prove $\postPred x {\outputRunFunc x}$, we do a case analysis on $x$.

\begin{itemize}
\item Case $x = \mcasAlg(\listvar u)$. 
Identical to the $\mcasAlg(\listvar u)$ case in the proof for Lemma \ref{lem::appendix::impl::postcondition-predicate-holds-mcas}, but
using Lemma \ref{lem::appendix::opor::visibility-lemma-mcas} to conclude that in the $\outputRunFunc x = true$ case,
for every $i \in \listvar u$,
there is $z$ such that $\visObsIndx {\pointerIndx {\pointGenEntry i}}{\descIndx} {z} x$ (hence 
$\visObs {\pointerIndx {\pointGenEntry i}} {z} x$)
and $\inputVal{z} {\pointGenEntry i} {\expGenEntry i}$; and in the $\outputRunFunc x = false$ case,
for some $i \in \listvar u$,
there is $z$ such that $\visObsIndx {\pointerIndx {\pointGenEntry i}}{\descIndx} {z} x$ (hence 
$\visObs {\pointerIndx {\pointGenEntry i}} {z} x$)
and $\inputVal{z} {\pointGenEntry i} v$ for some $v \neq {\expGenEntry i}$.

\item The cases $x = \mcasAllocAlg(v)$ and $x = \mcasWriteAlg(q,v)$ are identical as in the proof for 
Lemma \ref{lem::appendix::impl::postcondition-predicate-holds-mcas}.

\item Case $x = \rdcssReadAlg(q)$. We want to prove,
\begin{align*}
\exists z.\ \visObs {\pointerIndx q} z x \wedge \inputVal z q {\outputRunFunc x}
\end{align*}

Each case in the definition of $\runFunc x$
has a span $b$ such that $\firstRep b$ reads a descriptor or a non-descriptor value in $q$ at line \ref{appendix::alg-access-Read-Oport-MCAS}.

If $\firstRep b$ reads a non-descriptor value, then $b$ is a d-span. 
Therefore, by Lemma \ref{lem::appendix::opor::visibility-lemma-mcas},
there is $z$ such that $\visObsIndx {\pointerIndx q}{\descIndx} {z} x$
(hence $\visObs {\pointerIndx q} {z} x$)
and $\inputVal z q {\outputRunFunc x}$.

If $\firstRep b$ reads a descriptor $desc$, then $b$ is an o-span, and $q = \pointGenEntry j$ for some 
$j \in \entriesProp {desc}$. 
\begin{itemize}
\item Case $\outputProp {\lastRep b} = \SUCCEEDED$.
Then $\outputRunFunc x = \newGenEntry j$
and by Lemma \ref{lem::appendix::opor::visibility-lemma-mcas}, there is $z$ 
such that $\visObsIndx {\pointerIndx q}{\opporIndx} {z} x$
(hence $\visObs {\pointerIndx q} {z} x$)
and $\inputVal z q {\outputRunFunc x}$.

\item Case $\outputProp {\lastRep b} \neq \SUCCEEDED$.
Then $\outputRunFunc x = \expGenEntry j$
and by Lemma \ref{lem::appendix::opor::visibility-lemma-mcas}, there is $z$ 
such that $\visObsIndx {\pointerIndx q}{\opporIndx} {z} x$
(hence $\visObs {\pointerIndx q} {z} x$)
and $\inputVal z q {\outputRunFunc x}$.
\end{itemize} 
\end{itemize}
\end{prf}

\begin{lem}
\label{lem::appendix::opor::different-span-kinds-do-not-share-reps-mcas}
Axiom
\axiomORef{opor::different-span-kinds-do-not-share-reps} holds.
\end{lem}

\begin{prf}
By definition of the d-spans and o-spans, they do not share rep events. 
For example, spans in $\opportunisticSpans p$ have as first rep event
the read at line \ref{appendix::alg-access-Read-Oport-MCAS} that always 
returns a descriptor. While d-spans in $\descriptorSpans p$ never return
a descriptor in their first rep event.
\end{prf}

\begin{lem}
Axiom
\axiomORef{opor::runs-have-at-most-one-type-of-span} holds.
\end{lem}

\begin{prf}
An inspection of the cases in $\runFuncSymbol$ shows that d-spans and o-spans
do not get mixed in the output sets.
\end{prf}

\begin{lem}
Axiom
\axiomORef{opor::opportunism-is-unique} holds.
\end{lem}

\begin{prf}
$\mcasReadAlg$ is the only event in the definition of $\runFuncSymbol$ that can produce o-spans. 
Notice that the cases for $\mcasReadAlg$ produce only one o-span. Therefore,
any two o-spans in the denotation of an $\mcasReadAlg$ event will be equal.
\end{prf}

\begin{lem}
Axiom
\axiomORef{opor::writers-have-writer-blocks} holds.
\end{lem}

\begin{prf}
Suppose $\hspans{}{x} \neq \emptyset$. We do a case analysis on $x$.
\begin{itemize}
\item Case $x = \mcasAlg(\listvar u)$.

$\Longrightarrow$. By definition of $\writesAbs p$, we have $p = \pointGenEntry j$ for some $j \in \listvar u$ and,
\begin{align}
\label{eqn::appendix::opor::writers-have-writer-blocks-1-mcas}
\forall i \in \listvar u.\ \exists z.\ \visObs {\pointerIndx {\pointGenEntry i}} z {x} \wedge \inputVal z {\pointGenEntry i} {\expGenEntry i}
\end{align}

Hence, for every $i \in \listvar u$, there is $z$ such that $\visObsIndx {\pointerIndx {\pointGenEntry i}}{\descIndx} z {x}$ or
$\visObsIndx {\pointerIndx {\pointGenEntry i}}{\opporIndx} z {x}$.
If $\visObsIndx {\pointerIndx {\pointGenEntry i}}{\opporIndx} z {x}$ holds,
then $x$ would have o-spans in its denotation, which is impossible. So, we can assume 
$\visObsIndx {\pointerIndx {\pointGenEntry i}}{\descIndx} z {x}$.

By Lemma \ref{lem::appendix::opor::visibility-lemma-mcas}, for every $i \in \listvar u$, there are
$b_i \in \hspans {\pointGenEntry i} x$ such that
$\firstRep {b_i}$ reads value $\expGenEntry i$.

The only case in the denotation that matches this conditions is (1).(a), i.e.,
the successful denotation. In particular, for $j \in \listvar u$,
$b_j \in \hspans{}{x} \cap \writesSpans {\pointGenEntry j}$.

$\Longleftarrow$. Identical to the proof of Lemma 
\ref{lem::appendix::impl::writers-have-writer-blocks-mcas}, but using
Lemma \ref{lem::appendix::opor::visibility-lemma-mcas} instead.

\item The cases for $x = \mcasAllocAlg(v)$, $x = \mcasWriteAlg(q,v)$, and
$x = \mcasReadAlg(q)$ are similar to the proof of 
Lemma \ref{lem::appendix::impl::writers-have-writer-blocks-mcas}.
\end{itemize}
\end{prf}

\begin{lem}
Axiom
\axiomORef{opor::allocs-have-alloc-blocks} holds.
\end{lem}

\begin{prf}
The proof is very similar to the proof of Lemma \ref{lem::appendix::impl::allocs-have-alloc-blocks-mcas},
but using code line \ref{appendix::alloc-data-Alloc-Oport-MCAS}.
\end{prf}

\begin{lem}
	\label{lem::appendix::opor::spans-contained-in-event-mcas}
Axiom
\axiomORef{opor::blocks-contained-in-abstract-time-interval} holds.
\end{lem}

\begin{prf}
We prove each item.

\begin{itemize}
\item (i).
Let us focus on the $\mcasReadAlg$ case, since the rest of cases are as 
in the proof of Lemma \ref{lem::appendix::impl::blocks-contained-in-abstract-time-interval-mcas}.

All the three cases for $\mcasReadAlg$ contain a single span $b$ such that $\firstRep b$ is
executed by the same thread that invoked $x$. Therefore, $b$ starts after $x$ starts.

\item (ii).
Let us focus on the $\mcasReadAlg$ case, since the rest of cases are as 
in the proof of Lemma \ref{lem::appendix::impl::blocks-contained-in-abstract-time-interval-mcas}.

All the three cases for $\mcasReadAlg$ contain a single span $b$ such that $\lastRep b$ is
executed by the same thread that invoked $x$. Therefore, we can choose $i \defini \lastRep b$,
since it is guarantied that it will finish before $x$ finishes.
\end{itemize}
\end{prf}

\begin{thm}
\label{thm::appendix::opor::opor-axioms-satisfied-mcas}
$\genStructName{\MCAS}(\visObsSymbol {\pointerIndx p}, \visSepSymbol{\pointerIndx p})$ 
is implemented by
opportunistic structure $\genOportSpanStructName{\MCAS}$.
\end{thm}

\begin{prf}
All opportunism axioms hold from Lemma \ref{lem::appendix::opor::non-interference-of-lifespans-mcas} to 
Lemma \ref{lem::appendix::opor::spans-contained-in-event-mcas}.
\end{prf}

\begin{thm}
\label{thm::appendix::opor::mcas-is-linearizable}
The MCAS implementation of Figure \ref{appendix::alg-MCAS-opor} is linearizable.
\end{thm}

\begin{prf}
By Theorem \ref{thm::appendix::lin::rdcss-linearizability-from-vis-structure}, it
  suffices to show that
  $\genStructName{\MCAS}(\visObsSymbol {\pointerIndx p},
  \visSepSymbol{\pointerIndx p})$ is valid.  But by
  Theorem \ref{thm::appendix::opor::opor-axioms-imply-visibility}, it suffices that
  $\genStructName{\MCAS}(\visObsSymbol {\pointerIndx p},
  \visSepSymbol{\pointerIndx p})$ is implemented by opportunistic
  structure $\genOportSpanStructName{\MCAS}$. This is given by Theorem
  \ref{thm::appendix::opor::opor-axioms-satisfied-mcas}.
\end{prf}

\end{document}